\documentclass{bmcart}

\usepackage{amsthm,amsmath}
\usepackage[utf8]{inputenc} 
\usepackage[numbers]{natbib}
\usepackage{soul}
\usepackage{url}
\usepackage{breakurl}

\usepackage{graphicx}
\usepackage{amsmath}
\usepackage{booktabs}
\usepackage{tabularx}
\usepackage{multirow}
\usepackage{times}
\usepackage{color}
\usepackage{float}
\usepackage{adjustbox}
\usepackage{longtable}
\usepackage{rotating}
\usepackage{lscape}
\usepackage{threeparttable}
\usepackage{rotating}

\newtheorem{definition}{Definition}

\newcommand{\eg}{\textit{e}.\textit{g}.,}

\newcommand{\fan}[1]{\textcolor{black}{#1}}

\newcommand{\rfan}[1]{\textcolor{black}{#1}}
\newcommand{\sfan}[1]{\textcolor{black}{#1}}
\newcommand{\ssfan}[1]{\textcolor{black}{#1}}
\newcommand{\revfan}[1]{\textcolor{black}{#1}}
\newcommand{\revfantwo}[1]{\textcolor{black}{#1}}
\newcommand{\revfanthree}[1]{\textcolor{black}{#1}}

\startlocaldefs
\endlocaldefs

\begin{document}

\begin{frontmatter}

\begin{fmbox}
\dochead{Review}


\title{Cryptocurrency Trading: A Comprehensive Survey}

%
\author[
  addressref={aff1},                   
  corref={aff1},                       
  email={fan.fang@kcl.ac.uk}   
]{\inits{F.F.}\fnm{Fan} \snm{Fang}}
\author[
  addressref={aff1},
  email={carmine.ventre@kcl.ac.uk}
]{\inits{C.V.}\fnm{Carmine} \snm{Ventre}}
\author[
  addressref={aff2},
  email={mike@turintech.ai}
]{\inits{M.B.}\fnm{Michail} \snm{Basios}}
\author[
  addressref={aff2},
  email={leslie@turintech.ai}
]{\inits{L.K.}\fnm{Leslie} \snm{Kanthan}}
\author[
  addressref={aff2},
  email={david@turintech.ai}
]{\inits{D.M.R.}\fnm{David} \snm{Martinez-Rego}}
\author[
  addressref={aff2},
  email={fan@turintech.ai}
]{\inits{F.W.}\fnm{Fan} \snm{Wu}}
\author[
  addressref={aff2},
  corref={aff2},
  email={lingbo@turintech.ai}
]{\inits{L.L.}\fnm{Lingbo} \snm{Li}}


\address[id=aff1]{
  \orgdiv{Department of Informatics},             
  \orgname{King's College London},          
  \city{London},                              
  \cny{UK}                                    
}
\address[id=aff2]{%
  \orgdiv{},
  \orgname{Turing Intelligence Technology Limited},
  \street{},
  \postcode{}
  \city{London},
  \cny{UK}
}



\end{fmbox}


\begin{abstractbox}

\begin{abstract} 
In recent years, 
the tendency of the number of financial institutions \revfantwo{to include} \fan{cryptocurrencies} in their portfolios has accelerated. Cryptocurrencies are the first pure digital assets to be included by asset managers. 
\revfan{Although they have some commonalities with more traditional assets, they have their own separate nature and their behaviour as an asset is still in the process of being understood.}
It is therefore important to summarise existing research papers and results on cryptocurrency trading, including available trading platforms, trading signals, trading strategy research and risk management. 
This paper provides a comprehensive survey of cryptocurrency trading research, by covering \revfan{146} research papers on various aspects of cryptocurrency trading (\eg{} cryptocurrency trading system\fan{s}, bubble and extreme condition, prediction of volatility and return, \fan{crypto-assets portfolio construction} and crypto-assets, technical trading and others). 
This paper also analyses datasets, research trends and distribution among research objects (contents/properties) and technologies, concluding with some promising opportunities that remain open in cryptocurrency trading. 
\end{abstract}




\end{abstractbox}
%

\end{frontmatter}



\section{Introduction}\label{introduction}
Cryptocurrencies have experienced broad market acceptance and fast development despite their recent conception. Many hedge funds and asset managers have begun 
to include cryptocurrency-related assets into their portfolios and trading strategies. The academic community has similarly spent considerable efforts in researching cryptocurrency trading. This paper seeks to provide a comprehensive survey of the research on cryptocurrency trading, by which we mean any study aimed at facilitating and building strategies to trade cryptocurrencies.

\begin{figure}
\centering
\includegraphics[width=1.0\columnwidth]{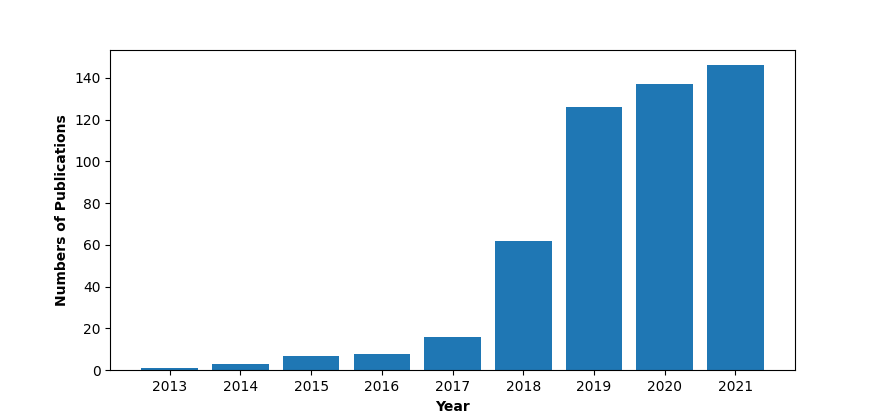}
\caption{Cryptocurrency Trading Publications (cumulative) during 2013-2021(June 2021)}
\label{fig:accu}
\end{figure}

As an emerging market and research direction, cryptocurrencies and cryptocurrency trading have seen considerable progress and a notable upturn in interest and activity~\cite{farell2015analysis}.
From Figure \ref{fig:accu}, we observe over 85\% of papers have appeared since 2018, demonstrating the emergence of cryptocurrency trading as a new research area in financial trading. \revfan{The sampling interval of this survey is from 2013 to June 2021.}

The literature is organised according to six distinct aspects of cryptocurrency trading: 
\begin{itemize}
	\item Cryptocurrency trading software system\fan{s} (i.e., real-time trading systems, turtle trading systems, arbitrage trading systems);
	\item Systematic trading including technical analysis, pairs trading and other systematic trading methods;
	\item Emergent trading technologies including econometric methods, machine learning technology and other emergent trading methods;
	\item Portfolio and cryptocurrency assets including research among cryptocurrency co-movements and crypto-asset portfolio research;
	\item Market condition research including bubbles~\cite{flood1986evaluation} or crash analysis and extreme conditions;
	\item Other Miscellaneous cryptocurrency trading research.
\end{itemize}

In this survey we aim at compiling the most relevant research in these areas and extract a set of descriptive indicators that can give an idea of the level of maturity research in this area has achieved.

We also summarise research distribution (among \fan{research} properties and categories/\fan{research} technologies). 
\revfan{The distribution among properties defines the classification of research objectives and content. The distribution among technologies defines the classification of methods or technological approaches to the study of cryptocurrency trading.}
Specifically, we subdivide research distribution among categories/technologies into statistical methods and machine learning technologies. 
Moreover, We identify datasets and opportunities (potential research directions) that have appeared in the cryptocurrency trading area. To ensure that our survey is self-contained, we aim to provide sufficient material to adequately guide financial trading researchers who are interested in cryptocurrency trading. 

There has been related work that discussed or partially surveyed the literature related to cryptocurrency trading. \revfan{Kyriazis et al.~\citep{kyriazis2019survey} investigated the efficiency and profitable trading opportunities in the cryptocurrency market. Ahamad et al.~\citep{ahamad2013survey} and Sharma et al.~\citep{sharma2017survey} gave a brief survey on cryptocurrencies, merits of cryptocurrencies compared to fiat currencies and compared different cryptocurrencies that are proposed in the literature.} Ujan et al.~\citep{mukhopadhyay2016brief} gave a brief survey of cryptocurrency systems.
\ssfan{Ignasi et al.~\citep{merediz2019bibliometric} performed a bibliometric analysis of bitcoin literature.
The outcomes of this \revfan{related} work focused on specific area in cryptocurrency, including cryptocurrencies and cryptocurrency market introduction, cryptocurrency systems / platforms, bitcoin literature review, etc.}
To the best of our knowledge, no previous work has provided a comprehensive survey particularly focused on cryptocurrency trading. 

In summary, the paper makes the following contributions:

\begin{description}
	\item[Definition.] This paper defines cryptocurrency trading and categorises it into: cryptocurrency markets, cryptocurrency trading models, and cryptocurrency trading strategies. The core content of this survey is trading strategies for cryptocurrencies while we cover all aspects of it.
	\item[Multidisciplinary Survey.] The paper provides a comprehensive survey of \revfan{146} cryptocurrency trading papers, across different academic disciplines such as finance and economics, artificial intelligence and computer science. Some papers may cover multiple aspects and will be surveyed for each category.
	\item[Analysis.] The paper analyses the research distribution, datasets and trends that characterise the cryptocurrency trading literature. 
	\item[Horizons.] The paper identifies challenges, promising research directions in cryptocurrency trading, aimed to promote and facilitate further research. 
\end{description}

\begin{figure} 
\centering
\includegraphics[width=1.0\columnwidth]{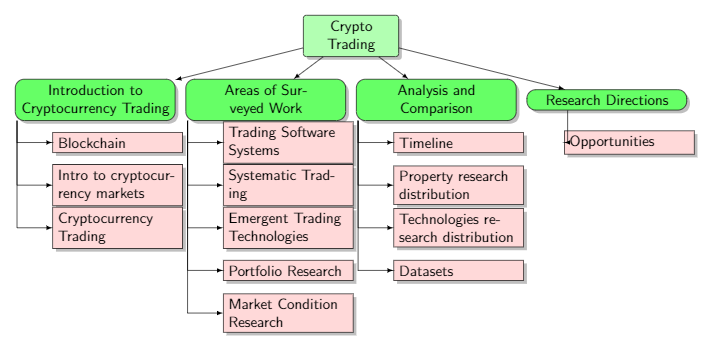}
\caption{Tree structure of the contents in this paper}
\label{fig:cont}
\end{figure}

Figure \ref{fig:cont} depicts the paper structure, which is informed by the review schema adopted. More details about this can be found in Section \ref{papercollection}. 


\section{Cryptocurrency Trading}\label{cryptotrading}
This section provides an introduction to cryptocurrency trading. We will discuss \textbf{Blockchain}, as the enabling technology, \textbf{cryptocurrency markets} and \textbf{cryptocurrency trading strategies}.

\subsection{Blockchain}\label{blockchain}
\subsubsection{Blockchain Technology Introduction}
\begin{figure} 
\centering
\includegraphics[width=.98\columnwidth]{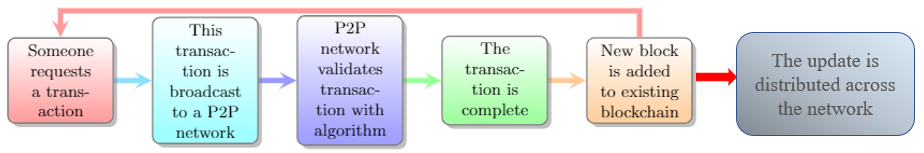}
\caption{Workflow of Blockchain transaction}
\label{fig:blkchain}
\end{figure}

\textbf{Blockchain} is a digital ledger of economic transactions that can be used to record not just financial transactions, but any object with an intrinsic value.
~\citep{tapscott2016blockchain}.
In its simplest form, a Blockchain is a series of immutable data records with timestamps,
which are managed by a cluster of machines that do not belong to any single entity. Each of these data \emph{block}s is protected by cryptographic principle and bound to each other in a \emph{chain} (cf. Figure~\ref{fig:blkchain} for the workflow).

Cryptocurrencies like Bitcoin are \revfan{conducted} on a peer-to-peer network structure. Each peer has a complete history of all transactions, thus recording the balance of each account. For example, a transaction is a file that says ``A pays X Bitcoins to B'' that is signed by A using its private key. This is basic public-key cryptography, but also the building block on which cryptocurrencies are based. After being signed, the transaction is broadcast on the network. When a peer discovers a new transaction, it checks to make sure that the signature is valid (\revfan{this is equivalent to using the signer's public key,} denoted as \ssfan{the} algorithm in Figure~\ref{fig:blkchain}). 
If the verification is valid then the block is added to the chain; all other blocks added after it will ``confirm'' that transaction. For example, if a transaction is contained in block 502 and the length of the blockchain is 507 blocks, it means that the transaction has 5 confirmations (507-502)~\citep{blockchainblog}.


\subsubsection{From Blockchain to cryptocurrencies}
Confirmation is a critical concept in cryptocurrencies; only miners can confirm transactions. Miners add blocks to the Blockchain; they retrieve transactions in the previous block and combine it with the hash of the preceding block to obtain its hash, and then store the derived hash into the current block. Miners in Blockchain accept transactions, mark them as legitimate and broadcast them across the network.  After the miner confirms the transaction, each node must add it to its database. In layman terms, it has become part of the Blockchain and miners undertake this work to obtain cryptocurrency tokens, such as Bitcoin. 
In contrast to Blockchain, cryptocurrencies are related to the use of tokens based on distributed ledger technology. 
Any transaction involving purchase, sale, investment, etc. involves a Blockchain native token or sub-token. Blockchain is a platform that drives cryptocurrency and is a technology that acts as a distributed ledger for the network. The network creates \ssfan{a} means of transaction and \ssfan{enables} the transfer of value and information. Cryptocurrencies are the tokens used in these networks to send value and pay for these transactions. They can be thought of as tools on the Blockchain, and in some cases can also function as resources or utilities. \ssfan{In other} instances, they are used to \ssfan{digitise} the value of assets. 
\revfan{In summary, cryptocurrencies are part of an ecosystem based on Blockchain technology.}
\subsection{Introduction of cryptocurrency market}
\subsubsection{What is cryptocurrency?}
\textbf{Cryptocurrency} is a decentralised medium of exchange which uses cryptographic functions to conduct financial transactions~\citep{doran2014forensic}. Cryptocurrencies leverage the Blockchain technology to gain decentralisation, transparency, and immutability~\citep{meunier2018blockchain}. In the above, we have discussed how Blockchain technology is implemented for cryptocurrencies.

In general, the security of cryptocurrencies is built on cryptography, neither by people nor on trust~\citep{narayanan2016bitcoin}. For example, Bitcoin uses a method called ``Elliptic Curve Cryptography'' to ensure that transactions involving Bitcoin are secure~\citep{wang2017designated}. Elliptic curve cryptography is a type of public-key cryptography that relies on mathematics to ensure the security of transactions.
When someone attempts to circumvent the aforesaid encryption scheme by brute force, 
it takes them \ssfan{one-tenth} the age of the universe to find a value match when trying 250 billion possibilities every second~\citep{elliptic}. 
\revfan{Regarding its use as a currency, cryptocurrency has properties similar to fiat currencies.}
It has a controlled supply. 
\revfan{Most cryptocurrencies limit the availability of their currency volumes.}
E.g. for Bitcoin, the supply will decrease over time and will reach its final quantity sometime around 2,140.  All cryptocurrencies control the supply of tokens through a timetable encoded in the Blockchain. 

One of the most important features of cryptocurrencies is the exclusion of financial institution intermediaries~\citep{harwick2016cryptocurrency}.  The absence of a ``middleman” lowers transaction costs for traders. For comparison, if a bank's database is hacked or damaged, the bank will rely entirely on its backup to recover any information that is lost or compromised. 
With cryptocurrencies, even if part of the network is compromised, the rest will continue to be able to verify transactions correctly. Cryptocurrencies also have the important feature of not being controlled by any central authority~\citep{rose2015evolution}: the decentralised nature of the Blockchain ensures cryptocurrencies are theoretically immune to government control and interference.

\revfan{The pure digital asset is anything that exists in a digital format and carries with it the right to use it. Currently, digital assets include digital documents, motion picture and so on; the market for digital assets has in fact evolved since its inception in 2009, with the first digital asset ``Bitcoin"~\citep{kaal2020digital}. For this reason, we call the cryptocurrency the ``first pure digital asset".}

\begin{figure} 
\centering
\includegraphics[width=.95\columnwidth]{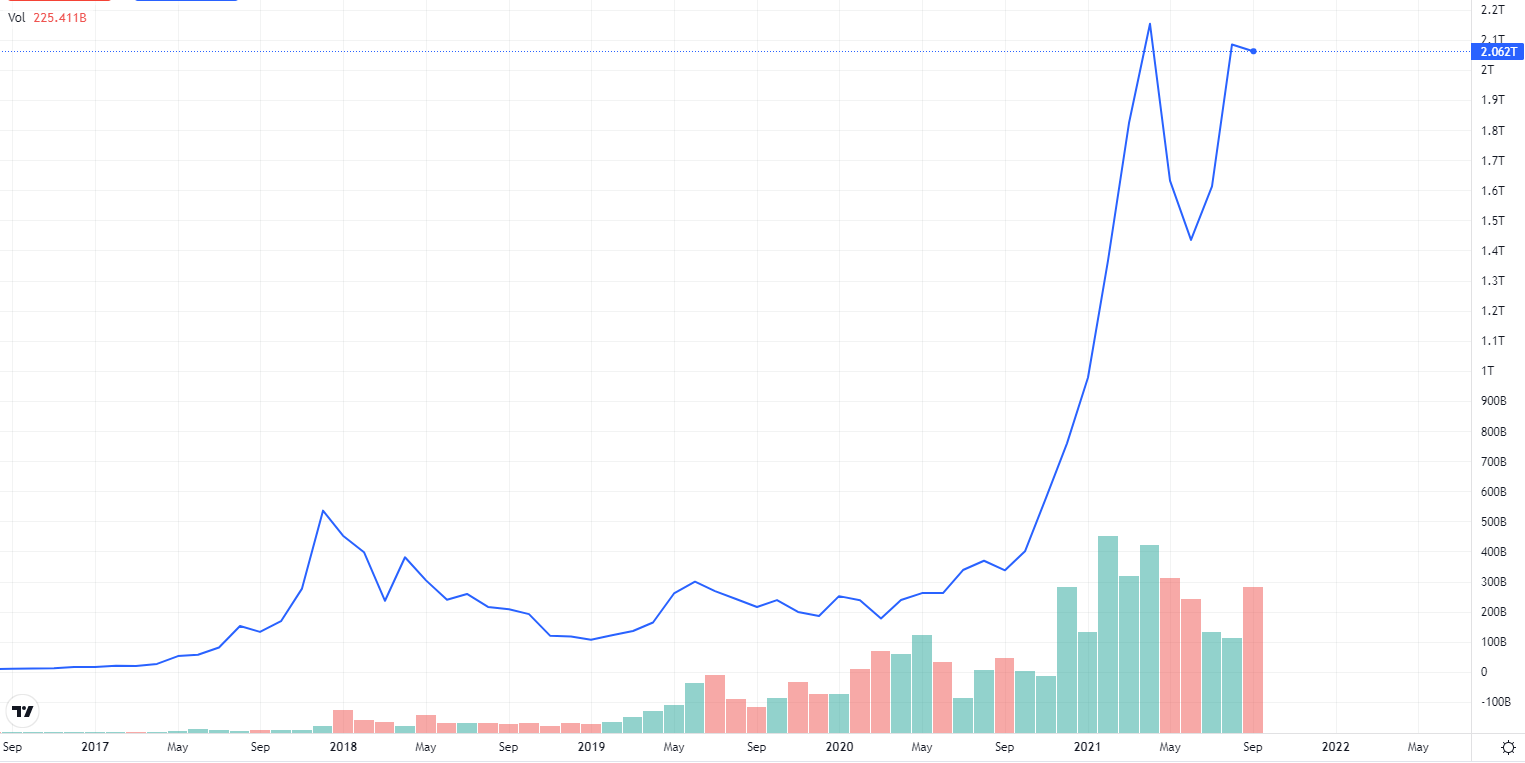}
\caption{Total Market Capitalization and Volume of cryptocurrency market, USD~\citep{cryptohistory}}
\label{fig:history}
\end{figure}

As of December 20, 2019, there exist 4,950 cryptocurrencies and 20,325 cryptocurrency markets; the market cap is around 190 billion dollars~\citep{cryptorank}. 
Figure~\ref{fig:history} shows historical data on global market capitalisation and 24-hour trading volume~\citep{cryptohistory}. \revfan{The blue line is the total cryptocurrency market capitalization and green/red histogram is the total cryptocurrency market volume.} The total market cap is calculated by aggregating the dollar market cap of all cryptocurrencies. From the figure, we can observe how cryptocurrencies experience exponential growth in 2017 and a large bubble burst in early 2018. \revfan{In the wake of the pandemic, cryptocurrencies raised dramatically in value in 2020. In 2021, the market value of cryptocurrencies has been very volatile but consistently at historically high levels.}

\revfan{There are three mainstream cryptocurrencies~\citep{mainstream}: Bitcoin (BTC), Ethereum (ETH), and Litecoin (LTC).}
Bitcoin was created in 2009 and garnered massive popularity. On October 31, 2008, an individual or group of individuals operating under the pseudonym Satoshi Nakamoto released the Bitcoin white paper and described it as: ''A pure peer-to-peer version of electronic cash that can be sent online for payment from one party to another without going through a \ssfan{counterparty}, ie. a financial institution.''~\citep{nakano2018bitcoin}
Launched by Vitalik Buterin in 2015, Ethereum is a special Blockchain with a special token called Ether (ETH symbol in exchanges). A very important feature of Ethereum is the ability to create new tokens on the Ethereum Blockchain. The Ethereum network went live on July 30, 2015, and pre-mined 72 million Ethereum.
Litecoin is a peer-to-peer cryptocurrency created by Charlie Lee. It was created according to the Bitcoin protocol, but it uses a different hashing algorithm. \revfan{Litecoin uses a memory-intensive proof-of-work algorithm, Scrypt.}

\begin{figure} 
\centering
\includegraphics[width=.95\columnwidth]{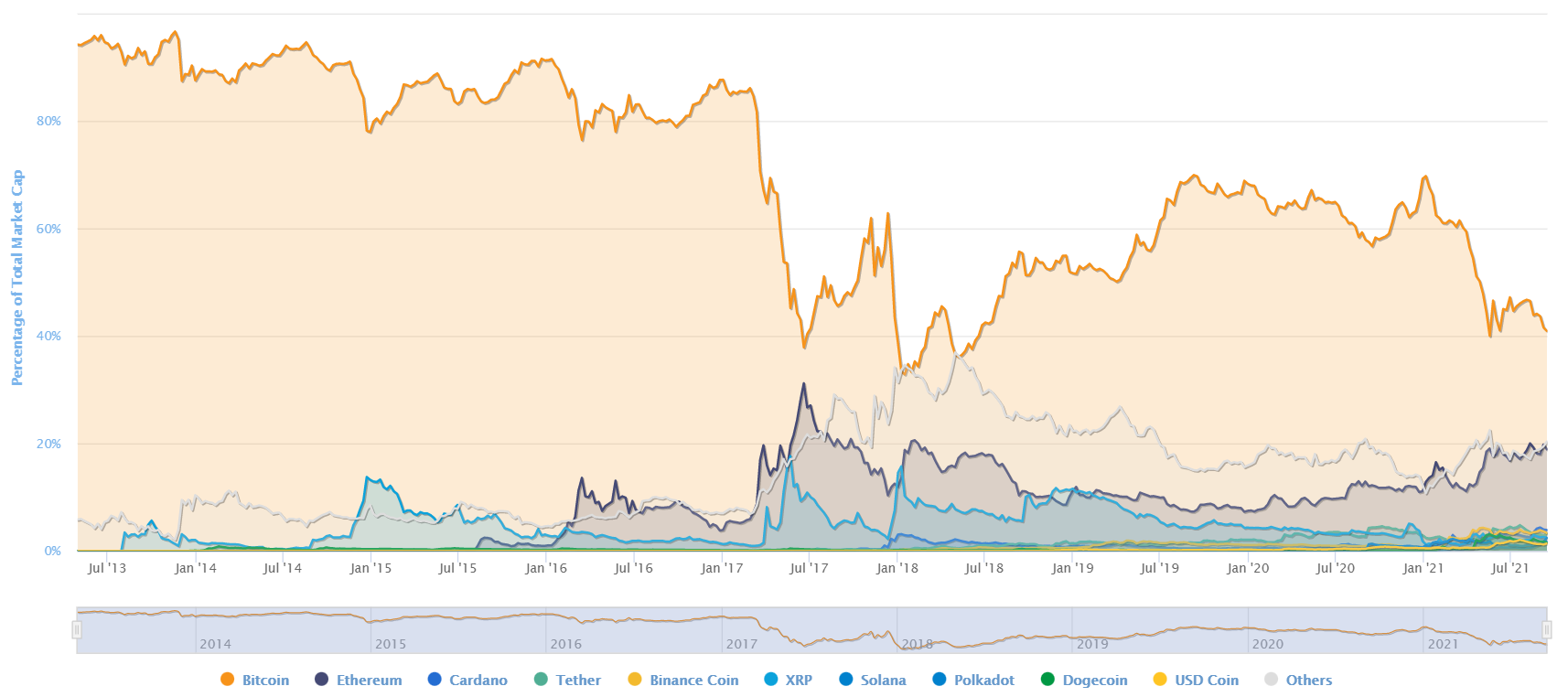}
\caption{Percentage of Total Market Capitalisation~\citep{crymarketper}}
\label{fig:percentmarket}
\end{figure}

Figure~\ref{fig:percentmarket} shows percentages of total cryptocurrency market capitalisation; \revfan{Bitcoin and Ethereum account for the majority of the total market capitalisation (data collected on 14 September 2021).}

\subsubsection{Cryptocurrency Exchanges}
A cryptocurrency exchange or digital currency exchange (DCE) is a business that allows customers to trade cryptocurrencies. Cryptocurrency exchanges can be market makers, usually using the bid-ask spread as a commission for services, or as a matching platform, by simply charging fees.
\revfan{A cryptocurrency exchange or digital currency exchange (DCE) is a place that allows customers to trade cryptocurrencies. Cryptocurrency exchanges can be market makers (usually using the bid-ask spread as a commission for services) or a matching platform (simply charging fees).}

Table~\ref{tbl:exchanges} shows the top or classical cryptocurrency exchanges according to the rank list, by volume, compiled on ``nomics'' website~\citep{exchanges}. Chicago Mercantile Exchange (CME), Chicago Board Options Exchange (CBOE) as well as BAKKT (backed by New York Stock Exchange) are regulated cryptocurrency exchanges. 
Fiat currency data also comes from ``nomics'' website~\citep{exchanges}. 
Regulatory authority and supported currencies of listed exchanges are collected from official websites or blogs. 

\begin{table*} 
\centering
\caption{Cryptocurrency exchanges Lists}
\resizebox{1\textwidth}{!}{
\begin{tabular}{llllll}
\hline
Exchanges & Category & Supported currencies & Fiat Currency & Registration country & Regulatory authority~                                      \\
\hline
CME                      & Derivatives    &   BTC and Ethereum~\citep{CMEcrypto}        & USD & USA~\citep{CMEcountry}  & CFTC~\citep{CMEregu}                                              \\
CBOE                     & Derivatives    &    BTC~\citep{CBOEcrypto}       & USD        & USA~\citep{CBOEcountry}        &  CFTC~\citep{CBOEregu}                                \\
BAKKT (NYSE)                     & Derivatives    &    BTC~\citep{BAKKTregu}       & USD & USA~\citep{BAKKTcountry}    & CFTC~\citep{BAKKTregu}                                     \\
BitMex                   & Derivatives             & 12 cryptocurrencies~\citep{Bitmexcrypto} & USD & Seychelles~\citep{Bitmexcountry} & -    \\
Binance                  & Spot             & 98 cryptocurrencies~\citep{Binancecrypto} & \textcolor[rgb]{0.18,0.18,0.18}{EUR, NGN, RUB, TRY}
& Malta~\citep{Binancecountry} & FATF~\citep{Binanceregu}\\
Coinbase                 & Spot             & 28 cryptocurrencies~\citep{Coinbasecrypto}  & \textcolor[rgb]{0.18,0.18,0.18}{EUR, GBP, USD}    & USA~\citep{Coinbasecountry} & SEC~\citep{Coinbaseregu}  \\
Bitfinex                 & Spot             & $>100$ cryptocurrencies~\citep{Bitfinexcrypto}  & \textcolor[rgb]{0.18,0.18,0.18}{EUR, GBP, JPY, USD} & British Virgin Islands~\citep{Bitfinexcountry} & NYAG~\citep{Bitfinexregu} \\
Bitstamp                 & Spot             & 5 cryptocurrencies~\citep{Bitstampcrypto}   & \textcolor[rgb]{0.18,0.18,0.18}{EUR, USD}        & Luxembourg~\citep{Bitstampcountry} & CSSF~\citep{Bitstampregu}   \\
Poloniex                 & Spot          & 23 cryptocurrencies~\citep{Poloniexcrypto}   &    USD      & USA~\citep{Poloniexcrypto}    & -        \\
\hline
\end{tabular}}
\label{tbl:exchanges}
\end{table*}

\subsection{Cryptocurrency Trading}

\subsubsection{Definition}
First we give a definition of \textit{cryptocurrency trading}.

\begin{definition}
Cryptocurrency trading is the act of buying and selling of cryptocurrencies with the intention of making a profit. 
\end{definition}
The definition of cryptocurrency trading can be broken down into three aspects: object, operation mode and trading strategy. The object of cryptocurrency trading is the asset being traded, which is ``cryptocurrency''. The operation mode of cryptocurrency trading depends on the means of transaction in the cryptocurrency market, which can be classified into ``trading of cryptocurrency Contract for Differences (CFD)'' (\fan{The contract between the two parties, often referred to as the ``buyer'' and ``seller'', stipulates that the buyer will pay the seller the difference \sfan{between themselves when the position closes}~\citep{CDFreference}}) and ``buying and selling cryptocurrencies via an exchange''. A trading strategy in cryptocurrency trading, formulated by an investor, is an algorithm that defines a set of predefined rules to buy and sell on cryptocurrency markets. 

\subsubsection{Advantages of Trading Cryptocurrency}
The benefits of cryptocurrency trading include:
\begin{description}
	\item[Drastic fluctuations.] The volatility of cryptocurrencies are often likely to attract speculative interest and investors. The rapid fluctuations of intraday prices can provide traders with great money-earning opportunities, but it also includes more risk.
	\item[24-hour market.] The cryptocurrency market is available 24 hours a day, 7 days a week because it is a decentralised market. Unlike buying and selling stocks and commodities, the cryptocurrency market is not traded physically from a single location.
	Cryptocurrency transactions can take place between individuals, in different venues across the world.
	\item[Near Anonymity.] Buying goods and services using cryptocurrencies is done online and does not require to make one's own identity public. With increasing concerns over identity theft and privacy, cryptocurrencies can thus provide users with some advantages regarding privacy. 
	\revfan{Different exchanges have specific Know-Your-Customer (KYC) measures for identifying users or customers~\citep{cryptoKYC}. }
	The KYC undertook in the exchanges allows financial institutions to reduce the financial risk while maximising the wallet owner's anonymity.
	\item[Peer-to-peer transactions.] One of the biggest benefits of cryptocurrencies is that they do not involve financial institution intermediaries. As mentioned above, this can reduce transaction costs. Moreover, this feature might appeal \ssfan{to} users who distrust traditional systems.
    Over-the-counter (OTC) cryptocurrency markets offer, in this context, peer-to-peer transactions on the Blockchain. 
	The most famous cryptocurrency OTC market is ``LocalBitcoin~\citep{localbtc}''.
	\item[Programmable ``smart'' capabilities.]  Some cryptocurrencies can bring other benefits to holders, including limited ownership and voting rights. Cryptocurrencies may also include \ssfan{a} partial ownership interest in physical assets such as artwork or real estate.
\end{description}

\revfan{\subsubsection{Disadvantages of Trading Cryptocurrency}
The disadvantages of cryptocurrency trading include:
\begin{description}
    \item[Scalability Problem.] Before the massive expansion of the technology infrastructure, the number of transactions and the speed of transactions cannot compete with traditional currency trading. Scalability issues led to a multi-day trading backlog in March 2020, affecting traders looking to move cryptocurrencies from their personal wallets to exchanges~\citep{priceplunge}.
    \item[Cybersecurity Issues.] As a digital technology, cryptocurrencies are subject to cyber security breaches and can fall into the hands of hackers. Recently, over \$600 million of ethereum and other cryptocurrencies were stolen in August 2021 in blockchain-based platform Poly Network~\citep{cryptostolen}. Mitigating this situation requires ongoing maintenance of the security infrastructure and the use of enhanced cyber security measures that go beyond those used in traditional banking\revfanthree{~\citep{kou2021fintech}}.
    \item[Regulations.] Authorities around the world face challenging questions about the nature and regulation of cryptocurrency as some parts of the system and its associated risks are largely unknown. There are currently three types of regulatory systems used to control digital currencies, they include: closed system for the Chinese market, open and liberal for the Swiss market,and open and strict system for the US market~\citep{cryptoregul}. At the same time, we notice that some countries such as India is not at par in using the cryptocurrency. As Buffett said, ``It doesn’t make sense. This thing is not regulated. It’s not under control. It’s not under the supervision of $[\ldots]$ United States Federal Reserve or any other central bank~\citep{buffett}."
\end{description}
}

\section{Cryptocurrency Trading Strategy}
Cryptocurrency trading strategy is the main focus of this survey. 
There are many trading strategies, which can be broadly divided into two main categories: technical and fundamental. 
\revfan{Technical and fundamental trading are two main trading analysis thoughts when it comes to analyzing the financial markets. Most traders use these two analysis methods or both~\citep{oberlechner2001importance}. From a survey on stock prediction, we in fact know that 66\% of the relevant research work was based on technical analysis; while 23\% and 11\% were based on fundamental analysis and general analysis, respectively~\citep{nti2020systematic}. Cryptocurrency trading can draw on the experience of stock market trading in most scenarios. So we divide trading strategies into two main categories: technical and fundamental trading.}

They are similar in the sense that they both rely on quantifiable information that can be backtested against historical data to verify their performance.
In recent years, a third \fan{kind of} trading strategy, \ssfan{which} we call \revfan{programmatic trading}, has received increasing attention. 
Such a trading strategy is similar to a technical trading strategy because it uses trading activity information on the exchange to make buying or selling decisions. 
\revfan{programmatic} traders build trading strategies with quantitative data, which is mainly derived from price, volume, technical indicators or ratios to take advantage of inefficiencies in the market and are executed automatically by trading software.
Cryptocurrency market is different from traditional markets as there are more arbitrage opportunities, higher fluctuation 
and 
transparency.
Due to these characteristics, most traders and analysts prefer using \revfan{programmatic trading} in cryptocurrency markets.

\subsection{Cryptocurrency Trading Software System}
Software trading systems allow international transactions, process customer accounts and information, and accept and execute transaction orders~\citep{calo2002global}.
A \textbf{cryptocurrency trading system} is a set of principles and procedures that are pre-programmed to \fan{allow} trade between cryptocurrencies and between fiat currencies and cryptocurrencies.
Cryptocurrency trading systems are built to overcome price manipulation, cybercriminal activities and transaction delays~\citep{bauriya2019real}. 
When developing a cryptocurrency trading system, we must consider \ssfan{the} capital market, base asset, investment plan and strategies~\citep{Julian2019}. Strategies are the most important part of an effective cryptocurrency trading system and they will be introduced below. There exist several cryptocurrency trading systems that are available commercially, for example, Capfolio, 3Commas, CCXT, Freqtrade and Ctubio. From these cryptocurrency trading systems, investors can obtain professional trading strategy support, fairness and transparency from \ssfan{the} professional third-party consulting \ssfan{companies} and fast customer services.

\subsection{Systematic Trading}
\textbf{Systematic Trading} is a way to define trading goals, risk controls and rules. In general, systematic trading includes high frequency trading and slower investment types like systematic trend tracking.
In this survey, we divide systematic cryptocurrency trading into technical analysis, pairs trading and others.
Technical analysis in cryptocurrency trading is the act of using historical patterns of transaction data to assist a trader in assessing current and projecting future market conditions for the purpose of making profitable trades.
Price and volume charts summarise all trading activity made by market participants in an exchange and affect their decisions.
Some experiments showed that the use of specific technical trading rules allows generating excess returns, which is useful to cryptocurrency traders and investors in making optimal trading and investment decisions~\citep{gerritsen2019profitability}.
Pairs trading is a systematic trading strategy \ssfan{that} considers two similar \fan{assets} with slightly different spreads.
If the spread widens, short the high \revfan{cryptocurrencies} and buy the low \revfan{cryptocurrencies}. When the spread narrows again to a certain equilibrium value, a profit is generated~\citep{elliott2005pairs}.
Papers shown in this section involve the analysis and comparison of technical indicators, pairs and informed trading, amongst other strategies.

\subsection{\revfan{Tools for building automated trading systems}}
\revfan{Tools for building automated trading systems in cryptocurrency market are those emergent trading strategies for cryptocurrency. These include strategies that are based on econometrics and machine learning technologies.}

\subsubsection{Econometrics on Cryptocurrency}
\textbf{Econometric} methods apply a combination of statistical and economic theories to estimate economic variables and predict their values~\citep{vogelvang2005econometrics}. 
\textbf{Statistical models} use mathematical equations to encode information extracted from the data~\citep{kaufman2013trading}. 
In some cases, statistical modeling techniques can quickly provide sufficiently accurate models~\citep{ben2002hybrid}.
Other methods might be used, such as sentiment-based prediction and long-and-short-term volatility classification based prediction~\citep{chang2015sophistication}. 
The prediction of volatility can be used to judge the price fluctuation of cryptocurrencies, which is also valuable for the pricing of cryptocurrency-related derivatives~\citep{kat1994volatility}. 

When studying cryptocurrency trading using econometrics, researchers apply statistical models on time-series data 
like generalised autoregressive conditional heteroskedasticity (GARCH) and 
BEKK (named after Baba, Engle, Kraft and Kroner, 1995~\citep{engle1995multivariate}) models to evaluate the fluctuation of cryptocurrencies~\citep{caporin2012we}. 
A \textbf{linear statistical model} is a method to evaluate the linear relationship between prices and an explanatory variable~\citep{neter1996applied}. 
When there exists more than one explanatory variable, we can model the linear relationship between explanatory (independent) and response (dependent) variables with multiple linear models.
The common linear statistical model used in \ssfan{the} time-series analysis is \ssfan{the} autoregressive moving average (ARMA) model~\citep{choi2012arma}. 


\subsubsection{Machine Learning Technology}
Machine learning is an efficient tool for developing Bitcoin and other cryptocurrency trading strategies ~\citep{mcnally2018predicting} because it can 
infer data relationships that are often not directly observable by humans. 
From the most basic perspective, Machine Learning relies on the definition of two main components: input features and objective function. 
The definition of Input Features (data sources) is where knowledge of fundamental and technical analysis comes into play. 
We may divide the input into several groups of features, for example, those based on Economic indicators (\fan{such as, gross domestic product indicator}, interest rates, etc.), Social indicators (Google Trends, Twitter, etc.), Technical indicators (price, volume, etc.) and other Seasonal indicators (time of day, day of \ssfan{the} week, etc.).
The objective function defines the fitness criteria one uses to judge if the Machine Learning model has learnt the task at hand. 
Typical predictive models try to anticipate numeric (e.g., price) or categorical (e.g., trend) unseen outcomes. The machine learning model is trained by using historic input data (sometimes called in-sample) to \emph{generalise} patterns therein to unseen (out-of-sample) data to (approximately) achieve the goal defined by the objective function. Clearly, in the case of trading, the goal is to infer trading signals from market indicators which help to anticipate asset future returns. 

\begin{figure} 
\centering
\includegraphics[width=1\columnwidth]{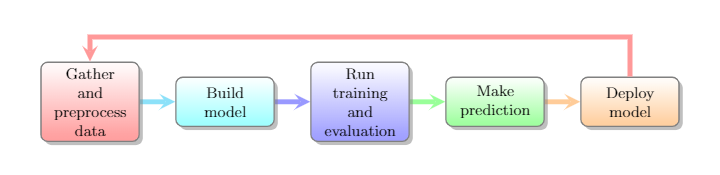}
\caption{Process of machine learning in predicting cryptocurrency}
\label{fig:machi}
\end{figure}

Generalisation error is a pervasive concern in the application of Machine Learning to real applications, and of utmost importance in Financial applications. We need to use statistical approaches, such as cross validation, to validate the model before we actually use it to make predictions. In machine learning, this is typically called ``validation''.
The process of using machine learning technology to predict cryptocurrency is shown in Figure~\ref{fig:machi}.

\begin{table} 
\centering
\caption{Comparison among Different machine learning methods}
\resizebox{.90\textwidth}{!}{
\begin{tabular}{llll}
                 & Supervised Learning                                                                 & Unsupervised Learning                                                                                       & Reinforcement Learning                                                                                                                       \\ \hline
Definition       & \begin{tabular}[c]{@{}l@{}}The machine learns by \\ using labeled data\end{tabular} & \begin{tabular}[c]{@{}l@{}}Machine training through \\ unlabelled data without \\ any guidance\end{tabular} & \begin{tabular}[c]{@{}l@{}}Agents interact with their environment\\ by performing actions and learning from\\ errors or rewards\end{tabular} \\ \hline
Type of problems & Rgression or classification                                                         & Association or clustering                                                                                   & Reward-based                                                                                                                                 \\ \hline
Type of data     & Labeled adata                                                                       & Unlabeled data                                                                                              & No predefined data                                                                                                                           \\ \hline
Training         & External supervision                                                                & No supervision                                                                                              & No supervision                                                                                                                               \\ \hline
Approach         & \begin{tabular}[c]{@{}l@{}}Mapping tagged inputs to\\ unknown outputs\end{tabular}  & \begin{tabular}[c]{@{}l@{}}Understanding patterns or\\ finding outputs\end{tabular}                         & \begin{tabular}[c]{@{}l@{}}Follow the trail-and-error\\ method\end{tabular}                                                                  \\ \hline
\end{tabular}
}
\label{tbl:mlcompare}
\end{table}

Depending on the formulation of the main learning loop, we can classify Machine Learning approaches into three categories: Supervised learning, Unsupervised learning and Reinforcement learning. \revfan{We list a general comparison~\citep{machinediffe} among these three machine learning methods in Table~\ref{tbl:mlcompare}.} \textbf{Supervised learning} is used to derive a predictive function from \ssfan{labeled} training data. Labeled training data means that each training instance includes inputs and expected outputs. Usually, these expected \ssfan{outputs} are produced by a supervisor and represent the expected behaviour of the model. The most used labels in trading are derived from in sample future returns of assets. \textbf{Unsupervised learning} tries to infer structure from unlabeled training data and it can be used during exploratory data analysis to discover hidden patterns or to group data according to any pre-defined similarity metrics.
\textbf{Reinforcement learning} utilises software agents trained to maximise a utility function, which defines their  
objective; this is flexible enough to allow 
agents 
to exchange short term returns for future ones.
In \ssfan{the} financial sector, some trading challenges can be expressed as a game in which an agent aims at maximising \ssfan{the} return at the end of \ssfan{the} period.

The use of machine learning in cryptocurrency trading research encompasses the connection between data sources' understanding and machine learning model research. Further concrete examples are shown in \ssfan{a} later section.

\subsection{Portfolio Research}
\textbf{Portfolio theory} advocates diversification of investments to maximize returns for a given level of risk by allocating assets strategically. The celebrated mean-variance optimisation is a prominent example of this approach \cite{MVO}. Generally, \textbf{crypto asset} denotes a digital asset (i.e., cryptocurrencies and derivatives).
There are some common ways to build a diversified portfolio in crypto assets. 
The first method is to diversify across markets, which is to mix a wide variety of investments within a portfolio of \ssfan{the} cryptocurrency market. 
The second method is to consider the industry sector, which is to avoid investing too much money in any one category. 
Diversified investment of portfolio in \ssfan{the} cryptocurrency market includes portfolio across cryptocurrencies~\citep{liu2019portfolio} and portfolio across \ssfan{the} global market including stocks and futures~\citep{kajtazi2019role}. 

\subsection{Market Condition Research}
Market condition research appears especially important for cryptocurrencies. A \textbf{financial bubble} is a significant increase in the price of an asset without changes in its intrinsic value\revfanthree{~\citep{brunnermeier2013bubbles,kou2021bankruptcy}}. Many experts pinpoint a cryptocurrency bubble in 2017 when the prices of cryptocurrencies grew by 900$\%$. In 2018, Bitcoin 
faced a collapse in its value. This significant fluctuation inspired researchers to study bubbles and extreme conditions in cryptocurrency trading. 
\revfan{The cryptocurrency market has experienced a near continuous bull market since the fall of 2020, with the value of Bitcoin soaring from \$10,645 on October 7, 2020 to an all-time high of \$63,346 on April 15, 2021. This represents a gain of approximately +600\% in just six months~\citep{cryptro2021}.}
\revfantwo{Some experts believe that the extreme volatility of exchange rates means that cryptocurrency exposure should be kept at a low percentage of your portfolio. ``I understand if you want to buy it because you believe the price will rise, but make sure it's only a small part of your portfolio, maybe 1\% or 2\%!" says Thanos Papasavvas, founder of research group ABP Invest, who has a 20-year background in asset management~\citep{cryptoftsurvey}. In any case, bubbles and crash analysis is an important researching area in cryptocurrency trading.}

\section{Paper Collection and Review Schema}\label{papercollection}
The section introduces the scope and approach of our paper collection, a basic analysis, and the structure of our survey.

\subsection{Survey Scope}\label{surveyscope}
We adopt a bottom-up approach to the research in cryptocurrency trading, starting from the systems up to risk management techniques. For the underlying trading system, the focus is \ssfan{on} the optimisation of trading platforms structure and improvements of computer science technologies.

At a higher level, researchers focus on the design of models to predict return or volatility in cryptocurrency markets. These techniques become useful to the generation of trading signals. 
on the next level above predictive models, researchers discuss technical trading methods to trade in real cryptocurrency markets. 
Bubbles and extreme conditions are hot topics in cryptocurrency trading because, as discussed above, these markets \fan{have shown to be highly volatile (whilst volatility went down after crashes)}. Portfolio and cryptocurrency asset management are effective methods to control risk. We group these two areas in risk management research. 
Other papers included in this survey include topics like pricing rules, dynamic market analysis, regulatory implications, and so on. Table~\ref{tbl:scop} shows the general scope of cryptocurrency trading included in this survey. 

Since many trading strategies and methods in cryptocurrency trading are closely related to stock trading, some researchers migrate or use the research results for the latter to the former.
When conducting this research, we only consider those papers whose research \ssfan{focuses} on cryptocurrency markets or a comparison of trading in those and other financial markets.

\begin{table} 
\centering
\caption{Survey scope table}
\resizebox{.55\textwidth}{!}{
\begin{tabular}{ll} 
\hline
\multirow{4}{*}{\textcolor[rgb]{0.2,0.2,0.2}{Trading (bottom up)} } & \textcolor[rgb]{0.2,0.2,0.2}{Trading System}                     \\ 
\cline{2-2}
                                                                    & \textcolor[rgb]{0.2,0.2,0.2}{Prediction (return)}                \\
                                                                    & \textcolor[rgb]{0.2,0.2,0.2}{Prediction (volatility)}            \\ 
\cline{2-2}
                                                                    & \textcolor[rgb]{0.2,0.2,0.2}{Technical trading methods}          \\ 
\hline
{\textcolor[rgb]{0.2,0.2,0.2}{Risk management} }     & \textcolor[rgb]{0.2,0.2,0.2}{Bubble and extreme condition}       \\ 
\cline{2-2}
                                                                    & \textcolor[rgb]{0.2,0.2,0.2}{Porfolio and Cryptocurrency asset}  \\ 
\hline
\multicolumn{2}{l}{\textcolor[rgb]{0.2,0.2,0.2}{Others} }                                                                              \\
\hline
\end{tabular}
}
\label{tbl:scop}
\end{table}

Specifically, we apply the following criteria when collecting papers related to cryptocurrency trading:
\begin{enumerate}
	\item The paper introduces or discusses the general idea of cryptocurrency trading or one of the related aspects of cryptocurrency trading. 
	\item The paper proposes an approach, study or framework that targets optimised efficiency or accuracy of cryptocurrency trading. 
	\item The paper compares different approaches or perspectives in trading cryptocurrency. 
\end{enumerate}
By ``cryptocurrency trading'' here, we mean one of the terms listed in Table \ref{tbl:scop} and discussed above.

Some researchers gave a brief survey of cryptocurrency~\citep{ahamad2013survey,sharma2017survey}, cryptocurrency systems~\citep{mukhopadhyay2016brief} and cryptocurrency trading opportunities~\citep{kyriazis2019survey}. These surveys are rather limited in scope as compared to ours, which also includes a discussion on the latest papers in the area; we want to remark that this is a fast-moving research field.

\subsection{Paper Collection Methodology}
To collect the papers in different areas or platforms, we used keyword searches on Google Scholar and arXiv, two of the most popular scientific databases. 
\ssfan{We also choose other public repositories like SSRN but we find that almost all academic papers in these platforms can also be retrieved via Google Scholar; consequently, in our statistical analysis, we count those as Google Scholar hits. 
We choose arXiv as another source since it allows this survey to be contemporary with all the most recent findings in the area. The interested reader is warned that these papers have not undergone formal peer review.} The keywords used for searching and collecting are listed below. [Crypto] means \ssfan{the} cryptocurrency market, which is our research interest because methods might be different among different markets. 
\revfan{We conducted 6 searches across the two repositories until July 1, 2021.}

\begin{enumerate}
\item[-] [Crypto] + Trading
\item[-] [Crypto] + Trading system
\item[-] [Crypto] + Prediction
\item[-] [Crypto] + Trading strategy
\item[-] [Crypto] + Risk Management 
\item[-] [Crypto] + Portfolio
\end{enumerate}

To ensure high coverage, 
we adopted the so-called \textbf{snowballing}\fan{~\citep{wohlin2014guidelines}} method on each paper found through these keywords. We checked papers added from snowballing methods that satisfy the criteria introduced above until we reached closure. 


\subsection{Collection Results}
Table~\ref{tbl:query} shows the details of the results from our paper collection. 
Keyword searches and snowballing resulted in \revfan{146} papers across the six research areas of interest in Section~\ref{surveyscope}. 

\begin{table} 
\centering
\caption{Paper query results. \#Hits, \#Title, and \#Body denote the number of papers returned by the search, left after title filtering, and left after body filtering, respectively.}
\resizebox{.58\textwidth}{!}{
\begin{tabular}{llll} 
\hline
Key Words                                                                                      & \#Hits & \#Title & \#Body  \\ 
\hline
{[}Crypto] + Trading                                                                           & 612  & 60    & 41    \\
{[}Crypto] + Trading System                                                                    & 4    & 3     & 2     \\
{[}Crypto] + Prediction                                                                        & 40   & 20    & 18    \\
{[}Crypto] + Trading Strategy                                                                  & 23   & 10     & 9     \\
\begin{tabular}[c]{@{}l@{}}{[}Crypto] + Risk Management /\\{[}Crypto] + Portfolio\end{tabular} & 128  & 20    & 16    \\
\hline
Query                                                                                          & -    & -     & 86    \\
Snowball                                                                                       & -    & -     & 60    \\
Overall                                                                                        & -    & -     & 146   \\
\hline
\end{tabular}
}
\label{tbl:query}
\end{table}

\begin{figure} 
\centering
\includegraphics[width=.95\columnwidth]{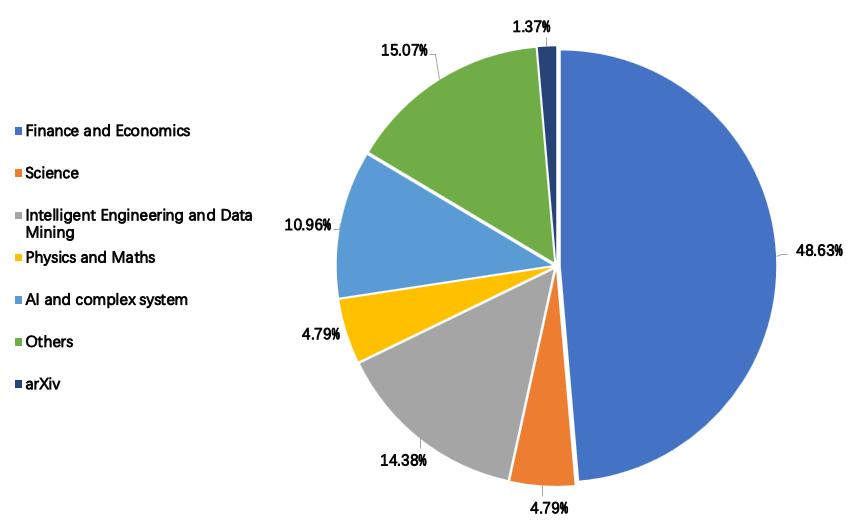}
\caption{Publication Venue Distribution}
\label{fig:publ}
\end{figure}

\revfan{Figure~\ref{fig:publ} shows the distribution of papers published at different research sites.
Among all the papers, 
48.63\% papers are published in Finance and Economics venues such as Journal of Financial Economics (JFE), Cambridge Centre for Alternative Finance (CCAF), Finance Research Letters, Centre for Economic Policy Research (CEPR), Finance Research Letters (FRL), Journal of Risk and Financial Management (JRFM) and some other high impact financial journals; 
4.79\% papers are published in Science venues such as Public Library Of Science one (PLOS one), Royal Society open science and SAGE; 
14.38\% papers are published in Intelligent Engineering and Data Mining venues such as Symposium Series on Computational Intelligence (SSCI), Intelligent Systems Conference (IntelliSys), Intelligent Data Engineering and Automated Learning (IDEAL) and International Conference on Data Mining (ICDM); 
4.79\% papers are published in Physics / Physicians venues (mostly in Physics venue) such as Physica A and Maths venue like Journal of Mathematics; 
10.96\% papers are published in AI and complex system venues such as Complexity and International Federation for Information Processing (IFIP);  
15.07\% papers are published in Others venues which contains independently published papers and dissertations; 1.37\% papers are published on arXiv. 
The distribution of different venues shows that cryptocurrency trading is mostly published in Finance and Economics venues, but with a wide diversity otherwise.}

\subsection{Survey Organisation}
We discuss the contributions of the collected papers and a statistical analysis of these papers in the remainder of the paper, according to Table~\ref{tbl:review}.

\begin{table*} 
\centering
\caption{Review Schema}
\resizebox{1\textwidth}{!}{
\begin{tabular}{lll}
\hline
Classification                                          & Sec  & Topic                                                    \\ 
\hline
\multirow{4}{*}{Cryptocurrency Trading Software System} & \ref{infrastructure}  & Trading Infrastructure System                  \\
                                                        & \ref{realtime}  & Real-time Cryptocurrency Trading System           \\
                                                        & \ref{turtle}  & Turtle trading system in Cryptocurrency market           \\
                                                        & \ref{arbitrage}  & Arbitrage Trading Systems for Cryptocurrencies           \\ 
                              & \ref{comparisontradingsys}    & Comparison of three cryptocurrency trading systems
                              \\
\hline
\multirow{3}{*}{Systematic Trading} & \ref{techanalysis}  & Technical Analysis                  \\
                                                        & \ref{pairstrading}  & Pairs Trading          \\
                                                        & \ref{systematicothers}  & Others           \\ 
\hline
\multirow{3}{*}{Emergent Trading Technologies}          & \ref{econometrics}  & Econometrics on cryptocurrency                           \\
                                                        & \ref{machinelearning}  & Machine learning technology                              \\
                                                        & \ref{emergentothers}  & Others                                                   \\ 
\hline
\multirow{4}{*}{Portfolio, Cryptocurrency Assets and Market condition}    & \ref{relatedfactor}  & Research among cryptocurrency pairs and related factors  \\
                                                        & \ref{portfolio}  & Crypto-asset portfolio research                          \\ 
             & \ref{bubbles}  & Bubbles and crash analysis                               \\
                                                        & \ref{extremecondition}  & Extreme condition                                        \\ 
\hline
Others                                                  & \ref{othersrelatedtrading}   & Others related to Cryptocurrency Trading                 \\ 
\hline
\multirow{4}{*}{Summary Analysis of Literature Review}  & \ref{timeline} & Timeline                                                 \\
                                                        & \ref{distributionpro} & Research distribution among properties                   \\
                                                        & \ref{distributiontech} & Research distribution among categories and technologies  \\
                                                        & \ref{datasetssec} & Datasets used in cryptocurrency trading                  \\
\hline
\end{tabular}}
\label{tbl:review}
\end{table*}  

The papers in our collection are organised and presented from six angles. We introduce the work about several different cryptocurrency trading software systems in Section \ref{tradingsoftware}. Section \ref{systematictradingsec} introduces systematic trading applied to cryptocurrency trading. In Section \ref{emergenttradingsec}, we introduce some emergent trading technologies including econometrics on cryptocurrencies, machine learning technologies and other emergent trading technologies in \ssfan{the} cryptocurrency market. Section \ref{portfoliosec} introduces research on cryptocurrency pairs and related factors and crypto-asset portfolios research. In Section \ref{bubbles} and Section \ref{extremecondition} we discuss cryptocurrency market condition research, including bubbles, crash analysis, and extreme conditions. Section \ref{othersrelatedtrading} introduces other research included in cryptocurrency trading not covered above.

We would like to emphasize that the six headings above focus on a particular aspect of cryptocurrency trading; we give a complete organisation of the papers collected under each heading. This implies that those papers covering more than one aspect will be discussed in different sections, once from each angle. 

We analyse and compare the number of research papers on different cryptocurrency trading properties and technologies in Section~\ref{summarysec}, where we also summarise the datasets and the timeline of research in cryptocurrency trading. 

We build upon this review to conclude in Section~\ref{opportunitiessec} with some opportunities for future research.

\section{Cryptocurrency Trading Software Systems}\label{tradingsoftware}

\subsection{Trading Infrastructure Systems}\label{infrastructure}
Following the development of computer science and cryptocurrency trading, many cryptocurrency trading systems/bots have been developed. Table~\ref{tbl:sysexist} compares the cryptocurrency trading systems existing in the market. The table is sorted based on URL types (GitHub or Official website) and GitHub stars (if appropriate). 

\textbf{Capfolio} is a proprietary payable cryptocurrency trading system which is a professional analysis platform and has \ssfan{an} advanced backtesting engine~\citep{capfolio}. 
It supports five different cryptocurrency exchanges. 

\textbf{3 Commas} is a proprietary payable cryptocurrency trading system platform \ssfan{that} can take profit and stop-loss orders at the same time~\citep{3commas}. 
Twelve different cryptocurrency exchanges are compatible with this system.

\textbf{CCXT} is a cryptocurrency trading system with a unified API out of the box and optional normalized data and supports many Bitcoin / Ether / Altcoin exchange markets and merchant APIs.
Any trader or developer can create a trading strategy based on this data and access public transactions through the APIs~\citep{CCXT}.
The CCXT library is used to connect and trade with cryptocurrency exchanges and payment processing services worldwide.
It provides quick access to market data for storage, analysis, visualisation,
indicator development, algorithmic trading, strategy backtesting, automated code generation 
and related software engineering. 
It is designed for coders, skilled traders, data scientists and financial analysts to build trading algorithms.
Current CCXT features include:  
\begin{itemize}
    \item Support for many cryptocurrency exchanges;
    \item Fully implemented public and private APIs;
    \item Optional normalized data for cross-exchange analysis and arbitrage;
    \item Out-of-the-box unified API, very easy to integrate.
\end{itemize}

\textbf{Blackbird} Bitcoin Arbitrage is a C++ trading system that automatically executes long / short arbitrage between Bitcoin exchanges. It can generate market-neutral strategies \ssfan{that} do not transfer funds between exchanges~\citep{Blackbird}.
The motivation behind Blackbird is to naturally profit from these temporary price differences between different exchanges while being market neutral.
Unlike other Bitcoin arbitrage systems, Blackbird does not sell but actually short sells Bitcoin on the short exchange. 
This feature offers two important advantages. 
Firstly, the strategy is always market agnostic: fluctuations (rising or falling) in the Bitcoin market will not affect the strategy returns.
This eliminates the huge risks of this strategy. 
Secondly, this strategy does not require transferring funds (USD or BTC) between Bitcoin exchanges.
Buy and sell transactions are conducted in parallel on two different exchanges.
There is no need to deal with transmission delays.

\textbf{StockSharp} is an open-source trading platform for trading at any market of the world including 48 cryptocurrency exchanges~\citep{stocksharp}. 
It has a free C\# library and free trading charting application.
Manual or automatic trading (algorithmic trading robot, regular or HFT) can be run on this platform.
StockSharp consists of five components that offer different features:
\begin{itemize}
    \item S\#.Designer - Free universal algorithm strategy app, easy to create strategies;
    \item S\#.Data - free software that can automatically load and store market data;
    \item S\#.Terminal - free trading chart application (trading terminal);
    \item S\#.Shell - ready-made graphics framework that can be changed according to needs and has \ssfan{a} fully open source in C\#;
    \item S\#.API - a free C\# library for programmers using Visual Studio. Any trading strategies can be created in S\#.API.
\end{itemize}

\textbf{Freqtrade} is a free and open-source cryptocurrency trading robot system written in Python. It is designed to support all major exchanges and is controlled by telegram. It contains backtesting, mapping and money management tools, and strategy optimization through machine learning~\citep{freqtrade}.
Freqtrade has \ssfan{the} following features:
\begin{itemize}
    \item Persistence: Persistence is achieved through \ssfan{SQLite} technology;
    \item Strategy optimization through machine learning: Use machine learning to optimize your trading strategy parameters with real trading data;
    \item Marginal Position Size: Calculates winning rate, risk-return ratio, optimal stop loss and adjusts position size, and then trades positions for each specific market;
    \item Telegram management: use telegram to manage the robot.
    \item Dry run: Run the robot without spending money;
\end{itemize}

\textbf{CryptoSignal} is a professional technical analysis cryptocurrency trading system~\citep{cryptosignal}. 
Investors can track over 500 coins of Bittrex, Bitfinex, GDAX, Gemini and more. 
Automated technical analysis \ssfan{includes} momentum, RSI, Ichimoku Cloud, MACD, etc. 
The system gives alerts including Email, Slack, Telegram, etc.
CryptoSignal has two primary features. First of all, it offers modular code for easy implementation of trading strategies; Secondly, it is easy to install with Docker.

\textbf{Ctubio} is a C++ based low latency (high frequency) cryptocurrency trading system~\citep{ctubio}. 
This trading system can place or cancel orders through supported cryptocurrency exchanges in less than a few milliseconds. 
Moreover, it provides a charting system that can visualise the trading account status including trades completed, target position for fiat currency, etc.

\textbf{Catalyst} is an analysis and visualization of \ssfan{the} cryptocurrency trading system~\citep{catalyst}. 
It makes trading strategies easy to express and \ssfan{backtest them on} historical data (daily and minute resolution), providing analysis and insights into the performance of specific strategies. 
Catalyst allows users to share and organise data and build profitable, data-driven investment strategies.
Catalyst not only supports the trading execution but also offers historical price data of all crypto assets (from minute to daily resolution). Catalyst also has backtesting and real-time trading capabilities, which enables users to seamlessly transit between the two different trading modes.
Lastly, Catalyst integrates statistics and machine learning libraries (such as matplotlib, scipy, statsmodels and sklearn) to support the development, analysis and visualization of the latest trading systems.

\textbf{Golang Crypto Trading Bot} is a Go based cryptocurrency trading system~\citep{golang}. 
Users can test the strategy in sandbox environment simulation. 
If simulation mode is enabled, a fake balance for each coin must be specified for each exchange.

\begin{table*}
\centering
\caption{Comparison of existing cryptocurrency trading systems. \#Exchange, Language, and \#Popularity denote the number of the exchanges that are supported by this software, programming language used, and the popularity of the software (number of the stars in Github).}
\resizebox{1\textwidth}{!}{
\begin{tabular}{@{}lllllll} 
\cline{1-7}
Name & Features                                                                                                                                                          & \#Exchange & Language   & Open-Source & URL                            & \#Popularity  \\ 
\cline{1-7}
Capfolio                                              & Professional analysis platform,                                                                                                                                   & 5                                                                     & {Not mentioned}                                                 & {No}                                          & {Official website~\citep{capfolio}} & {}             \\ 
                                                                       & Advanced backtesting engine                                                                                                                                       &                                                                                        &                                                                                &                                                              &                                   &                               \\
3 Commas                                              & Simultaneous take profit and                                                                                                                                      & {12}                                                                    & {Not mentioned}                                                 & {No}                                          & {Official website~\citep{3commas}} & {}             \\ 
                                                                       & stop loss orders                                                                                                                                                  &                                                                                        &                                                                                &                                                              &                                   &                               \\ 
CCXT                                                  & \textcolor[rgb]{0.141,0.161,0.18}{An out of the box unified API,}                                                                                                 & {10}                                                                    & {\textcolor[rgb]{0.141,0.161,0.18}{JavaScript / Python / PHP} } & {Yes}                                         & {GitHub~\citep{CCXT}}           & {13k}          \\
                                                                       & \textcolor[rgb]{0.141,0.161,0.18}{optional normalized data}                                                                                                       &                                                                                        &                                                                                &                                                              &                                   &                               \\
BlackBird                                             & \textcolor[rgb]{0.141,0.161,0.18}{Strategy is market-neutral}                                                                                                     & {8}                                                                     & {C++}                                                           & {Yes}                                         & {GitHub~\citep{Blackbird}}           & {4.7k}         \\
                                                                       & \textcolor[rgb]{0.141,0.161,0.18}{strategy not transfer funds between exchanges}                                                                                  &                                                                                        &                                                                                &                                                              &                                   &                               \\
StockSharp                                            & Free C\# library,                                                                                                                                                 & {48}                                                                    & {C\#}                                                           & {Yes}                                         & {GitHub~\citep{stocksharp}}           & {2.6k}         \\
                                                                       & \textcolor[rgb]{0.141,0.161,0.18}{free}\textcolor[rgb]{0.141,0.161,0.18}{~trading charting application }                                                          &                                                                                        &                                                                                &                                                              &                                   &                               \\
Freqtrade                                             & \textcolor[rgb]{0.141,0.161,0.18}{Strategy Optimization by machine learning,}                                                                                     & {2}                                                                     & {Python}                                                        & {Yes}                                         & {GitHub~\citep{freqtrade}}           & {2.4k}         \\
                                                                       & \textcolor[rgb]{0.141,0.161,0.18}{Calculate edge position sizing}                                                                                                 &                                                                                        &                                                                                &                                                              &                                   &                               \\
CryptoSignal                                                           & Technical analysis trading system                                                                                                                                 & 4                                                                                      & Python                                                                         & Yes                                                          & GitHub~\citep{cryptosignal}                            & 1.9k                          \\
Ctubio                                                                 & \textcolor[rgb]{0.141,0.161,0.18}{Low latency}                                                                                                                    & 1                                                                                      & C\textcolor[rgb]{0.133,0.133,0.133}{+}\textcolor[rgb]{0.133,0.133,0.133}{+}    & Yes                                                          & GitHub~\citep{ctubio}                            & 1.7k                          \\
Catalyst                                              & Analysis and~\textcolor[rgb]{0.141,0.161,0.18}{visualization of system}                                                                                           & {4}                                                                     & {Python}                                                        & {Yes}                                         & {GitHub~\citep{catalyst}}           & \multirow{2}{*}{1.7k}         \\
                                                                       & \begin{tabular}[c]{@{}l@{}}\textcolor[rgb]{0.141,0.161,0.18}{seamless transition between live }\\\textcolor[rgb]{0.141,0.161,0.18}{and~back-testing}\end{tabular} &                                                                                        &                                                                                &                                                              &                                   &                               \\
GoLang                                                                 & \textcolor[rgb]{0.141,0.161,0.18}{Sandbox environment simulation}                                                                                                 & 7                                                                                      & Go                                                                             & Yes                                                          & GitHub~\citep{golang}                            & 277                           \\
\hline
\end{tabular}}
\label{tbl:sysexist}
\end{table*}


\subsection{Real-time Cryptocurrency Trading Systems}\label{realtime}
Amit et al.~\citep{bauriya2019real} developed a real-time Cryptocurrency Trading System. 
A real-time cryptocurrency trading system is composed of clients, servers and databases. Traders use a web-application to login \ssfan{to the} server to buy/sell crypto assets. The server collects cryptocurrency market data by creating a script \ssfan{that} uses the Coinmarket API. 
Finally, the database collects balances, trades and order book information from \ssfan{the} server. 
The authors tested the system with an experiment that demonstrates user-friendly and secure experiences for traders in \ssfan{the} cryptocurrency exchange platform. 

\subsection{Turtle trading system in Cryptocurrency market}\label{turtle}
The original Turtle Trading system is a trend following trading system developed in the 1970s. The idea is to generate buy and sell signals on stock for short-term and long-term breakouts and its cut-loss condition which is measured by Average true range (ATR)~\citep{kamrat2018technical}. 
The trading system will adjust the size of assets based on their volatility. Essentially, if a turtle accumulates a position in a highly volatile market, it will be offset by a low volatility position.
Extended Turtle Trading system is improved with smaller time interval spans and introduces a new rule by using exponential moving average (EMA). Three EMA values are used to trigger \ssfan{the} ``buy'' signal: 30EMA (Fast), 60EMA (Slow), 100EMA (Long).
The author of~\citep{kamrat2018technical} performed backtesting and comparing both trading systems (Original Turtle and Extended Turtle) on 8 prominent cryptocurrencies. 
Through the experiment, Original Turtle Trading System achieved \ssfan{an} 18.59\% average net profit margin (percentage of net profit over total revenue) and 35.94\% average profitability (percentage of winning trades over total numbers of trades) in 87 trades through nearly one year. Extended Turtle Trading System achieved 114.41\% average net profit margin and 52.75\% average profitability in 41 trades through the same time interval. 
This research showed how Extended Turtle Trading System compared can improve over Original Turtle Trading System in trading cryptocurrencies. 

\subsection{Arbitrage Trading Systems for Cryptocurrencies}\label{arbitrage}
Christian~\citep{puauna2018arbitrage} introduced arbitrage trading systems for cryptocurrencies. 
Arbitrage trading aims to spot the differences in price that can occur when there are discrepancies in the levels of supply and demand across multiple exchanges. As a result, a trader could realise a quick and low-risk profit by buying from one exchange and selling at a higher price on a different exchange. Arbitrage trading signals are caught by automated trading software. The technical differences between data sources impose a server process to be organised for each data source. Relational databases and SQL are reliable solution due to \ssfan{the} large \ssfan{amounts} of relational data.
The author used the system to catch arbitrage opportunities on 25 May 2018 among 787 cryptocurrencies on 7 different exchanges. 
The research paper~\citep{puauna2018arbitrage} listed the best ten trading signals made by this system from 186 available found signals. The results showed that the system caught \ssfan{the} trading signal of ``BTG-BTC'' to get a profit \ssfan{of} up to 495.44\% when arbitraging to buy in Cryptopia exchange and sell in Binance exchange.
Another three well-traded arbitrage signals (profit expectation around 20\% mentioned by the author) were found on 25 May 2018. Arbitrage Trading Software System introduced in that paper presented general principles and implementation of arbitrage trading system in \ssfan{the} cryptocurrency market.

\subsection{Characteristics of three cryptocurrency trading systems}\label{comparisontradingsys}
Real-time trading systems use real-time \ssfan{functions} to collect data and generate trading algorithms. 
Turtle trading system and arbitrage trading system have shown a sharp contrast in their profit and risk behaviour. Using Turtle trading system in cryptocurrency markets got high returns with high risk. Arbitrage trading system is inferior in terms of revenue but also has a lower risk. One feature that turtle trading system and arbitrage trading system have in common is they performed well in capturing alpha.

\section{Systematic Trading}\label{systematictradingsec}

\subsection{Technical Analysis}\label{techanalysis}
Many researchers have focused on technical indicators (patterns) analysis for trading on cryptocurrency markets. Examples of studies with this approach include ``Turtle Soup pattern strategy''~\citep{guides2018eos}, ``Nem 
(XEM) strategy''~\citep{guides2018nem}, ``Amazing Gann Box strategy''~\citep{guides2018free}, ``Busted Double Top Pattern strategy''~\citep{guides2018iota}, and ``Bottom Rotation Trading strategy''~\citep{guides2018tether}. 
Table~\ref{tbl:techni} shows the comparison among these five classical technical trading strategies using technical indicators.
``Turtle soup pattern strategy''~\citep{guides2018eos} used \ssfan{a} 2-day breakout of price in predicting price trends of cryptocurrencies. This strategy is a kind of chart trading pattern. 
``Nem (XEM) strategy'' combined Rate of Change (ROC) indicator and Relative Strength Index (RSI) in predicting price trends~\citep{guides2018nem}.
``Amazing Gann Box'' predicted exact points of increase and decrease in Gann Box which are used to catch explosive trends of cryptocurrency price~\citep{guides2018free}. 
Technical analysis tools such as candlestick and box charts with Fibonacci Retracement based on golden ratio are used in this technical analysis. 
Fibonacci Retracement uses horizontal lines to indicate where possible support and resistance levels are in the market.
``Busted Double Top Pattern'' used \ssfan{a} Bearish reversal trading pattern which generates a sell signal to predict price trends~\citep{guides2018iota}. 
``Bottom Rotation Trading'' is a technical analysis method \ssfan{that} picks the bottom before the reversal happens. This strategy used \ssfan{a} price chart pattern and box chart as technical analysis tools.

\begin{table*} 
\centering
\caption{Comparison among five classical technical trading strategies}
\resizebox{1\textwidth}{!}{
\begin{tabular}{@{}lll} 
\hline
Technical trading strategy & Core Methods                                                                                                         & Tecchnical tools/patterns                                                                             \\ 
\hline
Turtle Soup pattern~\citep{guides2018eos}        & 2-daybreakout of price~                                                                                              & Chart trading patterns                                                                                \\
Nem (XEM)~\citep{guides2018nem}                  & Price trends combined ROC \& RSI                                                                                    & \begin{tabular}[c]{@{}l@{}}Rate of Change indictor (ROC)\\Relative strength index (RSI)\end{tabular}  \\
Amazing Gann Box~\citep{guides2018free}           & \begin{tabular}[c]{@{}l@{}}Predict exact points of rises and~falls~\\in Gann Box (catch explosive trends)\end{tabular} & \begin{tabular}[c]{@{}l@{}}Candlestick, boxcharts with\\Fibonacci Retracement\end{tabular}             \\
Busted Double Top Pattern~\citep{guides2018iota}  & \begin{tabular}[c]{@{}l@{}}Bearish reversal trading pattern that\\generates a sell signal\end{tabular}               & Price chart pattern                                                                                   \\
Bottom Rotation Trading~\citep{guides2018tether}    & \begin{tabular}[c]{@{}l@{}}Pick the bottom before the reversal\\happens\end{tabular}                                 & Price chart pattern, box chart                                                                        \\
\hline
\end{tabular}}
\label{tbl:techni}
\end{table*}
    
Sungjoo et al.~\citep{ha2018finding} investigated using genetic programming (GP) to find attractive technical patterns in the cryptocurrency market. Over 12 technical indicators including Moving Average (MA) and Stochastic oscillator were used in experiments; adjusted gain, match count, relative market pressure and diversity measures have been used to quantify the attractiveness of technical patterns. With extended experiments, the GP system is shown to find successfully attractive technical patterns, which are useful for portfolio optimization. 
Hudson et al.~\citep{hudson2019technical} applied almost $15,000$ \ssfan{to} technical trading rules (classified into MA rules, filter rules, support resistance rules, oscillator rules and channel breakout rules). This comprehensive study found that technical trading rules provide investors with significant predictive power and profitability.
\ssfan{Corbet} et al.~\citep{corbet2019effectiveness} analysed various technical trading rules in the form of the moving average-oscillator
and trading range break-out strategies to generate higher returns in cryptocurrency markets. By using one-minute dollar-denominated Bitcoin close-price data, the backtest showed variable-length moving average (VMA) rule performs best considering it generates \ssfan{the} most useful signals in high frequency trading.

\revfan{Grovys et al.~\citep{grobys2020technical} examined a simple moving average trading strategy using daily price data for the 11 most traded cryptocurrencies over the period 2016-2018. The results showed that, excluding Bitcoin, technical trading rules produced an annualised excess return of 8.76\% after controlling for average market returns. The analysis also suggests that cryptocurrency markets are inefficient.
AI-Yahyaee et al.~\citep{al2020cryptocurrency} examined multiple fractals, long memory processes and efficiency assumptions of major cryptocurrencies using Hurst exponents, time-rolling MF-DFA and quantile regression methods. The results showed that all markets provide evidence of long-term memory properties and multiple fractals. Furthermore, the inefficiency of cryptocurrency markets is time-varying. The researchers concluded that high liquidity with low volatility facilitates arbitrage opportunities for active traders.}


\subsection{Pairs Trading}\label{pairstrading}
Pairs trading is a trading strategy that attempts to exploit \ssfan{the} mean-reversion between the prices of certain securities.
Miroslav~\citep{fil2019pairs} investigated \ssfan{the} applicability of standard pairs trading approaches on cryptocurrency data with the benchmarks of Gatev et al.~\citep{gatev2006pairs}.
The pairs trading strategy is constructed in two steps. Firstly, suitable pairs with a stable long-run relationship are identified. Secondly, the long-run equilibrium is calculated and pairs trading strategy is defined by the spread based on the values. 
The research also extended intra-day pairs trading using high frequency data.
Overall, the model was able to achieve \ssfan{a} 3\% monthly profit in Miroslav's experiments~\citep{fil2019pairs}.
Broek ~\citep{van2018cointegration} applied pairs trading based on cointegration in cryptocurrency trading and 31 pairs were found to be significantly cointegrated (within sector and cross-sector). By selecting four pairs and testing over a 60-day trading period, the pairs trading strategy got its profitability from arbitrage opportunities, which rejected the Efficient-market hypothesis (EMH) for the cryptocurrency market.      
\ssfan{Lintihac et al~\citep{lintilhac2017model} proposed an optimal dynamic pair trading strategy model for a portfolio of assets. The experiment used stochastic control techniques to calculate optimal portfolio weights and correlated the results with several other strategies commonly used by practitioners including static dual-threshold strategies.
Thomas et al.~\citep{li2016optimal} proposed a pairwise trading model incorporating time-varying volatility with constant elasticity of variance type. The experiment calculated the best pair strategy by using a finite difference method and estimated parameters by generalised moment method.}
\subsection{Others}\label{systematicothers}
Other systematic trading methods in cryptocurrency trading mainly include informed trading.
Using USD / BTC exchange rate trading data, Feng et al.~\citep{feng2018informed} found evidence of informed trading in the Bitcoin market in \ssfan{those} quantiles of the order sizes of buyer-initiated (seller-initiated) orders are abnormally
high before large positive (negative) events, compared to the quantiles of seller-initiated (buyer-initiated) orders; this study adopts a new indicator inspired by the volume imbalance indicator~\citep{easley2008time}.
The evidence of informed trading in \ssfan{the} Bitcoin market suggests that investors profit on their private information when they get information before it is widely available.

\section{Emergent Trading Technologies}\label{emergenttradingsec}
\subsection{Econometrics on cryptocurrency}\label{econometrics}
Copula-quantile causality analysis and Granger-causality analysis are methods to investigate causality in cryptocurrency trading analysis.
Bouri et al.~\citep{bouri2019trading} applied a copula-quantile causality approach on volatility in the cryptocurrency market. The approach of the experiment extended \ssfan{the} Copula-Granger-causality in distribution (CGCD) method of Lee and Yang~\citep{lee2014granger} in 2014. The experiment constructed two tests of CGCD using copula functions. 
\rfan{The parametric test employed six parametric copula functions to discover dependency density between variables. The performance matrix of these functions varies with independent copula density.} 
Three distribution regions are the focus of this research: left tail (1\%, 5\%, 10\% quantile), central region (40\%, 60\% quantile and median) and right tail (90\%, 95\%, 99\% quantile).
The study provided significant evidence of Granger causality from trading volume to the returns of seven large cryptocurrencies on both left and right tails. 
\ssfan{Elie et al.~\citep{bouri2020volatility} examined the causal linkages among the volatility of leading cryptocurrencies via the frequency-domain test of Bodart and Candelon~\citep{bodart2009evidence} and distinguished between temporary and permanent causation. The results showed that permanent shocks are more important in explaining Granger causality whereas transient shocks dominate the causality of smaller cryptocurrencies in the long term.}
Badenhorst~\citep{badenhorst2019effect} attempted to reveal whether spot and derivative market volumes affect Bitcoin price volatility with \ssfan{the} Granger-causality method and ARCH (1,1).
The result shows spot trading volumes have a significant positive effect on price volatility while \ssfan{the} relationship between cryptocurrency volatility and \ssfan{the} derivative market is uncertain. 
\ssfan{Elie et al.~\citep{bouri2020return} used a dynamic equicorrelation (DECO) model and reported evidence that the average earnings equilibrium correlation changes over time between the 12 leading cryptocurrencies. The results showed increased cryptocurrency market consolidation despite significant price declined in 2018. Furthermore, measurement of trading volume and uncertainty are key determinants of integration.}

Several econometrics methods in time-series research, such as GARCH and BEKK, have been used in the literature on cryptocurrency trading.
Conrad et al.~\citep{conrad2018long} used the GARCH-MIDAS model to extract long and short-term volatility components of \ssfan{the} Bitcoin market. 
The technical details of this model decomposed the conditional variance into \ssfan{the} low-frequency and high-frequency \ssfan{components}. 
The results identified that S\&P 500 realized volatility has a negative and highly significant effect on long-term Bitcoin volatility and S\&P 500 volatility risk premium has a significantly positive effect on long-term Bitcoin volatility.
\sfan{Ardia et al.~\citep{ardia2019regime} used the Markov Switching GARCH (MSGARCH) model to test the existence of institutional changes in the GARCH volatility dynamics of Bitcoin's logarithmic returns. Moreover, a Bayesian method was used for estimating model parameters and calculating VaR prediction. The results showed that MSGARCH models clearly outperform single-regime GARCH for Value-at-Risk forecasting.
Troster et al.~\citep{troster2019bitcoin} performed general GARCH and GAS (Generalized Auto-regressive Score) analysis to model and predict Bitcoin's returns and risks. The experiment found that the GAS model with heavy-tailed distribution can provide the best out-of-sample prediction and goodness-of-fit attributes for Bitcoin's return and risk modeling. The results also illustrated the importance of \ssfan{modeling} excess kurtosis for Bitcoin returns.}

\sfan{Charles et al.~\citep{charles2019volatility} studied four cryptocurrency markets including Bitcoin, Dash, Litecoin and Ripple. Results showed cryptocurrency returns are strongly characterised by the presence of jumps as well as structural breaks except \ssfan{the} Dash market. Four GARCH-type models (i.e., GARCH, APARCH, IGARCH and FIGARCH) and three return types with structural breaks (original returns, jump-filtered returns, and jump-filtered returns) are considered. The research indicated the importance of jumps in cryptocurrency volatility and structural breakthroughs.}
\revfan{Malladi et al.~\citep{malladi2021time} examined the time series analysis of Bitcoin and Ripple's returns and volatility to examine the dependence of their prices in part on global equity indices, gold prices and fear indicators such as volatility indices and US economic policy uncertainty indices. Autoregressive-moving-average model with exogenous inputs model (ARMAX), GARCH, VAR and Granger causality tests are used in the experiments. The results showed that there is no causal relationship between global stock market and gold returns on bitcoin returns, but a causal relationship between ripple returns on bitcoin prices is found.}

Some researchers focused on long memory methods for volatility in cryptocurrency markets. Long memory methods focused on long-range dependence and significant long-term correlations among fluctuations on markets.
Chaim et al.~\citep{chaim2019nonlinear} estimated a multivariate stochastic volatility model with discontinuous jumps in cryptocurrency markets. 
The results showed that permanent volatility appears to be driven by major market developments and popular interest \ssfan{levels}.
Caporale et al.~\citep{caporale2018persistence} examined persistence in \ssfan{the} cryptocurrency market by Rescaled range (R/S) analysis and fractional integration. 
The results of the study indicated that the market is persistent (there is a positive correlation between its past and future values) and that its level changes over time.
Khuntin et al.~\citep{khuntia2018adaptive} applied the adaptive market hypothesis (AMH) in the predictability of Bitcoin evolving returns.
\rfan{The consistent test of Dominguez and Lobato~\citep{dominguez2003testing}, generalized spectral (GS) of Escanciano and Velasco~\citep{escanciano2006generalized} are applied in capturing time-varying linear and nonlinear dependence in bitcoin returns.}
The results verified Evolving Efficiency in Bitcoin price changes and evidence of dynamic efficiency in line with AMH's claims.
\revfan{Gradojevic et al.~\citep{gradojevic2021volatility} examined volatility cascades across multiple trading ranges in the cryptocurrency market. Using a wavelet Hidden Markov Tree model, authors estimated the transition probability of propagating high or low volatility at one time scale (range) to high or low volatility at the next time scale. The results showed that the volatility cascade tends to be symmetrical when moving from long to short term. In contrast, when moving from short to long term, the volatility cascade is very asymmetric.}

\revfan{Nikolova et al.~\citep{nikolova2020novel} provided a new method to calculate the probability of volatility clusters, especially for cryptocurrencies (high volatility of their exchange rates). The authors used the FD4 method to calculate the Hurst index of a volatility series and describe explicit criteria for determining the existence of fixed size volatility clusters by calculation. The results showed that the volatility of cryptocurrencies changes more rapidly than that of traditional assets, and much more rapidly than that of Bitcoin/USD, Ethereum/USD, and Ripple/USD pairs.
Ma et al.~\citep{ma2020cryptocurrency} investigated whether a new Markov Regime Transformation Mixed Data Sampling (MRS-MIADS) model can improve the prediction accuracy of Bitcoin's Realised Variance (RV). The results showed that the proposed new MRS-MIDAS model exhibits statistically significant improvements in predicting the RV of Bitcoin. At the same time, the occurrence of jumps significantly increases the persistence of high volatility and switches between high and low volatility.}

Katsiampa et al.~\citep{katsiampa2018volatility} applied three pair-wise bivariate BEKK models to examine the conditional volatility dynamics along with interlinkages and conditional correlations between three pairs of cryptocurrencies in 2018.
More specifically, the BEKK-MGARCH methodology also captured cross-market effects of shocks and volatility, which are also known as shock transmission effects and volatility spillover effects.
The experiment found evidence of bi-directional shock transmission effects between Bitcoin and both Ether and Litcoin.
In particular, bi-directional shock spillover effects are identified between three pairs (Bitcoin, Ether and Litcoin) and time-varying conditional correlations exist with positive correlations mostly prevailing.
In 2019, Katsiampa~\citep{katsiampa2019empirical} further researched an asymmetric diagonal BEKK model to examine conditional variances of five cryptocurrencies that are significantly affected by both previous squared errors and past conditional volatility. The experiment tested the null hypothesis of the unit root against \ssfan{the} stationarity hypothesis. Once stationarity is ensured, ARCH LM is tested for ARCH effects to examine \ssfan{the} requirement of volatility \ssfan{modeling} in return series. 
Moreover, volatility co-movements among cryptocurrency pairs are also tested by \ssfan{the} multivariate GARCH model. 
The results confirmed the non-normality and heteroskedasticity of price returns in cryptocurrency markets. 
The finding also identified the effects of cryptocurrencies' volatility dynamics due to major news.

Hultman~\citep{hultman2018volatility} set out to examine GARCH (1,1), bivariate-BEKK (1,1) and a standard stochastic model to forecast the volatility of Bitcoin. A rolling window approach is used in these experiments.
Mean absolute error (MAE), Mean squared error (MSE) and Root-mean-square deviation (RMSE) are three loss criteria adopted to evaluate the degree of error between predicted and true values.
The result shows the following rank of loss functions: GARCH (1,1) \textgreater\  bivariate-BEKK (1,1) \textgreater\ Standard stochastic for all the three different loss criteria; in other words, GARCH(1,1) appeared best in predicting the volatility of Bitcoin. 
Wavelet time-scale persistence analysis is also applied in \ssfan{the} prediction and research of volatility in cryptocurrency markets~\citep{omane2019wavelet}. The results showed that information efficiency (efficiency) and volatility persistence in \ssfan{the} cryptocurrency market are highly sensitive to time scales, measures of returns and volatility, and institutional changes.
Adjepong et al.~\citep{omane2019wavelet} connected with similar research \ssfan{by} Corbet et al.~\citep{corbet2018exploring} and showed that GARCH is quicker than BEKK to absorb new information regarding the data.

\revfan{Zhang et al.~\citep{zhang2020idiosyncratic} examined how to price exceptional volatility in a cross-section of cryptocurrency returns. Using portfolio-level analysis and Fama-MacBeth regression analysis, the authors demonstrated that idiosyncratic volatility is positively correlated with expected returns on cryptocurrencies.}




\subsection{Machine Learning Technology}\label{machinelearning}
As we have previously stated, Machine learning technology constructs computer algorithms that automatically improve themselves by finding patterns in existing data without explicit instructions~\citep{holmes1994weka}. 
The rapid development of machine learning in recent years has promoted its application to cryptocurrency trading, especially in \ssfan{the} prediction of cryptocurrency returns.
\revfan{Some ML algorithms solve both classification and regression problems from a methodological point of view. For clearer classification, we focus on the application of these ML algorithms in cryptocurrency trading. For example, Decision Tree (DT) can solve both classification and regression problems. But in cryptocurrency trading, researchers focus more on using DT in solving classification problems. Here we classify DT as ``Classification Algorithms".}

\subsubsection{Common Machine Learning Technology in this survey}
Several machine learning technologies are applied in cryptocurrency trading. 
We distinguish these by the objective set to the algorithm: classification, clustering, regression, reinforcement learning. We have separated a section specifically on deep learning due to its intrinsic variation of techniques and wide adoption.

\textbf{Classification Algorithms}.
Classification in machine learning \ssfan{has} the objective of categorising incoming objects into different categories as needed, where we can assign labels to each category (e.g., up and down).
Naive Bayes (NB)~\citep{rish2001empirical}, Support Vector Machine (SVM)~\citep{wang2005support}, K-Nearest Neighbours (KNN)~\citep{wang2005support}, Decision Tree (DT)~\citep{friedl1997decision}, Random Forest (RF)~\citep{liaw2002classification} and Gradient Boosting (GB)~\citep{friedman2001greedy} algorithms habe been used in cryptocurrency trading based on papers we collected.
NB is a probabilistic classifier based on Bayes' theorem with strong (naive) conditional independence assumptions between features~\citep{rish2001empirical}.
SVM is \ssfan{a} supervised learning model that aims at achieving high margin classifiers connecting to learning bounds theory~\citep{zemmal2016adaptive}. 
SVMs \ssfan{assign} new examples to one category or another, making it a non-probabilistic binary linear classifier~\citep{wang2005support}, although some corrections can make a probabilistic interpretation of their output~\citep{keerthi2001improvements}.
KNN is a memory-based or lazy learning algorithm, where the function is only approximated locally, and all calculations are \ssfan{being} postponed to inference time ~\citep{wang2005support}.
DT is a decision support tool algorithm that uses a tree-like decision graph or model to segment input patterns into regions to then assign an associated label to each region~\citep{friedl1997decision,fang2020better}. 
RF is an ensemble learning method. 
The algorithm operates by constructing a large number of decision trees during training and outputting the average consensus as predicted class in the case of classification or mean prediction value in the case of regression ~\citep{liaw2002classification}.
GB produces a prediction model in the form of an ensemble of weak prediction models~\citep{friedman2001greedy}.

\textbf{Clustering Algorithms}.
Clustering is a machine learning technique that involves grouping data points in a way that each group shows some regularity ~\citep{jianliang2009application}. 
K-Means is a vector quantization used for clustering analysis in data mining. 
K-means stores the $k$-centroids used to define the clusters; a point is considered to be in a particular cluster if it is closer to the cluster's centroid than any other centroid~\citep{wagstaff2001constrained}. 
K-Means is one of the most used clustering algorithms used in cryptocurrency trading according to the papers we collected. 
\revfanthree{Clustering algorithms have been successfully applied in many financial applications, such as fraud detection, rejection inference and credit assessment. Automated detection clusters are critical as they help to understand sub-patterns of data that can be used to infer user behaviour and identify potential risks~\citep{li2021integrated,kou2014evaluation}.}

\textbf{Regression Algorithms}.
We have defined regression as any statistical technique that aims at estimating a continuous value~\citep{kutner2005applied}.
Linear Regression (LR) and Scatterplot Smoothing are common techniques used in solving regression problems in cryptocurrency trading.
LR is a linear method used to model the relationship between a scalar response (or dependent variable) and one or more explanatory variables (or independent variables)~\citep{kutner2005applied}.
Scatterplot Smoothing is a technology to fit functions through scatter plots to best represent relationships between variables~\citep{friedman1984monotone}.

\textbf{Deep Learning Algorithms}.
Deep learning is a modern take on artificial neural networks (ANNs)~\citep{10.1145/3285029}, made possible by the advances in computational power. An ANN is a computational system inspired by the natural neural networks that make up the animal's brain. The system ``learns'' to perform tasks including \ssfan{the} prediction by considering examples. Deep learning's superior accuracy comes from high computational complexity cost. Deep learning algorithms are currently the basis for many modern artificial intelligence applications~\citep{sze2017efficient}.
Convolutional neural networks (CNNs)~\citep{lawrence1997face}, Recurrent neural networks (RNNs)~\citep{mikolov2011extensions},
Gated recurrent units (GRUs)~\citep{chung2014empirical}, Multilayer perceptron (MLP) and Long short-term memory (LSTM)~\citep{cheng2016long} networks are \ssfan{the} most common deep learning technologies used in cryptocurrency trading.
A CNN is a specific type of neural network layer commonly used for supervised learning. CNNs have found their best success in image processing and natural language processing problems. An attempt to use CNNs in cryptocurrency can be shown in ~\citep{kalchbrenner2014convolutional}.
An RNN is a type of artificial neural network in which connections between nodes form a directed graph with possible loops. This structure of RNNs makes them suitable for processing time-series data~\citep{mikolov2011extensions} due to the introduction of memory in the recurrent connections. They face nevertheless for the vanishing gradients problem~\citep{pascanu2013difficulty} and so different variations have been recently proposed.
LSTM~\citep{cheng2016long} is a particular RNN architecture widely used. LSTMs have shown to be superior to nongated RNNs on financial time-series problems because they have the ability to selectively remember patterns for a long time.
A GRU~\citep{chung2014empirical} is another gated version of the standard RNN which has been used in crypto trading~\citep{dutta2020gated}.
Another deep learning technology used in cryptocurrency trading is Seq2seq, which is a specific implementation of the Encoder–Decoder architecture~\citep{xu2017seq2seq}. 
Seq2seq was first aimed at solving natural language processing problems but has been also applied it in cryptocurrency trend predictions in~\citep{sriram2017cold}. 

\textbf{Reinforcement Learning Algorithms}.
Reinforcement learning (RL) is an area of machine learning leveraging the idea that software agents act in the environment to maximize a cumulative reward~\citep{sutton1998introduction}.
Deep Q-Learning (DQN)~\citep{gu2016continuous} and Deep Boltzmann Machine (DBM)~\citep{salakhutdinov2009deep} are common technologies used in cryptocurrency trading using RL.
Deep Q learning uses neural networks to approximate Q-value functions. A state is given as input, and Q values for all possible actions are generated as outputs~\citep{gu2016continuous}.
DBM is a type of binary paired Markov random field (undirected probability graphical model) with multiple layers of hidden random variables~\citep{salakhutdinov2009deep}. It is a network of randomly coupled random binary units.

\subsubsection{Research on Machine Learning Models}
In the development of machine learning trading \ssfan{signals}, technical indicators have usually been used as input features.  
Nakano et al.~\citep{nakano2018bitcoin} explored Bitcoin intraday technical trading based on ANNs for return prediction. 
The experiment obtained medium frequency price and volume data (time interval of data is 15min) of Bitcoin from a cryptocurrency exchange. An ANN predicts the price trends (up and down) in \ssfan{the} next period from the input data. Data is preprocessed to construct a training dataset \ssfan{that} contains a matrix of technical patterns including EMA, Emerging Markets Small Cap (EMSD), relative strength index (RSI), etc. Their numerical experiments contain different research aspects including base ANN research, effects of different layers, effects of different activation functions, different outputs, different inputs and effects of additional technical indicators.
The results have shown that the use of various technical indicators possibly prevents over-fitting in the classification of non-stationary financial time-series data, which enhances trading performance compared to \ssfan{the} primitive technical trading strategy. (Buy-and-Hold is the benchmark strategy in this experiment.)

Some classification and regression machine learning models are applied in cryptocurrency trading by predicting price trends. Most researchers have focused on the comparison of different classification and regression machine learning methods.
Sun et al.~\citep{sun2019using} used random forests (RFs) with factors in Alpha01~\citep{kakushadze2016101} (capturing features from \ssfan{the} history of \ssfan{the} cryptocurrency market) to build a prediction model.
The experiment collected data from API in cryptocurrency exchanges and selected 5-minute frequency data for backtesting.
The results showed that the performances are proportional to the amount of data (more data, more accurate) and the factors used in the RF model appear to have different importance. For example, ``Alpha024'' and ``Alpha032'' features appeared as the most important in the model adopted. (The alpha features come from paper ``101 Formulaic Alphas"~\citep{kakushadze2016101}.) 
Vo et al.~\citep{vo2018high} applied RFs in High-Frequency cryptocurrency Trading (HFT) and compared it with deep learning models. Minute-level data is collected when utilising a forward fill imputation method to replace the NULL value (i.e., a missing value). 
Different periods and RF trees are tested in the experiments. The authors also compared F-1 precision and recall metrics between RF and Deep Learning (DL). The results showed that RF is effective despite multicollinearity occurring in ML features, \fan{the lack of model identification also} potentially leading to model identification issues; this research also attempted to create an HFT strategy for Bitcoin using RF. 

Maryna et al.~\citep{zenkova2019robustness} investigated the profitability of an algorithmic trading strategy based on training an SVM model to identify cryptocurrencies with high or low predicted returns. The results showed that the performance of the SVM strategy was the fourth being better only than S\&P B\&H strategy, which simply buys-and-hold the S\&P index. (There are other 4 benchmark strategies in this research.)
The authors observed that SVM needs a large number of parameters and so is very prone to overfitting, which caused its bad performance. 
Barnwal et al.~\citep{barnwal2019stacking} used generative and discriminative classifiers to create a stacking model, particularly 3 generative and 6 discriminative classifiers combined by a one-layer Neural Network, to predict the direction of cryptocurrency price. 
A discriminative classifier directly \ssfan{models} the relationship between unknown and known data, while 
generative classifiers model the prediction indirectly through the data generation distribution~\citep{ng2002discriminative}. 
Technical indicators including trend, momentum, volume and volatility, are collected as features of the model. The authors discussed how different classifiers and features affect the prediction.   
Attanasio et al.~\citep{attanasio2019quantitative} compared a variety of classification algorithms including SVM, NB and RF in predicting next-day price trends of a given cryptocurrency. 
The results showed that due to the heterogeneity and volatility of cryptocurrencies' financial instruments, forecasting models based on a series of forecasts appeared better than a single classification technology in trading cryptocurrencies.
Madan et al.~\citep{madan2015automated} \ssfan{modeled the} Bitcoin price prediction problem as a binomial classification task, experimenting with a custom algorithm that leverages both random forests and
generalized linear models. 
Daily data, 10-minute data and 10-second data are used in the experiments.
The experiments showed that 10-minute data gave a better sensitivity and specificity ratio than 10-second data (10-second prediction achieved around 10\% accuracy). Considering predictive trading, 10-minute data helped show clearer trends in the experiment compared to 10-second backtesting. 
Similarly, Virk~\citep{virkprediction} compared RF, SVM, GB and LR to predict \ssfan{the} price of Bitcoin. 
The results showed that SVM achieved the highest accuracy of 62.31\% and precision value 0.77 among binomial classification machine learning algorithms.

Different deep learning models have been used in finding patterns of price movements in cryptocurrency markets. 
Zhengy et al.~\citep{zhengyang2019prediction} implemented two machine learning models, fully-connected ANN and LSTM to predict cryptocurrency price dynamics. The results showed that ANN, \ssfan{in general,} outperforms LSTM although theoretically, LSTM is more suitable than ANN in terms of modeling time series dynamics; the performance measures considered are MAE and RMSE in joint prediction (five cryptocurrencies daily prices prediction). The findings show that \ssfan{the} future state of a time series for cryptocurrencies is highly dependent on its historic evolution.
Kwon et al.~\citep{kwon2019time} used an LSTM model, with a three-dimensional price tensor representing the past price changes of cryptocurrencies as input. 
This model outperforms the GB model in terms of F1-score. Specifically, it has a performance improvement of about $7\%$ over the GB model in 10-minute price prediction. In particular, the experiments showed that LSTM is more suitable when classifying cryptocurrency data with high volatility. 

Alessandretti et al.~\citep{alessandretti2018anticipating} tested Gradient boosting decision trees (including single regression and XGBoost-augmented regression) and \ssfan{the} LSTM model on forecasting daily cryptocurrency prices. They found methods based on gradient boosting decision trees worked best when predictions were based on short-term windows of 5/10 days while LSTM  worked best when predictions were based on 50 days of data. The relative importance of the features in both models are compared and an optimised portfolio composition (based on geometric mean return and Sharpe ratio) is discussed in this paper. Phaladisailoed et al.~\citep{phaladisailoed2018machine} chose regression models (Theil-Sen Regression and Huber Regression) and deep learning-based models (LSTM and GRU) to compare the performance of predicting the rise and fall of Bitcoin price. In terms of two common measure metrics, MSE and R-Square (R$^2$), GRU shows the best accuracy.


Researchers have also focused on comparing classical statistical models and machine/deep learning models.
Rane et al.~\citep{rane2019systematic} described classical time series prediction methods and machine learning algorithms used for predicting Bitcoin price. Statistical models such as Autoregressive Integrated Moving Average models (ARIMA), Binomial Generalized Linear Model and GARCH are compared with machine learning models such as SVM, LSTM and Non-linear Auto-Regressive with Exogenous Input Model (NARX). The observation and results showed that \ssfan{the} NARX model is the best model with nearly 52\% predicting accuracy based on 10 seconds interval.  
Rebane et al.~\citep{rebane2018seq2seq} compared traditional models like ARIMA with \ssfan{a} modern popular model like seq2seq 
in predicting cryptocurrency returns. 
The result showed that the seq2seq model exhibited demonstrable improvement over the ARIMA model for Bitcoin-USD prediction but the seq2seq model showed very poor performance in extreme cases. The authors proposed performing additional investigations, such as the use of LSTM instead of GRU units to improve the performance. 
Similar models were also compared by Stuerner et al.~\citep{stuerner2019algorithmic} who explored the superiority of automated investment approach in trend following and technical analysis in cryptocurrency trading.
Samuel et al.~\citep{persson2018hybrid} explored \ssfan{the} vector autoregressive model (VAR model), a more complex RNN, and a hybrid of the two in residual recurrent neural networks (R2N2) in predicting cryptocurrency returns. The RNN with ten hidden layers is optimised for the setting and the neural network augmented by VAR allows the network to be shallower, quicker and to have a better prediction than \ssfan{an} RNN. RNN, VAR and R2N2 models are compared. The results showed that the VAR model has phenomenal test period performance and thus props up the R2N2 model, while the RNN performs poorly. This research is an attempt \ssfan{at} optimisation of model design and applying to \ssfan{the} prediction on cryptocurrency returns. 

\revfantwo{\subsubsection{Deep Neural Network}
Deep Neural Network architectures play important roles in forecasting. Researchers had applied many advanced deep neural network models in cryptocurrency trading like stacking (CNN + RNN) and Autoencoder-Decoder.
In this subsection, we describe the cutting edge Deep Neural Network researches in cryptocurrency trading. 
Recent studies show the productivity of using models based on such architectures for modeling and forecasting financial time series, including cryptocurrencies.
Livieris et al.~\citep{livieris2020cnn} proposed model called CNN-LSTM for accurate prediction of gold prices and movements. The first component of the model consists of a convolutional layer and a pooling layer, where complex mathematical operations are performed to develop the features of the input data. The second component uses the generated LSTM and the features of the dense layer. The results show that due to the sensitivity of the various hyperparameters of the proposed CNN-LSTM and its high complexity, additional optimisation configurations and major feature engineering have the potential to further improve the predictive power.
More Intelligent Evolutionary Optimisation (IEO) for hyperparameter optimisation is core problem when tuning the overall optimization process of machine learning models~\citep{huan2020ieo}.
Lu et al.~\citep{lu2020cnn} proposed a CNN-LSTM based method for stock price prediction. In In terms of MAE, RMSE and $R^2$ metrics, the experimental results showed that CNN-LSTM has the highest prediction accuracy and the best performance compared with MLP, CNN, RNN, LSTM, and CNN-RNN.}

\revfantwo{Fan et al.~\citep{fan2019deeplearning} applied an autoencoder-augmented LSTM structure in predicting the mid-price of 8 cryptocurrency pairs. Level-2 limit order book live data is collected and the experiment achieved 78\% accuracy of price movements prediction in high frequency trading (tick level). This research improved and verified the view of Sirignano et al.~\citep{sirignano2019universal} that universal models have better performance than currency-pair specific models for cryptocurrency markets. Moreover, ``Walkthrough'' (i.e., retrain the original deep learning model itself when it appears to no longer be valid) is proposed as a method to optimise the training of a deep learning model and shown to significantly improve the prediction accuracy.
Yao et al.~\citep{yao2018predictive} proposed a new method for predicting cryptocurrency prices based on deep learning techniques such as RNN and LSTM, taking into account various factors such as market capitalization, trading volume, circulating supply and maximum supply. The experimental results showed that the model performs well for a certain size of dataset.
Livieris et al.~\citep{livieris2020ensemble} combined three of the most widely used integration learning strategies: integrated averaging, bagging and stacking, with advanced deep learning models for predicting hourly prices of major cryptocurrencies. The proposed integrated model is evaluated using a state-of-the-art deep learning model as a component learner, which consists of a combination of LSTM, bidirectional LSTM and convolutional layers. The authors' detailed experimental analysis shows that integrated learning and deep learning can effectively reinforce each other to develop robust, stable and reliable predictive models.
Kumar et al.~\citep{kumar2020predicting} analyzed how deep learning techniques such as MLP and LSTM can help predict the price trend of Ethereum. By applying day/hour/minute historical data, the LSTM model is more robust and accurate to long-term dependencies than the MLP while LSTM outperformed the MLP marginally but not very significantly.
}

\subsubsection{Sentiment Analysis}
Sentiment analysis, a popular research topic in the age of social media, has also been adopted to improve predictions for cryptocurrency trading. This data source typically has to be combined with Machine Learning for the generation of trading signals.

Lamon et al.~\citep{lamon2017cryptocurrency} used daily news and social media data labeled on actual price changes, rather than on positive and negative sentiment. By this approach, the prediction on price is replaced with positive and negative sentiment. The experiment acquired cryptocurrency-related news article headlines from \ssfan{the} website like ``cryptocoinsnews'' and twitter API. Weights are taken in positive and negative words in \ssfan{the} cryptocurrency market. Authors compared Logistic Regression (LR), Linear Support Vector Machine (LSVM) and NB as classifiers and concluded that LR is the best classifier in daily price prediction with 43.9\% of price increases correctly predicted and 61.9\% of price decreases correctly forecasted.  
Smuts~\citep{smuts2019drives} conducted \ssfan{a} similar binary sentiment-based price prediction method with an LSTM model using Google Trends and Telegram sentiment. In detail, the sentiment was extracted from Telegram by using a novel measure called VADER
~\citep{hutto2014vader}. The backtesting reached 76\% accuracy on the test set during the first half of 2018 in predicting hourly prices. 

Nasir et al.~\citep{nasir2019forecasting} researched \ssfan{the} relationship between cryptocurrency returns and search engines. The experiment employed a rich set of established empirical approaches including VAR framework, copulas approach and non-parametric drawings of time series. The results found that Google searches exert significant influence on Bitcoin returns, especially in \ssfan{the} short-term \ssfan{intervals}.
Kristoufek~\citep{kristoufek2013bitcoin} discussed positive and negative feedback \ssfan{on} Google trends or daily views on Wikipedia. The author mentioned different methods including Cointegration, Vector autoregression and Vector error-correction model to find causal relationships between prices and searched terms in \ssfan{the} cryptocurrency market. The results indicated that search trends and cryptocurrency prices are connected. There is also a clear asymmetry between the effects of increased interest in currencies above or below their trend values from the experiment.
Young et al.~\citep{kim2016predicting} analysed user comments and replies in online communities and \ssfan{their} connection with cryptocurrency volatility. After crawling comments and replies in online communities, authors tagged the extent of positive and negative topics. Then \ssfan{the} relationship between price and \ssfan{the} number of \ssfan{transactions} of cryptocurrency is tested according to comments and replies to selected data. At last, a prediction model using machine learning based on selected data is created to predict fluctuations in \ssfan{the} cryptocurrency market. The results show the amount of accumulated data and animated community activities exerted a direct effect on fluctuation in the price and volume of \ssfan{a} cryptocurrency.

Phillips et al.~\citep{phillips2018mutual} applied dynamic topic \ssfan{modeling} and Hawkes model to decipher relationships between topics and cryptocurrency price movements. \ssfan{The authors} used Latent Dirichlet allocation (LDA) model for topic \ssfan{modeling}, which assumes each document contains multiple topics to different extents. The experiment showed that particular topics tend to precede certain types of price movements in \ssfan{the} cryptocurrency market and \ssfan{the} authors proposed the relationships could be built into real-time cryptocurrency trading. 
Li et al.~\citep{li2019sentiment} analysed Twitter sentiment and trading volume and an Extreme Gradient Boosting Regression Tree Model in \ssfan{the} prediction of ZClassic (ZCL) cryptocurrency market. Sentiment analysis using natural language processing from \ssfan{the} Python package ``Textblob'' assigns impactful words a polarity value. Values of weighted and unweighted sentiment indices are calculated on \ssfan{an} hourly basis by summing weights of coinciding tweets, which makes us compare this index to ZCL price data. The model achieved a Pearson correlation of 0.806 when applied to test data, yielding a statistical significance at the $p <  0.0001$ level. 
Flori~\citep{flori2019news} relied on a Bayesian framework that combines market-neutral information with subjective beliefs to construct diversified investment strategies in \ssfan{the} Bitcoin market. The result shows that news and media attention seem to contribute to influence the demand for Bitcoin and enlarge the perimeter of the potential investors, probably stimulating price euphoria and upwards-downwards market dynamics. \ssfan{The authors'} research highlighted the importance of news in guiding portfolio re-balancing.
\ssfan{Elie et al.~\citep{bouri2019predicting} compared the ability of newspaper-based metrics and internet search-based uncertainty metrics in predicting bitcoin returns. The predictive power of Internet-based economic uncertainty-related query indices is statistically stronger than that of newspapers in predicting bitcoin returns.}

Similarly, Colianni et al.~\citep{colianni2015algorithmic}, Garcia et al.~\citep{garcia2015social}, Zamuda et al.~\citep{zamuda2019forecasting} et al. used sentiment analysis technology applying it in the cryptocurrency trading area and had similar results. 
Colianni et al.~\citep{colianni2015algorithmic} cleaned data and applied supervised machine learning algorithms such as logistic regression, Naive Bayes and support vector machines, etc. \ssfan{on} Twitter Sentiment Analysis for cryptocurrency trading. 
Garcia et al.~\citep{garcia2015social} applied multidimensional analysis and impulse analysis in social signals of sentiment effects and algorithmic trading of Bitcoin. The results verified the long-standing assumption that transaction-based social media sentiment has the potential to generate a positive return on investment.
Zamuda et al.~\citep{zamuda2019forecasting} adopted new sentiment analysis indicators and used multi-target portfolio selection to avoid risks in cryptocurrency trading. The perspective is rationalized based on the elastic demand for computing resources of the cloud infrastructure. \ssfan{A} general model evaluating \ssfan{the} influence between user's network Action-Reaction-Influence-Model (ARIM) is mentioned in this research.
Bartolucci et al.~\citep{bartoluccibutterfly} researched cryptocurrency prices with the ``Butterfly effect'', which means ``issues'' of \ssfan{the} open-source project provides insights to improve prediction of cryptocurrency prices. Sentiment, politeness, emotions analysis of GitHub comments are applied in Ethereum and Bitcoin markets. The results showed that these metrics have predictive power on cryptocurrency prices.

\subsubsection{Reinforcement Learning}
Deep reinforcement algorithms bypass prediction and go straight to market management actions to achieve high cumulated profit~\citep{henderson2018deep,liu2021agent}.
Bu et al.~\citep{bu2018learning} proposed a combination of double Q-network and unsupervised pre-training using DBM to generate and enhance the optimal Q-function in cryptocurrency trading. The trading model contains agents in series in the form of two neural networks, unsupervised learning modules and environments. The input market state connects \ssfan{an} encoding network which includes spectral feature extraction (convolution-pooling module) and temporal feature extraction (LSTM module). A double-Q network follows the encoding network and actions are generated from this network. Compared to existing deep learning models (LSTM, CNN, MLP, etc.), this model achieved \ssfan{the} highest profit even facing an extreme market situation (recorded 24\% of \ssfan{the} profit while cryptocurrency market price drops by -64\%). 
Juchli~\citep{juchli2018limit} applied two implementations of reinforcement learning agents, a Q-Learning agent, which serves as the learner when no market variables are provided, and a DQN agent which was developed to handle the features previously mentioned. The DQN agent was backtested under the application of two different neural network architectures. 
The results showed that the DQN-CNN agent (convolutional neural network) is superior to the DQN-MLP agent (multilayer perceptron) in backtesting prediction.
Lucarelli et al.~\citep{lucarelli2019deep} focused on improving automated cryptocurrency trading with a deep reinforcement learning approach. Double and Dueling double deep Q-learning networks are compared for 4 years. By setting rewards functions as Sharpe ratio and profit, the double Q-learning method demonstrated to be the most profitable approach in trading cryptocurrency.
\revfan{Sattarov et al.~\citep{sattarov2020recommending} applied deep reinforcement learning and used historical data from BTC, LTC and ETH to observe historical price movements and acted on real-time prices. The model proposed by the authors helped traders to correctly choose one of the following three actions: buy, sell and hold stocks and get advice on the correct option. Experiments applying BTC via deep reinforcement learning showed that investors made a net profit of 14.4\% in one month. Similarly, tests on LTC and ETH ended with 74\% and 41\% profits respectively.
Koker et al.~\citep{koker2020cryptocurrency} pointed out direct reinforcement (DR) based model for active trading. Within the model, the authors attempt to estimate the parameters of the non-linear autoregressive model to achieve maximum risk-adjusted returns. Traders can take long or short positions in each of our sampled cryptocurrency markets, establish or hold them at the end of time interval $t$, and re-evaluate at the end of $t+1$. The results provide some preliminary evidence that cryptocurrency prices may not follow a purely random wandering process.}

\subsection{Others}\label{emergentothers}
Atsalakis et al.~\citep{atsalakis2019bitcoin} proposes a computational
intelligence technique that uses a hybrid Neuro-Fuzzy controller, namely PATSOS, to forecast the direction in the change of the daily price of Bitcoin. The proposed methodology outperforms two other computational intelligence models, the first being developed with a simpler neuro-fuzzy approach, and the second being developed with artificial neural networks. According to the signals of the proposed model, the investment return obtained through trading simulation is 71.21\% higher than the investment return obtained through a simple buy and hold strategy. This application is proposed for the first time in \ssfan{the} forecasting of Bitcoin price movements. 
Topological data analysis is applied to forecasting price trends of cryptocurrency markets in~\citep{kim2018time}. The approach is to harness topological features of attractors of dynamical systems for arbitrary temporal data. The results showed that the method can effectively separate important topological patterns and sampling noise (like bid–ask bounces, discreteness of price changes, differences in trade sizes or informational content of price changes, etc.) by providing theoretical results.
Kurbucz~\citep{kurbucz2019predicting} designed a complex method consisting of single-hidden layer feedforward neural networks (SLFNs) to (i) determine the predictive power of the most frequent edges of the transaction network (a public ledger that records all Bitcoin transactions) on the future
price of Bitcoin; and, (ii) to provide an efficient technique for applying this untapped dataset in day trading. The research found a significantly high accuracy (60.05\%) for the price movement classifications base on information \ssfan{that} can be obtained using a small subset of edges (approximately 0.45\% of all unique edges).
\sfan{It is worth noting that, Kondor et al.~\citep{kondor2014inferring,kondor2014rich} firstly published some papers giving analysis on transaction networks on cryptocurrency markets and applied related research in identifying Bitcoin users~\citep{juhasz2018bayesian}.}

\sfan{Abay et al.~\citep{abay2019chainnet} attempted 
to understand the network dynamics behind the Blockchain graphs using topological features. The results showed that standard graph features such as the degree distribution of transaction graphs may not be sufficient to capture network dynamics and their potential impact on Bitcoin price fluctuations.}
Maurice et al~\citep{omane2019wavelet} applied wavelet time-scale persistence in analysing returns and volatility in cryptocurrency markets.
The experiment examined \ssfan{the} long-memory and market efficiency characteristics in cryptocurrency markets using daily data for more than two years. The authors employed a log-periodogram regression method in researching stationarity in \ssfan{the} cryptocurrency market and used ARFIMA-FIGARCH class of models in examining long-memory behaviour of cryptocurrencies across time and scale. 
In general, experiments indicated that heterogeneous memory behaviour existed in eight cryptocurrency markets using daily data over the full-time period and across scales (August 25, 2015 to March 13, 2018).
\section{Portfolio, Cryptocurrency Assets and Market Condition Research}\label{portfoliosec}
\subsection{Research among cryptocurrency pairs and related factors}\label{relatedfactor}
Ji et al.~\citep{ji2019dynamic} examined connectedness via return and volatility spillovers across six large cryptocurrencies (collected from coinmarketcap lists from August 7 2015 to February 22 2018) and found Litecoin and Bitcoin to have the most effect on other cryptocurrencies. The authors followed methods of Diebold et al.~\citep{diebold2014network} and built positive/negative returns and volatility connectedness networks. Furthermore, the regression model is used to identify drivers of various cryptocurrency integration levels. Further analysis revealed that the relationship between each cryptocurrency in terms of return and volatility is not necessarily due to its market size. 
Adjepong et al.~\citep{omane2019multiresolution} explored market coherence and volatility causal linkages of seven leading cryptocurrencies. Wavelet-based methods are used to examine market connectedness. Parametric and nonparametric tests are employed to investigate directions of volatility spillovers of the assets. Experiments revealed from diversification benefits to linkages of connectedness and volatility in cryptocurrency markets.
\ssfan{Elie et al.~\citep{bouri2020bitcoin} found the presence of jumps was detected in a series of 12 cryptocurrency returns, and significant jumping activity was found in all cases. More results underscore the importance of the jump in trading volume for the formation of cryptocurrency leapfrogging.}
\revfan{Stanislaw et al.~\citep{drozdz2020competition} examined the correlation of daily exchange rate fluctuations within a basket of the 100 highest market capitalization cryptocurrencies for the period October 1, 2015 to March 31, 2019. The corresponding dynamics mainly involve one of the leading eigenvalues of the correlation matrix, while the others are mainly consistent with the eigenvalues of the Wishart random matrix. The study shows that Bitcoin (BTC) was dominant during the period under consideration, signalling exchange rate dynamics at least as influential as the US dollar (USD).}

Some researchers explored \ssfan{the} relationship between cryptocurrency and different factors, including futures, gold, etc.
Hale et al.~\citep{hale2018futures} suggested that Bitcoin prices rise and fall rapidly after CME issues futures consistent with pricing dynamics. Specifically, the authors pointed out that the rapid rise and subsequent decline in prices after the introduction of futures
is consistent with trading behaviour in \ssfan{the} cryptocurrency market. 
\ssfan{Werner et al.~\citep{kristjanpoller2020cryptocurrencies} focused on the asymmetric interrelationships between major currencies and cryptocurrencies. The results of multiple fractal asymmetric de-trending cross-correlation analysis show evidence of significant persistence and asymmetric multiplicity in the cross-correlation between most cryptocurrency pairs and ETF pairs.}
Bai et al.~\citep{bai2019automated} studied a trading algorithm for foreign exchange on a cryptocurrency Market using \ssfan{the} Automated Triangular Arbitrage method. Implementing \ssfan{a} pricing strategy, implementing trading algorithms and developing a given trading simulation are three problems solved by this research.
Kang et al.~\citep{kang2019co} examined the hedging and diversification properties of gold futures versus Bitcoin prices by using dynamic conditional correlations (DCCs) and wavelet coherence. DCC-GARCH model~\citep{engle2002dynamic} is used to estimate the time-varying correlation between Bitcoin and gold futures by \ssfan{modeling} the variance and the co-variance but also \fan{this two} flexibility. Wavelet coherence method focused more on co-movement between Bitcoin and gold futures. From experiments, the wavelet coherence results indicated volatility persistence, causality and phase difference between Bitcoin and gold.
\revfan{Qiao et al~\citep{qiao2020time} used wavelet coherence and relevance networks to investigate synergistic motion between Bitcoin and other cryptocurrencies. The authors then tested the hedging effect of bitcoin on others at different time frequencies by risk reduction and downside risk reduction. The empirical results provide evidence of linkage and hedging effects. Bitcoin's returns and volatility are ahead of other cryptocurrencies at low frequencies from the analysis, and in the long run, Bitcoin has a more pronounced hedging effect on other cryptocurrencies.}
Dyhrberg et al~\citep{dyhrberg2016bitcoin} applied \ssfan{the} GARCH model and the exponential GARCH model in analysing similarities between Bitcoin, gold and the US dollar. The experiments showed that Bitcoin, gold and the US dollar have \ssfan{similarities} with the variables of the GARCH model, have similar hedging capabilities and react symmetrically to good and bad news. The authors observed that Bitcoin can combine some advantages of commodities and currencies in financial markets to be a tool for portfolio management.

Baur et al.~\citep{baur2018Bitcoin} extended the research of Dyhrberg et al.; \ssfan{the} same data and sample periods are tested~\citep{dyhrberg2016bitcoin} with GARCH and EGARCH-(1,1) models but the experiments reached different conclusions. Baur et al. found that Bitcoin has unique risk-return characteristics compared with other assets. They noticed that Bitcoin excess returns and volatility resemble a rather highly speculative asset with respect to gold or the US dollar.
Bouri et al.~\citep{bouri2017Bitcoin} studied \ssfan{the} relationship between Bitcoin and energy commodities by applying DCCs and GARCH (1,1) \ssfan{models}. In particular, the results showed that Bitcoin is a strong hedge and safe haven for energy commodities.
Kakushadze~\citep{kakushadze2018cryptoasset} proposed factor models for the cross-section of daily cryptoasset returns and provided source code for data downloads, computing risk factors and backtesting for all cryptocurrencies and a host of various other digital assets. The results showed that cross-sectional statistical arbitrage trading may be possible for cryptoassets subject to efficient executions and shorting.
Beneki et al.~\citep{beneki2019investigating} tested hedging abilities between Bitcoin and Ethereum by a multivariate BEKK-GARCH methodology and impulse response analysis within VAR model. The results indicated a volatility transaction from Ethereum to Bitcoin, which implied possible profitable trading strategies on the cryptocurrency derivatives market.
Guglielmo et al.~\citep{caporale2018day} examined \ssfan{the} week effect in cryptocurrency markets and explored the feasibility of this indicator \fan{in trading practice}. Student $t$-test, ANOVA, Kruskal–Wallis and Mann–Whitney tests were carried out for cryptocurrency data in order to compare time periods that may be characterised by anomalies with other time periods. When \ssfan{an} anomaly is detected, an algorithm was established to exploit profit opportunities (MetaTrader terminal in MQL4 is mentioned in this research). The results showed evidence of anomaly (abnormal positive returns on Mondays) in \ssfan{the} Bitcoin market by backtesting in 2013-2016.

\revfan{A number of special research methods have proven to be relevant to cryptocurrency pairs, which is reflected in cryptocurrency trading. Delfabbro et al.~\citep{delfabbro2021cryptocurrency} pointed out that cryprocurrency trading have similarities to gambling. Decisions are often based on limited information, short-term profit motives, and highly volatile and uncertain outcomes. The authors examined whether gambling and problem gambling are reliable predictors of reported cryptocurrency trading strength. Results showed that problem gambling scores (PGSI) and engaging in stock trading were significantly correlated with measures of cryptocurrency trading intensity based on time spent per day, number of trades and level of expenditure.
In further research, Delfabbro et al.~\citep{delfabbro2021psychology} reviewed the specific structural features of cryptocurrency trading and its potential to give rise to excessive or harmful behaviour, including over-spending and compulsive checking. There are some similarities noted between online sports betting and day trading, but there are also some important differences. These include the 24/7 nature of trading, the global nature of the market and the powerful role of social media, social influences and non-balance sheet related events as determinants of price movement.
Cheng et al.~\citep{cheng2020relationship} investigated whether the economic policy uncertainty (EPU) index provided by Baker et al.~\citep{baker2016measuring} can predict the returns of cryptocurrencies. The results suggest that China's EPU Index can predict monthly returns for Bitcoin, whereas the EPU Index for the US or other Asian countries has no predictive power. In addition, China's ban on cryptocurrency trading only affects bitcoin returns among major cryptocurrencies.
Leirvik~\citep{leirvik2021cryptocurrency} analysed the relationship between the particular volatility of market liquidity and the returns of the five largest cryptocurrencies by market capitalisation. The results showed that in general there is a positive correlation between the volatility of liquidity and the returns of large-cap cryptocurrencies. For the most liquid and popular cryptocurrencies, this effect does not exist: Bitcoin. Moreover, the liquidity of cryptocurrencies increases over time, but varies greatly over time.}

\subsection{Crypto-asset Portfolio Research}\label{portfolio}
Some researchers applied portfolio theory for crypto assets. 
Corbet et al.~\citep{corbet2019cryptocurrencies} gave a systematic analysis of cryptocurrencies as financial assets. 
Brauneis et al.~\citep{brauneis2019cryptocurrency} applied the Markowitz mean-variance framework in order to assess \ssfan{the} risk-return benefits of cryptocurrency portfolios. In an out-of-sample analysis accounting for transaction cost, they found that combining cryptocurrencies enriches the set of `low’-risk cryptocurrency investment opportunities. In terms of the Sharpe ratio and certainty equivalent returns, the $1/N$-portfolio \fan{(i.e., ``naive'' strategies, such as equally dividing amongst asset classes)} outperformed single cryptocurrencies and more than 75\% \fan{in terms of the Sharpe ratio and certainty equivalent returns} of mean-variance optimal portfolios.
Castro et al.~\citep{castro2019crypto} developed a portfolio optimisation model based on the Omega measure which is more comprehensive than the Markowitz model and applied this to four crypto-asset investment portfolios by means of a numerical application. Experiments showed crypto-assets improves the return of the portfolios, but on the other hand, also increase the risk exposure. 

Bedi et al.~\citep{bedi2020investment} examined diversification capabilities of Bitcoin for a global portfolio spread across six asset classes from the standpoint of investors dealing in five major fiat currencies, namely US Dollar, Great Britain Pound, Euro, Japanese Yen and Chinese Yuan. They employed modified Conditional Value-at-Risk and standard deviation as measures of risk to perform portfolio optimisations across three asset allocation strategies and provided insights into \ssfan{the} sharp disparity in Bitcoin trading volumes across national currencies from a portfolio theory perspective.
Similar research has been done by Antipova et al.~\citep{antipova2019building}, which explored the possibility of establishing and optimizing a global portfolio by diversifying investments using one or more cryptocurrencies, and assessing returns to investors in terms of risks and returns.
Fantazzini et al.~\citep{fantazzini2019multivariate} proposed a set of models \ssfan{that} can be used to estimate the market risk for a portfolio of crypto-currencies, and simultaneously estimate their credit risk using the Zero Price Probability (ZPP) model. The results revealed the superiority of the t-copula/skewed-t GARCH model for market risk, and the ZPP-based models for credit risk. 
\ssfan{Qiang et al.~\citep{ji2019realised} examined the common dynamics of bitcoin exchanges. Using a connectivity metric based on the actual daily volatility of the bitcoin price, they found that Coinbase is undoubtedly the market leader, while Binance performance is surprisingly weak. The results also suggested that safer asset extraction is more important for volatility linkages between Bitcoin exchanges relative to trading volumes.}
\revfan{Fasanya et al.~\citep{fasanya2020returns} quantified returns and volatility transmission between cryptocurrency portfolios by using a spillover approach and rolling sample analysis. The results showed that there is a significant difference between the behaviour of cryptocurrency portfolio returns and the volatility spillover index over time. Given the spillover index, the authors found evidence of interdependence between cryptocurrency portfolios, with the spillover index showing an increased degree of integration between cryptocurrency portfolios.}

Trucios et al.~\citep{trucios2019value} proposed a methodology based on vine copulas and robust volatility models to estimate the Value-at-Risk (VaR) and Expected Shortfall (ES) of cryptocurrency portfolios. The proposed algorithm displayed good performance in estimating both VaR and ES.
Hrytsiuk et al.~\citep{hrytsiuk2019cryptocurrency} showed that the cryptocurrency returns can be described by the Cauchy distribution and obtained the analytical expressions for VaR risk measures and performed calculations accordingly. As a result of the optimisation, the sets of optimal cryptocurrency portfolios were built in their experiments. 

Jiang et al.~\citep{jiang2017cryptocurrency} proposed a two-hidden-layer CNN that takes the historical price of a group of \fan{cryptocurrency} assets as an input and outputs the weight of the group of \fan{cryptocurrency} assets. \sfan{This research focused on portfolio research in cryptocurrency assets using emerging technologies like CNN.} Training is conducted in an intensive manner to maximise cumulative returns, which is considered a reward function of the CNN network. The performance of the CNN strategy is compared with the three benchmarks and the other three portfolio management algorithms (buy and hold strategy, Uniform Constant Rebalanced Portfolio and \fan{Universal Portfolio with Online Newton Step and Passive Aggressive Mean Reversion}); the results are positive in that the model is only second to the Passive Aggressive Mean Reversion algorithm (PAMR). 
Estalayo et al.~\citep{estalayo2019return} reported initial findings around the combination of DL models and Multi-Objective Evolutionary Algorithms (MOEAs) for allocating cryptocurrency portfolios. Technical rationale and details were given on the design of a stacked DL recurrent neural network, and how its predictive power can be exploited for yielding accurate \ssfan{ex-ante} estimates of the return and risk of the portfolio. Results obtained for a set of experiments carried out with real cryptocurrency data have verified the superior performance of their designed deep learning model with respect to other regression techniques.



\subsection{Bubbles and Crash Analysis}\label{bubbles}
\revfantwo{Bubbles and crash analysis is an important researching area in cryptocurrency trading.}
Phillips and Yu proposed a methodology to test for the presence of cryptocurrency bubble~\citep{cheung2015crypto}, which is extended by Shaen et al.~\citep{corbet2018datestamping}. 
The method is based on supremum Augmented Dickey–Fuller (SADF) to test for the bubble through the inclusion of a sequence of \ssfan{forwarding} recursive right-tailed ADF unit root tests. An extended methodology generalised SADF (GSAFD), is also tested for bubbles within cryptocurrency data. 
The research concluded that there is no clear evidence of a persistent bubble in cryptocurrency markets including Bitcoin or Ethereum. 
Bouri et al.~\citep{bouri2019co} date-stamped price explosiveness in seven large cryptocurrencies and revealed evidence of multiple periods of explosivity in all cases. GSADF is used to identify multiple explosiveness periods and logistic regression is employed to uncover evidence of co-explosivity across cryptocurrencies. The results showed that the likelihood of explosive periods in one cryptocurrency generally depends on the presence of explosivity in other cryptocurrencies and points toward a contemporaneous co-explosivity that does not necessarily depend on the size of each cryptocurrency.

Extended research by Phillips et al.~\citep{phillips2015testinga,phillips2015testingb} (who applied a recursive augmented Dickey-Fuller algorithm, which is called PSY test) and Landsnes et al.~\citep{enoksen2019can} studied possible predictors of bubble periods of certain cryptocurrencies. 
The evaluation includes multiple bubble periods in all cryptocurrencies. The result shows that higher volatility and trading volume is positively associated with the presence of bubbles across cryptocurrencies. 
In terms of bubble prediction, \ssfan{the} authors found the probit model to perform better than the linear models.

Phillips et al.~\citep{phillips2017predicting} used Hidden Markov Model (HMM) and Superiority and Inferiority Ranking (SIR) method to identify bubble-like behaviour in cryptocurrency time series. 
Considering HMM and SIR method, \ssfan{an} epidemic detection mechanism is used in social media to predict cryptocurrency price bubbles, which classify bubbles through epidemic and non-epidemic labels. 
Experiments have demonstrated a strong relationship between Reddit 
usage and cryptocurrency prices.
This work also provides some empirical evidence that bubbles mirror the social epidemic-like spread of an investment idea.
Guglielmo et al.~\citep{caporale2019price} examined the price overreactions in the case of cryptocurrency trading. 
\fan{Some parametric and non-parametric tests confirmed \ssfan{the} presence of price patterns after overreactions, which identified that the next-day price changes in both directions are bigger than after ``normal'' days. The results also showed that the overreaction detected in the cryptocurrency market would not give available profit opportunities (possibly due to transaction costs) that cannot be considered as evidence of the EMH.}
Chaim et al.~\citep{chaim2018volatility} analysed \ssfan{the} high unconditional volatility of cryptocurrency from a standard log-normal stochastic volatility model to discontinuous jumps of volatility and returns. The experiment indicated the importance of incorporating permanent jumps to volatility in cryptocurrency markets.

\revfan{J.L. Cross et al.~\citep{cross2021returns} investigated the existence and nature of the interdependence of bitcoin, ethereum, litecoin and ripple during the cryptocurrency bubble of 2017-18. A generalized time-varying asset pricing model approach is proposed. The results showed that the negative news impact of the boom period in 2017 for LiteCoin and Ripple, which incurred a risk premium for investors, could explain the returns of cryptocurrencies during the 2018 crash.}

\subsection{Extreme condition}\label{extremecondition}
Differently from traditional fiat currencies, cryptocurrencies are risky and exhibit heavier tail behaviour. 
Paraskevi et al.~\citep{katsiampa2018cryptocurrency} found extreme dependence between returns and trading volumes. 
Evidence of asymmetric return-volume relationship in the cryptocurrency market was also found by \ssfan{the} experiment, as a result of discrepancies in the correlation between positive and negative return exceedances across all the cryptocurrencies.


There has been a price crash in late 2017 to early 2018 in cryptocurrency~\citep{yaya2018persistent}. Yaya et al.~\citep{yaya2018persistent} researched \ssfan{the} persistence and dependence of Bitcoin on other popular alternative coins before and after \ssfan{the} 2017/18 crash in cryptocurrency markets. The result showed that higher persistence of shocks is expected after the crash due to speculations in the mind of cryptocurrency traders, and more \ssfan{evidence} of non-mean reversions, implying chances of further price fall in cryptocurrencies.

\revfantwo{Manahov~\citep{manahov2021cryptocurrency} obtained millisecond data for major cryptocurrencies as well as the cryptocurrency indices Cryptocurrency IndeX (CRIX) and Cryptocurrencies Index 30 (CCI30) to investigate the relationship between cryptocurrency liquidity, herding behaviour and profitability during extreme price movements (EPM). Millisecond data was obtained for major cryptocurrencies as well as the cryptocurrency indices CRIX and CCI30 to investigate the relationship between cryptocurrency liquidity, herding behaviour and profitability during EPM. The experiments demonstrate that cryptocurrency traders (CTs) can promote EPM and demand liquidity even during periods of maximum EPM. The authors' robustness checks suggest that herding behaviour follows a dynamic pattern with decreasing magnitude over time.
Shahzad et al.~\citep{shahzad2021extreme} investigated the interdependence of median-based and tail-based returns between cryptocurrencies under normal and extreme market conditions. The experiment used daily data and combines LASSO techniques with quantile regression within a network analysis framework. The main results showed that the interdependence of the tails is higher than the median, especially in the right tail. Fluctuations in market, size and momentum drive return connectivity and clustering coefficients under both normal and extreme market conditions.
Chan et al.~\citep{chan2022extreme} examined the extreme dependence and correlation between high-frequency cryptocurrency (Bitcoin and Ethereum, relative to the euro and the US dollar) returns and trading volumes in the extreme tails associated with booms and busts in cryptocurrency markets. Experiments with extreme value theory methods highlight how these results can help traders and practitioners who rely on technical indicators in their trading strategies - especially in times of extreme market volatility or irrational market booms.}

\section{Others related to Cryptocurrency Trading}\label{othersrelatedtrading}
Some other research papers related to cryptocurrency trading treat distributed in market behaviour, regulatory mechanisms and benchmarks.

Krafft et al.~\citep{krafft2018experimental} and Yang~\citep{yang2018behavioral} analysed market dynamics and behavioural anomalies respectively to understand effects of market behaviour in \ssfan{the} cryptocurrency market. 
Krafft et al. discussed potential ultimate causes, potential behavioural mechanisms and potential moderating contextual factors to enumerate possible influence of GUI and API on cryptocurrency markets. Then they highlighted \ssfan{the} potential social and economic impact of human-computer interaction in digital agency design.
Yang, on the other hand, applied behavioural theories of asset pricing anomalies in testing 20 market anomalies using cryptocurrency trading data. The results showed that anomaly research focused more on the role of speculators, which gave a new idea to research the momentum and reversal in \ssfan{the} cryptocurrency market.
\sfan{Cocco et al.~\citep{cocco2016modeling} implemented a mechanism to form a Bitcoin price and specific behaviour for each type of trader including the initial wealth distribution following Pareto's law, order-based transaction and price settlement mechanism. Specifically, the model reproduced the unit root attributes of the price series, the fat tail phenomenon, the volatility clustering of price returns, the generation of Bitcoins, hashing power and power consumption.}

Leclair~\citep{leclair2018herding} and Vidal-Thom\'as et al.~\citep{vidal2019herding} analysed the existence of herding in the cryptocurrency market. 
Leclair applied herding methods of Huang and Salmon~\citep{hwang2004market} in estimating the market herd dynamics in the CAPM framework.
Vidal-Thom\'as et al. analyse the existence of herds in the cryptocurrency market by returning \ssfan{the} cross-sectional standard (absolute) deviations.
Both their findings showed significant evidence of market herding in \ssfan{the} cryptocurrency market. 
Makarov et al.~\citep{makarov2019trading} studied price impact and arbitrage dynamics in the cryptocurrency market and found that 85\% of the variations in
bitcoin returns and the idiosyncratic components of order flow~\citep{liu2021call} play an important role in explaining the size of the arbitrage spreads between exchanges.
\revfan{King et al.~\citep{king2021herding} examined the extent to which herding and feedback trading behaviour drive the price dynamics of nine major cryptocurrencies. The study documented heterogeneity in the types of feedback trading strategies used by investors in different markets and evidence of herding or ``trend chasing" behaviour in some cryptocurrency markets.}

 In November 2019, Griffin et al. put forward a paper on the thesis of unsupported digital money inflating cryptocurrency prices~\citep{griffin2019bitcoin}, which caused a great stir in the academic circle and public opinion. Using algorithms to analyse Blockchain data, they found that purchases with Tether are timed following market downturns and result in sizeable increases in Bitcoin prices. By mapping the blockchains of Bitcoin and Tether, they were able to establish that one large player on Bitfinex uses Tether to purchase large amounts of Bitcoin when prices are falling and following the prod of Tether.  

More researches involved benchmark and development in cryptocurrency market
~\citep{hileman2017global,zhou2018algorithmic}, regulatory framework analysis~\citep{shanaev2020taming,feinstein2021impact}, data mining technology in cryptocurrency trading~\citep{patil2018study}, application of efficient market hypothesis in \ssfan{the} cryptocurrency market~\citep{sigaki2019clustering}, \revfan{Decentralized Exchanges (DEXs)} and artificial financial markets for studying a cryptocurrency market~\citep{cocco2017using}.
Hileman et al.~\citep{hileman2017global} segmented the cryptocurrency industry into four key sectors: exchanges,
wallets, payments and mining. They gave a benchmarking study of individuals, data, regulation, compliance practices, costs of firms and \ssfan{a} global map of mining in \ssfan{the} cryptocurrency market in 2017.
Zhou et al.~\citep{zhou2018algorithmic} discussed the status and future of computer trading in the largest group of Asia-Pacific economies and then considered algorithmic and high frequency trading in cryptocurrency markets as well.
Shanaev et al.~\citep{shanaev2020taming} used data on 120 regulatory events to study the implications of cryptocurrency regulation and the results showed that stricter regulation of cryptocurrency is not desirable.
\revfan{Feinstein et al.~\citep{feinstein2021impact} collected raw data on global cryptocurrency regulations and used them to empirically test the trading activity of many exchanges against key regulatory announcements. No systematic evidence has been found that regulatory measures cause traders to flee or enter the affected regional jurisdictions according to authors' analysis.}
Akhilesh et al.~\citep{patil2018study} used the average absolute error calculated between the actual and predicted values \fan{of the market sentiment of different cryptocurrencies on that day} as a method for quantifying the uncertainty. They used the comparison of uncertainty quantification methods and opinion mining to analyse current market conditions.
Sigaki et al.~\citep{sigaki2019clustering} used permutation entropy and statistical complexity on the sliding time window returned by the price log to quantify the dynamic efficiency of more than four hundred cryptocurrencies. As a result, the cryptocurrency market showed significant compliance with efficient market assumptions.
\revfan{Aspris et al.~\citep{aspris2021decentralized} surveyed the rapid rise of DEXs, including automated market makers. The study demonstrated the significant differences in the listing and trading characteristics of these tokens compared to their centralised equivalents.}
Cocco et al.~\citep{cocco2017using} described an agent-based artificial cryptocurrency market in which heterogeneous agents buy or sell cryptocurrencies. The proposed simulator is able to reproduce some real statistical properties of price returns observed in the Bitcoin real market.
Marko~\citep{ogorevc2019cryptocurrency} considered the future use of cryptocurrencies as money based on the long-term value of cryptocurrencies. 
Neil et al.~\citep{gandal2014competition} analysed the influence of network effect on the competition of new cryptocurrency markets.
\sfan{Bariviera and Merediz-Sola~\citep{bariviera2020we} gave a survey based on hybrid analysis, which proposed a methodological hybrid method for \ssfan{a} comprehensive literature review and provided the latest technology in the cryptocurrency economics literature.}

There also exists some research and papers introducing the basic process and rules of cryptocurrency trading including findings of Hansel et al.~\citep{hansel2018cryptocurrency}, Kate~\citep{kate2018cryptocurrency}, Garza et al.~\citep{garza2019formal}, Ward et al.~\citep{ward2018algorithmic} and \sfan{Fantazzini et al.~\citep{fantazzini2019quantitative}}.
Hansel et al.~\citep{hansel2018cryptocurrency} introduced \ssfan{the} basics of cryptocurrency, Bitcoin and Blockchain, ways to identify \ssfan{the} profitable \ssfan{trends} in the market, ways to use Altcoin trading platforms such as GDAX and Coinbase, methods of using a crypto wallet to store and protect the coins in their book.
Kate et al.~\citep{kate2018cryptocurrency} set six steps to show how to start an investment without any technical skills in \ssfan{the} cryptocurrency market. This book is an entry-level trading manual for starters learning cryptocurrency trading.
Garza et al.~\citep{garza2019formal} simulated \ssfan{an} automatic cryptocurrency trading system, which helps investors limit systemic risks and improve market returns. This paper is an example to start designing an automatic cryptocurrency trading system.
Ward et al.~\citep{ward2018algorithmic} discussed algorithmic cryptocurrency trading using several general algorithms, and
modifications thereof including adjusting the parameters used in each strategy, as well as mixing multiple strategies or dynamically changing between strategies. This paper is an example to start algorithmic trading in cryptocurrency market.
\sfan{Fantazzini et al.~\citep{fantazzini2019quantitative} introduced the R packages Bitcoin-Finance and bubble, including financial analysis of cryptocurrency markets including Bitcoin.}

\sfan{A community resource, that is, a platform for scholarly communication, about cryptocurrencies and Blockchains is ``Blockchain research network", see~\citep{blockchaincommunity}.}

\section{Summary Analysis of Literature Review}\label{summarysec}
This section analyses the timeline, the research distribution among technology and methods, the research distribution among properties. It also summarises the datasets that have been used in cryptocurrency trading research.

\subsection{Timeline}\label{timeline}
\begin{figure} 
\centering
\includegraphics[width=1\columnwidth]{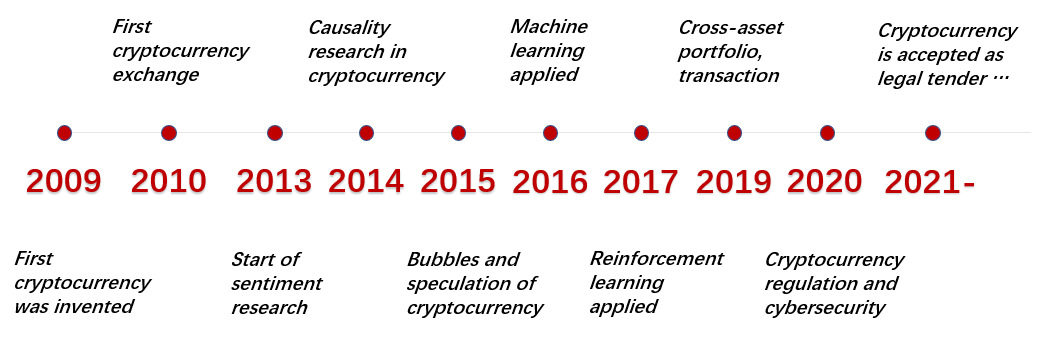}
\caption{Timeline of cryptocurrency trading research}
\label{fig:time}
\end{figure}

Figure~\ref{fig:time} shows several major events in cryptocurrency trading. The timeline contains milestone events in cryptocurrency trading and important scientific breakthroughs in this area.

As early as 2009, Satoshi Nakamoto proposed and invented \ssfan{the} first decentralised cryptocurrency, Bitcoin~\citep{nakamoto2009bitcoin}. It is considered to be the start of cryptocurrency. 
In 2010, the first cryptocurrency exchange was founded, which means cryptocurrency would not be an OTC market but traded on exchanges based on \ssfan{an} auction market system.

In 2013, Kristoufek~\citep{kristoufek2013bitcoin} concluded that there is a strong correlation between Bitcoin price and the frequency of ``Bitcoin'' search queries in Google Trends and Wikipedia. 
In 2014, Lee and Yang~\citep{lee2014granger} firstly proposed to check causality from copula-based causality in the quantile method from trading volumes of seven major cryptocurrencies to returns and volatility.

In 2015, Cheah et al.~\citep{cheah2015speculative} discussed \ssfan{the} bubble and speculation of Bitcoin and cryptocurrencies. 
In 2016, 
Dyhrberg explored Bitcoin volatility using GARCH models combined with gold and US dollars~\citep{dyhrberg2016bitcoin}.

From late 2016 to 2017, machine learning and deep learning technology were applied in \ssfan{the} prediction of cryptocurrency return. In 2016, McNally et al.~\citep{mcnally2016predicting} predicted Bitcoin price using \ssfan{the} LSTM algorithm. Bell and Zbikowski et al.~\citep{bell2016bitcoin,zbikowski2016application} applied SVM algorithm to predict trends of cryptocurrency price. 
In 2017, Jiang et al.~\citep{jiang2017cryptocurrency} used double Q-network and pre-trained it using DBM for the prediction of cryptocurrencies \fan{portfolio weights}. 

\revfan{From 2019 to 2020, several research directions including cross asset portfolios~\citep{bedi2020investment,castro2019crypto,brauneis2019cryptocurrency}, transaction network applications~\citep{kurbucz2019predicting,bouri2019co}, machine learning optimisation~\citep{rane2019systematic,atsalakis2019bitcoin,zhengyang2019prediction} have been considered in the cryptocurrency trading area.}

\revfan{In 2021, more regulation issues were put out the stage. On 18 May 2021, China banned financial institutions and payment companies from providing services related to cryptocurrency transactions, which led to a sharp drop in the price of bitcoin~\citep{cryptochina}. In June 2021, El Salvador becomes the first country to accept Bitcoin as legal tender~\citep{cryptolegal}.}

\subsection{Research Distribution among Properties}\label{distributionpro}
\begin{figure} 
\centering
\includegraphics[width=.95\columnwidth]{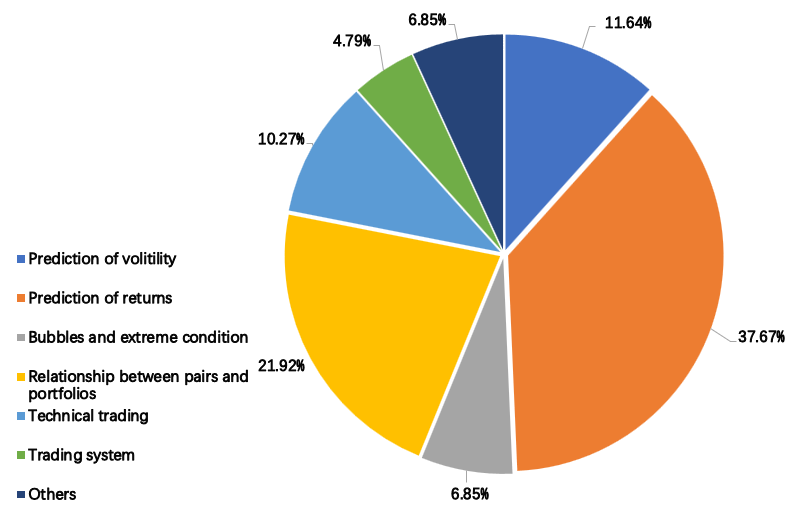}
\caption{Research distribution among cryptocurrency trading properties}
\label{fig:prop}
\end{figure}

We counted the number of papers covering different aspects of cryptocurrency trading. 
Figure \ref{fig:prop} shows the result. The attributes in the legend are ranked according to the number of papers that specifically test the attribute. 

\revfan{Over one-third (37.67\%) of the papers research prediction of returns. Another one-third of papers focus on researching bubbles and extreme conditions and the relationship between pairs and portfolios in cryptocurrency trading. The remaining researching topics (prediction of volatility, trading system, technical trading and others) have roughly one-third share.}

\subsection{Research Distribution among Categories and Technologies}\label{distributiontech}
This section introduces and compares categories and technologies in cryptocurrency trading. 
When papers cover multiple technologies or compare different methods, we draw statistics from different technical perspectives. 

Among all the \revfan{146} papers, \revfan{102} papers (\revfan{69.86}\%) cover statistical methods and machine learning categories. These 
papers basically research technical-level cryptocurrency trading including mathematical \ssfan{modeling} and 
statistics. 
Other papers related to trading systems on pure technical indicators 
and introducing 
the industry and its history are not included in this analysis.
Among all \revfan{102} papers, \revfan{88} papers (\revfan{86.28}\%) present statistical methods and technologies in cryptocurrency trading research and \revfan{13.72}\% papers research machine learning applied to cryptocurrency trading (cf. Figure~\ref{fig:tech}). It is interesting to mention that, there are \revfan{17} papers (\revfan{16.67}\%) applying and comparing more than one technique in cryptocurrency trading. 
More specifically, Bach et al.~\citep{bach2018machine}, Alessandretti et al.~\citep{alessandretti2018anticipating}, Vo et al.~\citep{vo2018high}, Phaladisailoed et al.~\citep{phaladisailoed2018machine},  Siaminos~\citep{siaminos2019predicting}, Rane et al.~\citep{rane2019systematic} used both statistical methods and machine learning methods in cryptocurrency trading. 

Table~\ref{tbl:hits} shows the results of search hits in all trading areas (not limited to cryptocurrencies). From the table, we can see that most research findings focused on statistical methods in trading, which means most of the research on traditional markets still focused on using statistical methods for trading. 
But we observed that machine learning in trading had a higher degree of attention. It might because \fan{the traditional technical and fundamental have been arbitraged, so the market has moved in recent years to find new anomalies to exploit}. 
Meanwhile, the results also showed there exist many opportunities for research in the widely studied areas of machine learning applied to \ssfan{trade} in cryptocurrency markets (cf. Section \ref{opportunitiessec}).

\begin{table} 
\centering
\caption{Search hits of research distribution in all trading areas}
\resizebox{.78\textwidth}{!}{
\begin{tabular}{lllll}
\hline
Technology Category      & Google Scholar hits & Google hits & Arxiv hits & SCOPUS Indexed citations  \\
\hline
Statistical methods      & 1.22M               & 62M         & 1240 & 2790        \\
Machine learning methods & 483K                & 150M        & 520  & 1,754        \\
\hline
\end{tabular}
}
\label{tbl:hits}
\end{table}

\subsubsection{Research Distribution among Statistical methods}
As from Figure~\ref{fig:tech}, we further classified the papers using statistical methods into 6 categories: (i) basic regression methods; (ii) linear classifiers and clustering; (iii) time-series analysis; (iv) decision trees and probabilistic classifiers; (v) modern portfolios theory; and, (vi) Others.

\textbf{Basic regression methods} include regression methods (Linear Regression), 
function estimation and CGCD 
method.
\textbf{Linear Classifiers and Clustering} include SVM and KNN algorithm.
\textbf{Time-series analysis} include GARCH model,
BEKK model, ARIMA model, Wavelet time-scale method.
\textbf{Decision Trees and probabilistic classifiers} include Boosting Tree, RF model.
\textbf{Modern portfolio theory} include Value-at-Risk (VaR) theory, expected-shortfall (ES), Markowitz mean-variance framework.
\textbf{Others} include industry, market data and research analysis in cryptocurrency market.

The figure shows that basic Regression methods and time-series analysis are the most commonly used methods in this area.

\subsubsection{Research Distribution among Machine Learning Categories}
\begin{figure}
\centering
\includegraphics[width=.95\columnwidth]{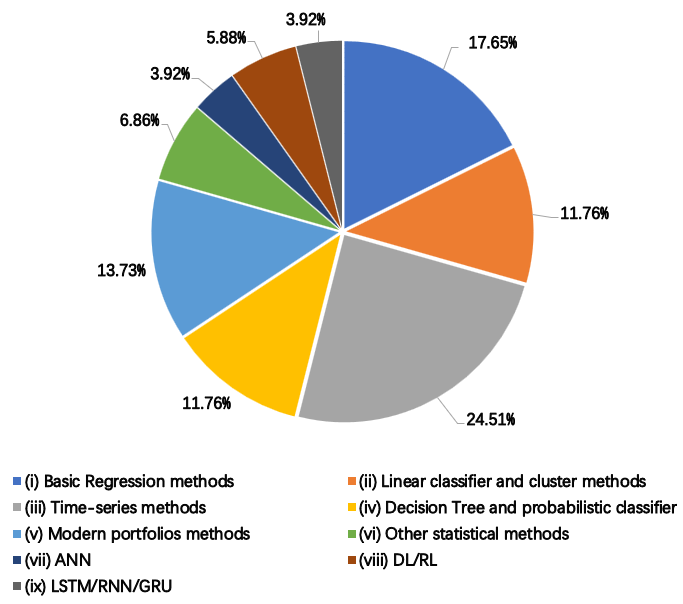}
\caption{Research distribution among cryptocurrency trading technologies and methods}
\label{fig:tech}
\end{figure}

Papers using machine learning account for \revfan{13.7} (c.f Figure~\ref{fig:tech}) of the total. We further classified these papers into three categories: (vii) ANNs, (viii) LSTM/RNN/GRUs, and (ix) DL/RL.

The figure also shows that methods based on LSTM, RNN and GRU are the most popular in this subfield.

\textbf{ANNs} \ssfan{contains} papers researching ANN applications in cryptocurrency trading such as back propagation (BP) NN.
\textbf{LSTM/RNN/GRUs} include papers using neural networks \ssfan{that} exploit the temporal structure of the data, a  technology especially suitable for time series prediction and financial trading. 
\textbf{DL/RL} \ssfan{includes} papers using Multilayer Neural Networks and Reinforcement Learning.
The difference between ANN and DL is that generally, DL refers to an ANN with multiple hidden layers while ANN refers to simple structure neural network contained input layer, hidden layer (one or multiple), and an output layer.

\subsection{Datasets used in Cryptocurrency Trading}\label{datasetssec}
Tables~\ref{tbl:dst1}--\ref{tbl:dst3} show the details for some representative datasets used in cryptocurrency trading research. Table~\ref{tbl:dst1} shows the market datasets. They mostly include price, trading volume, order-level information, collected from cryptocurrency exchanges. Table~\ref{tbl:dst2} shows the sentiment-based data. Most of \ssfan{the} datasets in this table contain market data and media/Internet data with emotional or statistical labels. Table~\ref{tbl:dst3} gives two examples of datasets used in the collected papers that are not covered in the first two tables.

The column ``Currency'' shows the types of cryptocurrencies included; this shows that Bitcoin is the most commonly used currency for cryptocurrency researches. The column ``Description'' shows \ssfan{a} general description and types of datasets. The column ``Data Resolution'' means latency of the data (e.g., used in the backtest) -- this is useful to distinguish between high-frequency trading and low-frequency trading. The column ``Time range'' shows the time span of datasets used in experiments; this is convenient to distinguish between the current performance in a specific time interval and the long-term effect. We also present how the dataset has been used (i.e., the task), cf. column ``Usage''. ``Data Sources'' gives details on where the data is retrieved from, including cryptocurrency exchanges, aggregated cryptocurrency index and user forums (for sentiment analysis).

\ssfan{Alexander et al.~\citep{alexander2020critical} made an investigation of cryptocurrency data as well. They summarised data collected from 152 published and SSRN discussion papers about cryptocurrencies and analysed their data quality. They found that less than half the cryptocurrency papers published since January 2017 employ correct data.}

\begin{table*} 
\caption{Datasets (1/3): Market Data}
\centering
\begin{turn}{90}
\resizebox{0.85\paperheight}{!}{
\begin{tabular}{@{}lllllll@{}}
\toprule
\textbf{Research Source} & \textbf{Description}                                                                                                                     & \textbf{Currency}                                                                        & \textbf{Data Resolution}                                                                               & \textbf{Time Range}                                                                                         & \textbf{Usage}
                                             &\textbf{Data Sources}        \\ \midrule
Bouri et al.~\citep{bouri2019trading}             & \begin{tabular}[t]{@{}l@{}}price, \\ volatility, \\ detrended volume data\end{tabular}                                                   & \begin{tabular}[t]{@{}l@{}}Bitcoin, \\ Ethereum,\\ 5 other cryptocurrencies\end{tabular} & daily                                                                                                & \begin{tabular}[t]{@{}l@{}}From: 2013/01/01\\ To: 2017/12/31\end{tabular}                                                                                    & Prediction of volatility/return     &
\begin{tabular}[t]{@{}l@{}}CoinMarketCap\end{tabular}
\\ \\
Nakano et al.~\citep{nakano2018bitcoin}            & \begin{tabular}[t]{@{}l@{}}high frequency price,\\ volume data\end{tabular}                                                              & Bitcoin                                                                                  & 15min                                                                                                & \begin{tabular}[t]{@{}l@{}}From: 2016/07/31\\ To: 2018/01/24\end{tabular}                                                                                     & Prediction of return
& \begin{tabular}[t]{@{}l@{}}Poloniex\end{tabular}\\ \\
Bu et al.~\citep{bu2018learning}                & \begin{tabular}[t]{@{}l@{}}three pieces time slice for \\ different r\textcolor[rgb]{0.133,0.133,0.133}{esearch objectives}\end{tabular} & Bitcoin and seven altcoins                                                               & Not mentioned                                                                                        & \begin{tabular}[t]{@{}l@{}}From: 2016/05/14\\ To: 2016/07/03\\From: 2018/01/01\\To: 2018/01/31\\From: 2017/07/01\\To: 2017/07/31\end{tabular} & Maximum profit with DRL    
& Not mentioned\\ \\
Sun et al.~\citep{sun2019using}               & price, volatility                                                                                                                        & \begin{tabular}[t]{@{}l@{}}ETC-USDT,\\ other 12 cryptocurrencies\end{tabular}            & \begin{tabular}[t]{@{}l@{}}1 minute, \\ 5 minutes,\\ 30 minutes, \\ one hour,\\ one day\end{tabular} & \begin{tabular}[t]{@{}l@{}}From: August 2017\\ To: December 2018\end{tabular}                                                                                  & Prediction of return 
& Binance, Bitfinex\\ \\
Guo et al.~\citep{guo2018Bitcoin}               & \begin{tabular}[t]{@{}l@{}}volatility, \\ order book data\end{tabular}                                                                                                           & Bitcoin                                                                                  & \begin{tabular}[t]{@{}l@{}}hourly volatility observations, \\ order book snapshots\end{tabular}      & \begin{tabular}[t]{@{}l@{}}From: September 2015\\ To: April 2017\end{tabular}                                                                                  & Prediction of volatility
& Not mentioned\\ \\
Vo et al.~\citep{vo2018high}                & \begin{tabular}[t]{@{}l@{}}timestamps,\\ the OHLC prices etc.\end{tabular}                                                & Bitcoin                                                                                  & 1minute                                                                                              & \begin{tabular}[t]{@{}l@{}}From: Starting 2015\\ To: End 2016\end{tabular}                                                                                    & Prediction of return 
&\begin{tabular}[t]{@{}l@{}} Bitstamp, Btce, Btcn,\\ Coinbase, Coincheck, and Kraken \end{tabular}\\ \\
Ross et al.~\citep{phillips2017predicting}              & price                                                                                                                                    & \begin{tabular}[t]{@{}l@{}}Bitcoin,\\ other 3 cryptocurrencies\end{tabular}              & daily                                                                                                & \begin{tabular}[t]{@{}l@{}}From: April 2015\\ To: September 2016\end{tabular}                                                                                 & Predicting bubbles   & CryptoCompare             \\ \\
Yaya et al.~\citep{yaya2018persistent}              & price                                                                                                                                    & \begin{tabular}[t]{@{}l@{}}Bitcoin,\\ other 12 cryptocurrencies\end{tabular}             & daily                                                                                                & \begin{tabular}[t]{@{}l@{}}From: 2015/08/07\\ To: 2018/11/28\end{tabular}                                                                           & Bubbles and crashes 
& Coin Metrics\\ \\
Brauneis et al.~\citep{brauneis2019cryptocurrency}          & \begin{tabular}[t]{@{}l@{}}individual price,\\trading volume\end{tabular}                                                           & \begin{tabular}[t]{@{}l@{}}500 most capitalized \\ Cryptocurrencies\end{tabular}         & daily                                                                                                & \begin{tabular}[t]{@{}l@{}}From: 2015/01/01\\ To: 2017/12/31\end{tabular}                                                                                  & Portfolios management 
& CoinMarketCap\\ \\
Feng et al.~\citep{feng2018informed}              & \begin{tabular}[t]{@{}l@{}}order-level USD/BTC \\ trading data\end{tabular}                                                                                                          & Bitcoin                                                                                  & order-level                                                                                          & \begin{tabular}[t]{@{}l@{}}From: 2011/09/13\\ To: 2017/07/17\end{tabular}                                                                                    & Trading strategy 
& Bitstamp\\ \\
Gradojevic et al.~\citep{gradojevic2021volatility}       &  \begin{tabular}[t]{@{}l@{}}price, volatility  \end{tabular}            
    &     \begin{tabular}[t]{@{}l@{}}BTC/USD\\ ETH/USD\\ XRP/USD \end{tabular}      & \begin{tabular}[t]{@{}l@{}}1-Minute\\ 1-Hour\\ 1-Day\\ 1-Month\end{tabular}   & \begin{tabular}[t]{@{}l@{}}From: 2015/01/01\\To: 2020/10/15\end{tabular}                       & Prediction of volatility   
    &   CryptoCompare  \\
\bottomrule
\end{tabular}
}
\end{turn}
\label{tbl:dst1}
\end{table*}

\begin{table*} 
\caption{Datasets (2/3): Sentiment-based data}
\begin{turn}{90}
\resizebox{0.85\paperheight}{!}{
\begin{tabular}{@{}llllll@{}}
\toprule
\textbf{Research Source} & \textbf{Description}                                                                                                & \textbf{Currency}                                    & \textbf{Time range}                                                                                                                                                                & \textbf{Usage}                                                     &\textbf{Data Sources}                                 \\ \midrule
Kim et al.~\citep{kim2016predicting}               & \begin{tabular}[t]{@{}l@{}}Online cryptocurrency communities data\\ and market data\end{tabular}                                                              &\begin{tabular}[t]{@{}l@{}} Bitcoin,Ethereum, Ripple \end{tabular}                            & \begin{tabular}[t]{@{}l@{}}From:~December 2013\\To:~August, 2016 (Bitcoin)\\From:~August 2015\\To:~August, 2016 (Ethereum)\\From: Creation\\To: August, 2016 (Ripple)\end{tabular} & \begin{tabular}[t]{@{}l@{}}Prediction of fluctuation  \end{tabular}                   & Each community’s HTML page                                                                    \\ \\
Phillips et al.~\citep{phillips2018mutual}          & \begin{tabular}[t]{@{}l@{}}Social media data and orice data \end{tabular}                                                                                   & \begin{tabular}[t]{@{}l@{}}Bitcoin and Ethereum \end{tabular}                                & \begin{tabular}[t]{@{}l@{}}From: 2016/08/30\\To: 2017/08/30\end{tabular}                                                                                                           &\begin{tabular}[t]{@{}l@{}}Predict~Mutual-Excitation of \\ Cryptocurrency Market Returns \end{tabular}   
& Reddit\\ \\
Smtus~\citep{smuts2019drives}                    & \begin{tabular}[t]{@{}l@{}}Hourly data on price and trading volume\\ and search terms from Google Trends\end{tabular} &\begin{tabular}[t]{@{}l@{}} Bitcoin, Ethereum\\ and their respective pricedrivers~\end{tabular} & \begin{tabular}[t]{@{}l@{}}From: 2017/12/01\\To: 2018/06/30\end{tabular}                                                                                                           & \begin{tabular}[t]{@{}l@{}}Prediction of price \end{tabular}                  & Google Trends, Telegram chat groups                                                                             \\ \\
Lamon et al.~\citep{lamon2017cryptocurrency}             & \begin{tabular}[t]{@{}l@{}}Daily price data and cryptocurrency\\related news article headlines\end{tabular}       & \begin{tabular}[t]{@{}l@{}}Bitcoin, Ethereum, Litecoin \end{tabular}                        & \begin{tabular}[t]{@{}l@{}}From: 2017/01/01\\To: 2017/11/30\end{tabular}                                                                                                           & \begin{tabular}[t]{@{}l@{}}Prediction of price \end{tabular}                              & Kaggle,  news headline                                                         \\ \\
Phillips et al.~\citep{phillips2018cryptocurrency}          & \begin{tabular}[t]{@{}l@{}}Price and social media factors from Reddit \end{tabular}                                                                          & \begin{tabular}[t]{@{}l@{}}Bitcoin, Ethereum, Monero \end{tabular}                            & \begin{tabular}[t]{@{}l@{}}From: 2010/09/10\\To: 2017/05/31 (Bitcoin)\\Others can reference the paper\end{tabular}                                                                 & \begin{tabular}[t]{@{}l@{}}Waveletcoherence analysis of price \end{tabular}               
& BraveNewCoin\\ \\
Kang et al.~\citep{kang2019whose}              &\begin{tabular}[t]{@{}l@{}}Market data and~posts and comments\\ including metadata  \end{tabular}                                                             & \begin{tabular}[t]{@{}l@{}}Bitcoin \end{tabular}                                             & \begin{tabular}[t]{@{}l@{}}From: 2009/11/22\\To: 2018/02/02\end{tabular}                                                                                                           & \begin{tabular}[t]{@{}l@{}}Relationships Between Bitcoin \\ Prices and User Groups in \\Online Community\end{tabular} 
& Bitcoin forum\\
\bottomrule
\end{tabular}}
\end{turn}
\label{tbl:dst2}
\end{table*}


\begin{table*} 
\caption{Datasets (3/3): Others}
\resizebox{\textwidth}{!}{
\begin{tabular}{@{}lllll@{}}
\toprule
\textbf{Research Source~} & \textbf{Description}                                                                                                                                                 & \textbf{Time range}                                                        & \textbf{Usage} &
\textbf{Data Sources}
\\ \midrule
Kurbucz~\citep{kurbucz2019predicting,Kondor_2014}                   & \begin{tabular}[t]{@{}l@{}}Raw and preprocessed data of all \\Bitcoin transactions and daily returns\end{tabular}                                                    & \begin{tabular}[t]{@{}l@{}}From: 2016/11/09\\To: 2018/02/05\end{tabular}   & \begin{tabular}[t]{@{}l@{}}Predicting the price of Bitcoin\\with~transaction network\end{tabular} 
& Bitcoin network dataset~\citep{senseable}\\ \\
Bedi et al.~\citep{bedi2020investment}               & \begin{tabular}[t]{@{}l@{}}A diversified portfolio including~equity,\\fixedincome, real estate, alternative \\investments, commodities and money market\end{tabular} & \begin{tabular}[t]{@{}l@{}}From:~July 2010\\To:~December 2018\end{tabular} & \begin{tabular}[t]{@{}l@{}}Cross-currency including cryptocurrency\\researching portfolios\end{tabular}
& Portfolio sources~\citep{bedi2020investment}\\
\bottomrule
\end{tabular}}
\label{tbl:dst3}
\end{table*}

\section{Opportunities in Cryptocurrency Trading}\label{opportunitiessec}
This section discusses potential opportunities for future research in cryptocurrency trading. 

\textbf{Sentiment-based research}. 
As discussed above, there is a substantial body of work, which uses natural language processing technology, for sentiment analysis with the ultimate goal of using news and media contents to improve the performance of cryptocurrency trading strategies. 

Possible research directions may lie in \ssfan{a} larger volume of media input (e.g., adding video sources) in sentiment analysis; updating baseline natural language processing model to perform more robust text preprocessing; applying neural networks in label training; extending samples in terms of holding period; transaction-fees; \revfanthree{opinion dynamics~\citep{zha2020opinion}} and, user reputation research.

\textbf{Long-and-short term trading research}.
There are significant differences between long and short time horizons in cryptocurrency trading. 
In long-term trading, investors might obtain greater profits but have more possibilities to control risk when managing a position for weeks or months. \sfan{It is mandatory to control for risk on long term strategies due to the increase in \ssfan{the} holding period, directly proportional to the risk incurred by the trader.} \fan{On the other hand, the longer the horizon, the higher the risk and the most important the risk control. The shorter the horizon, the higher the cost and the lower the risk, so cost takes over the design of a strategy.} In short-term trading, automated algorithmic trading can be applied when holding periods are less than a week. 
Researchers can differentiate between long-term and short-term trading in cryptocurrency trading by applying wavelet technology analysing bubble regimes~\citep{phillips2018cryptocurrency} and considering price explosiveness~\citep{bouri2019co} hypotheses for short-term and long-term research. 

The existing work is mainly about showing the differences between long and short-term cryptocurrency trading. Long-term trading means less time would cost in trend tracing and simple technical indicators in market analysis. Short-term trading can limit overall risk because small positions are used in every \ssfan{transaction}. But market noise (interference) and short transaction time might cause some stress in short term trading.
It might also be interesting to explore \ssfan{the} extraction of trading signals, time series research, application to portfolio management, \ssfan{the} relationship between \ssfan{a} huge market crash and small price drop, derivative pricing in cryptocurrency market, etc.

\textbf{Correlation between cryptocurrency and others}.
By the effects of monetary policy and business cycles 
that are not controlled by the central bank, cryptocurrency is always negatively correlated with overall financial market trends. There have been some studies discussing correlations between cryptocurrencies and other financial markets~\citep{kang2019co,castro2019crypto}, which can be used to predict the direction of the cryptocurrency market.

Considering the characteristics of cryptocurrency, \ssfan{the} correlation between cryptocurrency and other assets still requires further research. Possible breakthroughs might be achieved with principal component analysis, \ssfan{the} relationship between cryptocurrency and other currencies in extreme conditions (i.e., financial collapse). 

\textbf{Bubbles and crash research}.
To discuss the high volatility and return of cryptocurrencies, current research has focussed on bubbles of cryptocurrency markets~\citep{cheung2015crypto}, correlation between cryptocurrency bubbles and indicators like volatility index (VIX)~\citep{enoksen2019can} (which is a ``panic index'' to measure the implied volatility of S\&P500 Index Options), spillover effects in cryptocurrency market~\citep{luu2019spillover}.

Additional research for bubbles and crashes in cryptocurrency trading could include a connection between the process of bubble generation and financial collapse and conducting a coherent analysis (coherent process analysis from the formation of bubbles to aftermath analysis of bubble burst), analysis of bubble theory \fan{by Microeconomics}, trying other physical or industrial models in analysing bubbles in cryptocurrency market (i.e., \emph{Omori law}~\citep{weber2007relation}), discussing the supply and demand relationship of cryptocurrency in bubble analysis (like using supply and demand graph to simulate the generation of bubbles and \fan{simulate the bubble burst}).

\textbf{Game theory and agent-based analysis}.
Applying game theory or agent-based modelling in trading is a hot research direction in \ssfan{the} traditional financial market. It might also be interesting to apply this method to trading in cryptocurrency markets. 

\sfan{
\textbf{Public nature of Blockchain technology}.
Investigations on the connections between the formation of a given currency's transaction network and its price has increased rapidly in recent years; the growing attention on user identification ~\citep{juhasz2018bayesian} also strongly supports this direction. With an in-depth understanding of these networks, we may identify new features in price prediction and may be closer to understanding financial bubbles in cryptocurrency trading.
}

\sfan{
\textbf{Balance between the opening of trading research literature and the fading of alphas}.
Mclean et al.~\citep{mclean2016does} pointed \ssfan{out} that investors learn about mispricing in stock markets from academic publications. Similarly, cryptocurrency market predictability could also be affected by research papers in the area. A possible attempt is to try new pricing methods applying real-time market changes. Considering the proportion of informed traders increasing in \ssfan{the} cryptocurrency market in \ssfan{the} pricing process is another breaking point (looking for \ssfan{a} balance between alpha trading and trading research literature). 
}

\section{Conclusions}
We provided a comprehensive overview and analysis of \ssfan{the} research work \ssfan{on} cryptocurrency trading. 
This survey presented a nomenclature of the definitions and current state of the art. 
The paper provides a comprehensive survey of 146 cryptocurrency trading papers and analyses the research distribution that characterise the cryptocurrency trading literature. \revfan{Research distribution among properties and categories/technologies are analysed in this survey respectively.} We further summarised the datasets used for experiments and analysed the research trends and opportunities in cryptocurrency trading. \revfan{Future research directions and opportunities are discussed in Section ~\ref{opportunitiessec}.}

We expect this survey to be beneficial to academics (e.g., finance researchers) and quantitative traders alike. The survey represents a quick way to get familiar with the literature on cryptocurrency trading and can motivate more researchers to contribute to the pressing problems in the area, for example along the lines we have identified.


\begin{backmatter}


\bibliographystyle{bmc-mathphys} 
\bibliography{bmc_article}      


\begin{thebibliography}{311}
\ifx \bisbn   \undefined \def \bisbn  #1{ISBN #1}\fi
\ifx \binits  \undefined \def \binits#1{#1}\fi
\ifx \bauthor  \undefined \def \bauthor#1{#1}\fi
\ifx \batitle  \undefined \def \batitle#1{#1}\fi
\ifx \bjtitle  \undefined \def \bjtitle#1{#1}\fi
\ifx \bvolume  \undefined \def \bvolume#1{\textbf{#1}}\fi
\ifx \byear  \undefined \def \byear#1{#1}\fi
\ifx \bissue  \undefined \def \bissue#1{#1}\fi
\ifx \bfpage  \undefined \def \bfpage#1{#1}\fi
\ifx \blpage  \undefined \def \blpage #1{#1}\fi
\ifx \burl  \undefined \def \burl#1{\textsf{#1}}\fi
\ifx \doiurl  \undefined \def \doiurl#1{\textsf{#1}}\fi
\ifx \betal  \undefined \def \betal{\textit{et al.}}\fi
\ifx \binstitute  \undefined \def \binstitute#1{#1}\fi
\ifx \binstitutionaled  \undefined \def \binstitutionaled#1{#1}\fi
\ifx \bctitle  \undefined \def \bctitle#1{#1}\fi
\ifx \beditor  \undefined \def \beditor#1{#1}\fi
\ifx \bpublisher  \undefined \def \bpublisher#1{#1}\fi
\ifx \bbtitle  \undefined \def \bbtitle#1{#1}\fi
\ifx \bedition  \undefined \def \bedition#1{#1}\fi
\ifx \bseriesno  \undefined \def \bseriesno#1{#1}\fi
\ifx \blocation  \undefined \def \blocation#1{#1}\fi
\ifx \bsertitle  \undefined \def \bsertitle#1{#1}\fi
\ifx \bsnm \undefined \def \bsnm#1{#1}\fi
\ifx \bsuffix \undefined \def \bsuffix#1{#1}\fi
\ifx \bparticle \undefined \def \bparticle#1{#1}\fi
\ifx \barticle \undefined \def \barticle#1{#1}\fi
\ifx \bconfdate \undefined \def \bconfdate #1{#1}\fi
\ifx \botherref \undefined \def \botherref #1{#1}\fi
\ifx \url \undefined \def \url#1{\textsf{#1}}\fi
\ifx \bchapter \undefined \def \bchapter#1{#1}\fi
\ifx \bbook \undefined \def \bbook#1{#1}\fi
\ifx \bcomment \undefined \def \bcomment#1{#1}\fi
\ifx \oauthor \undefined \def \oauthor#1{#1}\fi
\ifx \citeauthoryear \undefined \def \citeauthoryear#1{#1}\fi
\ifx \endbibitem  \undefined \def \endbibitem {}\fi
\ifx \bconflocation  \undefined \def \bconflocation#1{#1}\fi
\ifx \arxivurl  \undefined \def \arxivurl#1{\textsf{#1}}\fi
\csname PreBibitemsHook\endcsname

\bibitem{farell2015analysis}
\begin{botherref}
\oauthor{\bsnm{Farell}, \binits{R.}}:
An analysis of the cryptocurrency industry
(2015)
\end{botherref}
\endbibitem

\bibitem{flood1986evaluation}
\begin{botherref}
\oauthor{\bsnm{Flood}, \binits{R.}},
\oauthor{\bsnm{Hodrick}, \binits{R.J.}},
\oauthor{\bsnm{Kaplan}, \binits{P.}}:
An evaluation of recent evidence on stock market bubbles.
National Bureau of Economic Research Cambridge, Mass., USA
(1986)
\end{botherref}
\endbibitem

\bibitem{kyriazis2019survey}
\begin{barticle}
\bauthor{\bsnm{Kyriazis}, \binits{N.A.}}:
\batitle{A survey on efficiency and profitable trading opportunities in
  cryptocurrency markets}.
\bjtitle{Journal of Risk and Financial Management}
\bvolume{12}(\bissue{2}),
\bfpage{67}
(\byear{2019})
\end{barticle}
\endbibitem

\bibitem{ahamad2013survey}
\begin{bchapter}
\bauthor{\bsnm{Ahamad}, \binits{S.}},
\bauthor{\bsnm{Nair}, \binits{M.}},
\bauthor{\bsnm{Varghese}, \binits{B.}}:
\bctitle{A survey on crypto currencies}.
In: \bbtitle{4th International Conference on Advances in Computer Science,
  AETACS},
pp. \bfpage{42}--\blpage{48}
(\byear{2013}).
\bcomment{Citeseer}
\end{bchapter}
\endbibitem

\bibitem{sharma2017survey}
\begin{barticle}
\bauthor{\bsnm{Sharma}, \binits{S.}},
\bauthor{\bsnm{Krishma}, \binits{N.}},
\bauthor{\bsnm{Raina}, \binits{E.}}:
\batitle{Survey paper on cryptocurrency}.
\bjtitle{International Journal of Scientific Research in Computer Science,
  Engineering and Information Technology Vol. 2 Issue}
\bvolume{3},
\bfpage{307}--\blpage{310}
(\byear{2017})
\end{barticle}
\endbibitem

\bibitem{mukhopadhyay2016brief}
\begin{bchapter}
\bauthor{\bsnm{Mukhopadhyay}, \binits{U.}},
\bauthor{\bsnm{Skjellum}, \binits{A.}},
\bauthor{\bsnm{Hambolu}, \binits{O.}},
\bauthor{\bsnm{Oakley}, \binits{J.}},
\bauthor{\bsnm{Yu}, \binits{L.}},
\bauthor{\bsnm{Brooks}, \binits{R.}}:
\bctitle{A brief survey of cryptocurrency systems}.
In: \bbtitle{2016 14th Annual Conference on Privacy, Security and Trust (PST)},
pp. \bfpage{745}--\blpage{752}
(\byear{2016}).
\bcomment{IEEE}
\end{bchapter}
\endbibitem

\bibitem{merediz2019bibliometric}
\begin{barticle}
\bauthor{\bsnm{Merediz-Sol{\`a}}, \binits{I.}},
\bauthor{\bsnm{Bariviera}, \binits{A.F.}}:
\batitle{A bibliometric analysis of bitcoin scientific production}.
\bjtitle{Research in International Business and Finance}
\bvolume{50},
\bfpage{294}--\blpage{305}
(\byear{2019})
\end{barticle}
\endbibitem

\bibitem{tapscott2016blockchain}
\begin{bbook}
\bauthor{\bsnm{Tapscott}, \binits{D.}},
\bauthor{\bsnm{Tapscott}, \binits{A.}}:
\bbtitle{Blockchain Revolution: How the Technology Behind Bitcoin Is Changing
  Money, Business, and the World}.
\bpublisher{Penguin}, \blocation{???}
(\byear{2016})
\end{bbook}
\endbibitem

\bibitem{blockchainblog}
\begin{botherref}
\oauthor{\bsnm{S.}, \binits{J.}}:
How do blockchain mining and transactions work explained in 7 simple steps.
\url{https://blog.goodaudience.com/how-a-miner-adds-transactions-to-the-blockchain-in-seven-steps-856053271476}.
[Online, Accessed: January 26, 2020]
(2018)
\end{botherref}
\endbibitem

\bibitem{doran2014forensic}
\begin{botherref}
\oauthor{\bsnm{Doran}, \binits{M.D.}}:
A forensic look at bitcoin cryptocurrency.
PhD thesis,
Utica College
(2014)
\end{botherref}
\endbibitem

\bibitem{meunier2018blockchain}
\begin{bchapter}
\bauthor{\bsnm{Meunier}, \binits{S.}}:
\bctitle{Blockchain 101: What is blockchain and how does this revolutionary
  technology work?}
In: \bbtitle{Transforming Climate Finance and Green Investment with
  Blockchains},
pp. \bfpage{23}--\blpage{34}.
\bpublisher{Elsevier}, \blocation{???}
(\byear{2018})
\end{bchapter}
\endbibitem

\bibitem{narayanan2016bitcoin}
\begin{bbook}
\bauthor{\bsnm{Narayanan}, \binits{A.}},
\bauthor{\bsnm{Bonneau}, \binits{J.}},
\bauthor{\bsnm{Felten}, \binits{E.}},
\bauthor{\bsnm{Miller}, \binits{A.}},
\bauthor{\bsnm{Goldfeder}, \binits{S.}}:
\bbtitle{Bitcoin and Cryptocurrency Technologies: a Comprehensive
  Introduction}.
\bpublisher{Princeton University Press}, \blocation{???}
(\byear{2016})
\end{bbook}
\endbibitem

\bibitem{wang2017designated}
\begin{botherref}
\oauthor{\bsnm{Wang}, \binits{H.}},
\oauthor{\bsnm{He}, \binits{D.}},
\oauthor{\bsnm{Ji}, \binits{Y.}}:
Designated-verifier proof of assets for bitcoin exchange using elliptic curve
  cryptography.
Future Generation Computer Systems
(2017)
\end{botherref}
\endbibitem

\bibitem{elliptic}
\begin{botherref}
\oauthor{\bsnm{Grayblock}}:
Elliptic-Curve Cryptography.
\url{https://medium.com/coinmonks/elliptic-curve-cryptography-6de8fc748b8b}.
[Online, Accessed December 29, 2019]
(2018)
\end{botherref}
\endbibitem

\bibitem{harwick2016cryptocurrency}
\begin{barticle}
\bauthor{\bsnm{Harwick}, \binits{C.}}:
\batitle{Cryptocurrency and the problem of intermediation}.
\bjtitle{The Independent Review}
\bvolume{20}(\bissue{4}),
\bfpage{569}--\blpage{588}
(\byear{2016})
\end{barticle}
\endbibitem

\bibitem{rose2015evolution}
\begin{barticle}
\bauthor{\bsnm{Rose}, \binits{C.}}:
\batitle{The evolution of digital currencies: Bitcoin, a cryptocurrency causing
  a monetary revolution}.
\bjtitle{International Business \& Economics Research Journal (IBER)}
\bvolume{14}(\bissue{4}),
\bfpage{617}--\blpage{622}
(\byear{2015})
\end{barticle}
\endbibitem

\bibitem{kaal2020digital}
\begin{botherref}
\oauthor{\bsnm{Kaal}, \binits{W.A.}}:
Digital asset market evolution.
Journal of Corporation Law,
20--02
(2020)
\end{botherref}
\endbibitem

\bibitem{cryptorank}
\begin{botherref}
\oauthor{\bsnm{CoinMaketCap}}:
Top 100 cryptocurrencies by market capitalization
(2019).
Accessed Accessed December 20, 2019
\end{botherref}
\endbibitem

\bibitem{cryptohistory}
\begin{botherref}
\oauthor{\bsnm{TradingView}}:
Total Crypto Market Capitalization and Volume.
\url{https://www.tradingview.com/markets/cryptocurrencies/global-charts/}.
[Online, Accessed September 10, 2021]
(2021)
\end{botherref}
\endbibitem

\bibitem{mainstream}
\begin{botherref}
\oauthor{\bsnm{Council}, \binits{F.B.}}:
The Main Roadblocks To Crypto Moving Mainstream.
\url{https://www.forbes.com/sites/forbesbusinesscouncil/2021/06/23/the-main-roadblocks-to-crypto-moving-mainstream/?sh=2e629de922b9}.
[Online, Accessed: June 23, 2021]
(2021)
\end{botherref}
\endbibitem

\bibitem{nakano2018bitcoin}
\begin{barticle}
\bauthor{\bsnm{Nakano}, \binits{M.}},
\bauthor{\bsnm{Takahashi}, \binits{A.}},
\bauthor{\bsnm{Takahashi}, \binits{S.}}:
\batitle{Bitcoin technical trading with artificial neural network}.
\bjtitle{Physica A: Statistical Mechanics and its Applications}
\bvolume{510},
\bfpage{587}--\blpage{609}
(\byear{2018})
\end{barticle}
\endbibitem

\bibitem{crymarketper}
\begin{botherref}
\oauthor{\bsnm{Coinmarketcap}}:
Percentage of Total Market Capitalization.
\url{https://coinmarketcap.com/charts/\#dominance-percentage}.
[Online, Accessed January 11, 2020]
(2020)
\end{botherref}
\endbibitem

\bibitem{exchanges}
\begin{botherref}
\oauthor{\bsnm{Nomics}}:
Top Cryptocurrency Exchanges List.
\url{https://nomics.com/exchanges}.
[Online, Accessed: January 11, 2020]
(2020)
\end{botherref}
\endbibitem

\bibitem{CMEcrypto}
\begin{botherref}
\oauthor{\bsnm{CME}}:
CME Cryptocurrency products.
\url{https://www.cmegroup.com/trading/cryptocurrency-indices.html}.
[Online, Accessed: February 11, 2020]
(2020)
\end{botherref}
\endbibitem

\bibitem{CMEcountry}
\begin{botherref}
\oauthor{\bsnm{CME}}:
CME groups overview.
\url{https://www.cmegroup.com/company/history/}.
[Online, Accessed: February 11, 2020]
(2020)
\end{botherref}
\endbibitem

\bibitem{CMEregu}
\begin{botherref}
\oauthor{\bsnm{CME}}:
CME Group Rules and Regulation Overview.
\url{https://www.cmegroup.com/education/courses/market-regulation/overview/cme-group-rules-and-regulation-overview.html}.
[Online, Accessed February 11, 2020]
(2020)
\end{botherref}
\endbibitem

\bibitem{CBOEcrypto}
\begin{botherref}
\oauthor{\bsnm{CBOE}}:
CBOE products.
\url{https://www.cboe.com}.
[Online, Accessed: February 11, 2020]
(2020)
\end{botherref}
\endbibitem

\bibitem{CBOEcountry}
\begin{botherref}
\oauthor{\bsnm{CBOE}}:
CBOE history.
\url{http://www.cboe.com/aboutcboe/history}.
[Online, Accessed: February 11, 2020]
(2020)
\end{botherref}
\endbibitem

\bibitem{CBOEregu}
\begin{botherref}
\oauthor{\bsnm{CBOE}}:
CFE Regulation.
\url{https://www.cboe.com/aboutcboe/legal-regulatory/departmental-overviews/cfe-regulation}.
[Online, Accessed February 11, 2020]
(2020)
\end{botherref}
\endbibitem

\bibitem{BAKKTregu}
\begin{botherref}
\oauthor{\bsnm{BAKKT}}:
BAKKT markets.
\url{https://www.bakkt.com/index}.
[Online, Accessed February 11, 2020]
(2020)
\end{botherref}
\endbibitem

\bibitem{BAKKTcountry}
\begin{botherref}
\oauthor{\bsnm{BAKKT}}:
BAKKT terms of use.
\url{https://www.bakkt.com/terms-of-use}.
[Online, Accessed: February 11, 2020]
(2020)
\end{botherref}
\endbibitem

\bibitem{Bitmexcrypto}
\begin{botherref}
\oauthor{\bsnm{Bitmex}}:
Beginner’s Guide to BitMEX: Complete Review.
\url{https://blockonomi.com/bitmex-review/}.
[Online, Accessed: February 11, 2020]
(2020)
\end{botherref}
\endbibitem

\bibitem{Bitmexcountry}
\begin{botherref}
\oauthor{\bsnm{Bitmex}}:
Bitmex.
\url{https://www.bitmex.com/register}.
[Online, Accessed: February 11, 2020]
(2020)
\end{botherref}
\endbibitem

\bibitem{Binancecrypto}
\begin{botherref}
\oauthor{\bsnm{Binance}}:
Binance Review 2020: Pros, Cons, Fees, Features, and Safety.
\url{https://insidebitcoins.com/cryptocurrency-exchanges/binance-review/}.
[Online, Accessed: February 11, 2020]
(2020)
\end{botherref}
\endbibitem

\bibitem{Binancecountry}
\begin{botherref}
\oauthor{\bsnm{Maltatoday}}:
Why world leader crypto exchange Binance moved to Malta.
\url{https://www.maltatoday.com.mt/business/business_news/93170/why_world_leader_crypto_exchange_binance_moved_to_malta\#.XlKZ8Gj7Q2x}.
[Online, Accessed: February 11, 2020]
(2020)
\end{botherref}
\endbibitem

\bibitem{Binanceregu}
\begin{botherref}
\oauthor{\bsnm{Binance}}:
Binance Partners with Coinfirm to Protect the Global Cryptocurrency Economy and
  Ensure Compliance with FATF AML Rules.
\url{https://www.binance.com/en/blog/386484403820867584/Binance-Partners-with-Coinfirm-to-Protect-the-Global-Cryptocurrency-Economy-and-Ensure-Compliance-with-FATF-AML-Rules}.
[Online, Accessed February 11, 2020]
(2020)
\end{botherref}
\endbibitem

\bibitem{Coinbasecrypto}
\begin{botherref}
\oauthor{\bsnm{Coinbase}}:
Coinbase Supported cryptocurrencies.
\url{https://help.coinbase.com/en/coinbase/getting-started/general-crypto-education/supported-cryptocurrencies.html}.
[Online, Accessed: February 11, 2020]
(2020)
\end{botherref}
\endbibitem

\bibitem{Coinbasecountry}
\begin{botherref}
\oauthor{\bsnm{Bloomberg}}:
Coinbase Inc.
\url{https://www.bloomberg.com/profile/company/0776164D:US}.
[Online, Accessed: February 11, 2020]
(2020)
\end{botherref}
\endbibitem

\bibitem{Coinbaseregu}
\begin{botherref}
\oauthor{\bsnm{Coinbase}}:
Our path to listing SEC-regulated crypto securities.
\url{https://blog.coinbase.com/our-path-to-listing-sec-regulated-crypto-securities-a1724e13bb5a}.
[Online, Accessed February 11, 2020]
(2020)
\end{botherref}
\endbibitem

\bibitem{Bitfinexcrypto}
\begin{botherref}
\oauthor{\bsnm{Bitfinex}}:
Bitfinex markets.
\url{https://www.bitfinex.com/}.
[Online, Accessed: February 11, 2020]
(2020)
\end{botherref}
\endbibitem

\bibitem{Bitfinexcountry}
\begin{botherref}
\oauthor{\bsnm{Bitfinex}}:
Bitfinex terms of service.
\url{https://www.bitfinex.com/legal/terms}.
[Online, Accessed: February 11, 2020]
(2020)
\end{botherref}
\endbibitem

\bibitem{Bitfinexregu}
\begin{botherref}
\oauthor{\bsnm{Bitfinex}}:
New York Court Rules That State Attorney Has Jurisdiction Over Bitfinex.
\url{https://cointelegraph.com/news/new-york-court-rules-that-state-attorney-has-jurisdiction-over-bitfinex}.
[Online, Accessed February 11, 2020]
(2020)
\end{botherref}
\endbibitem

\bibitem{Bitstampcrypto}
\begin{botherref}
\oauthor{\bsnm{Bitstamp}}:
Bitstamp Review 2020.
\url{https://www.fxempire.com/crypto/exchange/bitstamp/review}.
[Online, Accessed: February 11, 2020]
(2020)
\end{botherref}
\endbibitem

\bibitem{Bitstampcountry}
\begin{botherref}
\oauthor{\bsnm{Bitstamp}}:
Bitstamp who we are.
\url{https://www.bitstamp.net/about-us/}.
[Online, Accessed: February 11, 2020]
(2020)
\end{botherref}
\endbibitem

\bibitem{Bitstampregu}
\begin{botherref}
\oauthor{\bsnm{Bitstamp}}:
Terms of Use.
\url{https://www.bitstamp.net/terms-of-use/sa}.
[Online, Accessed February 11, 2020]
(2020)
\end{botherref}
\endbibitem

\bibitem{Poloniexcrypto}
\begin{botherref}
\oauthor{\bsnm{Poloniex}}:
Poloniex markets.
\url{https://poloniex.com/}.
[Online, Accessed: February 11, 2020]
(2020)
\end{botherref}
\endbibitem

\bibitem{CDFreference}
\begin{botherref}
\oauthor{\bsnm{Authority}, \binits{F.C.}}:
Contract for differences.
\url{https://www.fca.org.uk/firms/contracts-for-difference},
  \url{https://www:fca:org:uk/rms/contracts-for-dierence}.
[Online, Accessed January 29, 2020]
(2019)
\end{botherref}
\endbibitem

\bibitem{cryptoKYC}
\begin{botherref}
\oauthor{\bsnm{Adeyanju}, \binits{C.}}:
What crypto exchanges do to comply with KYC, AML and CFT regulations.
\url{https://cointelegraph.com/news/what-crypto-exchanges-do-to-comply-with-kyc-aml-and-cft-regulations}.
[Online, Accessed January 11, 2020]
(2019)
\end{botherref}
\endbibitem

\bibitem{localbtc}
\begin{botherref}
\oauthor{\bsnm{Localbtc}}:
Localbitcoins purchasing online.
\url{https://localbitcoins.com}.
[Online, Accessed: January 11, 2020]
(2020)
\end{botherref}
\endbibitem

\bibitem{priceplunge}
\begin{botherref}
\oauthor{\bsnm{Forbes}}:
Here’s What Caused Bitcoin’s ‘Extreme’ Price Plunge.
\url{https://www.forbes.com/sites/billybambrough/2020/03/19/major-bitcoin-exchange-bitmex-has-a-serious-problem/?sh=1be57a0d4f7d}.
[Online, Accessed: March 19, 2020]
(2021)
\end{botherref}
\endbibitem

\bibitem{cryptostolen}
\begin{botherref}
\oauthor{\bsnm{Forbes}}:
More Than \$600 Million Stolen In Ethereum And Other Cryptocurrencies—Marking
  One Of Crypto’s Biggest Hacks Ever.
\url{https://www.forbes.com/sites/jonathanponciano/2021/08/10/more-than-600-million-stolen-in-ethereum-and-other-cryptocurrencies-marking-one-of-cryptos-biggest-hacks-ever/?sh=502ce7387f62}.
[Online, Accessed: August 10, 2021]
(2021)
\end{botherref}
\endbibitem

\bibitem{kou2021fintech}
\begin{barticle}
\bauthor{\bsnm{Kou}, \binits{G.}},
\bauthor{\bsnm{Akdeniz}, \binits{{\"O}.O.}},
\bauthor{\bsnm{Din{\c{c}}er}, \binits{H.}},
\bauthor{\bsnm{Y{\"u}ksel}, \binits{S.}}:
\batitle{Fintech investments in european banks: a hybrid it2 fuzzy
  multidimensional decision-making approach}.
\bjtitle{Financial Innovation}
\bvolume{7}(\bissue{1}),
\bfpage{1}--\blpage{28}
(\byear{2021})
\end{barticle}
\endbibitem

\bibitem{cryptoregul}
\begin{botherref}
\oauthor{\bsnm{UKTN}}:
Bitcoin and the Challenges for Financial Regulation.
\url{https://www.uktech.news}.
[Online, Accessed: February 24, 2021]
(2021)
\end{botherref}
\endbibitem

\bibitem{buffett}
\begin{botherref}
\oauthor{\bsnm{Forbes}}:
Why Buffett Sees Bitcoin Bubble.
\url{https://www.forbes.com/sites/johnwasik/2017/11/06/why-buffett-sees-bitcoin-bubble/?sh=196c2a8062a8}.
[Online, Accessed: November 6, 2017]
(2017)
\end{botherref}
\endbibitem

\bibitem{oberlechner2001importance}
\begin{barticle}
\bauthor{\bsnm{Oberlechner}, \binits{T.}}:
\batitle{Importance of technical and fundamental analysis in the european
  foreign exchange market}.
\bjtitle{International Journal of Finance \& Economics}
\bvolume{6}(\bissue{1}),
\bfpage{81}--\blpage{93}
(\byear{2001})
\end{barticle}
\endbibitem

\bibitem{nti2020systematic}
\begin{barticle}
\bauthor{\bsnm{Nti}, \binits{I.K.}},
\bauthor{\bsnm{Adekoya}, \binits{A.F.}},
\bauthor{\bsnm{Weyori}, \binits{B.A.}}:
\batitle{A systematic review of fundamental and technical analysis of stock
  market predictions}.
\bjtitle{Artificial Intelligence Review}
\bvolume{53}(\bissue{4}),
\bfpage{3007}--\blpage{3057}
(\byear{2020})
\end{barticle}
\endbibitem

\bibitem{calo2002global}
\begin{botherref}
\oauthor{\bsnm{Calo}, \binits{B.}},
\oauthor{\bsnm{Johnson}, \binits{W.}}:
Global trading system.
Google Patents.
US Patent App. 09/769,036
(2002)
\end{botherref}
\endbibitem

\bibitem{bauriya2019real}
\begin{botherref}
\oauthor{\bsnm{Bauriya}, \binits{A.}},
\oauthor{\bsnm{Tikone}, \binits{A.}},
\oauthor{\bsnm{Nandgaonkar}, \binits{P.}},
\oauthor{\bsnm{Sakure}, \binits{K.S.}}:
Real-time cryptocurrency trading system.
International Research Journal of Engineering and Technology (IRJET)
\textbf{06}
(2019)
\end{botherref}
\endbibitem

\bibitem{Julian2019}
\begin{botherref}
\oauthor{\bsnm{Molina}, \binits{J.}}:
Develop your Crypto-Trading System Using Plain Logic, Part 1
(2019).
\url{https://medium.com/swlh/develop-your-crypto-trading-system-using-plain-logic-part-1-caac02f0a37d}
Accessed Sep 20, 2019
\end{botherref}
\endbibitem

\bibitem{gerritsen2019profitability}
\begin{botherref}
\oauthor{\bsnm{Gerritsen}, \binits{D.F.}},
\oauthor{\bsnm{Bouri}, \binits{E.}},
\oauthor{\bsnm{Ramezanifar}, \binits{E.}},
\oauthor{\bsnm{Roubaud}, \binits{D.}}:
The profitability of technical trading rules in the bitcoin market.
Finance Research Letters
(2019)
\end{botherref}
\endbibitem

\bibitem{elliott2005pairs}
\begin{barticle}
\bauthor{\bsnm{Elliott}, \binits{R.J.}},
\bauthor{\bsnm{Van Der~Hoek*}, \binits{J.}},
\bauthor{\bsnm{Malcolm}, \binits{W.P.}}:
\batitle{Pairs trading}.
\bjtitle{Quantitative Finance}
\bvolume{5}(\bissue{3}),
\bfpage{271}--\blpage{276}
(\byear{2005})
\end{barticle}
\endbibitem

\bibitem{vogelvang2005econometrics}
\begin{bbook}
\bauthor{\bsnm{Vogelvang}, \binits{B.}}:
\bbtitle{Econometrics: Theory and Applications with Eviews}.
\bpublisher{Pearson Education}, \blocation{???}
(\byear{2005})
\end{bbook}
\endbibitem

\bibitem{kaufman2013trading}
\begin{bbook}
\bauthor{\bsnm{Kaufman}, \binits{P.J.}}:
\bbtitle{Trading Systems and Methods,+ Website}
vol. \bseriesno{591}.
\bpublisher{John Wiley \& Sons}, \blocation{???}
(\byear{2013})
\end{bbook}
\endbibitem

\bibitem{ben2002hybrid}
\begin{barticle}
\bauthor{\bsnm{Ben-Akiva}, \binits{M.}},
\bauthor{\bsnm{McFadden}, \binits{D.}},
\bauthor{\bsnm{Train}, \binits{K.}},
\bauthor{\bsnm{Walker}, \binits{J.}},
\bauthor{\bsnm{Bhat}, \binits{C.}},
\bauthor{\bsnm{Bierlaire}, \binits{M.}},
\bauthor{\bsnm{Bolduc}, \binits{D.}},
\bauthor{\bsnm{Boersch-Supan}, \binits{A.}},
\bauthor{\bsnm{Brownstone}, \binits{D.}},
\bauthor{\bsnm{Bunch}, \binits{D.S.}}, \betal:
\batitle{Hybrid choice models: progress and challenges}.
\bjtitle{Marketing Letters}
\bvolume{13}(\bissue{3}),
\bfpage{163}--\blpage{175}
(\byear{2002})
\end{barticle}
\endbibitem

\bibitem{chang2015sophistication}
\begin{barticle}
\bauthor{\bsnm{Chang}, \binits{C.-C.}},
\bauthor{\bsnm{Hsieh}, \binits{P.-F.}},
\bauthor{\bsnm{Wang}, \binits{Y.-H.}}:
\batitle{Sophistication, sentiment, and misreaction}.
\bjtitle{Journal of Financial and Quantitative Analysis}
\bvolume{50}(\bissue{4}),
\bfpage{903}--\blpage{928}
(\byear{2015})
\end{barticle}
\endbibitem

\bibitem{kat1994volatility}
\begin{botherref}
\oauthor{\bsnm{Kat}, \binits{H.M.}},
\oauthor{\bsnm{Heynen}, \binits{R.C.}}:
Volatility prediction: A comparison of the stochastic volatility, garch (1, 1)
  and egarch (1, 1) models.
Journal of Derivatives
\textbf{2}(2)
(1994)
\end{botherref}
\endbibitem

\bibitem{engle1995multivariate}
\begin{barticle}
\bauthor{\bsnm{Engle}, \binits{R.F.}},
\bauthor{\bsnm{Kroner}, \binits{K.F.}}:
\batitle{Multivariate simultaneous generalized arch}.
\bjtitle{Econometric theory}
\bvolume{11}(\bissue{1}),
\bfpage{122}--\blpage{150}
(\byear{1995})
\end{barticle}
\endbibitem

\bibitem{caporin2012we}
\begin{barticle}
\bauthor{\bsnm{Caporin}, \binits{M.}},
\bauthor{\bsnm{McAleer}, \binits{M.}}:
\batitle{Do we really need both bekk and dcc? a tale of two multivariate garch
  models}.
\bjtitle{Journal of Economic Surveys}
\bvolume{26}(\bissue{4}),
\bfpage{736}--\blpage{751}
(\byear{2012})
\end{barticle}
\endbibitem

\bibitem{neter1996applied}
\begin{bbook}
\bauthor{\bsnm{Neter}, \binits{J.}},
\bauthor{\bsnm{Kutner}, \binits{M.H.}},
\bauthor{\bsnm{Nachtsheim}, \binits{C.J.}},
\bauthor{\bsnm{Wasserman}, \binits{W.}}:
\bbtitle{Applied Linear Statistical Models}
vol. \bseriesno{4}.
\bpublisher{Irwin Chicago}, \blocation{???}
(\byear{1996})
\end{bbook}
\endbibitem

\bibitem{choi2012arma}
\begin{bbook}
\bauthor{\bsnm{Choi}, \binits{B.}}:
\bbtitle{ARMA Model Identification}.
\bpublisher{Springer}, \blocation{???}
(\byear{2012})
\end{bbook}
\endbibitem

\bibitem{mcnally2018predicting}
\begin{bchapter}
\bauthor{\bsnm{McNally}, \binits{S.}},
\bauthor{\bsnm{Roche}, \binits{J.}},
\bauthor{\bsnm{Caton}, \binits{S.}}:
\bctitle{Predicting the price of bitcoin using machine learning}.
In: \bbtitle{2018 26th Euromicro International Conference on Parallel,
  Distributed and Network-based Processing (PDP)},
pp. \bfpage{339}--\blpage{343}
(\byear{2018}).
\bcomment{IEEE}
\end{bchapter}
\endbibitem

\bibitem{machinediffe}
\begin{botherref}
\oauthor{\bsnm{IntelliPaat}}:
Supervised Learning vs Unsupervised Learning vs Reinforcement Learning.
\url{https://intellipaat.com/blog/supervised-learning-vs-unsupervised-learning-vs-reinforcement-learning/}.
[Online, Accessed: September 14, 2021]
(2021)
\end{botherref}
\endbibitem

\bibitem{MVO}
\begin{barticle}
\bauthor{\bsnm{Markowitz}, \binits{H.}}:
\batitle{Portfolio selection}.
\bjtitle{The Journal of Finance}
\bvolume{7}(\bissue{1}),
\bfpage{77}--\blpage{91}
(\byear{1952})
\end{barticle}
\endbibitem

\bibitem{liu2019portfolio}
\begin{barticle}
\bauthor{\bsnm{Liu}, \binits{W.}}:
\batitle{Portfolio diversification across cryptocurrencies}.
\bjtitle{Finance Research Letters}
\bvolume{29},
\bfpage{200}--\blpage{205}
(\byear{2019})
\end{barticle}
\endbibitem

\bibitem{kajtazi2019role}
\begin{barticle}
\bauthor{\bsnm{Kajtazi}, \binits{A.}},
\bauthor{\bsnm{Moro}, \binits{A.}}:
\batitle{The role of bitcoin in well diversified portfolios: A comparative
  global study}.
\bjtitle{International Review of Financial Analysis}
\bvolume{61},
\bfpage{143}--\blpage{157}
(\byear{2019})
\end{barticle}
\endbibitem

\bibitem{brunnermeier2013bubbles}
\begin{bchapter}
\bauthor{\bsnm{Brunnermeier}, \binits{M.K.}},
\bauthor{\bsnm{Oehmke}, \binits{M.}}:
\bctitle{Bubbles, financial crises, and systemic risk}.
In: \bbtitle{Handbook of the Economics of Finance}
vol. \bseriesno{2},
pp. \bfpage{1221}--\blpage{1288}.
\bpublisher{Elsevier}, \blocation{???}
(\byear{2013})
\end{bchapter}
\endbibitem

\bibitem{kou2021bankruptcy}
\begin{barticle}
\bauthor{\bsnm{Kou}, \binits{G.}},
\bauthor{\bsnm{Xu}, \binits{Y.}},
\bauthor{\bsnm{Peng}, \binits{Y.}},
\bauthor{\bsnm{Shen}, \binits{F.}},
\bauthor{\bsnm{Chen}, \binits{Y.}},
\bauthor{\bsnm{Chang}, \binits{K.}},
\bauthor{\bsnm{Kou}, \binits{S.}}:
\batitle{Bankruptcy prediction for smes using transactional data and two-stage
  multiobjective feature selection}.
\bjtitle{Decision Support Systems}
\bvolume{140},
\bfpage{113429}
(\byear{2021})
\end{barticle}
\endbibitem

\bibitem{cryptro2021}
\begin{botherref}
\oauthor{\bsnm{Forbes}}:
Is The Crypto Market Maturing? An Analysis For Entrepreneurs.
\url{https://www.forbes.com/sites/theyec/2021/06/01/is-the-crypto-market-maturing-an-analysis-for-entrepreneurs/?sh=1170160bba22}.
[Online, Accessed: June 1, 2021]
(2021)
\end{botherref}
\endbibitem

\bibitem{cryptoftsurvey}
\begin{botherref}
\oauthor{\bsnm{FT}}:
Bitcoin: too good to miss or a bubble ready to burst?
\url{https://www.ft.com/crypto/}.
[Online, Accessed: November 9, 2021]
(2021)
\end{botherref}
\endbibitem

\bibitem{wohlin2014guidelines}
\begin{bchapter}
\bauthor{\bsnm{Wohlin}, \binits{C.}}:
\bctitle{Guidelines for snowballing in systematic literature studies and a
  replication in software engineering}.
In: \bbtitle{Proceedings of the 18th International Conference on Evaluation and
  Assessment in Software Engineering},
pp. \bfpage{1}--\blpage{10}
(\byear{2014})
\end{bchapter}
\endbibitem

\bibitem{capfolio}
\begin{botherref}
\oauthor{\bsnm{Capfolio}}:
Capfolio cryptocurrency trading platform.
\url{https://www.capfol.io/}.
[Online, Accessed: January 26, 2020]
(2020)
\end{botherref}
\endbibitem

\bibitem{3commas}
\begin{botherref}
\oauthor{\bsnm{3commas}}:
3Commas Smart Trading terminal and auto trading bots.
\url{https://3commas.io/}.
[Online, Accessed: January 26, 2020]
(2020)
\end{botherref}
\endbibitem

\bibitem{CCXT}
\begin{botherref}
\oauthor{\bsnm{Ccxt}}:
CCXT – CryptoCurrency eXchange Trading Library.
\url{https://github.com/ccxt/ccxt}.
[Online, Accessed: January 26, 2020]
(2020)
\end{botherref}
\endbibitem

\bibitem{Blackbird}
\begin{botherref}
\oauthor{\bsnm{Blackbird}}:
Blackbird Bitcoin Arbitrage: a long/short market-neutral strategy.
\url{https://github.com/butor/blackbird}.
[Online, Accessed: January 26, 2020]
(2020)
\end{botherref}
\endbibitem

\bibitem{stocksharp}
\begin{botherref}
\oauthor{\bsnm{Stocksharp}}:
StockSharp - trading platform.
\url{https://github.com/StockSharp/StockSharp}.
[Online, Accessed: January 26, 2020]
(2020)
\end{botherref}
\endbibitem

\bibitem{freqtrade}
\begin{botherref}
\oauthor{\bsnm{Fretrade}}:
Freqtrade.
\url{https://github.com/freqtrade/freqtrade}.
[Online, Accessed: January 26, 2020]
(2020)
\end{botherref}
\endbibitem

\bibitem{cryptosignal}
\begin{botherref}
\oauthor{\bsnm{Cryptosignal}}:
Automated Crypto Trading and Technical Analysis (TA) Bot.
\url{https://github.com/CryptoSignal/crypto-signal}.
[Online, Accessed: January 26, 2020]
(2020)
\end{botherref}
\endbibitem

\bibitem{ctubio}
\begin{botherref}
\oauthor{\bsnm{Ctubio}}:
Ctubio - Cryptocurrency trading bot.
\url{https://github.com/ctubio/Krypto-trading-bot}.
[Online, Accessed: January 26, 2020]
(2020)
\end{botherref}
\endbibitem

\bibitem{catalyst}
\begin{botherref}
\oauthor{\bsnm{Catalyst}}:
An Algorithmic Trading Library for Crypto-Assets in Python.
\url{https://github.com/enigmampc/catalyst}.
[Online, Accessed: January 26, 2020]
(2020)
\end{botherref}
\endbibitem

\bibitem{golang}
\begin{botherref}
\oauthor{\bsnm{Golang}}:
A golang implementation of a console-based trading bot for cryptocurrency
  exchanges.
\url{https://github.com/saniales/golang-crypto-trading-bot}.
[Online, Accessed: January 26, 2020]
(2020)
\end{botherref}
\endbibitem

\bibitem{kamrat2018technical}
\begin{bchapter}
\bauthor{\bsnm{Kamrat}, \binits{S.}},
\bauthor{\bsnm{Suesangiamsakul}, \binits{N.}},
\bauthor{\bsnm{Marukatat}, \binits{R.}}:
\bctitle{Technical analysis for cryptocurrency trading on mobile phones}.
In: \bbtitle{2018 3rd Technology Innovation Management and Engineering Science
  International Conference (TIMES-iCON)},
pp. \bfpage{1}--\blpage{4}
(\byear{2018}).
\bcomment{IEEE}
\end{bchapter}
\endbibitem

\bibitem{puauna2018arbitrage}
\begin{barticle}
\bauthor{\bsnm{P{\u{a}}una}, \binits{C.}}:
\batitle{Arbitrage trading systems for cryptocurrencies. design principles and
  server architecture}.
\bjtitle{Informatica Economica}
\bvolume{22}(\bissue{2}),
\bfpage{35}--\blpage{42}
(\byear{2018})
\end{barticle}
\endbibitem

\bibitem{guides2018eos}
\begin{botherref}
\oauthor{\bsnm{TradingstrategyGuides}}:
EOS Cryptocurrency Trading Strategy--Turtle Soup Pattern.
\url{https://tradingstrategyguides.com/eos-cryptocurrency-trading-strategy/}.
[Online, Accessed January 29, 2020]
(2019)
\end{botherref}
\endbibitem

\bibitem{guides2018nem}
\begin{botherref}
\oauthor{\bsnm{TradingstrategyGuides}}:
Nem (XEM) Cryptocurrency Strategy--Momentum Pinball Setup.
\url{https://tradingstrategyguides.com/nem-xem-cryptocurrency-strategy/}.
[Online, Accessed January 29, 2020]
(2019)
\end{botherref}
\endbibitem

\bibitem{guides2018free}
\begin{botherref}
\oauthor{\bsnm{TradingstrategyGuides}}:
Free OMNI Cryptocurrency Strategy--Amazing Gann Box.
\url{https://tradingstrategyguides.com/free-omni-cryptocurrency-strategy/}.
[Online, Accessed January 29, 2020]
(2019)
\end{botherref}
\endbibitem

\bibitem{guides2018iota}
\begin{botherref}
\oauthor{\bsnm{TradingstrategyGuides}}:
IOTA Cryptocurrency Strategy--Busted Double Top Pattern.
\url{https://tradingstrategyguides.com/iota-cryptocurrency-strategy/}.
[Online, Accessed January 29, 2020]
(2019)
\end{botherref}
\endbibitem

\bibitem{guides2018tether}
\begin{botherref}
\oauthor{\bsnm{TradingstrategyGuides}}:
Tether Trading Strategy--Bottom Rotation Trading.
\url{https://tradingstrategyguides.com/tether-trading-strategy/}.
[Online, Accessed January 29, 2020]
(2019)
\end{botherref}
\endbibitem

\bibitem{ha2018finding}
\begin{barticle}
\bauthor{\bsnm{Ha}, \binits{S.}},
\bauthor{\bsnm{Moon}, \binits{B.-R.}}:
\batitle{Finding attractive technical patterns in cryptocurrency markets}.
\bjtitle{Memetic Computing}
\bvolume{10}(\bissue{3}),
\bfpage{301}--\blpage{306}
(\byear{2018})
\end{barticle}
\endbibitem

\bibitem{hudson2019technical}
\begin{botherref}
\oauthor{\bsnm{Hudson}, \binits{R.}},
\oauthor{\bsnm{Urquhart}, \binits{A.}}:
Technical analysis and cryptocurrencies.
Available at SSRN 3387950
(2019)
\end{botherref}
\endbibitem

\bibitem{corbet2019effectiveness}
\begin{barticle}
\bauthor{\bsnm{Corbet}, \binits{S.}},
\bauthor{\bsnm{Eraslan}, \binits{V.}},
\bauthor{\bsnm{Lucey}, \binits{B.}},
\bauthor{\bsnm{Sensoy}, \binits{A.}}:
\batitle{The effectiveness of technical trading rules in cryptocurrency
  markets}.
\bjtitle{Finance Research Letters}
\bvolume{31},
\bfpage{32}--\blpage{37}
(\byear{2019})
\end{barticle}
\endbibitem

\bibitem{grobys2020technical}
\begin{barticle}
\bauthor{\bsnm{Grobys}, \binits{K.}},
\bauthor{\bsnm{Ahmed}, \binits{S.}},
\bauthor{\bsnm{Sapkota}, \binits{N.}}:
\batitle{Technical trading rules in the cryptocurrency market}.
\bjtitle{Finance Research Letters}
\bvolume{32},
\bfpage{101396}
(\byear{2020})
\end{barticle}
\endbibitem

\bibitem{al2020cryptocurrency}
\begin{barticle}
\bauthor{\bsnm{Al-Yahyaee}, \binits{K.H.}},
\bauthor{\bsnm{Mensi}, \binits{W.}},
\bauthor{\bsnm{Ko}, \binits{H.-U.}},
\bauthor{\bsnm{Yoon}, \binits{S.-M.}},
\bauthor{\bsnm{Kang}, \binits{S.H.}}:
\batitle{Why cryptocurrency markets are inefficient: The impact of liquidity
  and volatility}.
\bjtitle{The North American Journal of Economics and Finance}
\bvolume{52},
\bfpage{101168}
(\byear{2020})
\end{barticle}
\endbibitem

\bibitem{fil2019pairs}
\begin{botherref}
\oauthor{\bsnm{Fil}, \binits{M.}}:
Pairs trading in cryptocurrency markets.
Univerzita Karlova, Fakulta soci{\'a}ln{\'\i}ch v{\v{e}}d
(2019)
\end{botherref}
\endbibitem

\bibitem{gatev2006pairs}
\begin{barticle}
\bauthor{\bsnm{Gatev}, \binits{E.}},
\bauthor{\bsnm{Goetzmann}, \binits{W.N.}},
\bauthor{\bsnm{Rouwenhorst}, \binits{K.G.}}:
\batitle{Pairs trading: Performance of a relative-value arbitrage rule}.
\bjtitle{The Review of Financial Studies}
\bvolume{19}(\bissue{3}),
\bfpage{797}--\blpage{827}
(\byear{2006})
\end{barticle}
\endbibitem

\bibitem{van2018cointegration}
\begin{botherref}
\oauthor{\bparticle{van~den} \bsnm{Broek}, \binits{L.}},
\oauthor{\bsnm{Sharif}, \binits{Z.}}:
Cointegration-based pairs trading framework with application to the
  cryptocurrency market
(2018)
\end{botherref}
\endbibitem

\bibitem{lintilhac2017model}
\begin{barticle}
\bauthor{\bsnm{Lintilhac}, \binits{P.S.}},
\bauthor{\bsnm{Tourin}, \binits{A.}}:
\batitle{Model-based pairs trading in the bitcoin markets}.
\bjtitle{Quantitative Finance}
\bvolume{17}(\bissue{5}),
\bfpage{703}--\blpage{716}
(\byear{2017})
\end{barticle}
\endbibitem

\bibitem{li2016optimal}
\begin{barticle}
\bauthor{\bsnm{Li}, \binits{T.N.}},
\bauthor{\bsnm{Tourin}, \binits{A.}}:
\batitle{Optimal pairs trading with time-varying volatility}.
\bjtitle{International Journal of Financial Engineering}
\bvolume{3}(\bissue{03}),
\bfpage{1650023}
(\byear{2016})
\end{barticle}
\endbibitem

\bibitem{feng2018informed}
\begin{barticle}
\bauthor{\bsnm{Feng}, \binits{W.}},
\bauthor{\bsnm{Wang}, \binits{Y.}},
\bauthor{\bsnm{Zhang}, \binits{Z.}}:
\batitle{Informed trading in the bitcoin market}.
\bjtitle{Finance Research Letters}
\bvolume{26},
\bfpage{63}--\blpage{70}
(\byear{2018})
\end{barticle}
\endbibitem

\bibitem{easley2008time}
\begin{barticle}
\bauthor{\bsnm{Easley}, \binits{D.}},
\bauthor{\bsnm{Engle}, \binits{R.F.}},
\bauthor{\bsnm{O'Hara}, \binits{M.}},
\bauthor{\bsnm{Wu}, \binits{L.}}:
\batitle{Time-varying arrival rates of informed and uninformed trades}.
\bjtitle{Journal of Financial Econometrics}
\bvolume{6}(\bissue{2}),
\bfpage{171}--\blpage{207}
(\byear{2008})
\end{barticle}
\endbibitem

\bibitem{bouri2019trading}
\begin{barticle}
\bauthor{\bsnm{Bouri}, \binits{E.}},
\bauthor{\bsnm{Lau}, \binits{C.K.M.}},
\bauthor{\bsnm{Lucey}, \binits{B.}},
\bauthor{\bsnm{Roubaud}, \binits{D.}}:
\batitle{Trading volume and the predictability of return and volatility in the
  cryptocurrency market}.
\bjtitle{Finance Research Letters}
\bvolume{29},
\bfpage{340}--\blpage{346}
(\byear{2019})
\end{barticle}
\endbibitem

\bibitem{lee2014granger}
\begin{barticle}
\bauthor{\bsnm{Lee}, \binits{T.-H.}},
\bauthor{\bsnm{Yang}, \binits{W.}}:
\batitle{Granger-causality in quantiles between financial markets: Using copula
  approach}.
\bjtitle{International Review of Financial Analysis}
\bvolume{33},
\bfpage{70}--\blpage{78}
(\byear{2014})
\end{barticle}
\endbibitem

\bibitem{bouri2020volatility}
\begin{barticle}
\bauthor{\bsnm{Bouri}, \binits{E.}},
\bauthor{\bsnm{Lucey}, \binits{B.}},
\bauthor{\bsnm{Roubaud}, \binits{D.}}:
\batitle{The volatility surprise of leading cryptocurrencies: Transitory and
  permanent linkages}.
\bjtitle{Finance Research Letters}
\bvolume{33},
\bfpage{101188}
(\byear{2020})
\end{barticle}
\endbibitem

\bibitem{bodart2009evidence}
\begin{barticle}
\bauthor{\bsnm{Bodart}, \binits{V.}},
\bauthor{\bsnm{Candelon}, \binits{B.}}:
\batitle{Evidence of interdependence and contagion using a frequency domain
  framework}.
\bjtitle{Emerging markets review}
\bvolume{10}(\bissue{2}),
\bfpage{140}--\blpage{150}
(\byear{2009})
\end{barticle}
\endbibitem

\bibitem{badenhorst2019effect}
\begin{botherref}
\oauthor{\bsnm{Badenhorst}, \binits{J.J.}}, et al.:
Effect of bitcoin spot and derivative trading volumes on price volatility.
PhD thesis,
University of Pretoria
(2019)
\end{botherref}
\endbibitem

\bibitem{bouri2020return}
\begin{botherref}
\oauthor{\bsnm{Bouri}, \binits{E.}},
\oauthor{\bsnm{Vo}, \binits{X.V.}},
\oauthor{\bsnm{Saeed}, \binits{T.}}:
Return equicorrelation in the cryptocurrency market: Analysis and determinants.
Finance Research Letters,
101497
(2020)
\end{botherref}
\endbibitem

\bibitem{conrad2018long}
\begin{barticle}
\bauthor{\bsnm{Conrad}, \binits{C.}},
\bauthor{\bsnm{Custovic}, \binits{A.}},
\bauthor{\bsnm{Ghysels}, \binits{E.}}:
\batitle{Long-and short-term cryptocurrency volatility components: A
  garch-midas analysis}.
\bjtitle{Journal of Risk and Financial Management}
\bvolume{11}(\bissue{2}),
\bfpage{23}
(\byear{2018})
\end{barticle}
\endbibitem

\bibitem{ardia2019regime}
\begin{barticle}
\bauthor{\bsnm{Ardia}, \binits{D.}},
\bauthor{\bsnm{Bluteau}, \binits{K.}},
\bauthor{\bsnm{R{\"u}ede}, \binits{M.}}:
\batitle{Regime changes in bitcoin garch volatility dynamics}.
\bjtitle{Finance Research Letters}
\bvolume{29},
\bfpage{266}--\blpage{271}
(\byear{2019})
\end{barticle}
\endbibitem

\bibitem{troster2019bitcoin}
\begin{barticle}
\bauthor{\bsnm{Troster}, \binits{V.}},
\bauthor{\bsnm{Tiwari}, \binits{A.K.}},
\bauthor{\bsnm{Shahbaz}, \binits{M.}},
\bauthor{\bsnm{Macedo}, \binits{D.N.}}:
\batitle{Bitcoin returns and risk: A general garch and gas analysis}.
\bjtitle{Finance Research Letters}
\bvolume{30},
\bfpage{187}--\blpage{193}
(\byear{2019})
\end{barticle}
\endbibitem

\bibitem{charles2019volatility}
\begin{barticle}
\bauthor{\bsnm{Charles}, \binits{A.}},
\bauthor{\bsnm{Darn{\'e}}, \binits{O.}}, \betal:
\batitle{Volatility estimation for cryptocurrencies: Further evidence with
  jumps and structural breaks}.
\bjtitle{Economics Bulletin}
\bvolume{39}(\bissue{2}),
\bfpage{954}--\blpage{968}
(\byear{2019})
\end{barticle}
\endbibitem

\bibitem{malladi2021time}
\begin{barticle}
\bauthor{\bsnm{Malladi}, \binits{R.K.}},
\bauthor{\bsnm{Dheeriya}, \binits{P.L.}}:
\batitle{Time series analysis of cryptocurrency returns and volatilities}.
\bjtitle{Journal of Economics and Finance}
\bvolume{45}(\bissue{1}),
\bfpage{75}--\blpage{94}
(\byear{2021})
\end{barticle}
\endbibitem

\bibitem{chaim2019nonlinear}
\begin{barticle}
\bauthor{\bsnm{Chaim}, \binits{P.}},
\bauthor{\bsnm{Laurini}, \binits{M.P.}}:
\batitle{Nonlinear dependence in cryptocurrency markets}.
\bjtitle{The North American Journal of Economics and Finance}
\bvolume{48},
\bfpage{32}--\blpage{47}
(\byear{2019})
\end{barticle}
\endbibitem

\bibitem{caporale2018persistence}
\begin{barticle}
\bauthor{\bsnm{Caporale}, \binits{G.M.}},
\bauthor{\bsnm{Gil-Alana}, \binits{L.}},
\bauthor{\bsnm{Plastun}, \binits{A.}}:
\batitle{Persistence in the cryptocurrency market}.
\bjtitle{Research in International Business and Finance}
\bvolume{46},
\bfpage{141}--\blpage{148}
(\byear{2018})
\end{barticle}
\endbibitem

\bibitem{khuntia2018adaptive}
\begin{barticle}
\bauthor{\bsnm{Khuntia}, \binits{S.}},
\bauthor{\bsnm{Pattanayak}, \binits{J.}}:
\batitle{Adaptive market hypothesis and evolving predictability of bitcoin}.
\bjtitle{Economics Letters}
\bvolume{167},
\bfpage{26}--\blpage{28}
(\byear{2018})
\end{barticle}
\endbibitem

\bibitem{dominguez2003testing}
\begin{barticle}
\bauthor{\bsnm{Dom{\'\i}nguez}, \binits{M.A.}},
\bauthor{\bsnm{Lobato}, \binits{I.N.}}:
\batitle{Testing the martingale difference hypothesis}.
\bjtitle{Econometric Reviews}
\bvolume{22}(\bissue{4}),
\bfpage{351}--\blpage{377}
(\byear{2003})
\end{barticle}
\endbibitem

\bibitem{escanciano2006generalized}
\begin{barticle}
\bauthor{\bsnm{Escanciano}, \binits{J.C.}},
\bauthor{\bsnm{Velasco}, \binits{C.}}:
\batitle{Generalized spectral tests for the martingale difference hypothesis}.
\bjtitle{Journal of Econometrics}
\bvolume{134}(\bissue{1}),
\bfpage{151}--\blpage{185}
(\byear{2006})
\end{barticle}
\endbibitem

\bibitem{gradojevic2021volatility}
\begin{barticle}
\bauthor{\bsnm{Gradojevic}, \binits{N.}},
\bauthor{\bsnm{Tsiakas}, \binits{I.}}:
\batitle{Volatility cascades in cryptocurrency trading}.
\bjtitle{Journal of Empirical Finance}
\bvolume{62},
\bfpage{252}--\blpage{265}
(\byear{2021})
\end{barticle}
\endbibitem

\bibitem{nikolova2020novel}
\begin{barticle}
\bauthor{\bsnm{Nikolova}, \binits{V.}},
\bauthor{\bsnm{Trinidad~Segovia}, \binits{J.E.}},
\bauthor{\bsnm{Fern{\'a}ndez-Mart{\'\i}nez}, \binits{M.}},
\bauthor{\bsnm{S{\'a}nchez-Granero}, \binits{M.A.}}:
\batitle{A novel methodology to calculate the probability of volatility
  clusters in financial series: An application to cryptocurrency markets}.
\bjtitle{Mathematics}
\bvolume{8}(\bissue{8}),
\bfpage{1216}
(\byear{2020})
\end{barticle}
\endbibitem

\bibitem{ma2020cryptocurrency}
\begin{barticle}
\bauthor{\bsnm{Ma}, \binits{F.}},
\bauthor{\bsnm{Liang}, \binits{C.}},
\bauthor{\bsnm{Ma}, \binits{Y.}},
\bauthor{\bsnm{Wahab}, \binits{M.}}:
\batitle{Cryptocurrency volatility forecasting: A markov regime-switching midas
  approach}.
\bjtitle{Journal of Forecasting}
\bvolume{39}(\bissue{8}),
\bfpage{1277}--\blpage{1290}
(\byear{2020})
\end{barticle}
\endbibitem

\bibitem{katsiampa2018volatility}
\begin{botherref}
\oauthor{\bsnm{Katsiampa}, \binits{P.}},
\oauthor{\bsnm{Corbet}, \binits{S.}},
\oauthor{\bsnm{Lucey}, \binits{B.M.}}:
Volatility spillover effects in leading cryptocurrencies: A bekk-mgarch
  analysis.
Available at SSRN 3232912
(2018)
\end{botherref}
\endbibitem

\bibitem{katsiampa2019empirical}
\begin{botherref}
\oauthor{\bsnm{Katsiampa}, \binits{P.}}:
An empirical investigation of volatility dynamics in the cryptocurrency market.
Research in International Business and Finance
(2019)
\end{botherref}
\endbibitem

\bibitem{hultman2018volatility}
\begin{botherref}
\oauthor{\bsnm{Hultman}, \binits{H.}}:
Volatility forecasting an empirical study on bitcoin using garch and stochastic
  volatility models.
Master's thesis,
Lund University,
Sweden
(2018)
\end{botherref}
\endbibitem

\bibitem{omane2019wavelet}
\begin{barticle}
\bauthor{\bsnm{Omane-Adjepong}, \binits{M.}},
\bauthor{\bsnm{Alagidede}, \binits{P.}},
\bauthor{\bsnm{Akosah}, \binits{N.K.}}:
\batitle{Wavelet time-scale persistence analysis of cryptocurrency market
  returns and volatility}.
\bjtitle{Physica A: Statistical Mechanics and its Applications}
\bvolume{514},
\bfpage{105}--\blpage{120}
(\byear{2019})
\end{barticle}
\endbibitem

\bibitem{corbet2018exploring}
\begin{barticle}
\bauthor{\bsnm{Corbet}, \binits{S.}},
\bauthor{\bsnm{Meegan}, \binits{A.}},
\bauthor{\bsnm{Larkin}, \binits{C.}},
\bauthor{\bsnm{Lucey}, \binits{B.}},
\bauthor{\bsnm{Yarovaya}, \binits{L.}}:
\batitle{Exploring the dynamic relationships between cryptocurrencies and other
  financial assets}.
\bjtitle{Economics Letters}
\bvolume{165},
\bfpage{28}--\blpage{34}
(\byear{2018})
\end{barticle}
\endbibitem

\bibitem{zhang2020idiosyncratic}
\begin{barticle}
\bauthor{\bsnm{Zhang}, \binits{W.}},
\bauthor{\bsnm{Li}, \binits{Y.}}:
\batitle{Is idiosyncratic volatility priced in cryptocurrency markets?}
\bjtitle{Research in International Business and Finance}
\bvolume{54},
\bfpage{101252}
(\byear{2020})
\end{barticle}
\endbibitem

\bibitem{holmes1994weka}
\begin{bchapter}
\bauthor{\bsnm{Holmes}, \binits{G.}},
\bauthor{\bsnm{Donkin}, \binits{A.}},
\bauthor{\bsnm{Witten}, \binits{I.H.}}:
\bctitle{Weka: A machine learning workbench}.
In: \bbtitle{Proceedings of ANZIIS'94-Australian New Zealnd Intelligent
  Information Systems Conference},
pp. \bfpage{357}--\blpage{361}
(\byear{1994}).
\bcomment{IEEE}
\end{bchapter}
\endbibitem

\bibitem{rish2001empirical}
\begin{bchapter}
\bauthor{\bsnm{Rish}, \binits{I.}}, \betal:
\bctitle{An empirical study of the naive bayes classifier}.
In: \bbtitle{IJCAI 2001 Workshop on Empirical Methods in Artificial
  Intelligence},
vol. \bseriesno{3},
pp. \bfpage{41}--\blpage{46}
(\byear{2001})
\end{bchapter}
\endbibitem

\bibitem{wang2005support}
\begin{bbook}
\bauthor{\bsnm{Wang}, \binits{L.}}:
\bbtitle{Support Vector Machines: Theory and Applications}
vol. \bseriesno{177}.
\bpublisher{Springer}, \blocation{???}
(\byear{2005})
\end{bbook}
\endbibitem

\bibitem{friedl1997decision}
\begin{barticle}
\bauthor{\bsnm{Friedl}, \binits{M.A.}},
\bauthor{\bsnm{Brodley}, \binits{C.E.}}:
\batitle{Decision tree classification of land cover from remotely sensed data}.
\bjtitle{Remote sensing of environment}
\bvolume{61}(\bissue{3}),
\bfpage{399}--\blpage{409}
(\byear{1997})
\end{barticle}
\endbibitem

\bibitem{liaw2002classification}
\begin{barticle}
\bauthor{\bsnm{Liaw}, \binits{A.}},
\bauthor{\bsnm{Wiener}, \binits{M.}}, \betal:
\batitle{Classification and regression by randomforest}.
\bjtitle{R news}
\bvolume{2}(\bissue{3}),
\bfpage{18}--\blpage{22}
(\byear{2002})
\end{barticle}
\endbibitem

\bibitem{friedman2001greedy}
\begin{botherref}
\oauthor{\bsnm{Friedman}, \binits{J.H.}}:
Greedy function approximation: a gradient boosting machine.
Annals of statistics,
1189--1232
(2001)
\end{botherref}
\endbibitem

\bibitem{zemmal2016adaptive}
\begin{barticle}
\bauthor{\bsnm{Zemmal}, \binits{N.}},
\bauthor{\bsnm{Azizi}, \binits{N.}},
\bauthor{\bsnm{Dey}, \binits{N.}},
\bauthor{\bsnm{Sellami}, \binits{M.}}:
\batitle{Adaptive semi supervised support vector machine semi supervised
  learning with features cooperation for breast cancer classification}.
\bjtitle{Journal of Medical Imaging and Health Informatics}
\bvolume{6}(\bissue{1}),
\bfpage{53}--\blpage{62}
(\byear{2016})
\end{barticle}
\endbibitem

\bibitem{keerthi2001improvements}
\begin{barticle}
\bauthor{\bsnm{Keerthi}, \binits{S.S.}},
\bauthor{\bsnm{Shevade}, \binits{S.K.}},
\bauthor{\bsnm{Bhattacharyya}, \binits{C.}},
\bauthor{\bsnm{Murthy}, \binits{K.R.K.}}:
\batitle{Improvements to platt's smo algorithm for svm classifier design}.
\bjtitle{Neural computation}
\bvolume{13}(\bissue{3}),
\bfpage{637}--\blpage{649}
(\byear{2001})
\end{barticle}
\endbibitem

\bibitem{fang2020better}
\begin{botherref}
\oauthor{\bsnm{Fang}, \binits{F.}},
\oauthor{\bsnm{Ventre}, \binits{C.}},
\oauthor{\bsnm{Li}, \binits{L.}},
\oauthor{\bsnm{Kanthan}, \binits{L.}},
\oauthor{\bsnm{Wu}, \binits{F.}},
\oauthor{\bsnm{Basios}, \binits{M.}}:
Better model selection with a new definition of feature importance.
arXiv preprint arXiv:2009.07708
(2020)
\end{botherref}
\endbibitem

\bibitem{jianliang2009application}
\begin{bchapter}
\bauthor{\bsnm{Jianliang}, \binits{M.}},
\bauthor{\bsnm{Haikun}, \binits{S.}},
\bauthor{\bsnm{Ling}, \binits{B.}}:
\bctitle{The application on intrusion detection based on k-means cluster
  algorithm}.
In: \bbtitle{2009 International Forum on Information Technology and
  Applications},
vol. \bseriesno{1},
pp. \bfpage{150}--\blpage{152}
(\byear{2009}).
\bcomment{IEEE}
\end{bchapter}
\endbibitem

\bibitem{wagstaff2001constrained}
\begin{bchapter}
\bauthor{\bsnm{Wagstaff}, \binits{K.}},
\bauthor{\bsnm{Cardie}, \binits{C.}},
\bauthor{\bsnm{Rogers}, \binits{S.}},
\bauthor{\bsnm{Schr{\"o}dl}, \binits{S.}}, \betal:
\bctitle{Constrained k-means clustering with background knowledge}.
In: \bbtitle{Icml},
vol. \bseriesno{1},
pp. \bfpage{577}--\blpage{584}
(\byear{2001})
\end{bchapter}
\endbibitem

\bibitem{li2021integrated}
\begin{botherref}
\oauthor{\bsnm{Li}, \binits{T.}},
\oauthor{\bsnm{Kou}, \binits{G.}},
\oauthor{\bsnm{Peng}, \binits{Y.}},
\oauthor{\bsnm{Philip}, \binits{S.Y.}}:
An integrated cluster detection, optimization, and interpretation approach for
  financial data.
IEEE Transactions on Cybernetics
(2021)
\end{botherref}
\endbibitem

\bibitem{kou2014evaluation}
\begin{barticle}
\bauthor{\bsnm{Kou}, \binits{G.}},
\bauthor{\bsnm{Peng}, \binits{Y.}},
\bauthor{\bsnm{Wang}, \binits{G.}}:
\batitle{Evaluation of clustering algorithms for financial risk analysis using
  mcdm methods}.
\bjtitle{Information Sciences}
\bvolume{275},
\bfpage{1}--\blpage{12}
(\byear{2014})
\end{barticle}
\endbibitem

\bibitem{kutner2005applied}
\begin{bbook}
\bauthor{\bsnm{Kutner}, \binits{M.H.}},
\bauthor{\bsnm{Nachtsheim}, \binits{C.J.}},
\bauthor{\bsnm{Neter}, \binits{J.}},
\bauthor{\bsnm{Li}, \binits{W.}}, \betal:
\bbtitle{Applied Linear Statistical Models}
vol. \bseriesno{5}.
\bpublisher{McGraw-Hill Irwin New York}, \blocation{???}
(\byear{2005})
\end{bbook}
\endbibitem

\bibitem{friedman1984monotone}
\begin{barticle}
\bauthor{\bsnm{Friedman}, \binits{J.}},
\bauthor{\bsnm{Tibshirani}, \binits{R.}}:
\batitle{The monotone smoothing of scatterplots}.
\bjtitle{Technometrics}
\bvolume{26}(\bissue{3}),
\bfpage{243}--\blpage{250}
(\byear{1984})
\end{barticle}
\endbibitem

\bibitem{10.1145/3285029}
\begin{botherref}
\oauthor{\bsnm{Zhang}, \binits{S.}},
\oauthor{\bsnm{Yao}, \binits{L.}},
\oauthor{\bsnm{Sun}, \binits{A.}},
\oauthor{\bsnm{Tay}, \binits{Y.}}:
Deep learning based recommender system: A survey and new perspectives.
ACM Comput. Surv.
\textbf{52}(1)
(2019).
doi:\doiurl{10.1145/3285029}
\end{botherref}
\endbibitem

\bibitem{sze2017efficient}
\begin{barticle}
\bauthor{\bsnm{Sze}, \binits{V.}},
\bauthor{\bsnm{Chen}, \binits{Y.-H.}},
\bauthor{\bsnm{Yang}, \binits{T.-J.}},
\bauthor{\bsnm{Emer}, \binits{J.S.}}:
\batitle{Efficient processing of deep neural networks: A tutorial and survey}.
\bjtitle{Proceedings of the IEEE}
\bvolume{105}(\bissue{12}),
\bfpage{2295}--\blpage{2329}
(\byear{2017})
\end{barticle}
\endbibitem

\bibitem{lawrence1997face}
\begin{barticle}
\bauthor{\bsnm{Lawrence}, \binits{S.}},
\bauthor{\bsnm{Giles}, \binits{C.L.}},
\bauthor{\bsnm{Tsoi}, \binits{A.C.}},
\bauthor{\bsnm{Back}, \binits{A.D.}}:
\batitle{Face recognition: A convolutional neural-network approach}.
\bjtitle{IEEE transactions on neural networks}
\bvolume{8}(\bissue{1}),
\bfpage{98}--\blpage{113}
(\byear{1997})
\end{barticle}
\endbibitem

\bibitem{mikolov2011extensions}
\begin{bchapter}
\bauthor{\bsnm{Mikolov}, \binits{T.}},
\bauthor{\bsnm{Kombrink}, \binits{S.}},
\bauthor{\bsnm{Burget}, \binits{L.}},
\bauthor{\bsnm{{\v{C}}ernock{\`y}}, \binits{J.}},
\bauthor{\bsnm{Khudanpur}, \binits{S.}}:
\bctitle{Extensions of recurrent neural network language model}.
In: \bbtitle{2011 IEEE International Conference on Acoustics, Speech and Signal
  Processing (ICASSP)},
pp. \bfpage{5528}--\blpage{5531}
(\byear{2011}).
\bcomment{IEEE}
\end{bchapter}
\endbibitem

\bibitem{chung2014empirical}
\begin{botherref}
\oauthor{\bsnm{Chung}, \binits{J.}},
\oauthor{\bsnm{Gulcehre}, \binits{C.}},
\oauthor{\bsnm{Cho}, \binits{K.}},
\oauthor{\bsnm{Bengio}, \binits{Y.}}:
Empirical evaluation of gated recurrent neural networks on sequence modeling.
arXiv preprint arXiv:1412.3555
(2014)
\end{botherref}
\endbibitem

\bibitem{cheng2016long}
\begin{botherref}
\oauthor{\bsnm{Cheng}, \binits{J.}},
\oauthor{\bsnm{Dong}, \binits{L.}},
\oauthor{\bsnm{Lapata}, \binits{M.}}:
Long short-term memory-networks for machine reading.
arXiv preprint arXiv:1601.06733
(2016)
\end{botherref}
\endbibitem

\bibitem{kalchbrenner2014convolutional}
\begin{botherref}
\oauthor{\bsnm{Kalchbrenner}, \binits{N.}},
\oauthor{\bsnm{Grefenstette}, \binits{E.}},
\oauthor{\bsnm{Blunsom}, \binits{P.}}:
A convolutional neural network for modelling sentences.
arXiv preprint arXiv:1404.2188
(2014)
\end{botherref}
\endbibitem

\bibitem{pascanu2013difficulty}
\begin{bchapter}
\bauthor{\bsnm{Pascanu}, \binits{R.}},
\bauthor{\bsnm{Mikolov}, \binits{T.}},
\bauthor{\bsnm{Bengio}, \binits{Y.}}:
\bctitle{On the difficulty of training recurrent neural networks}.
In: \bbtitle{International Conference on Machine Learning},
pp. \bfpage{1310}--\blpage{1318}
(\byear{2013})
\end{bchapter}
\endbibitem

\bibitem{dutta2020gated}
\begin{barticle}
\bauthor{\bsnm{Dutta}, \binits{A.}},
\bauthor{\bsnm{Kumar}, \binits{S.}},
\bauthor{\bsnm{Basu}, \binits{M.}}:
\batitle{A gated recurrent unit approach to bitcoin price prediction}.
\bjtitle{Journal of Risk and Financial Management}
\bvolume{13}(\bissue{2}),
\bfpage{23}
(\byear{2020})
\end{barticle}
\endbibitem

\bibitem{xu2017seq2seq}
\begin{bchapter}
\bauthor{\bsnm{Xu}, \binits{Z.}},
\bauthor{\bsnm{Wang}, \binits{S.}},
\bauthor{\bsnm{Zhu}, \binits{F.}},
\bauthor{\bsnm{Huang}, \binits{J.}}:
\bctitle{Seq2seq fingerprint: An unsupervised deep molecular embedding for drug
  discovery}.
In: \bbtitle{Proceedings of the 8th ACM International Conference on
  Bioinformatics, Computational Biology, and Health Informatics},
pp. \bfpage{285}--\blpage{294}
(\byear{2017})
\end{bchapter}
\endbibitem

\bibitem{sriram2017cold}
\begin{botherref}
\oauthor{\bsnm{Sriram}, \binits{A.}},
\oauthor{\bsnm{Jun}, \binits{H.}},
\oauthor{\bsnm{Satheesh}, \binits{S.}},
\oauthor{\bsnm{Coates}, \binits{A.}}:
Cold fusion: Training seq2seq models together with language models.
arXiv preprint arXiv:1708.06426
(2017)
\end{botherref}
\endbibitem

\bibitem{sutton1998introduction}
\begin{bbook}
\bauthor{\bsnm{Sutton}, \binits{R.S.}},
\bauthor{\bsnm{Barto}, \binits{A.G.}}, \betal:
\bbtitle{Introduction to Reinforcement Learning}
vol. \bseriesno{135}.
\bpublisher{MIT press Cambridge}, \blocation{???}
(\byear{1998})
\end{bbook}
\endbibitem

\bibitem{gu2016continuous}
\begin{bchapter}
\bauthor{\bsnm{Gu}, \binits{S.}},
\bauthor{\bsnm{Lillicrap}, \binits{T.}},
\bauthor{\bsnm{Sutskever}, \binits{I.}},
\bauthor{\bsnm{Levine}, \binits{S.}}:
\bctitle{Continuous deep q-learning with model-based acceleration}.
In: \bbtitle{International Conference on Machine Learning},
pp. \bfpage{2829}--\blpage{2838}
(\byear{2016})
\end{bchapter}
\endbibitem

\bibitem{salakhutdinov2009deep}
\begin{bchapter}
\bauthor{\bsnm{Salakhutdinov}, \binits{R.}},
\bauthor{\bsnm{Hinton}, \binits{G.}}:
\bctitle{Deep boltzmann machines}.
In: \bbtitle{Artificial Intelligence and Statistics},
pp. \bfpage{448}--\blpage{455}
(\byear{2009})
\end{bchapter}
\endbibitem

\bibitem{sun2019using}
\begin{bchapter}
\bauthor{\bsnm{Sun}, \binits{J.}},
\bauthor{\bsnm{Zhou}, \binits{Y.}},
\bauthor{\bsnm{Lin}, \binits{J.}}:
\bctitle{Using machine learning for cryptocurrency trading}.
In: \bbtitle{2019 IEEE International Conference on Industrial Cyber Physical
  Systems (ICPS)},
pp. \bfpage{647}--\blpage{652}
(\byear{2019}).
\bcomment{IEEE}
\end{bchapter}
\endbibitem

\bibitem{kakushadze2016101}
\begin{barticle}
\bauthor{\bsnm{Kakushadze}, \binits{Z.}}:
\batitle{101 formulaic alphas}.
\bjtitle{Wilmott}
\bvolume{2016}(\bissue{84}),
\bfpage{72}--\blpage{81}
(\byear{2016})
\end{barticle}
\endbibitem

\bibitem{vo2018high}
\begin{botherref}
\oauthor{\bsnm{Vo}, \binits{A.}},
\oauthor{\bsnm{Yost-Bremm}, \binits{C.}}:
A high-frequency algorithmic trading strategy for cryptocurrency.
Journal of Computer Information Systems,
1--14
(2018)
\end{botherref}
\endbibitem

\bibitem{zenkova2019robustness}
\begin{barticle}
\bauthor{\bsnm{Ślepaczuk}, \binits{R.}},
\bauthor{\bsnm{Zenkova}, \binits{M.}}:
\batitle{Robustness of support vector machines in algorithmic trading on
  cryptocurrency market}.
\bjtitle{Central European Economic Journal}
\bvolume{5},
\bfpage{186}--\blpage{205}
(\byear{2018})
\end{barticle}
\endbibitem

\bibitem{barnwal2019stacking}
\begin{botherref}
\oauthor{\bsnm{Barnwal}, \binits{A.}},
\oauthor{\bsnm{Bharti}, \binits{H.}},
\oauthor{\bsnm{Ali}, \binits{A.}},
\oauthor{\bsnm{Singh}, \binits{V.}}:
Stacking with neural network for cryptocurrency investment.
arXiv preprint arXiv:1902.07855
(2019)
\end{botherref}
\endbibitem

\bibitem{ng2002discriminative}
\begin{bchapter}
\bauthor{\bsnm{Ng}, \binits{A.Y.}},
\bauthor{\bsnm{Jordan}, \binits{M.I.}}:
\bctitle{On discriminative vs. generative classifiers: A comparison of logistic
  regression and naive bayes}.
In: \bbtitle{Advances in Neural Information Processing Systems},
pp. \bfpage{841}--\blpage{848}
(\byear{2002})
\end{bchapter}
\endbibitem

\bibitem{attanasio2019quantitative}
\begin{bchapter}
\bauthor{\bsnm{Attanasio}, \binits{G.}},
\bauthor{\bsnm{Cagliero}, \binits{L.}},
\bauthor{\bsnm{Garza}, \binits{P.}},
\bauthor{\bsnm{Baralis}, \binits{E.}}:
\bctitle{Quantitative cryptocurrency trading: exploring the use of machine
  learning techniques}.
In: \bbtitle{Proceedings of the 5th Workshop on Data Science for Macro-modeling
  with Financial and Economic Datasets},
p. \bfpage{1}
(\byear{2019}).
\bcomment{ACM}
\end{bchapter}
\endbibitem

\bibitem{madan2015automated}
\begin{botherref}
\oauthor{\bsnm{Madan}, \binits{I.}},
\oauthor{\bsnm{Saluja}, \binits{S.}},
\oauthor{\bsnm{Zhao}, \binits{A.}}:
Automated bitcoin trading via machine learning algorithms.
URL: http://cs229. stanford. edu/proj2014/Isaac\% 20Madan
\textbf{20}
(2015)
\end{botherref}
\endbibitem

\bibitem{virkprediction}
\begin{botherref}
\oauthor{\bsnm{Virk}, \binits{D.S.}}:
Prediction of bitcoin price using data mining.
Master's thesis,
National College of Ireland
(2017)
\end{botherref}
\endbibitem

\bibitem{zhengyang2019prediction}
\begin{bchapter}
\bauthor{\bsnm{Zhengyang}, \binits{W.}},
\bauthor{\bsnm{Xingzhou}, \binits{L.}},
\bauthor{\bsnm{Jinjin}, \binits{R.}},
\bauthor{\bsnm{Jiaqing}, \binits{K.}}:
\bctitle{Prediction of cryptocurrency price dynamics with multiple machine
  learning techniques}.
In: \bbtitle{Proceedings of the 2019 4th International Conference on Machine
  Learning Technologies},
pp. \bfpage{15}--\blpage{19}
(\byear{2019}).
\bcomment{ACM}
\end{bchapter}
\endbibitem

\bibitem{kwon2019time}
\begin{botherref}
\oauthor{\bsnm{Kwon}, \binits{D.-H.}},
\oauthor{\bsnm{Kim}, \binits{J.-B.}},
\oauthor{\bsnm{Heo}, \binits{J.-S.}},
\oauthor{\bsnm{Kim}, \binits{C.-M.}},
\oauthor{\bsnm{Han}, \binits{Y.-H.}}:
Time series classification of cryptocurrency price trend based on a recurrent
  lstm neural network.
Journal of Information Processing Systems
\textbf{15}(3)
(2019)
\end{botherref}
\endbibitem

\bibitem{alessandretti2018anticipating}
\begin{botherref}
\oauthor{\bsnm{Alessandretti}, \binits{L.}},
\oauthor{\bsnm{ElBahrawy}, \binits{A.}},
\oauthor{\bsnm{Aiello}, \binits{L.M.}},
\oauthor{\bsnm{Baronchelli}, \binits{A.}}:
Anticipating cryptocurrency prices using machine learning.
Complexity
\textbf{2018}
(2018)
\end{botherref}
\endbibitem

\bibitem{phaladisailoed2018machine}
\begin{bchapter}
\bauthor{\bsnm{Phaladisailoed}, \binits{T.}},
\bauthor{\bsnm{Numnonda}, \binits{T.}}:
\bctitle{Machine learning models comparison for bitcoin price prediction}.
In: \bbtitle{2018 10th International Conference on Information Technology and
  Electrical Engineering (ICITEE)},
pp. \bfpage{506}--\blpage{511}
(\byear{2018}).
\bcomment{IEEE}
\end{bchapter}
\endbibitem

\bibitem{rane2019systematic}
\begin{bchapter}
\bauthor{\bsnm{Rane}, \binits{P.V.}},
\bauthor{\bsnm{Dhage}, \binits{S.N.}}:
\bctitle{Systematic erudition of bitcoin price prediction using machine
  learning techniques}.
In: \bbtitle{2019 5th International Conference on Advanced Computing \&
  Communication Systems (ICACCS)},
pp. \bfpage{594}--\blpage{598}
(\byear{2019}).
\bcomment{IEEE}
\end{bchapter}
\endbibitem

\bibitem{rebane2018seq2seq}
\begin{botherref}
\oauthor{\bsnm{Rebane}, \binits{J.}},
\oauthor{\bsnm{Karlsson}, \binits{I.}},
\oauthor{\bsnm{Denic}, \binits{S.}},
\oauthor{\bsnm{Papapetrou}, \binits{P.}}:
Seq2seq rnns and arima models for cryptocurrency prediction: A comparative
  study.
SIGKDD Fintech
\textbf{18}
(2018)
\end{botherref}
\endbibitem

\bibitem{stuerner2019algorithmic}
\begin{botherref}
\oauthor{\bsnm{Stuerner}, \binits{P.}}:
Algorithmic cryptocurrency trading.
PhD thesis,
Ulm University
(2019)
\end{botherref}
\endbibitem

\bibitem{persson2018hybrid}
\begin{botherref}
\oauthor{\bsnm{Persson}, \binits{S.}},
\oauthor{\bsnm{Slottje}, \binits{A.}},
\oauthor{\bsnm{Shaw}, \binits{I.}}:
Hybrid autoregressive-recurrent neural network architecture for algorithmic
  trading of cryptocurrencies.
Cs230 deep learning thesis,
Stanford University
(2018)
\end{botherref}
\endbibitem

\bibitem{livieris2020cnn}
\begin{barticle}
\bauthor{\bsnm{Livieris}, \binits{I.E.}},
\bauthor{\bsnm{Pintelas}, \binits{E.}},
\bauthor{\bsnm{Pintelas}, \binits{P.}}:
\batitle{A cnn--lstm model for gold price time-series forecasting}.
\bjtitle{Neural computing and applications}
\bvolume{32}(\bissue{23}),
\bfpage{17351}--\blpage{17360}
(\byear{2020})
\end{barticle}
\endbibitem

\bibitem{huan2020ieo}
\begin{botherref}
\oauthor{\bsnm{Huan}, \binits{Y.}},
\oauthor{\bsnm{Wu}, \binits{F.}},
\oauthor{\bsnm{Basios}, \binits{M.}},
\oauthor{\bsnm{Kanthan}, \binits{L.}},
\oauthor{\bsnm{Li}, \binits{L.}},
\oauthor{\bsnm{Xu}, \binits{B.}}:
Ieo: Intelligent evolutionary optimisation for hyperparameter tuning.
arXiv preprint arXiv:2009.06390
(2020)
\end{botherref}
\endbibitem

\bibitem{lu2020cnn}
\begin{botherref}
\oauthor{\bsnm{Lu}, \binits{W.}},
\oauthor{\bsnm{Li}, \binits{J.}},
\oauthor{\bsnm{Li}, \binits{Y.}},
\oauthor{\bsnm{Sun}, \binits{A.}},
\oauthor{\bsnm{Wang}, \binits{J.}}:
A cnn-lstm-based model to forecast stock prices.
Complexity
\textbf{2020}
(2020)
\end{botherref}
\endbibitem

\bibitem{fan2019deeplearning}
\begin{barticle}
\bauthor{\bsnm{Fang}, \binits{F.}},
\bauthor{\bsnm{Chung}, \binits{W.}},
\bauthor{\bsnm{Ventre}, \binits{C.}},
\bauthor{\bsnm{Basios}, \binits{M.}},
\bauthor{\bsnm{Kanthan}, \binits{L.}},
\bauthor{\bsnm{Li}, \binits{L.}},
\bauthor{\bsnm{Wu}, \binits{F.}}:
\batitle{Ascertaining price formation in cryptocurrency markets with machine
  learning}.
\bjtitle{The European Journal of Finance}
\bvolume{0}(\bissue{0}),
\bfpage{1}--\blpage{23}
(\byear{2021})
\end{barticle}
\endbibitem

\bibitem{sirignano2019universal}
\begin{barticle}
\bauthor{\bsnm{Sirignano}, \binits{J.}},
\bauthor{\bsnm{Cont}, \binits{R.}}:
\batitle{Universal features of price formation in financial markets:
  perspectives from deep learning}.
\bjtitle{Quantitative Finance}
\bvolume{19}(\bissue{9}),
\bfpage{1449}--\blpage{1459}
(\byear{2019})
\end{barticle}
\endbibitem

\bibitem{yao2018predictive}
\begin{barticle}
\bauthor{\bsnm{Yao}, \binits{Y.}},
\bauthor{\bsnm{Yi}, \binits{J.}},
\bauthor{\bsnm{Zhai}, \binits{S.}},
\bauthor{\bsnm{Lin}, \binits{Y.}},
\bauthor{\bsnm{Kim}, \binits{T.}},
\bauthor{\bsnm{Zhang}, \binits{G.}},
\bauthor{\bsnm{Lee}, \binits{L.Y.}}:
\batitle{Predictive analysis of cryptocurrency price using deep learning}.
\bjtitle{International Journal of Engineering \& Technology}
\bvolume{7}(\bissue{3.27}),
\bfpage{258}--\blpage{264}
(\byear{2018})
\end{barticle}
\endbibitem

\bibitem{livieris2020ensemble}
\begin{barticle}
\bauthor{\bsnm{Livieris}, \binits{I.E.}},
\bauthor{\bsnm{Pintelas}, \binits{E.}},
\bauthor{\bsnm{Stavroyiannis}, \binits{S.}},
\bauthor{\bsnm{Pintelas}, \binits{P.}}:
\batitle{Ensemble deep learning models for forecasting cryptocurrency
  time-series}.
\bjtitle{Algorithms}
\bvolume{13}(\bissue{5}),
\bfpage{121}
(\byear{2020})
\end{barticle}
\endbibitem

\bibitem{kumar2020predicting}
\begin{bchapter}
\bauthor{\bsnm{Kumar}, \binits{D.}},
\bauthor{\bsnm{Rath}, \binits{S.}}:
\bctitle{Predicting the trends of price for ethereum using deep learning
  techniques}.
In: \bbtitle{Artificial Intelligence and Evolutionary Computations in
  Engineering Systems},
pp. \bfpage{103}--\blpage{114}.
\bpublisher{Springer}, \blocation{???}
(\byear{2020})
\end{bchapter}
\endbibitem

\bibitem{lamon2017cryptocurrency}
\begin{barticle}
\bauthor{\bsnm{Lamon}, \binits{C.}},
\bauthor{\bsnm{Nielsen}, \binits{E.}},
\bauthor{\bsnm{Redondo}, \binits{E.}}:
\batitle{Cryptocurrency price prediction using news and social media
  sentiment}.
\bjtitle{SMU Data Sci. Rev}
\bvolume{1}(\bissue{3}),
\bfpage{1}--\blpage{22}
(\byear{2017})
\end{barticle}
\endbibitem

\bibitem{smuts2019drives}
\begin{barticle}
\bauthor{\bsnm{Smuts}, \binits{N.}}:
\batitle{What drives cryptocurrency prices?: An investigation of google trends
  and telegram sentiment}.
\bjtitle{ACM SIGMETRICS Performance Evaluation Review}
\bvolume{46}(\bissue{3}),
\bfpage{131}--\blpage{134}
(\byear{2019})
\end{barticle}
\endbibitem

\bibitem{hutto2014vader}
\begin{bchapter}
\bauthor{\bsnm{Hutto}, \binits{C.J.}},
\bauthor{\bsnm{Gilbert}, \binits{E.}}:
\bctitle{Vader: A parsimonious rule-based model for sentiment analysis of
  social media text}.
In: \bbtitle{Eighth International AAAI Conference on Weblogs and Social Media}
(\byear{2014})
\end{bchapter}
\endbibitem

\bibitem{nasir2019forecasting}
\begin{barticle}
\bauthor{\bsnm{Nasir}, \binits{M.A.}},
\bauthor{\bsnm{Huynh}, \binits{T.L.D.}},
\bauthor{\bsnm{Nguyen}, \binits{S.P.}},
\bauthor{\bsnm{Duong}, \binits{D.}}:
\batitle{Forecasting cryptocurrency returns and volume using search engines}.
\bjtitle{Financial Innovation}
\bvolume{5}(\bissue{1}),
\bfpage{2}
(\byear{2019})
\end{barticle}
\endbibitem

\bibitem{kristoufek2013bitcoin}
\begin{barticle}
\bauthor{\bsnm{Kristoufek}, \binits{L.}}:
\batitle{Bitcoin meets google trends and wikipedia: Quantifying the
  relationship between phenomena of the internet era}.
\bjtitle{Scientific reports}
\bvolume{3},
\bfpage{3415}
(\byear{2013})
\end{barticle}
\endbibitem

\bibitem{kim2016predicting}
\begin{barticle}
\bauthor{\bsnm{Kim}, \binits{Y.B.}},
\bauthor{\bsnm{Kim}, \binits{J.G.}},
\bauthor{\bsnm{Kim}, \binits{W.}},
\bauthor{\bsnm{Im}, \binits{J.H.}},
\bauthor{\bsnm{Kim}, \binits{T.H.}},
\bauthor{\bsnm{Kang}, \binits{S.J.}},
\bauthor{\bsnm{Kim}, \binits{C.H.}}:
\batitle{Predicting fluctuations in cryptocurrency transactions based on user
  comments and replies}.
\bjtitle{PloS one}
\bvolume{11}(\bissue{8}),
\bfpage{0161197}
(\byear{2016})
\end{barticle}
\endbibitem

\bibitem{phillips2018mutual}
\begin{bchapter}
\bauthor{\bsnm{Phillips}, \binits{R.C.}},
\bauthor{\bsnm{Gorse}, \binits{D.}}:
\bctitle{Mutual-excitation of cryptocurrency market returns and social media
  topics}.
In: \bbtitle{Proceedings of the 4th International Conference on Frontiers of
  Educational Technologies},
pp. \bfpage{80}--\blpage{86}
(\byear{2018}).
\bcomment{ACM}
\end{bchapter}
\endbibitem

\bibitem{li2019sentiment}
\begin{barticle}
\bauthor{\bsnm{Li}, \binits{T.R.}},
\bauthor{\bsnm{Chamrajnagar}, \binits{A.}},
\bauthor{\bsnm{Fong}, \binits{X.}},
\bauthor{\bsnm{Rizik}, \binits{N.}},
\bauthor{\bsnm{Fu}, \binits{F.}}:
\batitle{Sentiment-based prediction of alternative cryptocurrency price
  fluctuations using gradient boosting tree model}.
\bjtitle{Frontiers in Physics}
\bvolume{7},
\bfpage{98}
(\byear{2019})
\end{barticle}
\endbibitem

\bibitem{flori2019news}
\begin{barticle}
\bauthor{\bsnm{Flori}, \binits{A.}}:
\batitle{News and subjective beliefs: A bayesian approach to bitcoin
  investments}.
\bjtitle{Research in International Business and Finance}
\bvolume{50},
\bfpage{336}--\blpage{356}
(\byear{2019})
\end{barticle}
\endbibitem

\bibitem{bouri2019predicting}
\begin{botherref}
\oauthor{\bsnm{Bouri}, \binits{E.}},
\oauthor{\bsnm{Gupta}, \binits{R.}}:
Predicting bitcoin returns: Comparing the roles of newspaper-and internet
  search-based measures of uncertainty.
Finance Research Letters,
101398
(2019)
\end{botherref}
\endbibitem

\bibitem{colianni2015algorithmic}
\begin{botherref}
\oauthor{\bsnm{Colianni}, \binits{S.}},
\oauthor{\bsnm{Rosales}, \binits{S.}},
\oauthor{\bsnm{Signorotti}, \binits{M.}}:
Algorithmic trading of cryptocurrency based on twitter sentiment analysis.
CS229 Project,
1--5
(2015)
\end{botherref}
\endbibitem

\bibitem{garcia2015social}
\begin{barticle}
\bauthor{\bsnm{Garcia}, \binits{D.}},
\bauthor{\bsnm{Schweitzer}, \binits{F.}}:
\batitle{Social signals and algorithmic trading of bitcoin}.
\bjtitle{Royal Society open science}
\bvolume{2}(\bissue{9}),
\bfpage{150288}
(\byear{2015})
\end{barticle}
\endbibitem

\bibitem{zamuda2019forecasting}
\begin{bchapter}
\bauthor{\bsnm{Zamuda}, \binits{A.}},
\bauthor{\bsnm{Crescimanna}, \binits{V.}},
\bauthor{\bsnm{Burguillo}, \binits{J.C.}},
\bauthor{\bsnm{Dias}, \binits{J.M.}},
\bauthor{\bsnm{Wegrzyn-Wolska}, \binits{K.}},
\bauthor{\bsnm{Rached}, \binits{I.}},
\bauthor{\bsnm{Gonzlez}, \binits{H.}},
\bauthor{\bsnm{Senkerik}, \binits{R.}},
\bauthor{\bsnm{Pop}, \binits{C.}},
\bauthor{\bsnm{Cioara}, \binits{T.}}, \betal:
\bctitle{Forecasting cryptocurrency value by sentiment analysis: An
  hpc-oriented survey of the state-of-the-art in the cloud era}.
In: \bbtitle{High-Performance Modelling and Simulation for Big Data
  Applications},
pp. \bfpage{325}--\blpage{349}.
\bpublisher{Springer}, \blocation{???}
(\byear{2019})
\end{bchapter}
\endbibitem

\bibitem{bartoluccibutterfly}
\begin{botherref}
\oauthor{\bsnm{Bartolucci}, \binits{S.}},
\oauthor{\bsnm{Destefanis}, \binits{G.}},
\oauthor{\bsnm{Ortu}, \binits{M.}},
\oauthor{\bsnm{Uras}, \binits{N.}},
\oauthor{\bsnm{Marchesi}, \binits{M.}},
\oauthor{\bsnm{Tonelli}, \binits{R.}}:
The butterfly affect: Impact of development practices on cryptocurrency prices
(2019)
\end{botherref}
\endbibitem

\bibitem{henderson2018deep}
\begin{bchapter}
\bauthor{\bsnm{Henderson}, \binits{P.}},
\bauthor{\bsnm{Islam}, \binits{R.}},
\bauthor{\bsnm{Bachman}, \binits{P.}},
\bauthor{\bsnm{Pineau}, \binits{J.}},
\bauthor{\bsnm{Precup}, \binits{D.}},
\bauthor{\bsnm{Meger}, \binits{D.}}:
\bctitle{Deep reinforcement learning that matters}.
In: \bbtitle{Thirty-Second AAAI Conference on Artificial Intelligence}
(\byear{2018})
\end{bchapter}
\endbibitem

\bibitem{liu2021agent}
\begin{bchapter}
\bauthor{\bsnm{Liu}, \binits{B.}},
\bauthor{\bsnm{Polukarov}, \binits{M.}},
\bauthor{\bsnm{Ventre}, \binits{C.}},
\bauthor{\bsnm{Li}, \binits{L.}},
\bauthor{\bsnm{Kanthan}, \binits{L.}}:
\bctitle{Agent-based markets: Equilibrium strategies and robustness}.
In: \bbtitle{Proceedings of the 2nd ACM International Conference on AI in
  Finance}
(\byear{2021})
\end{bchapter}
\endbibitem

\bibitem{bu2018learning}
\begin{bchapter}
\bauthor{\bsnm{Bu}, \binits{S.-J.}},
\bauthor{\bsnm{Cho}, \binits{S.-B.}}:
\bctitle{Learning optimal q-function using deep boltzmann machine for reliable
  trading of cryptocurrency}.
In: \bbtitle{International Conference on Intelligent Data Engineering and
  Automated Learning},
pp. \bfpage{468}--\blpage{480}
(\byear{2018}).
\bcomment{Springer}
\end{bchapter}
\endbibitem

\bibitem{juchli2018limit}
\begin{botherref}
\oauthor{\bsnm{Juchli}, \binits{M.}}:
Limit order placement optimization with deep reinforcement learning: Learning
  from patterns in cryptocurrency market data.
Master's thesis,
TU Delft Electrical Engineering,
Netherlands
(2018)
\end{botherref}
\endbibitem

\bibitem{lucarelli2019deep}
\begin{bchapter}
\bauthor{\bsnm{Lucarelli}, \binits{G.}},
\bauthor{\bsnm{Borrotti}, \binits{M.}}:
\bctitle{A deep reinforcement learning approach for automated cryptocurrency
  trading}.
In: \bbtitle{IFIP International Conference on Artificial Intelligence
  Applications and Innovations},
pp. \bfpage{247}--\blpage{258}
(\byear{2019}).
\bcomment{Springer}
\end{bchapter}
\endbibitem

\bibitem{sattarov2020recommending}
\begin{barticle}
\bauthor{\bsnm{Sattarov}, \binits{O.}},
\bauthor{\bsnm{Muminov}, \binits{A.}},
\bauthor{\bsnm{Lee}, \binits{C.W.}},
\bauthor{\bsnm{Kang}, \binits{H.K.}},
\bauthor{\bsnm{Oh}, \binits{R.}},
\bauthor{\bsnm{Ahn}, \binits{J.}},
\bauthor{\bsnm{Oh}, \binits{H.J.}},
\bauthor{\bsnm{Jeon}, \binits{H.S.}}:
\batitle{Recommending cryptocurrency trading points with deep reinforcement
  learning approach}.
\bjtitle{Applied Sciences}
\bvolume{10}(\bissue{4}),
\bfpage{1506}
(\byear{2020})
\end{barticle}
\endbibitem

\bibitem{koker2020cryptocurrency}
\begin{barticle}
\bauthor{\bsnm{Koker}, \binits{T.E.}},
\bauthor{\bsnm{Koutmos}, \binits{D.}}:
\batitle{Cryptocurrency trading using machine learning}.
\bjtitle{Journal of Risk and Financial Management}
\bvolume{13}(\bissue{8}),
\bfpage{178}
(\byear{2020})
\end{barticle}
\endbibitem

\bibitem{atsalakis2019bitcoin}
\begin{barticle}
\bauthor{\bsnm{Atsalakis}, \binits{G.S.}},
\bauthor{\bsnm{Atsalaki}, \binits{I.G.}},
\bauthor{\bsnm{Pasiouras}, \binits{F.}},
\bauthor{\bsnm{Zopounidis}, \binits{C.}}:
\batitle{Bitcoin price forecasting with neuro-fuzzy techniques}.
\bjtitle{European Journal of Operational Research}
\bvolume{276}(\bissue{2}),
\bfpage{770}--\blpage{780}
(\byear{2019})
\end{barticle}
\endbibitem

\bibitem{kim2018time}
\begin{botherref}
\oauthor{\bsnm{Kim}, \binits{K.}},
\oauthor{\bsnm{Kim}, \binits{J.}},
\oauthor{\bsnm{Rinaldo}, \binits{A.}}:
Time series featurization via topological data analysis: an application to
  cryptocurrency trend forecasting.
arXiv preprint arXiv:1812.02987
(2018)
\end{botherref}
\endbibitem

\bibitem{kurbucz2019predicting}
\begin{barticle}
\bauthor{\bsnm{Kurbucz}, \binits{M.T.}}:
\batitle{Predicting the price of bitcoin by the most frequent edges of its
  transaction network}.
\bjtitle{Economics Letters}
\bvolume{184},
\bfpage{108655}
(\byear{2019})
\end{barticle}
\endbibitem

\bibitem{kondor2014inferring}
\begin{barticle}
\bauthor{\bsnm{Kondor}, \binits{D.}},
\bauthor{\bsnm{Csabai}, \binits{I.}},
\bauthor{\bsnm{Sz{\"u}le}, \binits{J.}},
\bauthor{\bsnm{P{\'o}sfai}, \binits{M.}},
\bauthor{\bsnm{Vattay}, \binits{G.}}:
\batitle{Inferring the interplay between network structure and market effects
  in bitcoin}.
\bjtitle{New Journal of Physics}
\bvolume{16}(\bissue{12}),
\bfpage{125003}
(\byear{2014})
\end{barticle}
\endbibitem

\bibitem{kondor2014rich}
\begin{botherref}
\oauthor{\bsnm{Kondor}, \binits{D.}},
\oauthor{\bsnm{P{\'o}sfai}, \binits{M.}},
\oauthor{\bsnm{Csabai}, \binits{I.}},
\oauthor{\bsnm{Vattay}, \binits{G.}}:
Do the rich get richer? an empirical analysis of the bitcoin transaction
  network.
PloS one
\textbf{9}(2)
(2014)
\end{botherref}
\endbibitem

\bibitem{juhasz2018bayesian}
\begin{botherref}
\oauthor{\bsnm{Juh{\'a}sz}, \binits{P.L.}},
\oauthor{\bsnm{St{\'e}ger}, \binits{J.}},
\oauthor{\bsnm{Kondor}, \binits{D.}},
\oauthor{\bsnm{Vattay}, \binits{G.}}:
A bayesian approach to identify bitcoin users.
PloS one
\textbf{13}(12)
(2018)
\end{botherref}
\endbibitem

\bibitem{abay2019chainnet}
\begin{bchapter}
\bauthor{\bsnm{Abay}, \binits{N.C.}},
\bauthor{\bsnm{Akcora}, \binits{C.G.}},
\bauthor{\bsnm{Gel}, \binits{Y.R.}},
\bauthor{\bsnm{Kantarcioglu}, \binits{M.}},
\bauthor{\bsnm{Islambekov}, \binits{U.D.}},
\bauthor{\bsnm{Tian}, \binits{Y.}},
\bauthor{\bsnm{Thuraisingham}, \binits{B.}}:
\bctitle{Chainnet: Learning on blockchain graphs with topological features}.
In: \bbtitle{2019 IEEE International Conference on Data Mining (ICDM)},
pp. \bfpage{946}--\blpage{951}
(\byear{2019}).
\bcomment{IEEE}
\end{bchapter}
\endbibitem

\bibitem{ji2019dynamic}
\begin{barticle}
\bauthor{\bsnm{Ji}, \binits{Q.}},
\bauthor{\bsnm{Bouri}, \binits{E.}},
\bauthor{\bsnm{Lau}, \binits{C.K.M.}},
\bauthor{\bsnm{Roubaud}, \binits{D.}}:
\batitle{Dynamic connectedness and integration in cryptocurrency markets}.
\bjtitle{International Review of Financial Analysis}
\bvolume{63},
\bfpage{257}--\blpage{272}
(\byear{2019})
\end{barticle}
\endbibitem

\bibitem{diebold2014network}
\begin{barticle}
\bauthor{\bsnm{Diebold}, \binits{F.X.}},
\bauthor{\bsnm{Y{\i}lmaz}, \binits{K.}}:
\batitle{On the network topology of variance decompositions: Measuring the
  connectedness of financial firms}.
\bjtitle{Journal of Econometrics}
\bvolume{182}(\bissue{1}),
\bfpage{119}--\blpage{134}
(\byear{2014})
\end{barticle}
\endbibitem

\bibitem{omane2019multiresolution}
\begin{barticle}
\bauthor{\bsnm{Omane-Adjepong}, \binits{M.}},
\bauthor{\bsnm{Alagidede}, \binits{I.P.}}:
\batitle{Multiresolution analysis and spillovers of major cryptocurrency
  markets}.
\bjtitle{Research in International Business and Finance}
\bvolume{49},
\bfpage{191}--\blpage{206}
(\byear{2019})
\end{barticle}
\endbibitem

\bibitem{bouri2020bitcoin}
\begin{barticle}
\bauthor{\bsnm{Bouri}, \binits{E.}},
\bauthor{\bsnm{Roubaud}, \binits{D.}},
\bauthor{\bsnm{Shahzad}, \binits{S.J.H.}}:
\batitle{Do bitcoin and other cryptocurrencies jump together?}
\bjtitle{The Quarterly Review of Economics and Finance}
\bvolume{76},
\bfpage{396}--\blpage{409}
(\byear{2020})
\end{barticle}
\endbibitem

\bibitem{drozdz2020competition}
\begin{barticle}
\bauthor{\bsnm{Dro{\.z}d{\.z}}, \binits{S.}},
\bauthor{\bsnm{Minati}, \binits{L.}},
\bauthor{\bsnm{O{\'s}wiecimka}, \binits{P.}},
\bauthor{\bsnm{Stanuszek}, \binits{M.}},
\bauthor{\bsnm{Watorek}, \binits{M.}}:
\batitle{Competition of noise and collectivity in global cryptocurrency
  trading: Route to a self-contained market}.
\bjtitle{Chaos: An Interdisciplinary Journal of Nonlinear Science}
\bvolume{30}(\bissue{2}),
\bfpage{023122}
(\byear{2020})
\end{barticle}
\endbibitem

\bibitem{hale2018futures}
\begin{botherref}
\oauthor{\bsnm{Hale}, \binits{G.}},
\oauthor{\bsnm{Krishnamurthy}, \binits{A.}},
\oauthor{\bsnm{Kudlyak}, \binits{M.}},
\oauthor{\bsnm{Shultz}, \binits{P.}}, et al.:
How futures trading changed bitcoin prices.
FRBSF Economic Letter
\textbf{12}
(2018)
\end{botherref}
\endbibitem

\bibitem{kristjanpoller2020cryptocurrencies}
\begin{barticle}
\bauthor{\bsnm{Kristjanpoller}, \binits{W.}},
\bauthor{\bsnm{Bouri}, \binits{E.}},
\bauthor{\bsnm{Takaishi}, \binits{T.}}:
\batitle{Cryptocurrencies and equity funds: Evidence from an asymmetric
  multifractal analysis}.
\bjtitle{Physica A: Statistical Mechanics and its Applications}
\bvolume{545},
\bfpage{123711}
(\byear{2020})
\end{barticle}
\endbibitem

\bibitem{bai2019automated}
\begin{botherref}
\oauthor{\bsnm{Bai}, \binits{S.}},
\oauthor{\bsnm{Robinson}, \binits{F.}}:
Automated triangular arbitrage:: A trading algorithm for foreign exchange on a
  cryptocurrency market.
Bachelor's thesis,
KTH Royal Institute of Technology,
Sweden
(2019)
\end{botherref}
\endbibitem

\bibitem{kang2019co}
\begin{botherref}
\oauthor{\bsnm{Kang}, \binits{S.H.}},
\oauthor{\bsnm{McIver}, \binits{R.P.}},
\oauthor{\bsnm{Hernandez}, \binits{J.A.}}:
Co-movements between bitcoin and gold: A wavelet coherence analysis.
Physica A: Statistical Mechanics and its Applications,
120888
(2019)
\end{botherref}
\endbibitem

\bibitem{engle2002dynamic}
\begin{barticle}
\bauthor{\bsnm{Engle}, \binits{R.}}:
\batitle{Dynamic conditional correlation: A simple class of multivariate
  generalized autoregressive conditional heteroskedasticity models}.
\bjtitle{Journal of Business \& Economic Statistics}
\bvolume{20}(\bissue{3}),
\bfpage{339}--\blpage{350}
(\byear{2002})
\end{barticle}
\endbibitem

\bibitem{qiao2020time}
\begin{barticle}
\bauthor{\bsnm{Qiao}, \binits{X.}},
\bauthor{\bsnm{Zhu}, \binits{H.}},
\bauthor{\bsnm{Hau}, \binits{L.}}:
\batitle{Time-frequency co-movement of cryptocurrency return and volatility:
  evidence from wavelet coherence analysis}.
\bjtitle{International Review of Financial Analysis}
\bvolume{71},
\bfpage{101541}
(\byear{2020})
\end{barticle}
\endbibitem

\bibitem{dyhrberg2016bitcoin}
\begin{barticle}
\bauthor{\bsnm{Dyhrberg}, \binits{A.H.}}:
\batitle{Bitcoin, gold and the dollar--a garch volatility analysis}.
\bjtitle{Finance Research Letters}
\bvolume{16},
\bfpage{85}--\blpage{92}
(\byear{2016})
\end{barticle}
\endbibitem

\bibitem{baur2018Bitcoin}
\begin{barticle}
\bauthor{\bsnm{Baur}, \binits{D.G.}},
\bauthor{\bsnm{Dimpfl}, \binits{T.}},
\bauthor{\bsnm{Kuck}, \binits{K.}}:
\batitle{Bitcoin, gold and the us dollar--a replication and extension}.
\bjtitle{Finance Research Letters}
\bvolume{25},
\bfpage{103}--\blpage{110}
(\byear{2018})
\end{barticle}
\endbibitem

\bibitem{bouri2017Bitcoin}
\begin{barticle}
\bauthor{\bsnm{Bouri}, \binits{E.}},
\bauthor{\bsnm{Jalkh}, \binits{N.}},
\bauthor{\bsnm{Moln{\'a}r}, \binits{P.}},
\bauthor{\bsnm{Roubaud}, \binits{D.}}:
\batitle{Bitcoin for energy commodities before and after the december 2013
  crash: diversifier, hedge or safe haven?}
\bjtitle{Applied Economics}
\bvolume{49}(\bissue{50}),
\bfpage{5063}--\blpage{5073}
(\byear{2017})
\end{barticle}
\endbibitem

\bibitem{kakushadze2018cryptoasset}
\begin{botherref}
\oauthor{\bsnm{Kakushadze}, \binits{Z.}}:
Cryptoasset factor models.
Algorithmic Finance
(Preprint),
1--18
(2018)
\end{botherref}
\endbibitem

\bibitem{beneki2019investigating}
\begin{barticle}
\bauthor{\bsnm{Beneki}, \binits{C.}},
\bauthor{\bsnm{Koulis}, \binits{A.}},
\bauthor{\bsnm{Kyriazis}, \binits{N.A.}},
\bauthor{\bsnm{Papadamou}, \binits{S.}}:
\batitle{Investigating volatility transmission and hedging properties between
  bitcoin and ethereum}.
\bjtitle{Research in International Business and Finance}
\bvolume{48},
\bfpage{219}--\blpage{227}
(\byear{2019})
\end{barticle}
\endbibitem

\bibitem{caporale2018day}
\begin{botherref}
\oauthor{\bsnm{Caporale}, \binits{G.M.}},
\oauthor{\bsnm{Plastun}, \binits{A.}}:
The day of the week effect in the cryptocurrency market.
Finance Research Letters
\textbf{31}
(2019)
\end{botherref}
\endbibitem

\bibitem{delfabbro2021cryptocurrency}
\begin{barticle}
\bauthor{\bsnm{Delfabbro}, \binits{P.}},
\bauthor{\bsnm{King}, \binits{D.}},
\bauthor{\bsnm{Williams}, \binits{J.}},
\bauthor{\bsnm{Georgiou}, \binits{N.}}:
\batitle{Cryptocurrency trading, gambling and problem gambling}.
\bjtitle{Addictive Behaviors}
\bvolume{122},
\bfpage{107021}
(\byear{2021})
\end{barticle}
\endbibitem

\bibitem{delfabbro2021psychology}
\begin{botherref}
\oauthor{\bsnm{Delfabbro}, \binits{P.}},
\oauthor{\bsnm{King}, \binits{D.L.}},
\oauthor{\bsnm{Williams}, \binits{J.}}:
The psychology of cryptocurrency trading: Risk and protective factors.
Journal of Behavioral Addictions
(2021)
\end{botherref}
\endbibitem

\bibitem{cheng2020relationship}
\begin{barticle}
\bauthor{\bsnm{Cheng}, \binits{H.-P.}},
\bauthor{\bsnm{Yen}, \binits{K.-C.}}:
\batitle{The relationship between the economic policy uncertainty and the
  cryptocurrency market}.
\bjtitle{Finance Research Letters}
\bvolume{35},
\bfpage{101308}
(\byear{2020})
\end{barticle}
\endbibitem

\bibitem{baker2016measuring}
\begin{barticle}
\bauthor{\bsnm{Baker}, \binits{S.R.}},
\bauthor{\bsnm{Bloom}, \binits{N.}},
\bauthor{\bsnm{Davis}, \binits{S.J.}}:
\batitle{Measuring economic policy uncertainty}.
\bjtitle{The quarterly journal of economics}
\bvolume{131}(\bissue{4}),
\bfpage{1593}--\blpage{1636}
(\byear{2016})
\end{barticle}
\endbibitem

\bibitem{leirvik2021cryptocurrency}
\begin{botherref}
\oauthor{\bsnm{Leirvik}, \binits{T.}}:
Cryptocurrency returns and the volatility of liquidity.
Finance Research Letters,
102031
(2021)
\end{botherref}
\endbibitem

\bibitem{corbet2019cryptocurrencies}
\begin{barticle}
\bauthor{\bsnm{Corbet}, \binits{S.}},
\bauthor{\bsnm{Lucey}, \binits{B.}},
\bauthor{\bsnm{Urquhart}, \binits{A.}},
\bauthor{\bsnm{Yarovaya}, \binits{L.}}:
\batitle{Cryptocurrencies as a financial asset: A systematic analysis}.
\bjtitle{International Review of Financial Analysis}
\bvolume{62},
\bfpage{182}--\blpage{199}
(\byear{2019})
\end{barticle}
\endbibitem

\bibitem{brauneis2019cryptocurrency}
\begin{barticle}
\bauthor{\bsnm{Brauneis}, \binits{A.}},
\bauthor{\bsnm{Mestel}, \binits{R.}}:
\batitle{Cryptocurrency-portfolios in a mean-variance framework}.
\bjtitle{Finance Research Letters}
\bvolume{28},
\bfpage{259}--\blpage{264}
(\byear{2019})
\end{barticle}
\endbibitem

\bibitem{castro2019crypto}
\begin{botherref}
\oauthor{\bsnm{Castro}, \binits{J.G.}},
\oauthor{\bsnm{Tito}, \binits{E.A.H.}},
\oauthor{\bsnm{Brand{\~a}o}, \binits{L.E.T.}},
\oauthor{\bsnm{Gomes}, \binits{L.L.}}:
Crypto-assets portfolio optimization under the omega measure.
The Engineering Economist,
1--21
(2019)
\end{botherref}
\endbibitem

\bibitem{bedi2020investment}
\begin{barticle}
\bauthor{\bsnm{Bedi}, \binits{P.}},
\bauthor{\bsnm{Nashier}, \binits{T.}}:
\batitle{On the investment credentials of bitcoin: A cross-currency
  perspective}.
\bjtitle{Research in International Business and Finance}
\bvolume{51},
\bfpage{101087}
(\byear{2020})
\end{barticle}
\endbibitem

\bibitem{antipova2019building}
\begin{botherref}
\oauthor{\bsnm{Antipova}, \binits{V.}}:
Building and testing global investment portfolios using alternative asset
  classes.
Master's thesis,
Vytautas Magnus University,
Lithuania
(2019)
\end{botherref}
\endbibitem

\bibitem{fantazzini2019multivariate}
\begin{barticle}
\bauthor{\bsnm{Fantazzini}, \binits{D.}},
\bauthor{\bsnm{Zimin}, \binits{S.}}:
\batitle{A multivariate approach for the simultaneous modelling of market risk
  and credit risk for cryptocurrencies}.
\bjtitle{Journal of Industrial and Business Economics}
\bvolume{47}(\bissue{1}),
\bfpage{19}--\blpage{69}
(\byear{2020})
\end{barticle}
\endbibitem

\bibitem{ji2019realised}
\begin{botherref}
\oauthor{\bsnm{Ji}, \binits{Q.}},
\oauthor{\bsnm{Bouri}, \binits{E.}},
\oauthor{\bsnm{Kristoufek}, \binits{L.}},
\oauthor{\bsnm{Lucey}, \binits{B.}}:
Realised volatility connectedness among bitcoin exchange markets.
Finance Research Letters,
101391
(2019)
\end{botherref}
\endbibitem

\bibitem{fasanya2020returns}
\begin{botherref}
\oauthor{\bsnm{Fasanya}, \binits{I.O.}},
\oauthor{\bsnm{Oyewole}, \binits{O.}},
\oauthor{\bsnm{Odudu}, \binits{T.}}:
Returns and volatility spillovers among cryptocurrency portfolios.
International Journal of Managerial Finance
(2020)
\end{botherref}
\endbibitem

\bibitem{trucios2019value}
\begin{botherref}
\oauthor{\bsnm{Truc{\'\i}os}, \binits{C.}},
\oauthor{\bsnm{Tiwari}, \binits{A.K.}},
\oauthor{\bsnm{Alqahtani}, \binits{F.}}:
Value-at-risk and expected shortfall in cryptocurrencies' portfolio: A vine
  copula-based approach.
Available at SSRN 3441892
(2019)
\end{botherref}
\endbibitem

\bibitem{hrytsiuk2019cryptocurrency}
\begin{bchapter}
\bauthor{\bsnm{Hrytsiuk}, \binits{P.}},
\bauthor{\bsnm{Babych}, \binits{T.}},
\bauthor{\bsnm{Bachyshyna}, \binits{L.}}:
\bctitle{Cryptocurrency portfolio optimization using value-at-risk measure}.
In: \bbtitle{Strategies, Models and Technologies of Economic Systems Management
  (SMTESM 2019)}
(\byear{2019}).
\bcomment{Atlantis Press}
\end{bchapter}
\endbibitem

\bibitem{jiang2017cryptocurrency}
\begin{bchapter}
\bauthor{\bsnm{Jiang}, \binits{Z.}},
\bauthor{\bsnm{Liang}, \binits{J.}}:
\bctitle{Cryptocurrency portfolio management with deep reinforcement learning}.
In: \bbtitle{2017 Intelligent Systems Conference (IntelliSys)},
pp. \bfpage{905}--\blpage{913}
(\byear{2017}).
\bcomment{IEEE}
\end{bchapter}
\endbibitem

\bibitem{estalayo2019return}
\begin{bchapter}
\bauthor{\bsnm{Estalayo}, \binits{I.}},
\bauthor{\bsnm{Del~Ser}, \binits{J.}},
\bauthor{\bsnm{Osaba}, \binits{E.}},
\bauthor{\bsnm{Bilbao}, \binits{M.N.}},
\bauthor{\bsnm{Muhammad}, \binits{K.}},
\bauthor{\bsnm{G{\'a}lvez}, \binits{A.}},
\bauthor{\bsnm{Iglesias}, \binits{A.}}:
\bctitle{Return, diversification and risk in cryptocurrency portfolios using
  deep recurrent neural networks and multi-objective evolutionary algorithms}.
In: \bbtitle{2019 IEEE Congress on Evolutionary Computation (CEC)},
pp. \bfpage{755}--\blpage{761}
(\byear{2019}).
\bcomment{IEEE}
\end{bchapter}
\endbibitem

\bibitem{cheung2015crypto}
\begin{barticle}
\bauthor{\bsnm{Cheung}, \binits{A.}},
\bauthor{\bsnm{Roca}, \binits{E.}},
\bauthor{\bsnm{Su}, \binits{J.-J.}}:
\batitle{Crypto-currency bubbles: an application of the phillips--shi--yu
  (2013) methodology on mt. gox bitcoin prices}.
\bjtitle{Applied Economics}
\bvolume{47}(\bissue{23}),
\bfpage{2348}--\blpage{2358}
(\byear{2015})
\end{barticle}
\endbibitem

\bibitem{corbet2018datestamping}
\begin{barticle}
\bauthor{\bsnm{Corbet}, \binits{S.}},
\bauthor{\bsnm{Lucey}, \binits{B.}},
\bauthor{\bsnm{Yarovaya}, \binits{L.}}:
\batitle{Datestamping the bitcoin and ethereum bubbles}.
\bjtitle{Finance Research Letters}
\bvolume{26},
\bfpage{81}--\blpage{88}
(\byear{2018})
\end{barticle}
\endbibitem

\bibitem{bouri2019co}
\begin{barticle}
\bauthor{\bsnm{Bouri}, \binits{E.}},
\bauthor{\bsnm{Shahzad}, \binits{S.J.H.}},
\bauthor{\bsnm{Roubaud}, \binits{D.}}:
\batitle{Co-explosivity in the cryptocurrency market}.
\bjtitle{Finance Research Letters}
\bvolume{29},
\bfpage{178}--\blpage{183}
(\byear{2019})
\end{barticle}
\endbibitem

\bibitem{phillips2015testinga}
\begin{barticle}
\bauthor{\bsnm{Phillips}, \binits{P.C.}},
\bauthor{\bsnm{Shi}, \binits{S.}},
\bauthor{\bsnm{Yu}, \binits{J.}}:
\batitle{Testing for multiple bubbles: Historical episodes of exuberance and
  collapse in the s\&p 500}.
\bjtitle{International economic review}
\bvolume{56}(\bissue{4}),
\bfpage{1043}--\blpage{1078}
(\byear{2015})
\end{barticle}
\endbibitem

\bibitem{phillips2015testingb}
\begin{barticle}
\bauthor{\bsnm{Phillips}, \binits{P.C.}},
\bauthor{\bsnm{Shi}, \binits{S.}},
\bauthor{\bsnm{Yu}, \binits{J.}}:
\batitle{Testing for multiple bubbles: Limit theory of real-time detectors}.
\bjtitle{International Economic Review}
\bvolume{56}(\bissue{4}),
\bfpage{1079}--\blpage{1134}
(\byear{2015})
\end{barticle}
\endbibitem

\bibitem{enoksen2019can}
\begin{botherref}
\oauthor{\bsnm{Enoksen}, \binits{F.A.}},
\oauthor{\bsnm{Landsnes}, \binits{C.J.}}:
What can predict bubbles in cryptocurrency prices?
Master's thesis,
University of Stavanger, Norway
(2019)
\end{botherref}
\endbibitem

\bibitem{phillips2017predicting}
\begin{bchapter}
\bauthor{\bsnm{Phillips}, \binits{R.C.}},
\bauthor{\bsnm{Gorse}, \binits{D.}}:
\bctitle{Predicting cryptocurrency price bubbles using social media data and
  epidemic modelling}.
In: \bbtitle{2017 IEEE Symposium Series on Computational Intelligence (SSCI)},
pp. \bfpage{1}--\blpage{7}
(\byear{2017}).
\bcomment{IEEE}
\end{bchapter}
\endbibitem

\bibitem{caporale2019price}
\begin{botherref}
\oauthor{\bsnm{Caporale}, \binits{G.M.}},
\oauthor{\bsnm{Plastun}, \binits{A.}}:
Price overreactions in the cryptocurrency market.
CESifo Working Paper 6861,
Munich
(2018).
\url{http://hdl.handle.net/10419/174984}
\end{botherref}
\endbibitem

\bibitem{chaim2018volatility}
\begin{barticle}
\bauthor{\bsnm{Chaim}, \binits{P.}},
\bauthor{\bsnm{Laurini}, \binits{M.P.}}:
\batitle{Volatility and return jumps in bitcoin}.
\bjtitle{Economics Letters}
\bvolume{173},
\bfpage{158}--\blpage{163}
(\byear{2018})
\end{barticle}
\endbibitem

\bibitem{cross2021returns}
\begin{botherref}
\oauthor{\bsnm{Cross}, \binits{J.L.}},
\oauthor{\bsnm{Hou}, \binits{C.}},
\oauthor{\bsnm{Trinh}, \binits{K.}}:
Returns, volatility and the cryptocurrency bubble of 2017-18.
Economic Modelling,
105643
(2021)
\end{botherref}
\endbibitem

\bibitem{katsiampa2018cryptocurrency}
\begin{botherref}
\oauthor{\bsnm{Katsiampa}, \binits{P.}},
\oauthor{\bsnm{Gkillas}, \binits{K.}},
\oauthor{\bsnm{Longin}, \binits{F.}}:
Cryptocurrency market activity during extremely volatile periods
(2018).
doi:\doiurl{10.2139/ssrn.3220781}
\end{botherref}
\endbibitem

\bibitem{yaya2018persistent}
\begin{botherref}
\oauthor{\bsnm{Yaya}, \binits{O.S.}},
\oauthor{\bsnm{Ogbonna}, \binits{E.A.}},
\oauthor{\bsnm{Olubusoye}, \binits{O.E.}}:
How persistent and dependent are pricing of bitcoin to other cryptocurrencies
  before and after 2017/18 crash?
(2018)
\end{botherref}
\endbibitem

\bibitem{manahov2021cryptocurrency}
\begin{barticle}
\bauthor{\bsnm{Manahov}, \binits{V.}}:
\batitle{Cryptocurrency liquidity during extreme price movements: is there a
  problem with virtual money?}
\bjtitle{Quantitative Finance}
\bvolume{21}(\bissue{2}),
\bfpage{341}--\blpage{360}
(\byear{2021})
\end{barticle}
\endbibitem

\bibitem{shahzad2021extreme}
\begin{botherref}
\oauthor{\bsnm{Shahzad}, \binits{S.J.H.}},
\oauthor{\bsnm{Bouri}, \binits{E.}},
\oauthor{\bsnm{Ahmad}, \binits{T.}},
\oauthor{\bsnm{Naeem}, \binits{M.A.}}:
Extreme tail network analysis of cryptocurrencies and trading strategies.
Finance Research Letters,
102106
(2021)
\end{botherref}
\endbibitem

\bibitem{chan2022extreme}
\begin{barticle}
\bauthor{\bsnm{Chan}, \binits{S.}},
\bauthor{\bsnm{Chu}, \binits{J.}},
\bauthor{\bsnm{Zhang}, \binits{Y.}},
\bauthor{\bsnm{Nadarajah}, \binits{S.}}:
\batitle{An extreme value analysis of the tail relationships between returns
  and volumes for high frequency cryptocurrencies}.
\bjtitle{Research in International Business and Finance}
\bvolume{59},
\bfpage{101541}
(\byear{2022})
\end{barticle}
\endbibitem

\bibitem{krafft2018experimental}
\begin{bchapter}
\bauthor{\bsnm{Krafft}, \binits{P.M.}},
\bauthor{\bsnm{Della~Penna}, \binits{N.}},
\bauthor{\bsnm{Pentland}, \binits{A.S.}}:
\bctitle{An experimental study of cryptocurrency market dynamics}.
In: \bbtitle{Proceedings of the 2018 CHI Conference on Human Factors in
  Computing Systems},
p. \bfpage{605}
(\byear{2018}).
\bcomment{ACM}
\end{bchapter}
\endbibitem

\bibitem{yang2018behavioral}
\begin{barticle}
\bauthor{\bsnm{Yang}, \binits{H.}}:
\batitle{Behavioral anomalies in cryptocurrency markets}.
\bjtitle{Available at SSRN 3174421}
(\byear{2018}).
doi:\doiurl{10.2139/ssrn.3174421}
\end{barticle}
\endbibitem

\bibitem{cocco2016modeling}
\begin{botherref}
\oauthor{\bsnm{Cocco}, \binits{L.}},
\oauthor{\bsnm{Marchesi}, \binits{M.}}:
Modeling and simulation of the economics of mining in the bitcoin market.
PloS one
\textbf{11}(10)
(2016)
\end{botherref}
\endbibitem

\bibitem{leclair2018herding}
\begin{botherref}
\oauthor{\bsnm{Leclair}, \binits{E.M.}}:
Herding in the cryptocurrency market.
Econ 5029 final research,
Carleton University,
Canada
(2018)
\end{botherref}
\endbibitem

\bibitem{vidal2019herding}
\begin{barticle}
\bauthor{\bsnm{Vidal-Tom{\'a}s}, \binits{D.}},
\bauthor{\bsnm{Ib{\'a}{\~n}ez}, \binits{A.M.}},
\bauthor{\bsnm{Farin{\'o}s}, \binits{J.E.}}:
\batitle{Herding in the cryptocurrency market: Cssd and csad approaches}.
\bjtitle{Finance Research Letters}
\bvolume{30},
\bfpage{181}--\blpage{186}
(\byear{2019})
\end{barticle}
\endbibitem

\bibitem{hwang2004market}
\begin{barticle}
\bauthor{\bsnm{Hwang}, \binits{S.}},
\bauthor{\bsnm{Salmon}, \binits{M.}}:
\batitle{Market stress and herding}.
\bjtitle{Journal of Empirical Finance}
\bvolume{11}(\bissue{4}),
\bfpage{585}--\blpage{616}
(\byear{2004})
\end{barticle}
\endbibitem

\bibitem{makarov2019trading}
\begin{barticle}
\bauthor{\bsnm{Makarov}, \binits{I.}},
\bauthor{\bsnm{Schoar}, \binits{A.}}:
\batitle{Trading and arbitrage in cryptocurrency markets}.
\bjtitle{Journal of Financial Economics}
\bvolume{135}(\bissue{2}),
\bfpage{293}--\blpage{319}
(\byear{2020})
\end{barticle}
\endbibitem

\bibitem{liu2021call}
\begin{bchapter}
\bauthor{\bsnm{Liu}, \binits{B.}},
\bauthor{\bsnm{Polukarov}, \binits{M.}},
\bauthor{\bsnm{Ventre}, \binits{C.}},
\bauthor{\bsnm{Li}, \binits{L.}},
\bauthor{\bsnm{Kanthan}, \binits{L.}}:
\bctitle{Call markets with adaptive clearing intervals}.
In: \bbtitle{Proceedings of the 20th International Conference on Autonomous
  Agents and MultiAgent Systems},
pp. \bfpage{1587}--\blpage{1589}
(\byear{2021})
\end{bchapter}
\endbibitem

\bibitem{king2021herding}
\begin{barticle}
\bauthor{\bsnm{King}, \binits{T.}},
\bauthor{\bsnm{Koutmos}, \binits{D.}}:
\batitle{Herding and feedback trading in cryptocurrency markets}.
\bjtitle{Annals of Operations Research}
\bvolume{300}(\bissue{1}),
\bfpage{79}--\blpage{96}
(\byear{2021})
\end{barticle}
\endbibitem

\bibitem{griffin2019bitcoin}
\begin{botherref}
\oauthor{\bsnm{Griffin}, \binits{J.M.}},
\oauthor{\bsnm{Shams}, \binits{A.}}:
Is bitcoin really un-tethered?
Available at SSRN 3195066
(2019)
\end{botherref}
\endbibitem

\bibitem{hileman2017global}
\begin{botherref}
\oauthor{\bsnm{Hileman}, \binits{G.}},
\oauthor{\bsnm{Rauchs}, \binits{M.}}:
Global cryptocurrency benchmarking study.
Cambridge Centre for Alternative Finance
\textbf{33}
(2017)
\end{botherref}
\endbibitem

\bibitem{zhou2018algorithmic}
\begin{barticle}
\bauthor{\bsnm{Zhou}, \binits{H.}},
\bauthor{\bsnm{Kalev}, \binits{P.S.}}:
\batitle{Algorithmic and high frequency trading in asia-pacific, now and the
  future}.
\bjtitle{Pacific-Basin Finance Journal}
\bvolume{53},
\bfpage{186}--\blpage{207}
(\byear{2019})
\end{barticle}
\endbibitem

\bibitem{shanaev2020taming}
\begin{barticle}
\bauthor{\bsnm{Shanaev}, \binits{S.}},
\bauthor{\bsnm{Sharma}, \binits{S.}},
\bauthor{\bsnm{Ghimire}, \binits{B.}},
\bauthor{\bsnm{Shuraeva}, \binits{A.}}:
\batitle{Taming the blockchain beast? regulatory implications for the
  cryptocurrency market}.
\bjtitle{Research in International Business and Finance}
\bvolume{51},
\bfpage{101080}
(\byear{2020})
\end{barticle}
\endbibitem

\bibitem{feinstein2021impact}
\begin{barticle}
\bauthor{\bsnm{Feinstein}, \binits{B.D.}},
\bauthor{\bsnm{Werbach}, \binits{K.}}:
\batitle{The impact of cryptocurrency regulation on trading markets}.
\bjtitle{Journal of Financial Regulation}
\bvolume{7}(\bissue{1}),
\bfpage{48}--\blpage{99}
(\byear{2021})
\end{barticle}
\endbibitem

\bibitem{patil2018study}
\begin{bchapter}
\bauthor{\bsnm{Patil}, \binits{A.P.}},
\bauthor{\bsnm{Akarsh}, \binits{T.}},
\bauthor{\bsnm{Parkavi}, \binits{A.}}:
\bctitle{A study of opinion mining and data mining techniques to analyse the
  cryptocurrency market}.
In: \bbtitle{2018 3rd International Conference on Computational Systems and
  Information Technology for Sustainable Solutions (CSITSS)},
pp. \bfpage{198}--\blpage{203}
(\byear{2018}).
\bcomment{IEEE}
\end{bchapter}
\endbibitem

\bibitem{sigaki2019clustering}
\begin{barticle}
\bauthor{\bsnm{Sigaki}, \binits{H.Y.}},
\bauthor{\bsnm{Perc}, \binits{M.}},
\bauthor{\bsnm{Ribeiro}, \binits{H.V.}}:
\batitle{Clustering patterns in efficiency and the coming-of-age of the
  cryptocurrency market}.
\bjtitle{Scientific reports}
\bvolume{9}(\bissue{1}),
\bfpage{1440}
(\byear{2019})
\end{barticle}
\endbibitem

\bibitem{cocco2017using}
\begin{barticle}
\bauthor{\bsnm{Cocco}, \binits{L.}},
\bauthor{\bsnm{Concas}, \binits{G.}},
\bauthor{\bsnm{Marchesi}, \binits{M.}}:
\batitle{Using an artificial financial market for studying a cryptocurrency
  market}.
\bjtitle{Journal of Economic Interaction and Coordination}
\bvolume{12}(\bissue{2}),
\bfpage{345}--\blpage{365}
(\byear{2017})
\end{barticle}
\endbibitem

\bibitem{aspris2021decentralized}
\begin{barticle}
\bauthor{\bsnm{Aspris}, \binits{A.}},
\bauthor{\bsnm{Foley}, \binits{S.}},
\bauthor{\bsnm{Svec}, \binits{J.}},
\bauthor{\bsnm{Wang}, \binits{L.}}:
\batitle{Decentralized exchanges: The “wild west” of cryptocurrency
  trading}.
\bjtitle{International Review of Financial Analysis}
\bvolume{77},
\bfpage{101845}
(\byear{2021})
\end{barticle}
\endbibitem

\bibitem{ogorevc2019cryptocurrency}
\begin{botherref}
\oauthor{\bsnm{Ogorevc}, \binits{M.}}:
Cryptocurrency as money: A trading strategy solution.
Available at SSRN 3436041
(2019)
\end{botherref}
\endbibitem

\bibitem{gandal2014competition}
\begin{botherref}
\oauthor{\bsnm{Gandal}, \binits{N.}},
\oauthor{\bsnm{Halaburda}, \binits{H.}}:
Competition in the cryptocurrency market.
(2014).
CEPR Discussion Paper No. DP10157
\end{botherref}
\endbibitem

\bibitem{bariviera2020we}
\begin{botherref}
\oauthor{\bsnm{Bariviera}, \binits{A.F.}},
\oauthor{\bsnm{Merediz-Sola}, \binits{I.}}:
Where do we stand in cryptocurrencies economic research? a survey based on
  hybrid analysis.
arXiv preprint arXiv:2003.09723
(2020)
\end{botherref}
\endbibitem

\bibitem{hansel2018cryptocurrency}
\begin{botherref}
\oauthor{\bsnm{Hansel}, \binits{D.}}:
Cryptocurrency trading: How to make money by trading bitcoin and other
  cryptocurrency (volume 2)
(2018)
\end{botherref}
\endbibitem

\bibitem{kate2018cryptocurrency}
\begin{botherref}
\oauthor{\bsnm{Kate}, \binits{C.}}:
Cryptocurrency trading for beginners: 6-steps action plan to your first
  investment
(2018)
\end{botherref}
\endbibitem

\bibitem{garza2019formal}
\begin{botherref}
\oauthor{\bsnm{Garza}, \binits{P.}}:
Formal automatic trading in the cryptocurrency era.
PhD thesis,
Politecnico di Torino
(2019)
\end{botherref}
\endbibitem

\bibitem{ward2018algorithmic}
\begin{botherref}
\oauthor{\bsnm{Ward}, \binits{M.}}:
Algorithmic trading for cryptocurrencies.
Departmental honors thesis,
UtahState University,
United States
(2018)
\end{botherref}
\endbibitem

\bibitem{fantazzini2019quantitative}
\begin{botherref}
\oauthor{\bsnm{Fantazzini}, \binits{D.}}:
Quantitative finance with r and cryptocurrencies.
Amazon KDP, ISBN-13,
978--1090685315
(2019)
\end{botherref}
\endbibitem

\bibitem{blockchaincommunity}
\begin{botherref}
\oauthor{\bparticle{research} \bsnm{network}, \binits{B.}}:
Platform for scholarly communication about cryptocurrencies and blockchains.
\url{https://www.blockchainresearchnetwork.org/}.
[Online, Accessed: April 17, 2020]
(2020)
\end{botherref}
\endbibitem

\bibitem{nakamoto2009bitcoin}
\begin{botherref}
\oauthor{\bsnm{Nakamoto}, \binits{S.}}:
Bitcoin open source implementation of p2p currency.
P2P foundation
\textbf{18}
(2009)
\end{botherref}
\endbibitem

\bibitem{cheah2015speculative}
\begin{barticle}
\bauthor{\bsnm{Cheah}, \binits{E.-T.}},
\bauthor{\bsnm{Fry}, \binits{J.}}:
\batitle{Speculative bubbles in bitcoin markets? an empirical investigation
  into the fundamental value of bitcoin}.
\bjtitle{Economics Letters}
\bvolume{130},
\bfpage{32}--\blpage{36}
(\byear{2015})
\end{barticle}
\endbibitem

\bibitem{mcnally2016predicting}
\begin{botherref}
\oauthor{\bsnm{McNally}, \binits{S.}}:
Predicting the price of bitcoin using machine learning.
PhD thesis,
Dublin, National College of Ireland
(2016)
\end{botherref}
\endbibitem

\bibitem{bell2016bitcoin}
\begin{botherref}
\oauthor{\bsnm{Bell}, \binits{T.}}:
Bitcoin trading agents.
University of Southampton
(2016)
\end{botherref}
\endbibitem

\bibitem{zbikowski2016application}
\begin{bchapter}
\bauthor{\bsnm{{\.Z}bikowski}, \binits{K.}}:
\bctitle{Application of machine learning algorithms for bitcoin automated
  trading}.
In: \bbtitle{Machine Intelligence and Big Data in Industry},
pp. \bfpage{161}--\blpage{168}.
\bpublisher{Springer}, \blocation{???}
(\byear{2016})
\end{bchapter}
\endbibitem

\bibitem{cryptochina}
\begin{botherref}
\oauthor{\bsnm{Reuters}}:
China bans financial, payment institutions from cryptocurrency business.
\url{https://www.reuters.com/technology/chinese-financial-payment-bodies-barred-cryptocurrency-business-2021-05-18/}.
[Online, Accessed: May 18, 2021]
(2021)
\end{botherref}
\endbibitem

\bibitem{cryptolegal}
\begin{botherref}
\oauthor{\bsnm{MercoPress}}:
Bitcoin legal tender in El Salvador, first country ever.
\url{https://en.mercopress.com/2021/06/10/bitcoin-legal-tender-in-el-salvador-first-country-ever}.
[Online, Accessed: June 10, 2021]
(2021)
\end{botherref}
\endbibitem

\bibitem{bach2018machine}
\begin{botherref}
\oauthor{\bsnm{Bach}, \binits{W.G.}},
\oauthor{\bsnm{Kasper}, \binits{L.}}:
On machine learning based cryptocurrency trading.
Master's thesis,
Aalborg University,
Denmark
(2018)
\end{botherref}
\endbibitem

\bibitem{siaminos2019predicting}
\begin{botherref}
\oauthor{\bsnm{Siaminos}, \binits{G.}}:
Predicting the value of cryptocurrencies using machine learning time series
  analysis time series analysis time
(2019)
\end{botherref}
\endbibitem

\bibitem{alexander2020critical}
\begin{barticle}
\bauthor{\bsnm{Alexander}, \binits{C.}},
\bauthor{\bsnm{Dakos}, \binits{M.}}:
\batitle{A critical investigation of cryptocurrency data and analysis}.
\bjtitle{Quantitative Finance}
\bvolume{20}(\bissue{2}),
\bfpage{173}--\blpage{188}
(\byear{2020})
\end{barticle}
\endbibitem

\bibitem{guo2018Bitcoin}
\begin{bchapter}
\bauthor{\bsnm{Guo}, \binits{T.}},
\bauthor{\bsnm{Bifet}, \binits{A.}},
\bauthor{\bsnm{Antulov-Fantulin}, \binits{N.}}:
\bctitle{Bitcoin volatility forecasting with a glimpse into buy and sell
  orders}.
In: \bbtitle{2018 IEEE International Conference on Data Mining (ICDM)},
pp. \bfpage{989}--\blpage{994}
(\byear{2018}).
\bcomment{IEEE}
\end{bchapter}
\endbibitem

\bibitem{phillips2018cryptocurrency}
\begin{barticle}
\bauthor{\bsnm{Phillips}, \binits{R.C.}},
\bauthor{\bsnm{Gorse}, \binits{D.}}:
\batitle{Cryptocurrency price drivers: Wavelet coherence analysis revisited}.
\bjtitle{PloS one}
\bvolume{13}(\bissue{4}),
\bfpage{0195200}
(\byear{2018})
\end{barticle}
\endbibitem

\bibitem{kang2019whose}
\begin{botherref}
\oauthor{\bsnm{Kang}, \binits{K.}},
\oauthor{\bsnm{Choo}, \binits{J.}},
\oauthor{\bsnm{Kim}, \binits{Y.}}:
Whose opinion matters? analyzing relationships between bitcoin prices and user
  groups in online community.
Social Science Computer Review,
0894439319840716
(2019)
\end{botherref}
\endbibitem

\bibitem{Kondor_2014}
\begin{barticle}
\bauthor{\bsnm{Kondor}, \binits{D.}},
\bauthor{\bsnm{Csabai}, \binits{I.}},
\bauthor{\bsnm{Szule}, \binits{J.}},
\bauthor{\bsnm{Psfai}, \binits{M.}},
\bauthor{\bsnm{Vattay}, \binits{G.}}:
\batitle{Inferring the interplay between network structure and market effects
  in bitcoin}.
\bjtitle{New Journal of Physics}
\bvolume{16}(\bissue{12}),
\bfpage{125003}
(\byear{2014}).
doi:\doiurl{10.1088/1367-2630/16/12/125003}
\end{barticle}
\endbibitem

\bibitem{senseable}
\begin{botherref}
\oauthor{\bsnm{MIT}}:
Bitcoin network dataset.
\url{https://senseable2015-6.mit.edu/bitcoin}.
[Online, Accessed January 11, 2020]
(2015)
\end{botherref}
\endbibitem

\bibitem{zha2020opinion}
\begin{barticle}
\bauthor{\bsnm{Zha}, \binits{Q.}},
\bauthor{\bsnm{Kou}, \binits{G.}},
\bauthor{\bsnm{Zhang}, \binits{H.}},
\bauthor{\bsnm{Liang}, \binits{H.}},
\bauthor{\bsnm{Chen}, \binits{X.}},
\bauthor{\bsnm{Li}, \binits{C.-C.}},
\bauthor{\bsnm{Dong}, \binits{Y.}}:
\batitle{Opinion dynamics in finance and business: a literature review and
  research opportunities}.
\bjtitle{Financial Innovation}
\bvolume{6}(\bissue{1}),
\bfpage{1}--\blpage{22}
(\byear{2020})
\end{barticle}
\endbibitem

\bibitem{luu2019spillover}
\begin{barticle}
\bauthor{\bsnm{Luu Duc~Huynh}, \binits{T.}}:
\batitle{Spillover risks on cryptocurrency markets: A look from var-svar
  granger causality and student copulas}.
\bjtitle{Journal of Risk and Financial Management}
\bvolume{12}(\bissue{2}),
\bfpage{52}
(\byear{2019})
\end{barticle}
\endbibitem

\bibitem{weber2007relation}
\begin{barticle}
\bauthor{\bsnm{Weber}, \binits{P.}},
\bauthor{\bsnm{Wang}, \binits{F.}},
\bauthor{\bsnm{Vodenska-Chitkushev}, \binits{I.}},
\bauthor{\bsnm{Havlin}, \binits{S.}},
\bauthor{\bsnm{Stanley}, \binits{H.E.}}:
\batitle{Relation between volatility correlations in financial markets and
  omori processes occurring on all scales}.
\bjtitle{Physical Review E}
\bvolume{76}(\bissue{1}),
\bfpage{016109}
(\byear{2007})
\end{barticle}
\endbibitem

\bibitem{mclean2016does}
\begin{barticle}
\bauthor{\bsnm{McLean}, \binits{R.D.}},
\bauthor{\bsnm{Pontiff}, \binits{J.}}:
\batitle{Does academic research destroy stock return predictability?}
\bjtitle{The Journal of Finance}
\bvolume{71}(\bissue{1}),
\bfpage{5}--\blpage{32}
(\byear{2016})
\end{barticle}
\endbibitem

\end{thebibliography}

\newcommand{\BMCxmlcomment}[1]{}

\BMCxmlcomment{

<refgrp>

<bibl id="B1">
  <title><p>An analysis of the cryptocurrency industry</p></title>
  <aug>
    <au><snm>Farell</snm><fnm>R</fnm></au>
  </aug>
  <pubdate>2015</pubdate>
</bibl>

<bibl id="B2">
  <title><p>An evaluation of recent evidence on stock market
  bubbles</p></title>
  <aug>
    <au><snm>Flood</snm><fnm>R</fnm></au>
    <au><snm>Hodrick</snm><fnm>RJ</fnm></au>
    <au><snm>Kaplan</snm><fnm>P</fnm></au>
  </aug>
  <source>National Bureau of Economic Research Cambridge, Mass., USA</source>
  <pubdate>1986</pubdate>
</bibl>

<bibl id="B3">
  <title><p>A survey on efficiency and profitable trading opportunities in
  cryptocurrency markets</p></title>
  <aug>
    <au><snm>Kyriazis</snm><fnm>NA</fnm></au>
  </aug>
  <source>Journal of Risk and Financial Management</source>
  <publisher>Multidisciplinary Digital Publishing Institute</publisher>
  <pubdate>2019</pubdate>
  <volume>12</volume>
  <issue>2</issue>
  <fpage>67</fpage>
</bibl>

<bibl id="B4">
  <title><p>A survey on crypto currencies</p></title>
  <aug>
    <au><snm>Ahamad</snm><fnm>S</fnm></au>
    <au><snm>Nair</snm><fnm>M</fnm></au>
    <au><snm>Varghese</snm><fnm>B</fnm></au>
  </aug>
  <source>4th International Conference on Advances in Computer Science,
  AETACS</source>
  <pubdate>2013</pubdate>
  <fpage>42</fpage>
  <lpage>-48</lpage>
</bibl>

<bibl id="B5">
  <title><p>Survey paper on cryptocurrency</p></title>
  <aug>
    <au><snm>Sharma</snm><fnm>S</fnm></au>
    <au><snm>Krishma</snm><fnm>NN</fnm></au>
    <au><snm>Raina</snm><fnm>EC</fnm></au>
  </aug>
  <source>International Journal of Scientific Research in Computer Science,
  Engineering and Information Technology Vol. 2 Issue</source>
  <pubdate>2017</pubdate>
  <volume>3</volume>
  <fpage>307</fpage>
  <lpage>-310</lpage>
</bibl>

<bibl id="B6">
  <title><p>A brief survey of cryptocurrency systems</p></title>
  <aug>
    <au><snm>Mukhopadhyay</snm><fnm>U</fnm></au>
    <au><snm>Skjellum</snm><fnm>A</fnm></au>
    <au><snm>Hambolu</snm><fnm>O</fnm></au>
    <au><snm>Oakley</snm><fnm>J</fnm></au>
    <au><snm>Yu</snm><fnm>L</fnm></au>
    <au><snm>Brooks</snm><fnm>R</fnm></au>
  </aug>
  <source>2016 14th annual conference on privacy, security and trust
  (PST)</source>
  <pubdate>2016</pubdate>
  <fpage>745</fpage>
  <lpage>-752</lpage>
</bibl>

<bibl id="B7">
  <title><p>A bibliometric analysis of bitcoin scientific
  production</p></title>
  <aug>
    <au><snm>Merediz Sol{\`a}</snm><fnm>I</fnm></au>
    <au><snm>Bariviera</snm><fnm>AF</fnm></au>
  </aug>
  <source>Research in International Business and Finance</source>
  <publisher>Elsevier</publisher>
  <pubdate>2019</pubdate>
  <volume>50</volume>
  <fpage>294</fpage>
  <lpage>-305</lpage>
</bibl>

<bibl id="B8">
  <title><p>Blockchain revolution: how the technology behind bitcoin is
  changing money, business, and the world</p></title>
  <aug>
    <au><snm>Tapscott</snm><fnm>D</fnm></au>
    <au><snm>Tapscott</snm><fnm>A</fnm></au>
  </aug>
  <publisher>Penguin</publisher>
  <pubdate>2016</pubdate>
</bibl>

<bibl id="B9">
  <title><p>How do blockchain mining and transactions work explained in 7
  simple steps</p></title>
  <aug>
    <au><snm>S.</snm><fnm>J</fnm></au>
  </aug>
  <source>\url{https://blog.goodaudience.com/how-a-miner-adds-transactions-to-the-blockchain-in-seven-steps-856053271476}</source>
  <pubdate>2018</pubdate>
  <note>[Online, Accessed: January 26, 2020]</note>
</bibl>

<bibl id="B10">
  <title><p>A forensic look at bitcoin cryptocurrency</p></title>
  <aug>
    <au><snm>Doran</snm><fnm>MD</fnm></au>
  </aug>
  <source>PhD thesis</source>
  <publisher>Utica College</publisher>
  <pubdate>2014</pubdate>
</bibl>

<bibl id="B11">
  <title><p>Blockchain 101: What is Blockchain and How Does This Revolutionary
  Technology Work?</p></title>
  <aug>
    <au><snm>Meunier</snm><fnm>S</fnm></au>
  </aug>
  <source>Transforming Climate Finance and Green Investment with
  Blockchains</source>
  <publisher>Elsevier</publisher>
  <pubdate>2018</pubdate>
  <fpage>23</fpage>
  <lpage>-34</lpage>
</bibl>

<bibl id="B12">
  <title><p>Bitcoin and cryptocurrency technologies: a comprehensive
  introduction</p></title>
  <aug>
    <au><snm>Narayanan</snm><fnm>A</fnm></au>
    <au><snm>Bonneau</snm><fnm>J</fnm></au>
    <au><snm>Felten</snm><fnm>E</fnm></au>
    <au><snm>Miller</snm><fnm>A</fnm></au>
    <au><snm>Goldfeder</snm><fnm>S</fnm></au>
  </aug>
  <publisher>Princeton University Press</publisher>
  <pubdate>2016</pubdate>
</bibl>

<bibl id="B13">
  <title><p>Designated-verifier proof of assets for bitcoin exchange using
  elliptic curve cryptography</p></title>
  <aug>
    <au><snm>Wang</snm><fnm>H</fnm></au>
    <au><snm>He</snm><fnm>D</fnm></au>
    <au><snm>Ji</snm><fnm>Y</fnm></au>
  </aug>
  <source>Future Generation Computer Systems</source>
  <publisher>Elsevier</publisher>
  <pubdate>2017</pubdate>
</bibl>

<bibl id="B14">
  <title><p>Elliptic-Curve Cryptography</p></title>
  <aug>
    <au><cnm>Grayblock</cnm></au>
  </aug>
  <source>\url{https://medium.com/coinmonks/elliptic-curve-cryptography-6de8fc748b8b}</source>
  <pubdate>2018</pubdate>
  <note>[Online, Accessed December 29, 2019]</note>
</bibl>

<bibl id="B15">
  <title><p>Cryptocurrency and the Problem of Intermediation</p></title>
  <aug>
    <au><snm>Harwick</snm><fnm>C</fnm></au>
  </aug>
  <source>The Independent Review</source>
  <publisher>JSTOR</publisher>
  <pubdate>2016</pubdate>
  <volume>20</volume>
  <issue>4</issue>
  <fpage>569</fpage>
  <lpage>-588</lpage>
</bibl>

<bibl id="B16">
  <title><p>The evolution of digital currencies: Bitcoin, A cryptocurrency
  causing A monetary revolution</p></title>
  <aug>
    <au><snm>Rose</snm><fnm>C</fnm></au>
  </aug>
  <source>International Business \& Economics Research Journal (IBER)</source>
  <pubdate>2015</pubdate>
  <volume>14</volume>
  <issue>4</issue>
  <fpage>617</fpage>
  <lpage>-622</lpage>
</bibl>

<bibl id="B17">
  <title><p>Digital Asset Market Evolution</p></title>
  <aug>
    <au><snm>Kaal</snm><fnm>WA</fnm></au>
  </aug>
  <source>Journal of Corporation Law</source>
  <pubdate>2020</pubdate>
  <fpage>20</fpage>
  <lpage>-02</lpage>
</bibl>

<bibl id="B18">
  <title><p>Top 100 Cryptocurrencies by Market Capitalization</p></title>
  <aug>
    <au><cnm>CoinMaketCap</cnm></au>
  </aug>
  <pubdate>2019</pubdate>
  <url>https://coinmarketcap.com/</url>
</bibl>

<bibl id="B19">
  <title><p>Total Crypto Market Capitalization and Volume</p></title>
  <aug>
    <au><cnm>TradingView</cnm></au>
  </aug>
  <source>\url{https://www.tradingview.com/markets/cryptocurrencies/global-charts/}</source>
  <pubdate>2021</pubdate>
  <note>[Online, Accessed September 10, 2021]</note>
</bibl>

<bibl id="B20">
  <title><p>The Main Roadblocks To Crypto Moving Mainstream</p></title>
  <aug>
    <au><snm>Council</snm><fnm>FB</fnm></au>
  </aug>
  <source>\url{https://www.forbes.com/sites/forbesbusinesscouncil/2021/06/23/the-main-roadblocks-to-crypto-moving-mainstream/?sh=2e629de922b9}</source>
  <pubdate>2021</pubdate>
  <note>[Online, Accessed: June 23, 2021]</note>
</bibl>

<bibl id="B21">
  <title><p>Bitcoin technical trading with artificial neural
  network</p></title>
  <aug>
    <au><snm>Nakano</snm><fnm>M</fnm></au>
    <au><snm>Takahashi</snm><fnm>A</fnm></au>
    <au><snm>Takahashi</snm><fnm>S</fnm></au>
  </aug>
  <source>Physica A: Statistical Mechanics and its Applications</source>
  <publisher>Elsevier</publisher>
  <pubdate>2018</pubdate>
  <volume>510</volume>
  <fpage>587</fpage>
  <lpage>-609</lpage>
</bibl>

<bibl id="B22">
  <title><p>Percentage of Total Market Capitalization</p></title>
  <aug>
    <au><cnm>Coinmarketcap</cnm></au>
  </aug>
  <source>\url{https://coinmarketcap.com/charts/\#dominance-percentage}</source>
  <pubdate>2020</pubdate>
  <note>[Online, Accessed January 11, 2020]</note>
</bibl>

<bibl id="B23">
  <title><p>Top Cryptocurrency Exchanges List</p></title>
  <aug>
    <au><cnm>Nomics</cnm></au>
  </aug>
  <source>\url{https://nomics.com/exchanges}</source>
  <pubdate>2020</pubdate>
  <note>[Online, Accessed: January 11, 2020]</note>
</bibl>

<bibl id="B24">
  <title><p>CME Cryptocurrency products</p></title>
  <aug>
    <au><cnm>CME</cnm></au>
  </aug>
  <source>\url{https://www.cmegroup.com/trading/cryptocurrency-indices.html}</source>
  <pubdate>2020</pubdate>
  <note>[Online, Accessed: February 11, 2020]</note>
</bibl>

<bibl id="B25">
  <title><p>CME groups overview</p></title>
  <aug>
    <au><cnm>CME</cnm></au>
  </aug>
  <source>\url{https://www.cmegroup.com/company/history/}</source>
  <pubdate>2020</pubdate>
  <note>[Online, Accessed: February 11, 2020]</note>
</bibl>

<bibl id="B26">
  <title><p>CME Group Rules and Regulation Overview</p></title>
  <aug>
    <au><cnm>CME</cnm></au>
  </aug>
  <source>\url{https://www.cmegroup.com/education/courses/market-regulation/overview/cme-group-rules-and-regulation-overview.html}</source>
  <pubdate>2020</pubdate>
  <note>[Online, Accessed February 11, 2020]</note>
</bibl>

<bibl id="B27">
  <title><p>CBOE products</p></title>
  <aug>
    <au><cnm>CBOE</cnm></au>
  </aug>
  <source>\url{https://www.cboe.com}</source>
  <pubdate>2020</pubdate>
  <note>[Online, Accessed: February 11, 2020]</note>
</bibl>

<bibl id="B28">
  <title><p>CBOE history</p></title>
  <aug>
    <au><cnm>CBOE</cnm></au>
  </aug>
  <source>\url{http://www.cboe.com/aboutcboe/history}</source>
  <pubdate>2020</pubdate>
  <note>[Online, Accessed: February 11, 2020]</note>
</bibl>

<bibl id="B29">
  <title><p>CFE Regulation</p></title>
  <aug>
    <au><cnm>CBOE</cnm></au>
  </aug>
  <source>\url{https://www.cboe.com/aboutcboe/legal-regulatory/departmental-overviews/cfe-regulation}</source>
  <pubdate>2020</pubdate>
  <note>[Online, Accessed February 11, 2020]</note>
</bibl>

<bibl id="B30">
  <title><p>BAKKT markets</p></title>
  <aug>
    <au><cnm>BAKKT</cnm></au>
  </aug>
  <source>\url{https://www.bakkt.com/index}</source>
  <pubdate>2020</pubdate>
  <note>[Online, Accessed February 11, 2020]</note>
</bibl>

<bibl id="B31">
  <title><p>BAKKT terms of use</p></title>
  <aug>
    <au><cnm>BAKKT</cnm></au>
  </aug>
  <source>\url{https://www.bakkt.com/terms-of-use}</source>
  <pubdate>2020</pubdate>
  <note>[Online, Accessed: February 11, 2020]</note>
</bibl>

<bibl id="B32">
  <title><p>Beginner’s Guide to BitMEX: Complete Review</p></title>
  <aug>
    <au><cnm>Bitmex</cnm></au>
  </aug>
  <source>\url{https://blockonomi.com/bitmex-review/}</source>
  <pubdate>2020</pubdate>
  <note>[Online, Accessed: February 11, 2020]</note>
</bibl>

<bibl id="B33">
  <title><p>Bitmex</p></title>
  <aug>
    <au><cnm>Bitmex</cnm></au>
  </aug>
  <source>\url{https://www.bitmex.com/register}</source>
  <pubdate>2020</pubdate>
  <note>[Online, Accessed: February 11, 2020]</note>
</bibl>

<bibl id="B34">
  <title><p>Binance Review 2020: Pros, Cons, Fees, Features, and
  Safety</p></title>
  <aug>
    <au><cnm>Binance</cnm></au>
  </aug>
  <source>\url{https://insidebitcoins.com/cryptocurrency-exchanges/binance-review/}</source>
  <pubdate>2020</pubdate>
  <note>[Online, Accessed: February 11, 2020]</note>
</bibl>

<bibl id="B35">
  <title><p>Why world leader crypto exchange Binance moved to Malta</p></title>
  <aug>
    <au><cnm>Maltatoday</cnm></au>
  </aug>
  <source>\url{https://www.maltatoday.com.mt/business/business_news/93170/why_world_leader_crypto_exchange_binance_moved_to_malta\#.XlKZ8Gj7Q2x}</source>
  <pubdate>2020</pubdate>
  <note>[Online, Accessed: February 11, 2020]</note>
</bibl>

<bibl id="B36">
  <title><p>Binance Partners with Coinfirm to Protect the Global Cryptocurrency
  Economy and Ensure Compliance with FATF AML Rules</p></title>
  <aug>
    <au><cnm>Binance</cnm></au>
  </aug>
  <source>\url{https://www.binance.com/en/blog/386484403820867584/Binance-Partners-with-Coinfirm-to-Protect-the-Global-Cryptocurrency-Economy-and-Ensure-Compliance-with-FATF-AML-Rules}</source>
  <pubdate>2020</pubdate>
  <note>[Online, Accessed February 11, 2020]</note>
</bibl>

<bibl id="B37">
  <title><p>Coinbase Supported cryptocurrencies</p></title>
  <aug>
    <au><cnm>Coinbase</cnm></au>
  </aug>
  <source>\url{https://help.coinbase.com/en/coinbase/getting-started/general-crypto-education/supported-cryptocurrencies.html}</source>
  <pubdate>2020</pubdate>
  <note>[Online, Accessed: February 11, 2020]</note>
</bibl>

<bibl id="B38">
  <title><p>Coinbase Inc.</p></title>
  <aug>
    <au><cnm>Bloomberg</cnm></au>
  </aug>
  <source>\url{https://www.bloomberg.com/profile/company/0776164D:US}</source>
  <pubdate>2020</pubdate>
  <note>[Online, Accessed: February 11, 2020]</note>
</bibl>

<bibl id="B39">
  <title><p>Our path to listing SEC-regulated crypto securities</p></title>
  <aug>
    <au><cnm>Coinbase</cnm></au>
  </aug>
  <source>\url{https://blog.coinbase.com/our-path-to-listing-sec-regulated-crypto-securities-a1724e13bb5a}</source>
  <pubdate>2020</pubdate>
  <note>[Online, Accessed February 11, 2020]</note>
</bibl>

<bibl id="B40">
  <title><p>Bitfinex markets</p></title>
  <aug>
    <au><cnm>Bitfinex</cnm></au>
  </aug>
  <source>\url{https://www.bitfinex.com/}</source>
  <pubdate>2020</pubdate>
  <note>[Online, Accessed: February 11, 2020]</note>
</bibl>

<bibl id="B41">
  <title><p>Bitfinex terms of service</p></title>
  <aug>
    <au><cnm>Bitfinex</cnm></au>
  </aug>
  <source>\url{https://www.bitfinex.com/legal/terms}</source>
  <pubdate>2020</pubdate>
  <note>[Online, Accessed: February 11, 2020]</note>
</bibl>

<bibl id="B42">
  <title><p>New York Court Rules That State Attorney Has Jurisdiction Over
  Bitfinex</p></title>
  <aug>
    <au><cnm>Bitfinex</cnm></au>
  </aug>
  <source>\url{https://cointelegraph.com/news/new-york-court-rules-that-state-attorney-has-jurisdiction-over-bitfinex}</source>
  <pubdate>2020</pubdate>
  <note>[Online, Accessed February 11, 2020]</note>
</bibl>

<bibl id="B43">
  <title><p>Bitstamp Review 2020</p></title>
  <aug>
    <au><cnm>Bitstamp</cnm></au>
  </aug>
  <source>\url{https://www.fxempire.com/crypto/exchange/bitstamp/review}</source>
  <pubdate>2020</pubdate>
  <note>[Online, Accessed: February 11, 2020]</note>
</bibl>

<bibl id="B44">
  <title><p>Bitstamp who we are</p></title>
  <aug>
    <au><cnm>Bitstamp</cnm></au>
  </aug>
  <source>\url{https://www.bitstamp.net/about-us/}</source>
  <pubdate>2020</pubdate>
  <note>[Online, Accessed: February 11, 2020]</note>
</bibl>

<bibl id="B45">
  <title><p>Terms of Use</p></title>
  <aug>
    <au><cnm>Bitstamp</cnm></au>
  </aug>
  <source>\url{https://www.bitstamp.net/terms-of-use/sa}</source>
  <pubdate>2020</pubdate>
  <note>[Online, Accessed February 11, 2020]</note>
</bibl>

<bibl id="B46">
  <title><p>Poloniex markets</p></title>
  <aug>
    <au><cnm>Poloniex</cnm></au>
  </aug>
  <source>\url{https://poloniex.com/}</source>
  <pubdate>2020</pubdate>
  <note>[Online, Accessed: February 11, 2020]</note>
</bibl>

<bibl id="B47">
  <title><p>Contract for differences</p></title>
  <aug>
    <au><snm>Authority</snm><fnm>FC</fnm></au>
  </aug>
  <source>\url{https://www.fca.org.uk/firms/contracts-for-difference},
  \url{https://www:fca:org:uk/rms/contracts-for-dierence}</source>
  <pubdate>2019</pubdate>
  <note>[Online, Accessed January 29, 2020]</note>
</bibl>

<bibl id="B48">
  <title><p>What crypto exchanges do to comply with KYC, AML and CFT
  regulations</p></title>
  <aug>
    <au><snm>Adeyanju</snm><fnm>C</fnm></au>
  </aug>
  <source>\url{https://cointelegraph.com/news/what-crypto-exchanges-do-to-comply-with-kyc-aml-and-cft-regulations}</source>
  <pubdate>2019</pubdate>
  <note>[Online, Accessed January 11, 2020]</note>
</bibl>

<bibl id="B49">
  <title><p>Localbitcoins purchasing online</p></title>
  <aug>
    <au><cnm>Localbtc</cnm></au>
  </aug>
  <source>\url{https://localbitcoins.com}</source>
  <pubdate>2020</pubdate>
  <note>[Online, Accessed: January 11, 2020]</note>
</bibl>

<bibl id="B50">
  <title><p>Here’s What Caused Bitcoin’s ‘Extreme’ Price
  Plunge</p></title>
  <aug>
    <au><cnm>Forbes</cnm></au>
  </aug>
  <source>\url{https://www.forbes.com/sites/billybambrough/2020/03/19/major-bitcoin-exchange-bitmex-has-a-serious-problem/?sh=1be57a0d4f7d}</source>
  <pubdate>2021</pubdate>
  <note>[Online, Accessed: March 19, 2020]</note>
</bibl>

<bibl id="B51">
  <title><p>More Than \$600 Million Stolen In Ethereum And Other
  Cryptocurrencies—Marking One Of Crypto’s Biggest Hacks Ever</p></title>
  <aug>
    <au><cnm>Forbes</cnm></au>
  </aug>
  <source>\url{https://www.forbes.com/sites/jonathanponciano/2021/08/10/more-than-600-million-stolen-in-ethereum-and-other-cryptocurrencies-marking-one-of-cryptos-biggest-hacks-ever/?sh=502ce7387f62}</source>
  <pubdate>2021</pubdate>
  <note>[Online, Accessed: August 10, 2021]</note>
</bibl>

<bibl id="B52">
  <title><p>Fintech investments in European banks: a hybrid IT2 fuzzy
  multidimensional decision-making approach</p></title>
  <aug>
    <au><snm>Kou</snm><fnm>G</fnm></au>
    <au><snm>Akdeniz</snm><fnm>{\"O}O</fnm></au>
    <au><snm>Din{\c{c}}er</snm><fnm>H</fnm></au>
    <au><snm>Y{\"u}ksel</snm><fnm>S</fnm></au>
  </aug>
  <source>Financial Innovation</source>
  <publisher>Springer</publisher>
  <pubdate>2021</pubdate>
  <volume>7</volume>
  <issue>1</issue>
  <fpage>1</fpage>
  <lpage>-28</lpage>
</bibl>

<bibl id="B53">
  <title><p>Bitcoin and the Challenges for Financial Regulation</p></title>
  <aug>
    <au><cnm>UKTN</cnm></au>
  </aug>
  <source>\url{https://www.uktech.news}</source>
  <pubdate>2021</pubdate>
  <note>[Online, Accessed: February 24, 2021]</note>
</bibl>

<bibl id="B54">
  <title><p>Why Buffett Sees Bitcoin Bubble</p></title>
  <aug>
    <au><cnm>Forbes</cnm></au>
  </aug>
  <source>\url{https://www.forbes.com/sites/johnwasik/2017/11/06/why-buffett-sees-bitcoin-bubble/?sh=196c2a8062a8}</source>
  <pubdate>2017</pubdate>
  <note>[Online, Accessed: November 6, 2017]</note>
</bibl>

<bibl id="B55">
  <title><p>Importance of technical and fundamental analysis in the European
  foreign exchange market</p></title>
  <aug>
    <au><snm>Oberlechner</snm><fnm>T</fnm></au>
  </aug>
  <source>International Journal of Finance \& Economics</source>
  <publisher>Wiley Online Library</publisher>
  <pubdate>2001</pubdate>
  <volume>6</volume>
  <issue>1</issue>
  <fpage>81</fpage>
  <lpage>-93</lpage>
</bibl>

<bibl id="B56">
  <title><p>A systematic review of fundamental and technical analysis of stock
  market predictions</p></title>
  <aug>
    <au><snm>Nti</snm><fnm>IK</fnm></au>
    <au><snm>Adekoya</snm><fnm>AF</fnm></au>
    <au><snm>Weyori</snm><fnm>BA</fnm></au>
  </aug>
  <source>Artificial Intelligence Review</source>
  <publisher>Springer</publisher>
  <pubdate>2020</pubdate>
  <volume>53</volume>
  <issue>4</issue>
  <fpage>3007</fpage>
  <lpage>-3057</lpage>
</bibl>

<bibl id="B57">
  <title><p>Global trading system</p></title>
  <aug>
    <au><snm>Calo</snm><fnm>B</fnm></au>
    <au><snm>Johnson</snm><fnm>W</fnm></au>
  </aug>
  <publisher>Google Patents</publisher>
  <pubdate>2002</pubdate>
  <note>US Patent App. 09/769,036</note>
</bibl>

<bibl id="B58">
  <title><p>Real-Time Cryptocurrency Trading System</p></title>
  <aug>
    <au><snm>Bauriya</snm><fnm>A</fnm></au>
    <au><snm>Tikone</snm><fnm>A</fnm></au>
    <au><snm>Nandgaonkar</snm><fnm>P</fnm></au>
    <au><snm>Sakure</snm><fnm>KS</fnm></au>
  </aug>
  <source>International Research Journal of Engineering and Technology
  (IRJET)</source>
  <pubdate>2019</pubdate>
  <volume>06</volume>
</bibl>

<bibl id="B59">
  <title><p>Develop your Crypto-Trading System Using Plain Logic, Part
  1</p></title>
  <aug>
    <au><snm>Molina</snm><fnm>J</fnm></au>
  </aug>
  <pubdate>2019</pubdate>
  <url>https://medium.com/swlh/develop-your-crypto-trading-system-using-plain-logic-part-1-caac02f0a37d</url>
</bibl>

<bibl id="B60">
  <title><p>The profitability of technical trading rules in the Bitcoin
  market</p></title>
  <aug>
    <au><snm>Gerritsen</snm><fnm>DF</fnm></au>
    <au><snm>Bouri</snm><fnm>E</fnm></au>
    <au><snm>Ramezanifar</snm><fnm>E</fnm></au>
    <au><snm>Roubaud</snm><fnm>D</fnm></au>
  </aug>
  <source>Finance Research Letters</source>
  <publisher>Elsevier</publisher>
  <pubdate>2019</pubdate>
</bibl>

<bibl id="B61">
  <title><p>Pairs trading</p></title>
  <aug>
    <au><snm>Elliott</snm><fnm>RJ</fnm></au>
    <au><snm>Van Der Hoek*</snm><fnm>J</fnm></au>
    <au><snm>Malcolm</snm><fnm>WP</fnm></au>
  </aug>
  <source>Quantitative Finance</source>
  <publisher>Taylor \& Francis</publisher>
  <pubdate>2005</pubdate>
  <volume>5</volume>
  <issue>3</issue>
  <fpage>271</fpage>
  <lpage>-276</lpage>
</bibl>

<bibl id="B62">
  <title><p>Econometrics: theory and applications with Eviews</p></title>
  <aug>
    <au><snm>Vogelvang</snm><fnm>B</fnm></au>
  </aug>
  <publisher>Pearson Education</publisher>
  <pubdate>2005</pubdate>
</bibl>

<bibl id="B63">
  <title><p>Trading Systems and Methods,+ Website</p></title>
  <aug>
    <au><snm>Kaufman</snm><fnm>PJ</fnm></au>
  </aug>
  <publisher>John Wiley \& Sons</publisher>
  <pubdate>2013</pubdate>
  <volume>591</volume>
</bibl>

<bibl id="B64">
  <title><p>Hybrid choice models: progress and challenges</p></title>
  <aug>
    <au><snm>Ben Akiva</snm><fnm>M</fnm></au>
    <au><snm>McFadden</snm><fnm>D</fnm></au>
    <au><snm>Train</snm><fnm>K</fnm></au>
    <au><snm>Walker</snm><fnm>J</fnm></au>
    <au><snm>Bhat</snm><fnm>C</fnm></au>
    <au><snm>Bierlaire</snm><fnm>M</fnm></au>
    <au><snm>Bolduc</snm><fnm>D</fnm></au>
    <au><snm>Boersch Supan</snm><fnm>A</fnm></au>
    <au><snm>Brownstone</snm><fnm>D</fnm></au>
    <au><snm>Bunch</snm><fnm>DS</fnm></au>
    <au><cnm>others</cnm></au>
  </aug>
  <source>Marketing Letters</source>
  <publisher>Springer</publisher>
  <pubdate>2002</pubdate>
  <volume>13</volume>
  <issue>3</issue>
  <fpage>163</fpage>
  <lpage>-175</lpage>
</bibl>

<bibl id="B65">
  <title><p>Sophistication, sentiment, and misreaction</p></title>
  <aug>
    <au><snm>Chang</snm><fnm>CC</fnm></au>
    <au><snm>Hsieh</snm><fnm>PF</fnm></au>
    <au><snm>Wang</snm><fnm>YH</fnm></au>
  </aug>
  <source>Journal of Financial and Quantitative Analysis</source>
  <publisher>Cambridge University Press</publisher>
  <pubdate>2015</pubdate>
  <volume>50</volume>
  <issue>4</issue>
  <fpage>903</fpage>
  <lpage>-928</lpage>
</bibl>

<bibl id="B66">
  <title><p>Volatility Prediction: A Comparison of the Stochastic Volatility,
  GARCH (1, 1) and Egarch (1, 1) Models.</p></title>
  <aug>
    <au><snm>Kat</snm><fnm>HM</fnm></au>
    <au><snm>Heynen</snm><fnm>RC</fnm></au>
  </aug>
  <source>Journal of Derivatives</source>
  <pubdate>1994</pubdate>
  <volume>2</volume>
  <issue>2</issue>
</bibl>

<bibl id="B67">
  <title><p>Multivariate simultaneous generalized ARCH</p></title>
  <aug>
    <au><snm>Engle</snm><fnm>RF</fnm></au>
    <au><snm>Kroner</snm><fnm>KF</fnm></au>
  </aug>
  <source>Econometric theory</source>
  <publisher>Cambridge University Press</publisher>
  <pubdate>1995</pubdate>
  <volume>11</volume>
  <issue>1</issue>
  <fpage>122</fpage>
  <lpage>-150</lpage>
</bibl>

<bibl id="B68">
  <title><p>Do we really need both BEKK and DCC? A tale of two multivariate
  GARCH models</p></title>
  <aug>
    <au><snm>Caporin</snm><fnm>M</fnm></au>
    <au><snm>McAleer</snm><fnm>M</fnm></au>
  </aug>
  <source>Journal of Economic Surveys</source>
  <publisher>Wiley Online Library</publisher>
  <pubdate>2012</pubdate>
  <volume>26</volume>
  <issue>4</issue>
  <fpage>736</fpage>
  <lpage>-751</lpage>
</bibl>

<bibl id="B69">
  <title><p>Applied linear statistical models</p></title>
  <aug>
    <au><snm>Neter</snm><fnm>J</fnm></au>
    <au><snm>Kutner</snm><fnm>MH</fnm></au>
    <au><snm>Nachtsheim</snm><fnm>CJ</fnm></au>
    <au><snm>Wasserman</snm><fnm>W</fnm></au>
  </aug>
  <publisher>Irwin Chicago</publisher>
  <pubdate>1996</pubdate>
  <volume>4</volume>
</bibl>

<bibl id="B70">
  <title><p>ARMA model identification</p></title>
  <aug>
    <au><snm>Choi</snm><fnm>B</fnm></au>
  </aug>
  <publisher>Springer Science \& Business Media</publisher>
  <pubdate>2012</pubdate>
</bibl>

<bibl id="B71">
  <title><p>Predicting the price of bitcoin using machine learning</p></title>
  <aug>
    <au><snm>McNally</snm><fnm>S</fnm></au>
    <au><snm>Roche</snm><fnm>J</fnm></au>
    <au><snm>Caton</snm><fnm>S</fnm></au>
  </aug>
  <source>2018 26th Euromicro International Conference on Parallel, Distributed
  and Network-based Processing (PDP)</source>
  <pubdate>2018</pubdate>
  <fpage>339</fpage>
  <lpage>-343</lpage>
</bibl>

<bibl id="B72">
  <title><p>Supervised Learning vs Unsupervised Learning vs Reinforcement
  Learning</p></title>
  <aug>
    <au><cnm>IntelliPaat</cnm></au>
  </aug>
  <source>\url{https://intellipaat.com/blog/supervised-learning-vs-unsupervised-learning-vs-reinforcement-learning/}</source>
  <pubdate>2021</pubdate>
  <note>[Online, Accessed: September 14, 2021]</note>
</bibl>

<bibl id="B73">
  <title><p>Portfolio Selection</p></title>
  <aug>
    <au><snm>Markowitz</snm><fnm>H</fnm></au>
  </aug>
  <source>The Journal of Finance</source>
  <pubdate>1952</pubdate>
  <volume>7</volume>
  <issue>1</issue>
  <fpage>77</fpage>
  <lpage>-91</lpage>
</bibl>

<bibl id="B74">
  <title><p>Portfolio diversification across cryptocurrencies</p></title>
  <aug>
    <au><snm>Liu</snm><fnm>W</fnm></au>
  </aug>
  <source>Finance Research Letters</source>
  <publisher>Elsevier</publisher>
  <pubdate>2019</pubdate>
  <volume>29</volume>
  <fpage>200</fpage>
  <lpage>-205</lpage>
</bibl>

<bibl id="B75">
  <title><p>The role of bitcoin in well diversified portfolios: A comparative
  global study</p></title>
  <aug>
    <au><snm>Kajtazi</snm><fnm>A</fnm></au>
    <au><snm>Moro</snm><fnm>A</fnm></au>
  </aug>
  <source>International Review of Financial Analysis</source>
  <publisher>Elsevier</publisher>
  <pubdate>2019</pubdate>
  <volume>61</volume>
  <fpage>143</fpage>
  <lpage>-157</lpage>
</bibl>

<bibl id="B76">
  <title><p>Bubbles, financial crises, and systemic risk</p></title>
  <aug>
    <au><snm>Brunnermeier</snm><fnm>MK</fnm></au>
    <au><snm>Oehmke</snm><fnm>M</fnm></au>
  </aug>
  <source>Handbook of the Economics of Finance</source>
  <publisher>Elsevier</publisher>
  <pubdate>2013</pubdate>
  <volume>2</volume>
  <fpage>1221</fpage>
  <lpage>-1288</lpage>
</bibl>

<bibl id="B77">
  <title><p>Bankruptcy prediction for SMEs using transactional data and
  two-stage multiobjective feature selection</p></title>
  <aug>
    <au><snm>Kou</snm><fnm>G</fnm></au>
    <au><snm>Xu</snm><fnm>Y</fnm></au>
    <au><snm>Peng</snm><fnm>Y</fnm></au>
    <au><snm>Shen</snm><fnm>F</fnm></au>
    <au><snm>Chen</snm><fnm>Y</fnm></au>
    <au><snm>Chang</snm><fnm>K</fnm></au>
    <au><snm>Kou</snm><fnm>S</fnm></au>
  </aug>
  <source>Decision Support Systems</source>
  <publisher>Elsevier</publisher>
  <pubdate>2021</pubdate>
  <volume>140</volume>
  <fpage>113429</fpage>
</bibl>

<bibl id="B78">
  <title><p>Is The Crypto Market Maturing? An Analysis For
  Entrepreneurs</p></title>
  <aug>
    <au><cnm>Forbes</cnm></au>
  </aug>
  <source>\url{https://www.forbes.com/sites/theyec/2021/06/01/is-the-crypto-market-maturing-an-analysis-for-entrepreneurs/?sh=1170160bba22}</source>
  <pubdate>2021</pubdate>
  <note>[Online, Accessed: June 1, 2021]</note>
</bibl>

<bibl id="B79">
  <title><p>Bitcoin: too good to miss or a bubble ready to burst?</p></title>
  <aug>
    <au><cnm>FT</cnm></au>
  </aug>
  <source>\url{https://www.ft.com/crypto/}</source>
  <pubdate>2021</pubdate>
  <note>[Online, Accessed: November 9, 2021]</note>
</bibl>

<bibl id="B80">
  <title><p>Guidelines for snowballing in systematic literature studies and a
  replication in software engineering</p></title>
  <aug>
    <au><snm>Wohlin</snm><fnm>C</fnm></au>
  </aug>
  <source>Proceedings of the 18th international conference on evaluation and
  assessment in software engineering</source>
  <pubdate>2014</pubdate>
  <fpage>1</fpage>
  <lpage>-10</lpage>
</bibl>

<bibl id="B81">
  <title><p>Capfolio cryptocurrency trading platform</p></title>
  <aug>
    <au><cnm>Capfolio</cnm></au>
  </aug>
  <source>\url{https://www.capfol.io/}</source>
  <pubdate>2020</pubdate>
  <note>[Online, Accessed: January 26, 2020]</note>
</bibl>

<bibl id="B82">
  <title><p>3Commas Smart Trading terminal and auto trading bots</p></title>
  <aug>
    <au><cnm>3commas</cnm></au>
  </aug>
  <source>\url{https://3commas.io/}</source>
  <pubdate>2020</pubdate>
  <note>[Online, Accessed: January 26, 2020]</note>
</bibl>

<bibl id="B83">
  <title><p>CCXT – CryptoCurrency eXchange Trading Library</p></title>
  <aug>
    <au><cnm>Ccxt</cnm></au>
  </aug>
  <source>\url{https://github.com/ccxt/ccxt}</source>
  <pubdate>2020</pubdate>
  <note>[Online, Accessed: January 26, 2020]</note>
</bibl>

<bibl id="B84">
  <title><p>Blackbird Bitcoin Arbitrage: a long/short market-neutral
  strategy</p></title>
  <aug>
    <au><cnm>Blackbird</cnm></au>
  </aug>
  <source>\url{https://github.com/butor/blackbird}</source>
  <pubdate>2020</pubdate>
  <note>[Online, Accessed: January 26, 2020]</note>
</bibl>

<bibl id="B85">
  <title><p>StockSharp - trading platform</p></title>
  <aug>
    <au><cnm>Stocksharp</cnm></au>
  </aug>
  <source>\url{https://github.com/StockSharp/StockSharp}</source>
  <pubdate>2020</pubdate>
  <note>[Online, Accessed: January 26, 2020]</note>
</bibl>

<bibl id="B86">
  <title><p>Freqtrade</p></title>
  <aug>
    <au><cnm>Fretrade</cnm></au>
  </aug>
  <source>\url{https://github.com/freqtrade/freqtrade}</source>
  <pubdate>2020</pubdate>
  <note>[Online, Accessed: January 26, 2020]</note>
</bibl>

<bibl id="B87">
  <title><p>Automated Crypto Trading and Technical Analysis (TA)
  Bot</p></title>
  <aug>
    <au><cnm>Cryptosignal</cnm></au>
  </aug>
  <source>\url{https://github.com/CryptoSignal/crypto-signal}</source>
  <pubdate>2020</pubdate>
  <note>[Online, Accessed: January 26, 2020]</note>
</bibl>

<bibl id="B88">
  <title><p>Ctubio - Cryptocurrency trading bot</p></title>
  <aug>
    <au><cnm>Ctubio</cnm></au>
  </aug>
  <source>\url{https://github.com/ctubio/Krypto-trading-bot}</source>
  <pubdate>2020</pubdate>
  <note>[Online, Accessed: January 26, 2020]</note>
</bibl>

<bibl id="B89">
  <title><p>An Algorithmic Trading Library for Crypto-Assets in
  Python</p></title>
  <aug>
    <au><cnm>Catalyst</cnm></au>
  </aug>
  <source>\url{https://github.com/enigmampc/catalyst}</source>
  <pubdate>2020</pubdate>
  <note>[Online, Accessed: January 26, 2020]</note>
</bibl>

<bibl id="B90">
  <title><p>A golang implementation of a console-based trading bot for
  cryptocurrency exchanges</p></title>
  <aug>
    <au><cnm>Golang</cnm></au>
  </aug>
  <source>\url{https://github.com/saniales/golang-crypto-trading-bot}</source>
  <pubdate>2020</pubdate>
  <note>[Online, Accessed: January 26, 2020]</note>
</bibl>

<bibl id="B91">
  <title><p>Technical Analysis for Cryptocurrency Trading on Mobile
  Phones</p></title>
  <aug>
    <au><snm>Kamrat</snm><fnm>S</fnm></au>
    <au><snm>Suesangiamsakul</snm><fnm>N</fnm></au>
    <au><snm>Marukatat</snm><fnm>R</fnm></au>
  </aug>
  <source>2018 3rd Technology Innovation Management and Engineering Science
  International Conference (TIMES-iCON)</source>
  <pubdate>2018</pubdate>
  <fpage>1</fpage>
  <lpage>-4</lpage>
</bibl>

<bibl id="B92">
  <title><p>Arbitrage Trading Systems for Cryptocurrencies. Design Principles
  and Server Architecture</p></title>
  <aug>
    <au><snm>P{\u{a}}una</snm><fnm>C</fnm></au>
  </aug>
  <source>Informatica Economica</source>
  <publisher>INFOREC Association</publisher>
  <pubdate>2018</pubdate>
  <volume>22</volume>
  <issue>2</issue>
  <fpage>35</fpage>
  <lpage>-42</lpage>
</bibl>

<bibl id="B93">
  <title><p>EOS Cryptocurrency Trading Strategy--Turtle Soup
  Pattern</p></title>
  <aug>
    <au><cnm>TradingstrategyGuides</cnm></au>
  </aug>
  <source>\url{https://tradingstrategyguides.com/eos-cryptocurrency-trading-strategy/}</source>
  <pubdate>2019</pubdate>
  <note>[Online, Accessed January 29, 2020]</note>
</bibl>

<bibl id="B94">
  <title><p>Nem (XEM) Cryptocurrency Strategy--Momentum Pinball
  Setup</p></title>
  <aug>
    <au><cnm>TradingstrategyGuides</cnm></au>
  </aug>
  <source>\url{https://tradingstrategyguides.com/nem-xem-cryptocurrency-strategy/}</source>
  <pubdate>2019</pubdate>
  <note>[Online, Accessed January 29, 2020]</note>
</bibl>

<bibl id="B95">
  <title><p>Free OMNI Cryptocurrency Strategy--Amazing Gann Box</p></title>
  <aug>
    <au><cnm>TradingstrategyGuides</cnm></au>
  </aug>
  <source>\url{https://tradingstrategyguides.com/free-omni-cryptocurrency-strategy/}</source>
  <pubdate>2019</pubdate>
  <note>[Online, Accessed January 29, 2020]</note>
</bibl>

<bibl id="B96">
  <title><p>IOTA Cryptocurrency Strategy--Busted Double Top Pattern</p></title>
  <aug>
    <au><cnm>TradingstrategyGuides</cnm></au>
  </aug>
  <source>\url{https://tradingstrategyguides.com/iota-cryptocurrency-strategy/}</source>
  <pubdate>2019</pubdate>
  <note>[Online, Accessed January 29, 2020]</note>
</bibl>

<bibl id="B97">
  <title><p>Tether Trading Strategy--Bottom Rotation Trading</p></title>
  <aug>
    <au><cnm>TradingstrategyGuides</cnm></au>
  </aug>
  <source>\url{https://tradingstrategyguides.com/tether-trading-strategy/}</source>
  <pubdate>2019</pubdate>
  <note>[Online, Accessed January 29, 2020]</note>
</bibl>

<bibl id="B98">
  <title><p>Finding attractive technical patterns in cryptocurrency
  markets</p></title>
  <aug>
    <au><snm>Ha</snm><fnm>S</fnm></au>
    <au><snm>Moon</snm><fnm>BR</fnm></au>
  </aug>
  <source>Memetic Computing</source>
  <publisher>Springer</publisher>
  <pubdate>2018</pubdate>
  <volume>10</volume>
  <issue>3</issue>
  <fpage>301</fpage>
  <lpage>-306</lpage>
</bibl>

<bibl id="B99">
  <title><p>Technical Analysis and Cryptocurrencies</p></title>
  <aug>
    <au><snm>Hudson</snm><fnm>R</fnm></au>
    <au><snm>Urquhart</snm><fnm>A</fnm></au>
  </aug>
  <source>Available at SSRN 3387950</source>
  <pubdate>2019</pubdate>
</bibl>

<bibl id="B100">
  <title><p>The effectiveness of technical trading rules in cryptocurrency
  markets</p></title>
  <aug>
    <au><snm>Corbet</snm><fnm>S</fnm></au>
    <au><snm>Eraslan</snm><fnm>V</fnm></au>
    <au><snm>Lucey</snm><fnm>B</fnm></au>
    <au><snm>Sensoy</snm><fnm>A</fnm></au>
  </aug>
  <source>Finance Research Letters</source>
  <publisher>Elsevier</publisher>
  <pubdate>2019</pubdate>
  <volume>31</volume>
  <fpage>32</fpage>
  <lpage>-37</lpage>
</bibl>

<bibl id="B101">
  <title><p>Technical trading rules in the cryptocurrency market</p></title>
  <aug>
    <au><snm>Grobys</snm><fnm>K</fnm></au>
    <au><snm>Ahmed</snm><fnm>S</fnm></au>
    <au><snm>Sapkota</snm><fnm>N</fnm></au>
  </aug>
  <source>Finance Research Letters</source>
  <publisher>Elsevier</publisher>
  <pubdate>2020</pubdate>
  <volume>32</volume>
  <fpage>101396</fpage>
</bibl>

<bibl id="B102">
  <title><p>Why cryptocurrency markets are inefficient: The impact of liquidity
  and volatility</p></title>
  <aug>
    <au><snm>Al Yahyaee</snm><fnm>KH</fnm></au>
    <au><snm>Mensi</snm><fnm>W</fnm></au>
    <au><snm>Ko</snm><fnm>HU</fnm></au>
    <au><snm>Yoon</snm><fnm>SM</fnm></au>
    <au><snm>Kang</snm><fnm>SH</fnm></au>
  </aug>
  <source>The North American Journal of Economics and Finance</source>
  <publisher>Elsevier</publisher>
  <pubdate>2020</pubdate>
  <volume>52</volume>
  <fpage>101168</fpage>
</bibl>

<bibl id="B103">
  <title><p>Pairs Trading in Cryptocurrency Markets</p></title>
  <aug>
    <au><snm>Fil</snm><fnm>M</fnm></au>
  </aug>
  <source>Univerzita Karlova, Fakulta soci{\'a}ln{\'\i}ch v{\v{e}}d</source>
  <pubdate>2019</pubdate>
</bibl>

<bibl id="B104">
  <title><p>Pairs trading: Performance of a relative-value arbitrage
  rule</p></title>
  <aug>
    <au><snm>Gatev</snm><fnm>E</fnm></au>
    <au><snm>Goetzmann</snm><fnm>WN</fnm></au>
    <au><snm>Rouwenhorst</snm><fnm>KG</fnm></au>
  </aug>
  <source>The Review of Financial Studies</source>
  <publisher>Oxford University Press</publisher>
  <pubdate>2006</pubdate>
  <volume>19</volume>
  <issue>3</issue>
  <fpage>797</fpage>
  <lpage>-827</lpage>
</bibl>

<bibl id="B105">
  <title><p>Cointegration-based pairs trading framework with application to the
  Cryptocurrency market</p></title>
  <aug>
    <au><snm>Broek</snm><fnm>L</fnm></au>
    <au><snm>Sharif</snm><fnm>Z</fnm></au>
  </aug>
  <pubdate>2018</pubdate>
</bibl>

<bibl id="B106">
  <title><p>Model-based pairs trading in the bitcoin markets</p></title>
  <aug>
    <au><snm>Lintilhac</snm><fnm>PS</fnm></au>
    <au><snm>Tourin</snm><fnm>A</fnm></au>
  </aug>
  <source>Quantitative Finance</source>
  <publisher>Taylor \& Francis</publisher>
  <pubdate>2017</pubdate>
  <volume>17</volume>
  <issue>5</issue>
  <fpage>703</fpage>
  <lpage>-716</lpage>
</bibl>

<bibl id="B107">
  <title><p>Optimal pairs trading with time-varying volatility</p></title>
  <aug>
    <au><snm>Li</snm><fnm>TN</fnm></au>
    <au><snm>Tourin</snm><fnm>A</fnm></au>
  </aug>
  <source>International Journal of Financial Engineering</source>
  <publisher>World Scientific</publisher>
  <pubdate>2016</pubdate>
  <volume>3</volume>
  <issue>03</issue>
  <fpage>1650023</fpage>
</bibl>

<bibl id="B108">
  <title><p>Informed trading in the Bitcoin market</p></title>
  <aug>
    <au><snm>Feng</snm><fnm>W</fnm></au>
    <au><snm>Wang</snm><fnm>Y</fnm></au>
    <au><snm>Zhang</snm><fnm>Z</fnm></au>
  </aug>
  <source>Finance Research Letters</source>
  <publisher>Elsevier</publisher>
  <pubdate>2018</pubdate>
  <volume>26</volume>
  <fpage>63</fpage>
  <lpage>-70</lpage>
</bibl>

<bibl id="B109">
  <title><p>Time-varying arrival rates of informed and uninformed
  trades</p></title>
  <aug>
    <au><snm>Easley</snm><fnm>D</fnm></au>
    <au><snm>Engle</snm><fnm>RF</fnm></au>
    <au><snm>O'Hara</snm><fnm>M</fnm></au>
    <au><snm>Wu</snm><fnm>L</fnm></au>
  </aug>
  <source>Journal of Financial Econometrics</source>
  <publisher>Oxford University Press</publisher>
  <pubdate>2008</pubdate>
  <volume>6</volume>
  <issue>2</issue>
  <fpage>171</fpage>
  <lpage>-207</lpage>
</bibl>

<bibl id="B110">
  <title><p>Trading volume and the predictability of return and volatility in
  the cryptocurrency market</p></title>
  <aug>
    <au><snm>Bouri</snm><fnm>E</fnm></au>
    <au><snm>Lau</snm><fnm>CKM</fnm></au>
    <au><snm>Lucey</snm><fnm>B</fnm></au>
    <au><snm>Roubaud</snm><fnm>D</fnm></au>
  </aug>
  <source>Finance Research Letters</source>
  <publisher>Elsevier</publisher>
  <pubdate>2019</pubdate>
  <volume>29</volume>
  <fpage>340</fpage>
  <lpage>-346</lpage>
</bibl>

<bibl id="B111">
  <title><p>Granger-causality in quantiles between financial markets: Using
  copula approach</p></title>
  <aug>
    <au><snm>Lee</snm><fnm>TH</fnm></au>
    <au><snm>Yang</snm><fnm>W</fnm></au>
  </aug>
  <source>International Review of Financial Analysis</source>
  <publisher>Elsevier</publisher>
  <pubdate>2014</pubdate>
  <volume>33</volume>
  <fpage>70</fpage>
  <lpage>-78</lpage>
</bibl>

<bibl id="B112">
  <title><p>The volatility surprise of leading cryptocurrencies: Transitory and
  permanent linkages</p></title>
  <aug>
    <au><snm>Bouri</snm><fnm>E</fnm></au>
    <au><snm>Lucey</snm><fnm>B</fnm></au>
    <au><snm>Roubaud</snm><fnm>D</fnm></au>
  </aug>
  <source>Finance Research Letters</source>
  <publisher>Elsevier</publisher>
  <pubdate>2020</pubdate>
  <volume>33</volume>
  <fpage>101188</fpage>
</bibl>

<bibl id="B113">
  <title><p>Evidence of interdependence and contagion using a frequency domain
  framework</p></title>
  <aug>
    <au><snm>Bodart</snm><fnm>V</fnm></au>
    <au><snm>Candelon</snm><fnm>B</fnm></au>
  </aug>
  <source>Emerging markets review</source>
  <publisher>Elsevier</publisher>
  <pubdate>2009</pubdate>
  <volume>10</volume>
  <issue>2</issue>
  <fpage>140</fpage>
  <lpage>-150</lpage>
</bibl>

<bibl id="B114">
  <title><p>Effect of Bitcoin spot and derivative trading volumes on price
  volatility</p></title>
  <aug>
    <au><snm>Badenhorst</snm><fnm>JJ</fnm></au>
    <au><cnm>others</cnm></au>
  </aug>
  <source>PhD thesis</source>
  <publisher>University of Pretoria</publisher>
  <pubdate>2019</pubdate>
</bibl>

<bibl id="B115">
  <title><p>Return equicorrelation in the cryptocurrency market: Analysis and
  determinants</p></title>
  <aug>
    <au><snm>Bouri</snm><fnm>E</fnm></au>
    <au><snm>Vo</snm><fnm>XV</fnm></au>
    <au><snm>Saeed</snm><fnm>T</fnm></au>
  </aug>
  <source>Finance Research Letters</source>
  <publisher>Elsevier</publisher>
  <pubdate>2020</pubdate>
  <fpage>101497</fpage>
</bibl>

<bibl id="B116">
  <title><p>Long-and short-term cryptocurrency volatility components: A
  GARCH-MIDAS analysis</p></title>
  <aug>
    <au><snm>Conrad</snm><fnm>C</fnm></au>
    <au><snm>Custovic</snm><fnm>A</fnm></au>
    <au><snm>Ghysels</snm><fnm>E</fnm></au>
  </aug>
  <source>Journal of Risk and Financial Management</source>
  <publisher>Multidisciplinary Digital Publishing Institute</publisher>
  <pubdate>2018</pubdate>
  <volume>11</volume>
  <issue>2</issue>
  <fpage>23</fpage>
</bibl>

<bibl id="B117">
  <title><p>Regime changes in Bitcoin GARCH volatility dynamics</p></title>
  <aug>
    <au><snm>Ardia</snm><fnm>D</fnm></au>
    <au><snm>Bluteau</snm><fnm>K</fnm></au>
    <au><snm>R{\"u}ede</snm><fnm>M</fnm></au>
  </aug>
  <source>Finance Research Letters</source>
  <publisher>Elsevier</publisher>
  <pubdate>2019</pubdate>
  <volume>29</volume>
  <fpage>266</fpage>
  <lpage>-271</lpage>
</bibl>

<bibl id="B118">
  <title><p>Bitcoin returns and risk: A general GARCH and GAS
  analysis</p></title>
  <aug>
    <au><snm>Troster</snm><fnm>V</fnm></au>
    <au><snm>Tiwari</snm><fnm>AK</fnm></au>
    <au><snm>Shahbaz</snm><fnm>M</fnm></au>
    <au><snm>Macedo</snm><fnm>DN</fnm></au>
  </aug>
  <source>Finance Research Letters</source>
  <publisher>Elsevier</publisher>
  <pubdate>2019</pubdate>
  <volume>30</volume>
  <fpage>187</fpage>
  <lpage>-193</lpage>
</bibl>

<bibl id="B119">
  <title><p>Volatility estimation for cryptocurrencies: Further evidence with
  jumps and structural breaks</p></title>
  <aug>
    <au><snm>Charles</snm><fnm>A</fnm></au>
    <au><snm>Darn{\'e}</snm><fnm>O</fnm></au>
    <au><cnm>others</cnm></au>
  </aug>
  <source>Economics Bulletin</source>
  <publisher>AccessEcon</publisher>
  <pubdate>2019</pubdate>
  <volume>39</volume>
  <issue>2</issue>
  <fpage>954</fpage>
  <lpage>-968</lpage>
</bibl>

<bibl id="B120">
  <title><p>Time series analysis of Cryptocurrency returns and
  volatilities</p></title>
  <aug>
    <au><snm>Malladi</snm><fnm>RK</fnm></au>
    <au><snm>Dheeriya</snm><fnm>PL</fnm></au>
  </aug>
  <source>Journal of Economics and Finance</source>
  <publisher>Springer</publisher>
  <pubdate>2021</pubdate>
  <volume>45</volume>
  <issue>1</issue>
  <fpage>75</fpage>
  <lpage>-94</lpage>
</bibl>

<bibl id="B121">
  <title><p>Nonlinear dependence in cryptocurrency markets</p></title>
  <aug>
    <au><snm>Chaim</snm><fnm>P</fnm></au>
    <au><snm>Laurini</snm><fnm>MP</fnm></au>
  </aug>
  <source>The North American Journal of Economics and Finance</source>
  <publisher>Elsevier</publisher>
  <pubdate>2019</pubdate>
  <volume>48</volume>
  <fpage>32</fpage>
  <lpage>-47</lpage>
</bibl>

<bibl id="B122">
  <title><p>Persistence in the cryptocurrency market</p></title>
  <aug>
    <au><snm>Caporale</snm><fnm>GM</fnm></au>
    <au><snm>Gil Alana</snm><fnm>L</fnm></au>
    <au><snm>Plastun</snm><fnm>A</fnm></au>
  </aug>
  <source>Research in International Business and Finance</source>
  <publisher>Elsevier</publisher>
  <pubdate>2018</pubdate>
  <volume>46</volume>
  <fpage>141</fpage>
  <lpage>-148</lpage>
</bibl>

<bibl id="B123">
  <title><p>Adaptive market hypothesis and evolving predictability of
  bitcoin</p></title>
  <aug>
    <au><snm>Khuntia</snm><fnm>S</fnm></au>
    <au><snm>Pattanayak</snm><fnm>JK</fnm></au>
  </aug>
  <source>Economics Letters</source>
  <publisher>Elsevier</publisher>
  <pubdate>2018</pubdate>
  <volume>167</volume>
  <fpage>26</fpage>
  <lpage>-28</lpage>
</bibl>

<bibl id="B124">
  <title><p>Testing the martingale difference hypothesis</p></title>
  <aug>
    <au><snm>Dom{\'\i}nguez</snm><fnm>MA</fnm></au>
    <au><snm>Lobato</snm><fnm>IN</fnm></au>
  </aug>
  <source>Econometric Reviews</source>
  <publisher>Taylor \& Francis</publisher>
  <pubdate>2003</pubdate>
  <volume>22</volume>
  <issue>4</issue>
  <fpage>351</fpage>
  <lpage>-377</lpage>
</bibl>

<bibl id="B125">
  <title><p>Generalized spectral tests for the martingale difference
  hypothesis</p></title>
  <aug>
    <au><snm>Escanciano</snm><fnm>JC</fnm></au>
    <au><snm>Velasco</snm><fnm>C</fnm></au>
  </aug>
  <source>Journal of Econometrics</source>
  <publisher>Elsevier</publisher>
  <pubdate>2006</pubdate>
  <volume>134</volume>
  <issue>1</issue>
  <fpage>151</fpage>
  <lpage>-185</lpage>
</bibl>

<bibl id="B126">
  <title><p>Volatility cascades in cryptocurrency trading</p></title>
  <aug>
    <au><snm>Gradojevic</snm><fnm>N</fnm></au>
    <au><snm>Tsiakas</snm><fnm>I</fnm></au>
  </aug>
  <source>Journal of Empirical Finance</source>
  <publisher>Elsevier</publisher>
  <pubdate>2021</pubdate>
  <volume>62</volume>
  <fpage>252</fpage>
  <lpage>-265</lpage>
</bibl>

<bibl id="B127">
  <title><p>A novel methodology to calculate the probability of volatility
  clusters in financial series: An application to cryptocurrency
  markets</p></title>
  <aug>
    <au><snm>Nikolova</snm><fnm>V</fnm></au>
    <au><snm>Trinidad Segovia</snm><fnm>JE</fnm></au>
    <au><snm>Fern{\'a}ndez Mart{\'\i}nez</snm><fnm>M</fnm></au>
    <au><snm>S{\'a}nchez Granero</snm><fnm>MA</fnm></au>
  </aug>
  <source>Mathematics</source>
  <publisher>Multidisciplinary Digital Publishing Institute</publisher>
  <pubdate>2020</pubdate>
  <volume>8</volume>
  <issue>8</issue>
  <fpage>1216</fpage>
</bibl>

<bibl id="B128">
  <title><p>Cryptocurrency volatility forecasting: A Markov regime-switching
  MIDAS approach</p></title>
  <aug>
    <au><snm>Ma</snm><fnm>F</fnm></au>
    <au><snm>Liang</snm><fnm>C</fnm></au>
    <au><snm>Ma</snm><fnm>Y</fnm></au>
    <au><snm>Wahab</snm><fnm>MIM</fnm></au>
  </aug>
  <source>Journal of Forecasting</source>
  <publisher>Wiley Online Library</publisher>
  <pubdate>2020</pubdate>
  <volume>39</volume>
  <issue>8</issue>
  <fpage>1277</fpage>
  <lpage>-1290</lpage>
</bibl>

<bibl id="B129">
  <title><p>Volatility Spillover Effects in Leading Cryptocurrencies: A
  BEKK-MGARCH Analysis</p></title>
  <aug>
    <au><snm>Katsiampa</snm><fnm>P</fnm></au>
    <au><snm>Corbet</snm><fnm>S</fnm></au>
    <au><snm>Lucey</snm><fnm>BM</fnm></au>
  </aug>
  <source>Available at SSRN 3232912</source>
  <pubdate>2018</pubdate>
</bibl>

<bibl id="B130">
  <title><p>An empirical investigation of volatility dynamics in the
  cryptocurrency market</p></title>
  <aug>
    <au><snm>Katsiampa</snm><fnm>P</fnm></au>
  </aug>
  <source>Research in International Business and Finance</source>
  <publisher>Elsevier</publisher>
  <pubdate>2019</pubdate>
</bibl>

<bibl id="B131">
  <title><p>Volatility Forecasting An Empirical Study on Bitcoin Using Garch
  and Stochastic Volatility models</p></title>
  <aug>
    <au><snm>Hultman</snm><fnm>H</fnm></au>
  </aug>
  <source>Master's thesis</source>
  <publisher>Lund University</publisher>
  <pubdate>2018</pubdate>
</bibl>

<bibl id="B132">
  <title><p>Wavelet time-scale persistence analysis of cryptocurrency market
  returns and volatility</p></title>
  <aug>
    <au><snm>Omane Adjepong</snm><fnm>M</fnm></au>
    <au><snm>Alagidede</snm><fnm>P</fnm></au>
    <au><snm>Akosah</snm><fnm>NK</fnm></au>
  </aug>
  <source>Physica A: Statistical Mechanics and its Applications</source>
  <publisher>Elsevier</publisher>
  <pubdate>2019</pubdate>
  <volume>514</volume>
  <fpage>105</fpage>
  <lpage>-120</lpage>
</bibl>

<bibl id="B133">
  <title><p>Exploring the dynamic relationships between cryptocurrencies and
  other financial assets</p></title>
  <aug>
    <au><snm>Corbet</snm><fnm>S</fnm></au>
    <au><snm>Meegan</snm><fnm>A</fnm></au>
    <au><snm>Larkin</snm><fnm>C</fnm></au>
    <au><snm>Lucey</snm><fnm>B</fnm></au>
    <au><snm>Yarovaya</snm><fnm>L</fnm></au>
  </aug>
  <source>Economics Letters</source>
  <publisher>Elsevier</publisher>
  <pubdate>2018</pubdate>
  <volume>165</volume>
  <fpage>28</fpage>
  <lpage>-34</lpage>
</bibl>

<bibl id="B134">
  <title><p>Is idiosyncratic volatility priced in cryptocurrency
  markets?</p></title>
  <aug>
    <au><snm>Zhang</snm><fnm>W</fnm></au>
    <au><snm>Li</snm><fnm>Y</fnm></au>
  </aug>
  <source>Research in International Business and Finance</source>
  <publisher>Elsevier</publisher>
  <pubdate>2020</pubdate>
  <volume>54</volume>
  <fpage>101252</fpage>
</bibl>

<bibl id="B135">
  <title><p>Weka: A machine learning workbench</p></title>
  <aug>
    <au><snm>Holmes</snm><fnm>G</fnm></au>
    <au><snm>Donkin</snm><fnm>A</fnm></au>
    <au><snm>Witten</snm><fnm>IH</fnm></au>
  </aug>
  <source>Proceedings of ANZIIS'94-Australian New Zealnd Intelligent
  Information Systems Conference</source>
  <pubdate>1994</pubdate>
  <fpage>357</fpage>
  <lpage>-361</lpage>
</bibl>

<bibl id="B136">
  <title><p>An empirical study of the naive Bayes classifier</p></title>
  <aug>
    <au><snm>Rish</snm><fnm>I</fnm></au>
    <au><cnm>others</cnm></au>
  </aug>
  <source>IJCAI 2001 workshop on empirical methods in artificial
  intelligence</source>
  <pubdate>2001</pubdate>
  <volume>3</volume>
  <issue>22</issue>
  <fpage>41</fpage>
  <lpage>-46</lpage>
</bibl>

<bibl id="B137">
  <title><p>Support vector machines: theory and applications</p></title>
  <aug>
    <au><snm>Wang</snm><fnm>L</fnm></au>
  </aug>
  <publisher>Springer Science \& Business Media</publisher>
  <pubdate>2005</pubdate>
  <volume>177</volume>
</bibl>

<bibl id="B138">
  <title><p>Decision tree classification of land cover from remotely sensed
  data</p></title>
  <aug>
    <au><snm>Friedl</snm><fnm>MA</fnm></au>
    <au><snm>Brodley</snm><fnm>CE</fnm></au>
  </aug>
  <source>Remote sensing of environment</source>
  <publisher>Elsevier</publisher>
  <pubdate>1997</pubdate>
  <volume>61</volume>
  <issue>3</issue>
  <fpage>399</fpage>
  <lpage>-409</lpage>
</bibl>

<bibl id="B139">
  <title><p>Classification and regression by randomForest</p></title>
  <aug>
    <au><snm>Liaw</snm><fnm>A</fnm></au>
    <au><snm>Wiener</snm><fnm>M</fnm></au>
    <au><cnm>others</cnm></au>
  </aug>
  <source>R news</source>
  <pubdate>2002</pubdate>
  <volume>2</volume>
  <issue>3</issue>
  <fpage>18</fpage>
  <lpage>-22</lpage>
</bibl>

<bibl id="B140">
  <title><p>Greedy function approximation: a gradient boosting
  machine</p></title>
  <aug>
    <au><snm>Friedman</snm><fnm>JH</fnm></au>
  </aug>
  <source>Annals of statistics</source>
  <publisher>JSTOR</publisher>
  <pubdate>2001</pubdate>
  <fpage>1189</fpage>
  <lpage>-1232</lpage>
</bibl>

<bibl id="B141">
  <title><p>Adaptive semi supervised support vector machine semi supervised
  learning with features cooperation for breast cancer
  classification</p></title>
  <aug>
    <au><snm>Zemmal</snm><fnm>N</fnm></au>
    <au><snm>Azizi</snm><fnm>N</fnm></au>
    <au><snm>Dey</snm><fnm>N</fnm></au>
    <au><snm>Sellami</snm><fnm>M</fnm></au>
  </aug>
  <source>Journal of Medical Imaging and Health Informatics</source>
  <publisher>American Scientific Publishers</publisher>
  <pubdate>2016</pubdate>
  <volume>6</volume>
  <issue>1</issue>
  <fpage>53</fpage>
  <lpage>-62</lpage>
</bibl>

<bibl id="B142">
  <title><p>Improvements to Platt's SMO algorithm for SVM classifier
  design</p></title>
  <aug>
    <au><snm>Keerthi</snm><fnm>SS</fnm></au>
    <au><snm>Shevade</snm><fnm>SK</fnm></au>
    <au><snm>Bhattacharyya</snm><fnm>C</fnm></au>
    <au><snm>Murthy</snm><fnm>KRK</fnm></au>
  </aug>
  <source>Neural computation</source>
  <publisher>MIT Press</publisher>
  <pubdate>2001</pubdate>
  <volume>13</volume>
  <issue>3</issue>
  <fpage>637</fpage>
  <lpage>-649</lpage>
</bibl>

<bibl id="B143">
  <title><p>Better Model Selection with a new Definition of Feature
  Importance</p></title>
  <aug>
    <au><snm>Fang</snm><fnm>F</fnm></au>
    <au><snm>Ventre</snm><fnm>C</fnm></au>
    <au><snm>Li</snm><fnm>L</fnm></au>
    <au><snm>Kanthan</snm><fnm>L</fnm></au>
    <au><snm>Wu</snm><fnm>F</fnm></au>
    <au><snm>Basios</snm><fnm>M</fnm></au>
  </aug>
  <source>arXiv preprint arXiv:2009.07708</source>
  <pubdate>2020</pubdate>
</bibl>

<bibl id="B144">
  <title><p>The application on intrusion detection based on k-means cluster
  algorithm</p></title>
  <aug>
    <au><snm>Jianliang</snm><fnm>M</fnm></au>
    <au><snm>Haikun</snm><fnm>S</fnm></au>
    <au><snm>Ling</snm><fnm>B</fnm></au>
  </aug>
  <source>2009 International Forum on Information Technology and
  Applications</source>
  <pubdate>2009</pubdate>
  <volume>1</volume>
  <fpage>150</fpage>
  <lpage>-152</lpage>
</bibl>

<bibl id="B145">
  <title><p>Constrained k-means clustering with background
  knowledge</p></title>
  <aug>
    <au><snm>Wagstaff</snm><fnm>K</fnm></au>
    <au><snm>Cardie</snm><fnm>C</fnm></au>
    <au><snm>Rogers</snm><fnm>S</fnm></au>
    <au><snm>Schr{\"o}dl</snm><fnm>S</fnm></au>
    <au><cnm>others</cnm></au>
  </aug>
  <source>Icml</source>
  <pubdate>2001</pubdate>
  <volume>1</volume>
  <fpage>577</fpage>
  <lpage>-584</lpage>
</bibl>

<bibl id="B146">
  <title><p>An Integrated Cluster Detection, Optimization, and Interpretation
  Approach for Financial Data</p></title>
  <aug>
    <au><snm>Li</snm><fnm>T</fnm></au>
    <au><snm>Kou</snm><fnm>G</fnm></au>
    <au><snm>Peng</snm><fnm>Y</fnm></au>
    <au><snm>Philip</snm><fnm>SY</fnm></au>
  </aug>
  <source>IEEE Transactions on Cybernetics</source>
  <publisher>IEEE</publisher>
  <pubdate>2021</pubdate>
</bibl>

<bibl id="B147">
  <title><p>Evaluation of clustering algorithms for financial risk analysis
  using MCDM methods</p></title>
  <aug>
    <au><snm>Kou</snm><fnm>G</fnm></au>
    <au><snm>Peng</snm><fnm>Y</fnm></au>
    <au><snm>Wang</snm><fnm>G</fnm></au>
  </aug>
  <source>Information Sciences</source>
  <publisher>Elsevier</publisher>
  <pubdate>2014</pubdate>
  <volume>275</volume>
  <fpage>1</fpage>
  <lpage>-12</lpage>
</bibl>

<bibl id="B148">
  <title><p>Applied linear statistical models</p></title>
  <aug>
    <au><snm>Kutner</snm><fnm>MH</fnm></au>
    <au><snm>Nachtsheim</snm><fnm>CJ</fnm></au>
    <au><snm>Neter</snm><fnm>J</fnm></au>
    <au><snm>Li</snm><fnm>W</fnm></au>
    <au><cnm>others</cnm></au>
  </aug>
  <publisher>McGraw-Hill Irwin New York</publisher>
  <pubdate>2005</pubdate>
  <volume>5</volume>
</bibl>

<bibl id="B149">
  <title><p>The monotone smoothing of scatterplots</p></title>
  <aug>
    <au><snm>Friedman</snm><fnm>J</fnm></au>
    <au><snm>Tibshirani</snm><fnm>R</fnm></au>
  </aug>
  <source>Technometrics</source>
  <publisher>Taylor \& Francis Group</publisher>
  <pubdate>1984</pubdate>
  <volume>26</volume>
  <issue>3</issue>
  <fpage>243</fpage>
  <lpage>-250</lpage>
</bibl>

<bibl id="B150">
  <title><p>Deep Learning Based Recommender System: A Survey and New
  Perspectives</p></title>
  <aug>
    <au><snm>Zhang</snm><fnm>S</fnm></au>
    <au><snm>Yao</snm><fnm>L</fnm></au>
    <au><snm>Sun</snm><fnm>A</fnm></au>
    <au><snm>Tay</snm><fnm>Y</fnm></au>
  </aug>
  <source>ACM Comput. Surv.</source>
  <publisher>New York, NY, USA: Association for Computing Machinery</publisher>
  <pubdate>2019</pubdate>
  <volume>52</volume>
  <issue>1</issue>
  <url>https://doi.org/10.1145/3285029</url>
</bibl>

<bibl id="B151">
  <title><p>Efficient processing of deep neural networks: A tutorial and
  survey</p></title>
  <aug>
    <au><snm>Sze</snm><fnm>V</fnm></au>
    <au><snm>Chen</snm><fnm>YH</fnm></au>
    <au><snm>Yang</snm><fnm>TJ</fnm></au>
    <au><snm>Emer</snm><fnm>JS</fnm></au>
  </aug>
  <source>Proceedings of the IEEE</source>
  <publisher>Ieee</publisher>
  <pubdate>2017</pubdate>
  <volume>105</volume>
  <issue>12</issue>
  <fpage>2295</fpage>
  <lpage>-2329</lpage>
</bibl>

<bibl id="B152">
  <title><p>Face recognition: A convolutional neural-network
  approach</p></title>
  <aug>
    <au><snm>Lawrence</snm><fnm>S</fnm></au>
    <au><snm>Giles</snm><fnm>CL</fnm></au>
    <au><snm>Tsoi</snm><fnm>AC</fnm></au>
    <au><snm>Back</snm><fnm>AD</fnm></au>
  </aug>
  <source>IEEE transactions on neural networks</source>
  <publisher>IEEE</publisher>
  <pubdate>1997</pubdate>
  <volume>8</volume>
  <issue>1</issue>
  <fpage>98</fpage>
  <lpage>-113</lpage>
</bibl>

<bibl id="B153">
  <title><p>Extensions of recurrent neural network language model</p></title>
  <aug>
    <au><snm>Mikolov</snm><fnm>T</fnm></au>
    <au><snm>Kombrink</snm><fnm>S</fnm></au>
    <au><snm>Burget</snm><fnm>L</fnm></au>
    <au><snm>{\v{C}}ernock{\`y}</snm><fnm>J</fnm></au>
    <au><snm>Khudanpur</snm><fnm>S</fnm></au>
  </aug>
  <source>2011 IEEE international conference on acoustics, speech and signal
  processing (ICASSP)</source>
  <pubdate>2011</pubdate>
  <fpage>5528</fpage>
  <lpage>-5531</lpage>
</bibl>

<bibl id="B154">
  <title><p>Empirical evaluation of gated recurrent neural networks on sequence
  modeling</p></title>
  <aug>
    <au><snm>Chung</snm><fnm>J</fnm></au>
    <au><snm>Gulcehre</snm><fnm>C</fnm></au>
    <au><snm>Cho</snm><fnm>K</fnm></au>
    <au><snm>Bengio</snm><fnm>Y</fnm></au>
  </aug>
  <source>arXiv preprint arXiv:1412.3555</source>
  <pubdate>2014</pubdate>
</bibl>

<bibl id="B155">
  <title><p>Long short-term memory-networks for machine reading</p></title>
  <aug>
    <au><snm>Cheng</snm><fnm>J</fnm></au>
    <au><snm>Dong</snm><fnm>L</fnm></au>
    <au><snm>Lapata</snm><fnm>M</fnm></au>
  </aug>
  <source>arXiv preprint arXiv:1601.06733</source>
  <pubdate>2016</pubdate>
</bibl>

<bibl id="B156">
  <title><p>A convolutional neural network for modelling sentences</p></title>
  <aug>
    <au><snm>Kalchbrenner</snm><fnm>N</fnm></au>
    <au><snm>Grefenstette</snm><fnm>E</fnm></au>
    <au><snm>Blunsom</snm><fnm>P</fnm></au>
  </aug>
  <source>arXiv preprint arXiv:1404.2188</source>
  <pubdate>2014</pubdate>
</bibl>

<bibl id="B157">
  <title><p>On the difficulty of training recurrent neural networks</p></title>
  <aug>
    <au><snm>Pascanu</snm><fnm>R</fnm></au>
    <au><snm>Mikolov</snm><fnm>T</fnm></au>
    <au><snm>Bengio</snm><fnm>Y</fnm></au>
  </aug>
  <source>International conference on machine learning</source>
  <pubdate>2013</pubdate>
  <fpage>1310</fpage>
  <lpage>-1318</lpage>
</bibl>

<bibl id="B158">
  <title><p>A Gated Recurrent Unit Approach to Bitcoin Price
  Prediction</p></title>
  <aug>
    <au><snm>Dutta</snm><fnm>A</fnm></au>
    <au><snm>Kumar</snm><fnm>S</fnm></au>
    <au><snm>Basu</snm><fnm>M</fnm></au>
  </aug>
  <source>Journal of Risk and Financial Management</source>
  <publisher>Multidisciplinary Digital Publishing Institute</publisher>
  <pubdate>2020</pubdate>
  <volume>13</volume>
  <issue>2</issue>
  <fpage>23</fpage>
</bibl>

<bibl id="B159">
  <title><p>Seq2seq fingerprint: An unsupervised deep molecular embedding for
  drug discovery</p></title>
  <aug>
    <au><snm>Xu</snm><fnm>Z</fnm></au>
    <au><snm>Wang</snm><fnm>S</fnm></au>
    <au><snm>Zhu</snm><fnm>F</fnm></au>
    <au><snm>Huang</snm><fnm>J</fnm></au>
  </aug>
  <source>Proceedings of the 8th ACM International Conference on
  Bioinformatics, Computational Biology, and Health Informatics</source>
  <pubdate>2017</pubdate>
  <fpage>285</fpage>
  <lpage>-294</lpage>
</bibl>

<bibl id="B160">
  <title><p>Cold fusion: Training seq2seq models together with language
  models</p></title>
  <aug>
    <au><snm>Sriram</snm><fnm>A</fnm></au>
    <au><snm>Jun</snm><fnm>H</fnm></au>
    <au><snm>Satheesh</snm><fnm>S</fnm></au>
    <au><snm>Coates</snm><fnm>A</fnm></au>
  </aug>
  <source>arXiv preprint arXiv:1708.06426</source>
  <pubdate>2017</pubdate>
</bibl>

<bibl id="B161">
  <title><p>Introduction to reinforcement learning</p></title>
  <aug>
    <au><snm>Sutton</snm><fnm>RS</fnm></au>
    <au><snm>Barto</snm><fnm>AG</fnm></au>
    <au><cnm>others</cnm></au>
  </aug>
  <publisher>MIT press Cambridge</publisher>
  <pubdate>1998</pubdate>
  <volume>135</volume>
</bibl>

<bibl id="B162">
  <title><p>Continuous deep q-learning with model-based
  acceleration</p></title>
  <aug>
    <au><snm>Gu</snm><fnm>S</fnm></au>
    <au><snm>Lillicrap</snm><fnm>T</fnm></au>
    <au><snm>Sutskever</snm><fnm>I</fnm></au>
    <au><snm>Levine</snm><fnm>S</fnm></au>
  </aug>
  <source>International Conference on Machine Learning</source>
  <pubdate>2016</pubdate>
  <fpage>2829</fpage>
  <lpage>-2838</lpage>
</bibl>

<bibl id="B163">
  <title><p>Deep boltzmann machines</p></title>
  <aug>
    <au><snm>Salakhutdinov</snm><fnm>R</fnm></au>
    <au><snm>Hinton</snm><fnm>G</fnm></au>
  </aug>
  <source>Artificial intelligence and statistics</source>
  <pubdate>2009</pubdate>
  <fpage>448</fpage>
  <lpage>-455</lpage>
</bibl>

<bibl id="B164">
  <title><p>Using machine learning for cryptocurrency trading</p></title>
  <aug>
    <au><snm>Sun</snm><fnm>J</fnm></au>
    <au><snm>Zhou</snm><fnm>Y</fnm></au>
    <au><snm>Lin</snm><fnm>J</fnm></au>
  </aug>
  <source>2019 IEEE International Conference on Industrial Cyber Physical
  Systems (ICPS)</source>
  <pubdate>2019</pubdate>
  <fpage>647</fpage>
  <lpage>-652</lpage>
</bibl>

<bibl id="B165">
  <title><p>101 formulaic alphas</p></title>
  <aug>
    <au><snm>Kakushadze</snm><fnm>Z</fnm></au>
  </aug>
  <source>Wilmott</source>
  <publisher>Wiley Online Library</publisher>
  <pubdate>2016</pubdate>
  <volume>2016</volume>
  <issue>84</issue>
  <fpage>72</fpage>
  <lpage>-81</lpage>
</bibl>

<bibl id="B166">
  <title><p>A High-Frequency Algorithmic Trading Strategy for
  Cryptocurrency</p></title>
  <aug>
    <au><snm>Vo</snm><fnm>A</fnm></au>
    <au><snm>Yost Bremm</snm><fnm>C</fnm></au>
  </aug>
  <source>Journal of Computer Information Systems</source>
  <publisher>Taylor \& Francis</publisher>
  <pubdate>2018</pubdate>
  <fpage>1</fpage>
  <lpage>-14</lpage>
</bibl>

<bibl id="B167">
  <title><p>Robustness of Support Vector Machines in Algorithmic Trading on
  Cryptocurrency Market</p></title>
  <aug>
    <au><snm>Ślepaczuk</snm><fnm>R</fnm></au>
    <au><snm>Zenkova</snm><fnm>M</fnm></au>
  </aug>
  <source>Central European Economic Journal</source>
  <pubdate>2018</pubdate>
  <volume>5</volume>
  <fpage>186</fpage>
  <lpage>205</lpage>
</bibl>

<bibl id="B168">
  <title><p>Stacking with Neural network for Cryptocurrency
  investment</p></title>
  <aug>
    <au><snm>Barnwal</snm><fnm>A</fnm></au>
    <au><snm>Bharti</snm><fnm>H</fnm></au>
    <au><snm>Ali</snm><fnm>A</fnm></au>
    <au><snm>Singh</snm><fnm>V</fnm></au>
  </aug>
  <source>arXiv preprint arXiv:1902.07855</source>
  <pubdate>2019</pubdate>
</bibl>

<bibl id="B169">
  <title><p>On discriminative vs. generative classifiers: A comparison of
  logistic regression and naive bayes</p></title>
  <aug>
    <au><snm>Ng</snm><fnm>AY</fnm></au>
    <au><snm>Jordan</snm><fnm>MI</fnm></au>
  </aug>
  <source>Advances in neural information processing systems</source>
  <pubdate>2002</pubdate>
  <fpage>841</fpage>
  <lpage>-848</lpage>
</bibl>

<bibl id="B170">
  <title><p>Quantitative cryptocurrency trading: exploring the use of machine
  learning techniques</p></title>
  <aug>
    <au><snm>Attanasio</snm><fnm>G</fnm></au>
    <au><snm>Cagliero</snm><fnm>L</fnm></au>
    <au><snm>Garza</snm><fnm>P</fnm></au>
    <au><snm>Baralis</snm><fnm>E</fnm></au>
  </aug>
  <source>Proceedings of the 5th Workshop on Data Science for Macro-modeling
  with Financial and Economic Datasets</source>
  <pubdate>2019</pubdate>
  <fpage>1</fpage>
</bibl>

<bibl id="B171">
  <title><p>Automated bitcoin trading via machine learning
  algorithms</p></title>
  <aug>
    <au><snm>Madan</snm><fnm>I</fnm></au>
    <au><snm>Saluja</snm><fnm>S</fnm></au>
    <au><snm>Zhao</snm><fnm>A</fnm></au>
  </aug>
  <source>URL: http://cs229. stanford. edu/proj2014/Isaac\% 20Madan</source>
  <pubdate>2015</pubdate>
  <volume>20</volume>
</bibl>

<bibl id="B172">
  <title><p>Prediction of Bitcoin Price using Data Mining</p></title>
  <aug>
    <au><snm>Virk</snm><fnm>DS</fnm></au>
  </aug>
  <source>Master's thesis</source>
  <publisher>National College of Ireland</publisher>
  <pubdate>2017</pubdate>
</bibl>

<bibl id="B173">
  <title><p>Prediction of Cryptocurrency Price Dynamics with Multiple Machine
  Learning Techniques</p></title>
  <aug>
    <au><snm>Zhengyang</snm><fnm>W</fnm></au>
    <au><snm>Xingzhou</snm><fnm>L</fnm></au>
    <au><snm>Jinjin</snm><fnm>R</fnm></au>
    <au><snm>Jiaqing</snm><fnm>K</fnm></au>
  </aug>
  <source>Proceedings of the 2019 4th International Conference on Machine
  Learning Technologies</source>
  <pubdate>2019</pubdate>
  <fpage>15</fpage>
  <lpage>-19</lpage>
</bibl>

<bibl id="B174">
  <title><p>Time Series Classification of Cryptocurrency Price Trend Based on a
  Recurrent LSTM Neural Network.</p></title>
  <aug>
    <au><snm>Kwon</snm><fnm>DH</fnm></au>
    <au><snm>Kim</snm><fnm>JB</fnm></au>
    <au><snm>Heo</snm><fnm>JS</fnm></au>
    <au><snm>Kim</snm><fnm>CM</fnm></au>
    <au><snm>Han</snm><fnm>YH</fnm></au>
  </aug>
  <source>Journal of Information Processing Systems</source>
  <pubdate>2019</pubdate>
  <volume>15</volume>
  <issue>3</issue>
</bibl>

<bibl id="B175">
  <title><p>Anticipating cryptocurrency prices using machine
  learning</p></title>
  <aug>
    <au><snm>Alessandretti</snm><fnm>L</fnm></au>
    <au><snm>ElBahrawy</snm><fnm>A</fnm></au>
    <au><snm>Aiello</snm><fnm>LM</fnm></au>
    <au><snm>Baronchelli</snm><fnm>A</fnm></au>
  </aug>
  <source>Complexity</source>
  <publisher>Hindawi</publisher>
  <pubdate>2018</pubdate>
  <volume>2018</volume>
</bibl>

<bibl id="B176">
  <title><p>Machine Learning Models Comparison for Bitcoin Price
  Prediction</p></title>
  <aug>
    <au><snm>Phaladisailoed</snm><fnm>T</fnm></au>
    <au><snm>Numnonda</snm><fnm>T</fnm></au>
  </aug>
  <source>2018 10th International Conference on Information Technology and
  Electrical Engineering (ICITEE)</source>
  <pubdate>2018</pubdate>
  <fpage>506</fpage>
  <lpage>-511</lpage>
</bibl>

<bibl id="B177">
  <title><p>Systematic Erudition of Bitcoin Price Prediction using Machine
  Learning Techniques</p></title>
  <aug>
    <au><snm>Rane</snm><fnm>PV</fnm></au>
    <au><snm>Dhage</snm><fnm>SN</fnm></au>
  </aug>
  <source>2019 5th International Conference on Advanced Computing \&
  Communication Systems (ICACCS)</source>
  <pubdate>2019</pubdate>
  <fpage>594</fpage>
  <lpage>-598</lpage>
</bibl>

<bibl id="B178">
  <title><p>Seq2Seq RNNs and ARIMA models for cryptocurrency prediction: A
  comparative study</p></title>
  <aug>
    <au><snm>Rebane</snm><fnm>J</fnm></au>
    <au><snm>Karlsson</snm><fnm>I</fnm></au>
    <au><snm>Denic</snm><fnm>S</fnm></au>
    <au><snm>Papapetrou</snm><fnm>P</fnm></au>
  </aug>
  <source>SIGKDD Fintech</source>
  <pubdate>2018</pubdate>
  <volume>18</volume>
</bibl>

<bibl id="B179">
  <title><p>Algorithmic Cryptocurrency Trading</p></title>
  <aug>
    <au><snm>Stuerner</snm><fnm>P</fnm></au>
  </aug>
  <source>PhD thesis</source>
  <publisher>Ulm University</publisher>
  <pubdate>2019</pubdate>
</bibl>

<bibl id="B180">
  <title><p>Hybrid Autoregressive-Recurrent Neural Network Architecture for
  Algorithmic Trading of Cryptocurrencies</p></title>
  <aug>
    <au><snm>Persson</snm><fnm>S</fnm></au>
    <au><snm>Slottje</snm><fnm>A</fnm></au>
    <au><snm>Shaw</snm><fnm>I</fnm></au>
  </aug>
  <source>Master's thesis</source>
  <publisher>Stanford University</publisher>
  <pubdate>2018</pubdate>
</bibl>

<bibl id="B181">
  <title><p>A CNN--LSTM model for gold price time-series
  forecasting</p></title>
  <aug>
    <au><snm>Livieris</snm><fnm>IE</fnm></au>
    <au><snm>Pintelas</snm><fnm>E</fnm></au>
    <au><snm>Pintelas</snm><fnm>P</fnm></au>
  </aug>
  <source>Neural computing and applications</source>
  <publisher>Springer</publisher>
  <pubdate>2020</pubdate>
  <volume>32</volume>
  <issue>23</issue>
  <fpage>17351</fpage>
  <lpage>-17360</lpage>
</bibl>

<bibl id="B182">
  <title><p>Ieo: Intelligent evolutionary optimisation for hyperparameter
  tuning</p></title>
  <aug>
    <au><snm>Huan</snm><fnm>Y</fnm></au>
    <au><snm>Wu</snm><fnm>F</fnm></au>
    <au><snm>Basios</snm><fnm>M</fnm></au>
    <au><snm>Kanthan</snm><fnm>L</fnm></au>
    <au><snm>Li</snm><fnm>L</fnm></au>
    <au><snm>Xu</snm><fnm>B</fnm></au>
  </aug>
  <source>arXiv preprint arXiv:2009.06390</source>
  <pubdate>2020</pubdate>
</bibl>

<bibl id="B183">
  <title><p>A CNN-LSTM-based model to forecast stock prices</p></title>
  <aug>
    <au><snm>Lu</snm><fnm>W</fnm></au>
    <au><snm>Li</snm><fnm>J</fnm></au>
    <au><snm>Li</snm><fnm>Y</fnm></au>
    <au><snm>Sun</snm><fnm>A</fnm></au>
    <au><snm>Wang</snm><fnm>J</fnm></au>
  </aug>
  <source>Complexity</source>
  <publisher>Hindawi</publisher>
  <pubdate>2020</pubdate>
  <volume>2020</volume>
</bibl>

<bibl id="B184">
  <title><p>Ascertaining price formation in cryptocurrency markets with machine
  learning</p></title>
  <aug>
    <au><snm>Fang</snm><fnm>F</fnm></au>
    <au><snm>Chung</snm><fnm>W</fnm></au>
    <au><snm>Ventre</snm><fnm>C</fnm></au>
    <au><snm>Basios</snm><fnm>M</fnm></au>
    <au><snm>Kanthan</snm><fnm>L</fnm></au>
    <au><snm>Li</snm><fnm>L</fnm></au>
    <au><snm>Wu</snm><fnm>F</fnm></au>
  </aug>
  <source>The European Journal of Finance</source>
  <pubdate>2021</pubdate>
  <volume>0</volume>
  <issue>0</issue>
  <fpage>1</fpage>
  <lpage>23</lpage>
</bibl>

<bibl id="B185">
  <title><p>Universal features of price formation in financial markets:
  perspectives from deep learning</p></title>
  <aug>
    <au><snm>Sirignano</snm><fnm>J</fnm></au>
    <au><snm>Cont</snm><fnm>R</fnm></au>
  </aug>
  <source>Quantitative Finance</source>
  <publisher>Taylor \& Francis</publisher>
  <pubdate>2019</pubdate>
  <volume>19</volume>
  <issue>9</issue>
  <fpage>1449</fpage>
  <lpage>-1459</lpage>
</bibl>

<bibl id="B186">
  <title><p>Predictive analysis of cryptocurrency price using deep
  learning</p></title>
  <aug>
    <au><snm>Yao</snm><fnm>Y</fnm></au>
    <au><snm>Yi</snm><fnm>J</fnm></au>
    <au><snm>Zhai</snm><fnm>S</fnm></au>
    <au><snm>Lin</snm><fnm>Y</fnm></au>
    <au><snm>Kim</snm><fnm>T</fnm></au>
    <au><snm>Zhang</snm><fnm>G</fnm></au>
    <au><snm>Lee</snm><fnm>LY</fnm></au>
  </aug>
  <source>International Journal of Engineering \& Technology</source>
  <pubdate>2018</pubdate>
  <volume>7</volume>
  <issue>3.27</issue>
  <fpage>258</fpage>
  <lpage>-264</lpage>
</bibl>

<bibl id="B187">
  <title><p>Ensemble deep learning models for forecasting cryptocurrency
  time-series</p></title>
  <aug>
    <au><snm>Livieris</snm><fnm>IE</fnm></au>
    <au><snm>Pintelas</snm><fnm>E</fnm></au>
    <au><snm>Stavroyiannis</snm><fnm>S</fnm></au>
    <au><snm>Pintelas</snm><fnm>P</fnm></au>
  </aug>
  <source>Algorithms</source>
  <publisher>Multidisciplinary Digital Publishing Institute</publisher>
  <pubdate>2020</pubdate>
  <volume>13</volume>
  <issue>5</issue>
  <fpage>121</fpage>
</bibl>

<bibl id="B188">
  <title><p>Predicting the trends of price for ethereum using deep learning
  techniques</p></title>
  <aug>
    <au><snm>Kumar</snm><fnm>D</fnm></au>
    <au><snm>Rath</snm><fnm>SK</fnm></au>
  </aug>
  <source>Artificial Intelligence and Evolutionary Computations in Engineering
  Systems</source>
  <publisher>Springer</publisher>
  <pubdate>2020</pubdate>
  <fpage>103</fpage>
  <lpage>-114</lpage>
</bibl>

<bibl id="B189">
  <title><p>Cryptocurrency price prediction using news and social media
  sentiment</p></title>
  <aug>
    <au><snm>Lamon</snm><fnm>C</fnm></au>
    <au><snm>Nielsen</snm><fnm>E</fnm></au>
    <au><snm>Redondo</snm><fnm>E</fnm></au>
  </aug>
  <source>SMU Data Sci. Rev</source>
  <pubdate>2017</pubdate>
  <volume>1</volume>
  <issue>3</issue>
  <fpage>1</fpage>
  <lpage>-22</lpage>
</bibl>

<bibl id="B190">
  <title><p>What Drives Cryptocurrency Prices?: An Investigation of Google
  Trends and Telegram Sentiment</p></title>
  <aug>
    <au><snm>Smuts</snm><fnm>N</fnm></au>
  </aug>
  <source>ACM SIGMETRICS Performance Evaluation Review</source>
  <publisher>ACM</publisher>
  <pubdate>2019</pubdate>
  <volume>46</volume>
  <issue>3</issue>
  <fpage>131</fpage>
  <lpage>-134</lpage>
</bibl>

<bibl id="B191">
  <title><p>Vader: A parsimonious rule-based model for sentiment analysis of
  social media text</p></title>
  <aug>
    <au><snm>Hutto</snm><fnm>CJ</fnm></au>
    <au><snm>Gilbert</snm><fnm>E</fnm></au>
  </aug>
  <source>Eighth international AAAI conference on weblogs and social
  media</source>
  <pubdate>2014</pubdate>
</bibl>

<bibl id="B192">
  <title><p>Forecasting cryptocurrency returns and volume using search
  engines</p></title>
  <aug>
    <au><snm>Nasir</snm><fnm>MA</fnm></au>
    <au><snm>Huynh</snm><fnm>TLD</fnm></au>
    <au><snm>Nguyen</snm><fnm>SP</fnm></au>
    <au><snm>Duong</snm><fnm>D</fnm></au>
  </aug>
  <source>Financial Innovation</source>
  <publisher>SpringerOpen</publisher>
  <pubdate>2019</pubdate>
  <volume>5</volume>
  <issue>1</issue>
  <fpage>2</fpage>
</bibl>

<bibl id="B193">
  <title><p>BitCoin meets Google Trends and Wikipedia: Quantifying the
  relationship between phenomena of the Internet era</p></title>
  <aug>
    <au><snm>Kristoufek</snm><fnm>L</fnm></au>
  </aug>
  <source>Scientific reports</source>
  <publisher>Nature Publishing Group</publisher>
  <pubdate>2013</pubdate>
  <volume>3</volume>
  <fpage>3415</fpage>
</bibl>

<bibl id="B194">
  <title><p>Predicting fluctuations in cryptocurrency transactions based on
  user comments and replies</p></title>
  <aug>
    <au><snm>Kim</snm><fnm>YB</fnm></au>
    <au><snm>Kim</snm><fnm>JG</fnm></au>
    <au><snm>Kim</snm><fnm>W</fnm></au>
    <au><snm>Im</snm><fnm>JH</fnm></au>
    <au><snm>Kim</snm><fnm>TH</fnm></au>
    <au><snm>Kang</snm><fnm>SJ</fnm></au>
    <au><snm>Kim</snm><fnm>CH</fnm></au>
  </aug>
  <source>PloS one</source>
  <publisher>Public Library of Science</publisher>
  <pubdate>2016</pubdate>
  <volume>11</volume>
  <issue>8</issue>
  <fpage>e0161197</fpage>
</bibl>

<bibl id="B195">
  <title><p>Mutual-excitation of cryptocurrency market returns and social media
  topics</p></title>
  <aug>
    <au><snm>Phillips</snm><fnm>RC</fnm></au>
    <au><snm>Gorse</snm><fnm>D</fnm></au>
  </aug>
  <source>Proceedings of the 4th International Conference on Frontiers of
  Educational Technologies</source>
  <pubdate>2018</pubdate>
  <fpage>80</fpage>
  <lpage>-86</lpage>
</bibl>

<bibl id="B196">
  <title><p>Sentiment-based prediction of alternative cryptocurrency price
  fluctuations using gradient boosting tree model</p></title>
  <aug>
    <au><snm>Li</snm><fnm>TR</fnm></au>
    <au><snm>Chamrajnagar</snm><fnm>A</fnm></au>
    <au><snm>Fong</snm><fnm>X</fnm></au>
    <au><snm>Rizik</snm><fnm>N</fnm></au>
    <au><snm>Fu</snm><fnm>F</fnm></au>
  </aug>
  <source>Frontiers in Physics</source>
  <publisher>Frontiers</publisher>
  <pubdate>2019</pubdate>
  <volume>7</volume>
  <fpage>98</fpage>
</bibl>

<bibl id="B197">
  <title><p>News and subjective beliefs: A Bayesian approach to Bitcoin
  investments</p></title>
  <aug>
    <au><snm>Flori</snm><fnm>A</fnm></au>
  </aug>
  <source>Research in International Business and Finance</source>
  <publisher>Elsevier</publisher>
  <pubdate>2019</pubdate>
  <volume>50</volume>
  <fpage>336</fpage>
  <lpage>-356</lpage>
</bibl>

<bibl id="B198">
  <title><p>Predicting Bitcoin returns: Comparing the roles of newspaper-and
  internet search-based measures of uncertainty</p></title>
  <aug>
    <au><snm>Bouri</snm><fnm>E</fnm></au>
    <au><snm>Gupta</snm><fnm>R</fnm></au>
  </aug>
  <source>Finance Research Letters</source>
  <publisher>Elsevier</publisher>
  <pubdate>2019</pubdate>
  <fpage>101398</fpage>
</bibl>

<bibl id="B199">
  <title><p>Algorithmic trading of cryptocurrency based on Twitter sentiment
  analysis</p></title>
  <aug>
    <au><snm>Colianni</snm><fnm>S</fnm></au>
    <au><snm>Rosales</snm><fnm>S</fnm></au>
    <au><snm>Signorotti</snm><fnm>M</fnm></au>
  </aug>
  <source>CS229 Project</source>
  <pubdate>2015</pubdate>
  <fpage>1</fpage>
  <lpage>-5</lpage>
</bibl>

<bibl id="B200">
  <title><p>Social signals and algorithmic trading of Bitcoin</p></title>
  <aug>
    <au><snm>Garcia</snm><fnm>D</fnm></au>
    <au><snm>Schweitzer</snm><fnm>F</fnm></au>
  </aug>
  <source>Royal Society open science</source>
  <publisher>The Royal Society Publishing</publisher>
  <pubdate>2015</pubdate>
  <volume>2</volume>
  <issue>9</issue>
  <fpage>150288</fpage>
</bibl>

<bibl id="B201">
  <title><p>Forecasting Cryptocurrency Value by Sentiment Analysis: An
  HPC-Oriented Survey of the State-of-the-Art in the Cloud Era</p></title>
  <aug>
    <au><snm>Zamuda</snm><fnm>A</fnm></au>
    <au><snm>Crescimanna</snm><fnm>V</fnm></au>
    <au><snm>Burguillo</snm><fnm>JC</fnm></au>
    <au><snm>Dias</snm><fnm>JM</fnm></au>
    <au><snm>Wegrzyn Wolska</snm><fnm>K</fnm></au>
    <au><snm>Rached</snm><fnm>I</fnm></au>
    <au><snm>Gonzlez</snm><fnm>H</fnm></au>
    <au><snm>Senkerik</snm><fnm>R</fnm></au>
    <au><snm>Pop</snm><fnm>C</fnm></au>
    <au><snm>Cioara</snm><fnm>T</fnm></au>
    <au><cnm>others</cnm></au>
  </aug>
  <source>High-Performance Modelling and Simulation for Big Data
  Applications</source>
  <publisher>Springer</publisher>
  <pubdate>2019</pubdate>
  <fpage>325</fpage>
  <lpage>-349</lpage>
</bibl>

<bibl id="B202">
  <title><p>The Butterfly Affect: Impact of Development Practices on
  Cryptocurrency Prices</p></title>
  <aug>
    <au><snm>Bartolucci</snm><fnm>S</fnm></au>
    <au><snm>Destefanis</snm><fnm>G</fnm></au>
    <au><snm>Ortu</snm><fnm>M</fnm></au>
    <au><snm>Uras</snm><fnm>N</fnm></au>
    <au><snm>Marchesi</snm><fnm>M</fnm></au>
    <au><snm>Tonelli</snm><fnm>R</fnm></au>
  </aug>
  <pubdate>2019</pubdate>
</bibl>

<bibl id="B203">
  <title><p>Deep reinforcement learning that matters</p></title>
  <aug>
    <au><snm>Henderson</snm><fnm>P</fnm></au>
    <au><snm>Islam</snm><fnm>R</fnm></au>
    <au><snm>Bachman</snm><fnm>P</fnm></au>
    <au><snm>Pineau</snm><fnm>J</fnm></au>
    <au><snm>Precup</snm><fnm>D</fnm></au>
    <au><snm>Meger</snm><fnm>D</fnm></au>
  </aug>
  <source>Thirty-Second AAAI Conference on Artificial Intelligence</source>
  <pubdate>2018</pubdate>
</bibl>

<bibl id="B204">
  <title><p>Agent-based Markets: Equilibrium Strategies and
  Robustness</p></title>
  <aug>
    <au><snm>Liu</snm><fnm>B</fnm></au>
    <au><snm>Polukarov</snm><fnm>M</fnm></au>
    <au><snm>Ventre</snm><fnm>C</fnm></au>
    <au><snm>Li</snm><fnm>L</fnm></au>
    <au><snm>Kanthan</snm><fnm>L</fnm></au>
  </aug>
  <source>Proceedings of the 2nd ACM International Conference on AI in
  Finance</source>
  <pubdate>2021</pubdate>
</bibl>

<bibl id="B205">
  <title><p>Learning optimal Q-function using deep Boltzmann machine for
  reliable trading of cryptocurrency</p></title>
  <aug>
    <au><snm>Bu</snm><fnm>SJ</fnm></au>
    <au><snm>Cho</snm><fnm>SB</fnm></au>
  </aug>
  <source>International Conference on Intelligent Data Engineering and
  Automated Learning</source>
  <pubdate>2018</pubdate>
  <fpage>468</fpage>
  <lpage>-480</lpage>
</bibl>

<bibl id="B206">
  <title><p>Limit order placement optimization with Deep Reinforcement
  Learning: Learning from patterns in cryptocurrency market data</p></title>
  <aug>
    <au><snm>Juchli</snm><fnm>M</fnm></au>
  </aug>
  <source>Master's thesis</source>
  <publisher>TU Delft Electrical Engineering</publisher>
  <pubdate>2018</pubdate>
</bibl>

<bibl id="B207">
  <title><p>A Deep Reinforcement Learning Approach for Automated Cryptocurrency
  Trading</p></title>
  <aug>
    <au><snm>Lucarelli</snm><fnm>G</fnm></au>
    <au><snm>Borrotti</snm><fnm>M</fnm></au>
  </aug>
  <source>IFIP International Conference on Artificial Intelligence Applications
  and Innovations</source>
  <pubdate>2019</pubdate>
  <fpage>247</fpage>
  <lpage>-258</lpage>
</bibl>

<bibl id="B208">
  <title><p>Recommending Cryptocurrency Trading Points with Deep Reinforcement
  Learning Approach</p></title>
  <aug>
    <au><snm>Sattarov</snm><fnm>O</fnm></au>
    <au><snm>Muminov</snm><fnm>A</fnm></au>
    <au><snm>Lee</snm><fnm>CW</fnm></au>
    <au><snm>Kang</snm><fnm>HK</fnm></au>
    <au><snm>Oh</snm><fnm>R</fnm></au>
    <au><snm>Ahn</snm><fnm>J</fnm></au>
    <au><snm>Oh</snm><fnm>HJ</fnm></au>
    <au><snm>Jeon</snm><fnm>HS</fnm></au>
  </aug>
  <source>Applied Sciences</source>
  <publisher>Multidisciplinary Digital Publishing Institute</publisher>
  <pubdate>2020</pubdate>
  <volume>10</volume>
  <issue>4</issue>
  <fpage>1506</fpage>
</bibl>

<bibl id="B209">
  <title><p>Cryptocurrency Trading Using Machine Learning</p></title>
  <aug>
    <au><snm>Koker</snm><fnm>TE</fnm></au>
    <au><snm>Koutmos</snm><fnm>D</fnm></au>
  </aug>
  <source>Journal of Risk and Financial Management</source>
  <publisher>Multidisciplinary Digital Publishing Institute</publisher>
  <pubdate>2020</pubdate>
  <volume>13</volume>
  <issue>8</issue>
  <fpage>178</fpage>
</bibl>

<bibl id="B210">
  <title><p>Bitcoin price forecasting with neuro-fuzzy techniques</p></title>
  <aug>
    <au><snm>Atsalakis</snm><fnm>GS</fnm></au>
    <au><snm>Atsalaki</snm><fnm>IG</fnm></au>
    <au><snm>Pasiouras</snm><fnm>F</fnm></au>
    <au><snm>Zopounidis</snm><fnm>C</fnm></au>
  </aug>
  <source>European Journal of Operational Research</source>
  <publisher>Elsevier</publisher>
  <pubdate>2019</pubdate>
  <volume>276</volume>
  <issue>2</issue>
  <fpage>770</fpage>
  <lpage>-780</lpage>
</bibl>

<bibl id="B211">
  <title><p>Time Series Featurization via Topological Data Analysis: an
  Application to Cryptocurrency Trend Forecasting</p></title>
  <aug>
    <au><snm>Kim</snm><fnm>K</fnm></au>
    <au><snm>Kim</snm><fnm>J</fnm></au>
    <au><snm>Rinaldo</snm><fnm>A</fnm></au>
  </aug>
  <source>arXiv preprint arXiv:1812.02987</source>
  <pubdate>2018</pubdate>
</bibl>

<bibl id="B212">
  <title><p>Predicting the price of Bitcoin by the most frequent edges of its
  transaction network</p></title>
  <aug>
    <au><snm>Kurbucz</snm><fnm>MT</fnm></au>
  </aug>
  <source>Economics Letters</source>
  <publisher>Elsevier</publisher>
  <pubdate>2019</pubdate>
  <volume>184</volume>
  <fpage>108655</fpage>
</bibl>

<bibl id="B213">
  <title><p>Inferring the interplay between network structure and market
  effects in Bitcoin</p></title>
  <aug>
    <au><snm>Kondor</snm><fnm>D</fnm></au>
    <au><snm>Csabai</snm><fnm>I</fnm></au>
    <au><snm>Sz{\"u}le</snm><fnm>J</fnm></au>
    <au><snm>P{\'o}sfai</snm><fnm>M</fnm></au>
    <au><snm>Vattay</snm><fnm>G</fnm></au>
  </aug>
  <source>New Journal of Physics</source>
  <publisher>IOP Publishing</publisher>
  <pubdate>2014</pubdate>
  <volume>16</volume>
  <issue>12</issue>
  <fpage>125003</fpage>
</bibl>

<bibl id="B214">
  <title><p>Do the rich get richer? An empirical analysis of the Bitcoin
  transaction network</p></title>
  <aug>
    <au><snm>Kondor</snm><fnm>D</fnm></au>
    <au><snm>P{\'o}sfai</snm><fnm>M</fnm></au>
    <au><snm>Csabai</snm><fnm>I</fnm></au>
    <au><snm>Vattay</snm><fnm>G</fnm></au>
  </aug>
  <source>PloS one</source>
  <publisher>Public Library of Science</publisher>
  <pubdate>2014</pubdate>
  <volume>9</volume>
  <issue>2</issue>
</bibl>

<bibl id="B215">
  <title><p>A bayesian approach to identify bitcoin users</p></title>
  <aug>
    <au><snm>Juh{\'a}sz</snm><fnm>PL</fnm></au>
    <au><snm>St{\'e}ger</snm><fnm>J</fnm></au>
    <au><snm>Kondor</snm><fnm>D</fnm></au>
    <au><snm>Vattay</snm><fnm>G</fnm></au>
  </aug>
  <source>PloS one</source>
  <publisher>Public Library of Science</publisher>
  <pubdate>2018</pubdate>
  <volume>13</volume>
  <issue>12</issue>
</bibl>

<bibl id="B216">
  <title><p>Chainnet: Learning on blockchain graphs with topological
  features</p></title>
  <aug>
    <au><snm>Abay</snm><fnm>NC</fnm></au>
    <au><snm>Akcora</snm><fnm>CG</fnm></au>
    <au><snm>Gel</snm><fnm>YR</fnm></au>
    <au><snm>Kantarcioglu</snm><fnm>M</fnm></au>
    <au><snm>Islambekov</snm><fnm>UD</fnm></au>
    <au><snm>Tian</snm><fnm>Y</fnm></au>
    <au><snm>Thuraisingham</snm><fnm>B</fnm></au>
  </aug>
  <source>2019 IEEE International Conference on Data Mining (ICDM)</source>
  <pubdate>2019</pubdate>
  <fpage>946</fpage>
  <lpage>-951</lpage>
</bibl>

<bibl id="B217">
  <title><p>Dynamic connectedness and integration in cryptocurrency
  markets</p></title>
  <aug>
    <au><snm>Ji</snm><fnm>Q</fnm></au>
    <au><snm>Bouri</snm><fnm>E</fnm></au>
    <au><snm>Lau</snm><fnm>CKM</fnm></au>
    <au><snm>Roubaud</snm><fnm>D</fnm></au>
  </aug>
  <source>International Review of Financial Analysis</source>
  <publisher>Elsevier</publisher>
  <pubdate>2019</pubdate>
  <volume>63</volume>
  <fpage>257</fpage>
  <lpage>-272</lpage>
</bibl>

<bibl id="B218">
  <title><p>On the network topology of variance decompositions: Measuring the
  connectedness of financial firms</p></title>
  <aug>
    <au><snm>Diebold</snm><fnm>FX</fnm></au>
    <au><snm>Y{\i}lmaz</snm><fnm>K</fnm></au>
  </aug>
  <source>Journal of Econometrics</source>
  <publisher>Elsevier</publisher>
  <pubdate>2014</pubdate>
  <volume>182</volume>
  <issue>1</issue>
  <fpage>119</fpage>
  <lpage>-134</lpage>
</bibl>

<bibl id="B219">
  <title><p>Multiresolution analysis and spillovers of major cryptocurrency
  markets</p></title>
  <aug>
    <au><snm>Omane Adjepong</snm><fnm>M</fnm></au>
    <au><snm>Alagidede</snm><fnm>IP</fnm></au>
  </aug>
  <source>Research in International Business and Finance</source>
  <publisher>Elsevier</publisher>
  <pubdate>2019</pubdate>
  <volume>49</volume>
  <fpage>191</fpage>
  <lpage>-206</lpage>
</bibl>

<bibl id="B220">
  <title><p>Do Bitcoin and other cryptocurrencies jump together?</p></title>
  <aug>
    <au><snm>Bouri</snm><fnm>E</fnm></au>
    <au><snm>Roubaud</snm><fnm>D</fnm></au>
    <au><snm>Shahzad</snm><fnm>SJH</fnm></au>
  </aug>
  <source>The Quarterly Review of Economics and Finance</source>
  <publisher>Elsevier</publisher>
  <pubdate>2020</pubdate>
  <volume>76</volume>
  <fpage>396</fpage>
  <lpage>-409</lpage>
</bibl>

<bibl id="B221">
  <title><p>Competition of noise and collectivity in global cryptocurrency
  trading: Route to a self-contained market</p></title>
  <aug>
    <au><snm>Dro{\.z}d{\.z}</snm><fnm>S</fnm></au>
    <au><snm>Minati</snm><fnm>L</fnm></au>
    <au><snm>O{\'s}wiecimka</snm><fnm>P</fnm></au>
    <au><snm>Stanuszek</snm><fnm>M</fnm></au>
    <au><snm>Watorek</snm><fnm>M</fnm></au>
  </aug>
  <source>Chaos: An Interdisciplinary Journal of Nonlinear Science</source>
  <publisher>AIP Publishing LLC</publisher>
  <pubdate>2020</pubdate>
  <volume>30</volume>
  <issue>2</issue>
  <fpage>023122</fpage>
</bibl>

<bibl id="B222">
  <title><p>How futures trading changed bitcoin prices</p></title>
  <aug>
    <au><snm>Hale</snm><fnm>G</fnm></au>
    <au><snm>Krishnamurthy</snm><fnm>A</fnm></au>
    <au><snm>Kudlyak</snm><fnm>M</fnm></au>
    <au><snm>Shultz</snm><fnm>P</fnm></au>
    <au><cnm>others</cnm></au>
  </aug>
  <source>FRBSF Economic Letter</source>
  <publisher>Federal Reserve Bank of San Francisco</publisher>
  <pubdate>2018</pubdate>
  <volume>12</volume>
</bibl>

<bibl id="B223">
  <title><p>Cryptocurrencies and equity funds: Evidence from an asymmetric
  multifractal analysis</p></title>
  <aug>
    <au><snm>Kristjanpoller</snm><fnm>W</fnm></au>
    <au><snm>Bouri</snm><fnm>E</fnm></au>
    <au><snm>Takaishi</snm><fnm>T</fnm></au>
  </aug>
  <source>Physica A: Statistical Mechanics and its Applications</source>
  <publisher>Elsevier</publisher>
  <pubdate>2020</pubdate>
  <volume>545</volume>
  <fpage>123711</fpage>
</bibl>

<bibl id="B224">
  <title><p>Automated Triangular Arbitrage:: A Trading Algorithm for Foreign
  Exchange on a Cryptocurrency Market</p></title>
  <aug>
    <au><snm>Bai</snm><fnm>S</fnm></au>
    <au><snm>Robinson</snm><fnm>F</fnm></au>
  </aug>
  <source>Master's thesis</source>
  <publisher>KTH Royal Institute of Technology</publisher>
  <pubdate>2019</pubdate>
</bibl>

<bibl id="B225">
  <title><p>Co-movements between Bitcoin and Gold: A wavelet coherence
  analysis</p></title>
  <aug>
    <au><snm>Kang</snm><fnm>SH</fnm></au>
    <au><snm>McIver</snm><fnm>RP</fnm></au>
    <au><snm>Hernandez</snm><fnm>JA</fnm></au>
  </aug>
  <source>Physica A: Statistical Mechanics and its Applications</source>
  <publisher>Elsevier</publisher>
  <pubdate>2019</pubdate>
  <fpage>120888</fpage>
</bibl>

<bibl id="B226">
  <title><p>Dynamic conditional correlation: A simple class of multivariate
  generalized autoregressive conditional heteroskedasticity models</p></title>
  <aug>
    <au><snm>Engle</snm><fnm>R</fnm></au>
  </aug>
  <source>Journal of Business \& Economic Statistics</source>
  <publisher>Taylor \& Francis</publisher>
  <pubdate>2002</pubdate>
  <volume>20</volume>
  <issue>3</issue>
  <fpage>339</fpage>
  <lpage>-350</lpage>
</bibl>

<bibl id="B227">
  <title><p>Time-frequency co-movement of cryptocurrency return and volatility:
  evidence from wavelet coherence analysis</p></title>
  <aug>
    <au><snm>Qiao</snm><fnm>X</fnm></au>
    <au><snm>Zhu</snm><fnm>H</fnm></au>
    <au><snm>Hau</snm><fnm>L</fnm></au>
  </aug>
  <source>International Review of Financial Analysis</source>
  <publisher>Elsevier</publisher>
  <pubdate>2020</pubdate>
  <volume>71</volume>
  <fpage>101541</fpage>
</bibl>

<bibl id="B228">
  <title><p>Bitcoin, gold and the dollar--A GARCH volatility
  analysis</p></title>
  <aug>
    <au><snm>Dyhrberg</snm><fnm>AH</fnm></au>
  </aug>
  <source>Finance Research Letters</source>
  <publisher>Elsevier</publisher>
  <pubdate>2016</pubdate>
  <volume>16</volume>
  <fpage>85</fpage>
  <lpage>-92</lpage>
</bibl>

<bibl id="B229">
  <title><p>Bitcoin, gold and the US dollar--A replication and
  extension</p></title>
  <aug>
    <au><snm>Baur</snm><fnm>DG</fnm></au>
    <au><snm>Dimpfl</snm><fnm>T</fnm></au>
    <au><snm>Kuck</snm><fnm>K</fnm></au>
  </aug>
  <source>Finance Research Letters</source>
  <publisher>Elsevier</publisher>
  <pubdate>2018</pubdate>
  <volume>25</volume>
  <fpage>103</fpage>
  <lpage>-110</lpage>
</bibl>

<bibl id="B230">
  <title><p>Bitcoin for energy commodities before and after the December 2013
  crash: diversifier, hedge or safe haven?</p></title>
  <aug>
    <au><snm>Bouri</snm><fnm>E</fnm></au>
    <au><snm>Jalkh</snm><fnm>N</fnm></au>
    <au><snm>Moln{\'a}r</snm><fnm>P</fnm></au>
    <au><snm>Roubaud</snm><fnm>D</fnm></au>
  </aug>
  <source>Applied Economics</source>
  <publisher>Taylor \& Francis</publisher>
  <pubdate>2017</pubdate>
  <volume>49</volume>
  <issue>50</issue>
  <fpage>5063</fpage>
  <lpage>-5073</lpage>
</bibl>

<bibl id="B231">
  <title><p>Cryptoasset factor models</p></title>
  <aug>
    <au><snm>Kakushadze</snm><fnm>Z</fnm></au>
  </aug>
  <source>Algorithmic Finance</source>
  <publisher>IOS Press</publisher>
  <pubdate>2018</pubdate>
  <issue>Preprint</issue>
  <fpage>1</fpage>
  <lpage>-18</lpage>
</bibl>

<bibl id="B232">
  <title><p>Investigating volatility transmission and hedging properties
  between Bitcoin and Ethereum</p></title>
  <aug>
    <au><snm>Beneki</snm><fnm>C</fnm></au>
    <au><snm>Koulis</snm><fnm>A</fnm></au>
    <au><snm>Kyriazis</snm><fnm>NA</fnm></au>
    <au><snm>Papadamou</snm><fnm>S</fnm></au>
  </aug>
  <source>Research in International Business and Finance</source>
  <publisher>Elsevier</publisher>
  <pubdate>2019</pubdate>
  <volume>48</volume>
  <fpage>219</fpage>
  <lpage>-227</lpage>
</bibl>

<bibl id="B233">
  <title><p>The day of the week effect in the cryptocurrency market</p></title>
  <aug>
    <au><snm>Caporale</snm><fnm>GM</fnm></au>
    <au><snm>Plastun</snm><fnm>A</fnm></au>
  </aug>
  <source>Finance Research Letters</source>
  <publisher>Elsevier</publisher>
  <pubdate>2019</pubdate>
  <volume>31</volume>
</bibl>

<bibl id="B234">
  <title><p>Cryptocurrency trading, gambling and problem gambling</p></title>
  <aug>
    <au><snm>Delfabbro</snm><fnm>P</fnm></au>
    <au><snm>King</snm><fnm>D</fnm></au>
    <au><snm>Williams</snm><fnm>J</fnm></au>
    <au><snm>Georgiou</snm><fnm>N</fnm></au>
  </aug>
  <source>Addictive Behaviors</source>
  <publisher>Elsevier</publisher>
  <pubdate>2021</pubdate>
  <volume>122</volume>
  <fpage>107021</fpage>
</bibl>

<bibl id="B235">
  <title><p>The psychology of cryptocurrency trading: Risk and protective
  factors</p></title>
  <aug>
    <au><snm>Delfabbro</snm><fnm>P</fnm></au>
    <au><snm>King</snm><fnm>DL</fnm></au>
    <au><snm>Williams</snm><fnm>J</fnm></au>
  </aug>
  <source>Journal of Behavioral Addictions</source>
  <publisher>Akad{\'e}miai Kiad{\'o} Budapest</publisher>
  <pubdate>2021</pubdate>
</bibl>

<bibl id="B236">
  <title><p>The relationship between the economic policy uncertainty and the
  cryptocurrency market</p></title>
  <aug>
    <au><snm>Cheng</snm><fnm>HP</fnm></au>
    <au><snm>Yen</snm><fnm>KC</fnm></au>
  </aug>
  <source>Finance Research Letters</source>
  <publisher>Elsevier</publisher>
  <pubdate>2020</pubdate>
  <volume>35</volume>
  <fpage>101308</fpage>
</bibl>

<bibl id="B237">
  <title><p>Measuring economic policy uncertainty</p></title>
  <aug>
    <au><snm>Baker</snm><fnm>SR</fnm></au>
    <au><snm>Bloom</snm><fnm>N</fnm></au>
    <au><snm>Davis</snm><fnm>SJ</fnm></au>
  </aug>
  <source>The quarterly journal of economics</source>
  <publisher>Oxford University Press</publisher>
  <pubdate>2016</pubdate>
  <volume>131</volume>
  <issue>4</issue>
  <fpage>1593</fpage>
  <lpage>-1636</lpage>
</bibl>

<bibl id="B238">
  <title><p>Cryptocurrency returns and the volatility of liquidity</p></title>
  <aug>
    <au><snm>Leirvik</snm><fnm>T</fnm></au>
  </aug>
  <source>Finance Research Letters</source>
  <publisher>Elsevier</publisher>
  <pubdate>2021</pubdate>
  <fpage>102031</fpage>
</bibl>

<bibl id="B239">
  <title><p>Cryptocurrencies as a financial asset: A systematic
  analysis</p></title>
  <aug>
    <au><snm>Corbet</snm><fnm>S</fnm></au>
    <au><snm>Lucey</snm><fnm>B</fnm></au>
    <au><snm>Urquhart</snm><fnm>A</fnm></au>
    <au><snm>Yarovaya</snm><fnm>L</fnm></au>
  </aug>
  <source>International Review of Financial Analysis</source>
  <publisher>Elsevier</publisher>
  <pubdate>2019</pubdate>
  <volume>62</volume>
  <fpage>182</fpage>
  <lpage>-199</lpage>
</bibl>

<bibl id="B240">
  <title><p>Cryptocurrency-portfolios in a mean-variance framework</p></title>
  <aug>
    <au><snm>Brauneis</snm><fnm>A</fnm></au>
    <au><snm>Mestel</snm><fnm>R</fnm></au>
  </aug>
  <source>Finance Research Letters</source>
  <publisher>Elsevier</publisher>
  <pubdate>2019</pubdate>
  <volume>28</volume>
  <fpage>259</fpage>
  <lpage>-264</lpage>
</bibl>

<bibl id="B241">
  <title><p>Crypto-assets portfolio optimization under the omega
  measure</p></title>
  <aug>
    <au><snm>Castro</snm><fnm>JG</fnm></au>
    <au><snm>Tito</snm><fnm>EAH</fnm></au>
    <au><snm>Brand{\~a}o</snm><fnm>LET</fnm></au>
    <au><snm>Gomes</snm><fnm>LL</fnm></au>
  </aug>
  <source>The Engineering Economist</source>
  <publisher>Taylor \& Francis</publisher>
  <pubdate>2019</pubdate>
  <fpage>1</fpage>
  <lpage>-21</lpage>
</bibl>

<bibl id="B242">
  <title><p>On the investment credentials of Bitcoin: A cross-currency
  perspective</p></title>
  <aug>
    <au><snm>Bedi</snm><fnm>P</fnm></au>
    <au><snm>Nashier</snm><fnm>T</fnm></au>
  </aug>
  <source>Research in International Business and Finance</source>
  <publisher>Elsevier</publisher>
  <pubdate>2020</pubdate>
  <volume>51</volume>
  <fpage>101087</fpage>
</bibl>

<bibl id="B243">
  <title><p>Building and testing global investment portfolios using alternative
  asset classes</p></title>
  <aug>
    <au><snm>Antipova</snm><fnm>V</fnm></au>
  </aug>
  <source>Master's thesis</source>
  <publisher>Vytautas Magnus University</publisher>
  <pubdate>2019</pubdate>
</bibl>

<bibl id="B244">
  <title><p>A multivariate approach for the simultaneous modelling of market
  risk and credit risk for cryptocurrencies</p></title>
  <aug>
    <au><snm>Fantazzini</snm><fnm>D</fnm></au>
    <au><snm>Zimin</snm><fnm>S</fnm></au>
  </aug>
  <source>Journal of Industrial and Business Economics</source>
  <publisher>Springer</publisher>
  <pubdate>2020</pubdate>
  <volume>47</volume>
  <issue>1</issue>
  <fpage>19</fpage>
  <lpage>-69</lpage>
</bibl>

<bibl id="B245">
  <title><p>Realised volatility connectedness among Bitcoin exchange
  markets</p></title>
  <aug>
    <au><snm>Ji</snm><fnm>Q</fnm></au>
    <au><snm>Bouri</snm><fnm>E</fnm></au>
    <au><snm>Kristoufek</snm><fnm>L</fnm></au>
    <au><snm>Lucey</snm><fnm>B</fnm></au>
  </aug>
  <source>Finance Research Letters</source>
  <publisher>Elsevier</publisher>
  <pubdate>2019</pubdate>
  <fpage>101391</fpage>
</bibl>

<bibl id="B246">
  <title><p>Returns and volatility spillovers among cryptocurrency
  portfolios</p></title>
  <aug>
    <au><snm>Fasanya</snm><fnm>IO</fnm></au>
    <au><snm>Oyewole</snm><fnm>O</fnm></au>
    <au><snm>Odudu</snm><fnm>T</fnm></au>
  </aug>
  <source>International Journal of Managerial Finance</source>
  <publisher>Emerald Publishing Limited</publisher>
  <pubdate>2020</pubdate>
</bibl>

<bibl id="B247">
  <title><p>Value-at-Risk and Expected Shortfall in Cryptocurrencies'
  Portfolio: A Vine Copula-based Approach</p></title>
  <aug>
    <au><snm>Truc{\'\i}os</snm><fnm>C</fnm></au>
    <au><snm>Tiwari</snm><fnm>AK</fnm></au>
    <au><snm>Alqahtani</snm><fnm>F</fnm></au>
  </aug>
  <source>Available at SSRN 3441892</source>
  <pubdate>2019</pubdate>
</bibl>

<bibl id="B248">
  <title><p>Cryptocurrency portfolio optimization using Value-at-Risk
  measure</p></title>
  <aug>
    <au><snm>Hrytsiuk</snm><fnm>P</fnm></au>
    <au><snm>Babych</snm><fnm>T</fnm></au>
    <au><snm>Bachyshyna</snm><fnm>L</fnm></au>
  </aug>
  <source>Strategies, Models and Technologies of Economic Systems Management
  (SMTESM 2019)</source>
  <pubdate>2019</pubdate>
</bibl>

<bibl id="B249">
  <title><p>Cryptocurrency portfolio management with deep reinforcement
  learning</p></title>
  <aug>
    <au><snm>Jiang</snm><fnm>Z</fnm></au>
    <au><snm>Liang</snm><fnm>J</fnm></au>
  </aug>
  <source>2017 Intelligent Systems Conference (IntelliSys)</source>
  <pubdate>2017</pubdate>
  <fpage>905</fpage>
  <lpage>-913</lpage>
</bibl>

<bibl id="B250">
  <title><p>Return, Diversification and Risk in Cryptocurrency Portfolios using
  Deep Recurrent Neural Networks and Multi-Objective Evolutionary
  Algorithms</p></title>
  <aug>
    <au><snm>Estalayo</snm><fnm>I</fnm></au>
    <au><snm>Del Ser</snm><fnm>J</fnm></au>
    <au><snm>Osaba</snm><fnm>E</fnm></au>
    <au><snm>Bilbao</snm><fnm>MN</fnm></au>
    <au><snm>Muhammad</snm><fnm>K</fnm></au>
    <au><snm>G{\'a}lvez</snm><fnm>A</fnm></au>
    <au><snm>Iglesias</snm><fnm>A</fnm></au>
  </aug>
  <source>2019 IEEE Congress on Evolutionary Computation (CEC)</source>
  <pubdate>2019</pubdate>
  <fpage>755</fpage>
  <lpage>-761</lpage>
</bibl>

<bibl id="B251">
  <title><p>Crypto-currency bubbles: an application of the Phillips--Shi--Yu
  (2013) methodology on Mt. Gox bitcoin prices</p></title>
  <aug>
    <au><snm>Cheung</snm><fnm>A</fnm></au>
    <au><snm>Roca</snm><fnm>E</fnm></au>
    <au><snm>Su</snm><fnm>JJ</fnm></au>
  </aug>
  <source>Applied Economics</source>
  <publisher>Taylor \& Francis</publisher>
  <pubdate>2015</pubdate>
  <volume>47</volume>
  <issue>23</issue>
  <fpage>2348</fpage>
  <lpage>-2358</lpage>
</bibl>

<bibl id="B252">
  <title><p>Datestamping the Bitcoin and Ethereum bubbles</p></title>
  <aug>
    <au><snm>Corbet</snm><fnm>S</fnm></au>
    <au><snm>Lucey</snm><fnm>B</fnm></au>
    <au><snm>Yarovaya</snm><fnm>L</fnm></au>
  </aug>
  <source>Finance Research Letters</source>
  <publisher>Elsevier</publisher>
  <pubdate>2018</pubdate>
  <volume>26</volume>
  <fpage>81</fpage>
  <lpage>-88</lpage>
</bibl>

<bibl id="B253">
  <title><p>Co-explosivity in the cryptocurrency market</p></title>
  <aug>
    <au><snm>Bouri</snm><fnm>E</fnm></au>
    <au><snm>Shahzad</snm><fnm>SJH</fnm></au>
    <au><snm>Roubaud</snm><fnm>D</fnm></au>
  </aug>
  <source>Finance Research Letters</source>
  <publisher>Elsevier</publisher>
  <pubdate>2019</pubdate>
  <volume>29</volume>
  <fpage>178</fpage>
  <lpage>-183</lpage>
</bibl>

<bibl id="B254">
  <title><p>Testing for multiple bubbles: Historical episodes of exuberance and
  collapse in the S&ampP 500</p></title>
  <aug>
    <au><snm>Phillips</snm><fnm>PC</fnm></au>
    <au><snm>Shi</snm><fnm>S</fnm></au>
    <au><snm>Yu</snm><fnm>J</fnm></au>
  </aug>
  <source>International economic review</source>
  <publisher>Wiley Online Library</publisher>
  <pubdate>2015</pubdate>
  <volume>56</volume>
  <issue>4</issue>
  <fpage>1043</fpage>
  <lpage>-1078</lpage>
</bibl>

<bibl id="B255">
  <title><p>Testing for multiple bubbles: Limit theory of real-time
  detectors</p></title>
  <aug>
    <au><snm>Phillips</snm><fnm>PC</fnm></au>
    <au><snm>Shi</snm><fnm>S</fnm></au>
    <au><snm>Yu</snm><fnm>J</fnm></au>
  </aug>
  <source>International Economic Review</source>
  <publisher>Wiley Online Library</publisher>
  <pubdate>2015</pubdate>
  <volume>56</volume>
  <issue>4</issue>
  <fpage>1079</fpage>
  <lpage>-1134</lpage>
</bibl>

<bibl id="B256">
  <title><p>What Can Predict Bubbles in Cryptocurrency Prices?</p></title>
  <aug>
    <au><snm>Enoksen</snm><fnm>FA</fnm></au>
    <au><snm>Landsnes</snm><fnm>CJ</fnm></au>
  </aug>
  <source>Master's thesis</source>
  <publisher>University of Stavanger, Norway</publisher>
  <pubdate>2019</pubdate>
</bibl>

<bibl id="B257">
  <title><p>Predicting cryptocurrency price bubbles using social media data and
  epidemic modelling</p></title>
  <aug>
    <au><snm>Phillips</snm><fnm>RC</fnm></au>
    <au><snm>Gorse</snm><fnm>D</fnm></au>
  </aug>
  <source>2017 IEEE Symposium Series on Computational Intelligence
  (SSCI)</source>
  <pubdate>2017</pubdate>
  <fpage>1</fpage>
  <lpage>-7</lpage>
</bibl>

<bibl id="B258">
  <title><p>Price Overreactions in the Cryptocurrency Market</p></title>
  <aug>
    <au><snm>Caporale</snm><fnm>GM</fnm></au>
    <au><snm>Plastun</snm><fnm>A</fnm></au>
  </aug>
  <source>CESifo Working Paper</source>
  <publisher>Munich: Center for Economic Studies and Ifo Institute
  (CESifo)</publisher>
  <pubdate>2018</pubdate>
  <issue>6861</issue>
  <url>http://hdl.handle.net/10419/174984</url>
</bibl>

<bibl id="B259">
  <title><p>Volatility and return jumps in bitcoin</p></title>
  <aug>
    <au><snm>Chaim</snm><fnm>P</fnm></au>
    <au><snm>Laurini</snm><fnm>MP</fnm></au>
  </aug>
  <source>Economics Letters</source>
  <publisher>Elsevier</publisher>
  <pubdate>2018</pubdate>
  <volume>173</volume>
  <fpage>158</fpage>
  <lpage>-163</lpage>
</bibl>

<bibl id="B260">
  <title><p>Returns, Volatility and the Cryptocurrency Bubble of
  2017-18</p></title>
  <aug>
    <au><snm>Cross</snm><fnm>JL</fnm></au>
    <au><snm>Hou</snm><fnm>C</fnm></au>
    <au><snm>Trinh</snm><fnm>K</fnm></au>
  </aug>
  <source>Economic Modelling</source>
  <publisher>Elsevier</publisher>
  <pubdate>2021</pubdate>
  <fpage>105643</fpage>
</bibl>

<bibl id="B261">
  <title><p>Cryptocurrency Market Activity During Extremely Volatile
  Periods</p></title>
  <aug>
    <au><snm>Katsiampa</snm><fnm>P</fnm></au>
    <au><snm>Gkillas</snm><fnm>K</fnm></au>
    <au><snm>Longin</snm><fnm>F</fnm></au>
  </aug>
  <pubdate>2018</pubdate>
  <url>https://ssrn.com/abstract=3220781</url>
</bibl>

<bibl id="B262">
  <title><p>How Persistent and Dependent are Pricing of Bitcoin to other
  Cryptocurrencies Before and After 2017/18 Crash?</p></title>
  <aug>
    <au><snm>Yaya</snm><fnm>OS</fnm></au>
    <au><snm>Ogbonna</snm><fnm>EA</fnm></au>
    <au><snm>Olubusoye</snm><fnm>OE</fnm></au>
  </aug>
  <pubdate>2018</pubdate>
</bibl>

<bibl id="B263">
  <title><p>Cryptocurrency liquidity during extreme price movements: is there a
  problem with virtual money?</p></title>
  <aug>
    <au><snm>Manahov</snm><fnm>V</fnm></au>
  </aug>
  <source>Quantitative Finance</source>
  <publisher>Taylor \& Francis</publisher>
  <pubdate>2021</pubdate>
  <volume>21</volume>
  <issue>2</issue>
  <fpage>341</fpage>
  <lpage>-360</lpage>
</bibl>

<bibl id="B264">
  <title><p>Extreme tail network analysis of cryptocurrencies and trading
  strategies</p></title>
  <aug>
    <au><snm>Shahzad</snm><fnm>SJH</fnm></au>
    <au><snm>Bouri</snm><fnm>E</fnm></au>
    <au><snm>Ahmad</snm><fnm>T</fnm></au>
    <au><snm>Naeem</snm><fnm>MA</fnm></au>
  </aug>
  <source>Finance Research Letters</source>
  <publisher>Elsevier</publisher>
  <pubdate>2021</pubdate>
  <fpage>102106</fpage>
</bibl>

<bibl id="B265">
  <title><p>An extreme value analysis of the tail relationships between returns
  and volumes for high frequency cryptocurrencies</p></title>
  <aug>
    <au><snm>Chan</snm><fnm>S</fnm></au>
    <au><snm>Chu</snm><fnm>J</fnm></au>
    <au><snm>Zhang</snm><fnm>Y</fnm></au>
    <au><snm>Nadarajah</snm><fnm>S</fnm></au>
  </aug>
  <source>Research in International Business and Finance</source>
  <publisher>Elsevier</publisher>
  <pubdate>2022</pubdate>
  <volume>59</volume>
  <fpage>101541</fpage>
</bibl>

<bibl id="B266">
  <title><p>An experimental study of cryptocurrency market dynamics</p></title>
  <aug>
    <au><snm>Krafft</snm><fnm>PM</fnm></au>
    <au><snm>Della Penna</snm><fnm>N</fnm></au>
    <au><snm>Pentland</snm><fnm>AS</fnm></au>
  </aug>
  <source>Proceedings of the 2018 CHI Conference on Human Factors in Computing
  Systems</source>
  <pubdate>2018</pubdate>
  <fpage>605</fpage>
</bibl>

<bibl id="B267">
  <title><p>Behavioral anomalies in cryptocurrency markets</p></title>
  <aug>
    <au><snm>Yang</snm><fnm>H</fnm></au>
  </aug>
  <source>Available at SSRN 3174421</source>
  <pubdate>2018</pubdate>
  <url>https://ssrn.com/abstract=3174421</url>
</bibl>

<bibl id="B268">
  <title><p>Modeling and Simulation of the Economics of Mining in the Bitcoin
  Market</p></title>
  <aug>
    <au><snm>Cocco</snm><fnm>L</fnm></au>
    <au><snm>Marchesi</snm><fnm>M</fnm></au>
  </aug>
  <source>PloS one</source>
  <publisher>Public Library of Science</publisher>
  <pubdate>2016</pubdate>
  <volume>11</volume>
  <issue>10</issue>
</bibl>

<bibl id="B269">
  <title><p>Herding in the cryptocurrency market</p></title>
  <aug>
    <au><snm>Leclair</snm><fnm>EM</fnm></au>
  </aug>
  <source>Master's thesis</source>
  <publisher>Carleton University</publisher>
  <pubdate>2018</pubdate>
</bibl>

<bibl id="B270">
  <title><p>Herding in the cryptocurrency market: CSSD and CSAD
  approaches</p></title>
  <aug>
    <au><snm>Vidal Tom{\'a}s</snm><fnm>D</fnm></au>
    <au><snm>Ib{\'a}{\~n}ez</snm><fnm>AM</fnm></au>
    <au><snm>Farin{\'o}s</snm><fnm>JE</fnm></au>
  </aug>
  <source>Finance Research Letters</source>
  <publisher>Elsevier</publisher>
  <pubdate>2019</pubdate>
  <volume>30</volume>
  <fpage>181</fpage>
  <lpage>-186</lpage>
</bibl>

<bibl id="B271">
  <title><p>Market stress and herding</p></title>
  <aug>
    <au><snm>Hwang</snm><fnm>S</fnm></au>
    <au><snm>Salmon</snm><fnm>M</fnm></au>
  </aug>
  <source>Journal of Empirical Finance</source>
  <publisher>Elsevier</publisher>
  <pubdate>2004</pubdate>
  <volume>11</volume>
  <issue>4</issue>
  <fpage>585</fpage>
  <lpage>-616</lpage>
</bibl>

<bibl id="B272">
  <title><p>Trading and arbitrage in cryptocurrency markets</p></title>
  <aug>
    <au><snm>Makarov</snm><fnm>I</fnm></au>
    <au><snm>Schoar</snm><fnm>A</fnm></au>
  </aug>
  <source>Journal of Financial Economics</source>
  <publisher>Elsevier</publisher>
  <pubdate>2020</pubdate>
  <volume>135</volume>
  <issue>2</issue>
  <fpage>293</fpage>
  <lpage>-319</lpage>
</bibl>

<bibl id="B273">
  <title><p>Call Markets with Adaptive Clearing Intervals</p></title>
  <aug>
    <au><snm>Liu</snm><fnm>B</fnm></au>
    <au><snm>Polukarov</snm><fnm>M</fnm></au>
    <au><snm>Ventre</snm><fnm>C</fnm></au>
    <au><snm>Li</snm><fnm>L</fnm></au>
    <au><snm>Kanthan</snm><fnm>L</fnm></au>
  </aug>
  <source>Proceedings of the 20th International Conference on Autonomous Agents
  and MultiAgent Systems</source>
  <pubdate>2021</pubdate>
  <fpage>1587</fpage>
  <lpage>-1589</lpage>
</bibl>

<bibl id="B274">
  <title><p>Herding and feedback trading in cryptocurrency markets</p></title>
  <aug>
    <au><snm>King</snm><fnm>T</fnm></au>
    <au><snm>Koutmos</snm><fnm>D</fnm></au>
  </aug>
  <source>Annals of Operations Research</source>
  <publisher>Springer</publisher>
  <pubdate>2021</pubdate>
  <volume>300</volume>
  <issue>1</issue>
  <fpage>79</fpage>
  <lpage>-96</lpage>
</bibl>

<bibl id="B275">
  <title><p>Is bitcoin really un-tethered?</p></title>
  <aug>
    <au><snm>Griffin</snm><fnm>JM</fnm></au>
    <au><snm>Shams</snm><fnm>A</fnm></au>
  </aug>
  <source>Available at SSRN 3195066</source>
  <pubdate>2019</pubdate>
</bibl>

<bibl id="B276">
  <title><p>Global cryptocurrency benchmarking study</p></title>
  <aug>
    <au><snm>Hileman</snm><fnm>G</fnm></au>
    <au><snm>Rauchs</snm><fnm>M</fnm></au>
  </aug>
  <source>Cambridge Centre for Alternative Finance</source>
  <pubdate>2017</pubdate>
  <volume>33</volume>
</bibl>

<bibl id="B277">
  <title><p>Algorithmic and high frequency trading in Asia-Pacific, now and the
  future</p></title>
  <aug>
    <au><snm>Zhou</snm><fnm>H</fnm></au>
    <au><snm>Kalev</snm><fnm>PS</fnm></au>
  </aug>
  <source>Pacific-Basin Finance Journal</source>
  <publisher>Elsevier</publisher>
  <pubdate>2019</pubdate>
  <volume>53</volume>
  <fpage>186</fpage>
  <lpage>-207</lpage>
</bibl>

<bibl id="B278">
  <title><p>Taming the blockchain beast? Regulatory implications for the
  cryptocurrency Market</p></title>
  <aug>
    <au><snm>Shanaev</snm><fnm>S</fnm></au>
    <au><snm>Sharma</snm><fnm>S</fnm></au>
    <au><snm>Ghimire</snm><fnm>B</fnm></au>
    <au><snm>Shuraeva</snm><fnm>A</fnm></au>
  </aug>
  <source>Research in International Business and Finance</source>
  <publisher>Elsevier</publisher>
  <pubdate>2020</pubdate>
  <volume>51</volume>
  <fpage>101080</fpage>
</bibl>

<bibl id="B279">
  <title><p>The impact of cryptocurrency regulation on trading
  markets</p></title>
  <aug>
    <au><snm>Feinstein</snm><fnm>BD</fnm></au>
    <au><snm>Werbach</snm><fnm>K</fnm></au>
  </aug>
  <source>Journal of Financial Regulation</source>
  <publisher>Oxford University Press</publisher>
  <pubdate>2021</pubdate>
  <volume>7</volume>
  <issue>1</issue>
  <fpage>48</fpage>
  <lpage>-99</lpage>
</bibl>

<bibl id="B280">
  <title><p>A Study of Opinion Mining and Data Mining Techniques to Analyse the
  Cryptocurrency Market</p></title>
  <aug>
    <au><snm>Patil</snm><fnm>AP</fnm></au>
    <au><snm>Akarsh</snm><fnm>TS</fnm></au>
    <au><snm>Parkavi</snm><fnm>A</fnm></au>
  </aug>
  <source>2018 3rd International Conference on Computational Systems and
  Information Technology for Sustainable Solutions (CSITSS)</source>
  <pubdate>2018</pubdate>
  <fpage>198</fpage>
  <lpage>-203</lpage>
</bibl>

<bibl id="B281">
  <title><p>Clustering patterns in efficiency and the coming-of-age of the
  cryptocurrency market</p></title>
  <aug>
    <au><snm>Sigaki</snm><fnm>HY</fnm></au>
    <au><snm>Perc</snm><fnm>M</fnm></au>
    <au><snm>Ribeiro</snm><fnm>HV</fnm></au>
  </aug>
  <source>Scientific reports</source>
  <publisher>Nature Publishing Group</publisher>
  <pubdate>2019</pubdate>
  <volume>9</volume>
  <issue>1</issue>
  <fpage>1440</fpage>
</bibl>

<bibl id="B282">
  <title><p>Using an artificial financial market for studying a cryptocurrency
  market</p></title>
  <aug>
    <au><snm>Cocco</snm><fnm>L</fnm></au>
    <au><snm>Concas</snm><fnm>G</fnm></au>
    <au><snm>Marchesi</snm><fnm>M</fnm></au>
  </aug>
  <source>Journal of Economic Interaction and Coordination</source>
  <publisher>Springer</publisher>
  <pubdate>2017</pubdate>
  <volume>12</volume>
  <issue>2</issue>
  <fpage>345</fpage>
  <lpage>-365</lpage>
</bibl>

<bibl id="B283">
  <title><p>Decentralized exchanges: The “wild west” of cryptocurrency
  trading</p></title>
  <aug>
    <au><snm>Aspris</snm><fnm>A</fnm></au>
    <au><snm>Foley</snm><fnm>S</fnm></au>
    <au><snm>Svec</snm><fnm>J</fnm></au>
    <au><snm>Wang</snm><fnm>L</fnm></au>
  </aug>
  <source>International Review of Financial Analysis</source>
  <publisher>Elsevier</publisher>
  <pubdate>2021</pubdate>
  <volume>77</volume>
  <fpage>101845</fpage>
</bibl>

<bibl id="B284">
  <title><p>Cryptocurrency As Money: A Trading Strategy Solution</p></title>
  <aug>
    <au><snm>Ogorevc</snm><fnm>M</fnm></au>
  </aug>
  <source>Available at SSRN 3436041</source>
  <pubdate>2019</pubdate>
</bibl>

<bibl id="B285">
  <title><p>Competition in the cryptocurrency market.</p></title>
  <aug>
    <au><snm>Gandal</snm><fnm>N</fnm></au>
    <au><snm>Halaburda</snm><fnm>H</fnm></au>
  </aug>
  <publisher>CEPR Discussion Paper No. DP10157</publisher>
  <pubdate>2014</pubdate>
  <note>CEPR Discussion Paper No. DP10157</note>
</bibl>

<bibl id="B286">
  <title><p>Where do we stand in cryptocurrencies economic research? A survey
  based on hybrid analysis</p></title>
  <aug>
    <au><snm>Bariviera</snm><fnm>AF</fnm></au>
    <au><snm>Merediz Sola</snm><fnm>I</fnm></au>
  </aug>
  <source>arXiv preprint arXiv:2003.09723</source>
  <pubdate>2020</pubdate>
</bibl>

<bibl id="B287">
  <title><p>Cryptocurrency Trading: How to Make Money by Trading Bitcoin and
  other Cryptocurrency (Volume 2)</p></title>
  <aug>
    <au><snm>Hansel</snm><fnm>D</fnm></au>
  </aug>
  <publisher>CreateSpace Independent Publishing Platform</publisher>
  <pubdate>2018</pubdate>
</bibl>

<bibl id="B288">
  <title><p>Cryptocurrency Trading For Beginners: 6-Steps Action Plan To Your
  First Investment</p></title>
  <aug>
    <au><snm>Kate</snm><fnm>C</fnm></au>
  </aug>
  <publisher>Independently published</publisher>
  <pubdate>2018</pubdate>
</bibl>

<bibl id="B289">
  <title><p>Formal automatic trading in the cryptocurrency era</p></title>
  <aug>
    <au><snm>Garza</snm><fnm>P</fnm></au>
  </aug>
  <source>PhD thesis</source>
  <publisher>Politecnico di Torino</publisher>
  <pubdate>2019</pubdate>
</bibl>

<bibl id="B290">
  <title><p>Algorithmic Trading for Cryptocurrencies</p></title>
  <aug>
    <au><snm>Ward</snm><fnm>M</fnm></au>
  </aug>
  <source>Master's thesis</source>
  <publisher>UtahState University</publisher>
  <pubdate>2018</pubdate>
</bibl>

<bibl id="B291">
  <title><p>Quantitative finance with R and cryptocurrencies</p></title>
  <aug>
    <au><snm>Fantazzini</snm><fnm>D</fnm></au>
  </aug>
  <source>Amazon KDP, ISBN-13</source>
  <pubdate>2019</pubdate>
  <fpage>978</fpage>
  <lpage>-1090685315</lpage>
</bibl>

<bibl id="B292">
  <title><p>Platform for scholarly communication about cryptocurrencies and
  blockchains</p></title>
  <aug>
    <au><snm>network</snm><fnm>B</fnm></au>
  </aug>
  <source>\url{https://www.blockchainresearchnetwork.org/}</source>
  <pubdate>2020</pubdate>
  <note>[Online, Accessed: April 17, 2020]</note>
</bibl>

<bibl id="B293">
  <title><p>Bitcoin open source implementation of P2P currency</p></title>
  <aug>
    <au><snm>Nakamoto</snm><fnm>S</fnm></au>
  </aug>
  <source>P2P foundation</source>
  <pubdate>2009</pubdate>
  <volume>18</volume>
</bibl>

<bibl id="B294">
  <title><p>Speculative bubbles in Bitcoin markets? An empirical investigation
  into the fundamental value of Bitcoin</p></title>
  <aug>
    <au><snm>Cheah</snm><fnm>ET</fnm></au>
    <au><snm>Fry</snm><fnm>J</fnm></au>
  </aug>
  <source>Economics Letters</source>
  <publisher>Elsevier</publisher>
  <pubdate>2015</pubdate>
  <volume>130</volume>
  <fpage>32</fpage>
  <lpage>-36</lpage>
</bibl>

<bibl id="B295">
  <title><p>Predicting the price of Bitcoin using Machine Learning</p></title>
  <aug>
    <au><snm>McNally</snm><fnm>S</fnm></au>
  </aug>
  <source>PhD thesis</source>
  <publisher>Dublin, National College of Ireland</publisher>
  <pubdate>2016</pubdate>
</bibl>

<bibl id="B296">
  <title><p>Bitcoin Trading Agents</p></title>
  <aug>
    <au><snm>Bell</snm><fnm>T</fnm></au>
  </aug>
  <source>University of Southampton</source>
  <pubdate>2016</pubdate>
</bibl>

<bibl id="B297">
  <title><p>Application of machine learning algorithms for bitcoin automated
  trading</p></title>
  <aug>
    <au><snm>{\.Z}bikowski</snm><fnm>K</fnm></au>
  </aug>
  <source>Machine Intelligence and Big Data in Industry</source>
  <publisher>Springer</publisher>
  <pubdate>2016</pubdate>
  <fpage>161</fpage>
  <lpage>-168</lpage>
</bibl>

<bibl id="B298">
  <title><p>China bans financial, payment institutions from cryptocurrency
  business</p></title>
  <aug>
    <au><cnm>Reuters</cnm></au>
  </aug>
  <source>\url{https://www.reuters.com/technology/chinese-financial-payment-bodies-barred-cryptocurrency-business-2021-05-18/}</source>
  <pubdate>2021</pubdate>
  <note>[Online, Accessed: May 18, 2021]</note>
</bibl>

<bibl id="B299">
  <title><p>Bitcoin legal tender in El Salvador, first country ever</p></title>
  <aug>
    <au><cnm>MercoPress</cnm></au>
  </aug>
  <source>\url{https://en.mercopress.com/2021/06/10/bitcoin-legal-tender-in-el-salvador-first-country-ever}</source>
  <pubdate>2021</pubdate>
  <note>[Online, Accessed: June 10, 2021]</note>
</bibl>

<bibl id="B300">
  <title><p>On Machine Learning Based Cryptocurrency Trading</p></title>
  <aug>
    <au><snm>Bach</snm><fnm>WG</fnm></au>
    <au><snm>Kasper</snm><fnm>LN</fnm></au>
  </aug>
  <source>Master's thesis</source>
  <publisher>Aalborg University</publisher>
  <pubdate>2018</pubdate>
</bibl>

<bibl id="B301">
  <title><p>Predicting the value of cryptocurrencies using machine learning
  time series analysis time series analysis time</p></title>
  <aug>
    <au><snm>Siaminos</snm><fnm>G</fnm></au>
  </aug>
  <pubdate>2019</pubdate>
</bibl>

<bibl id="B302">
  <title><p>A critical investigation of cryptocurrency data and
  analysis</p></title>
  <aug>
    <au><snm>Alexander</snm><fnm>C</fnm></au>
    <au><snm>Dakos</snm><fnm>M</fnm></au>
  </aug>
  <source>Quantitative Finance</source>
  <publisher>Taylor \& Francis</publisher>
  <pubdate>2020</pubdate>
  <volume>20</volume>
  <issue>2</issue>
  <fpage>173</fpage>
  <lpage>-188</lpage>
</bibl>

<bibl id="B303">
  <title><p>Bitcoin volatility forecasting with a glimpse into buy and sell
  orders</p></title>
  <aug>
    <au><snm>Guo</snm><fnm>T</fnm></au>
    <au><snm>Bifet</snm><fnm>A</fnm></au>
    <au><snm>Antulov Fantulin</snm><fnm>N</fnm></au>
  </aug>
  <source>2018 IEEE International Conference on Data Mining (ICDM)</source>
  <pubdate>2018</pubdate>
  <fpage>989</fpage>
  <lpage>-994</lpage>
</bibl>

<bibl id="B304">
  <title><p>Cryptocurrency price drivers: Wavelet coherence analysis
  revisited</p></title>
  <aug>
    <au><snm>Phillips</snm><fnm>RC</fnm></au>
    <au><snm>Gorse</snm><fnm>D</fnm></au>
  </aug>
  <source>PloS one</source>
  <publisher>Public Library of Science</publisher>
  <pubdate>2018</pubdate>
  <volume>13</volume>
  <issue>4</issue>
  <fpage>e0195200</fpage>
</bibl>

<bibl id="B305">
  <title><p>Whose Opinion Matters? Analyzing Relationships Between Bitcoin
  Prices and User Groups in Online Community</p></title>
  <aug>
    <au><snm>Kang</snm><fnm>K</fnm></au>
    <au><snm>Choo</snm><fnm>J</fnm></au>
    <au><snm>Kim</snm><fnm>Y</fnm></au>
  </aug>
  <source>Social Science Computer Review</source>
  <publisher>SAGE Publications Sage CA: Los Angeles, CA</publisher>
  <pubdate>2019</pubdate>
  <fpage>0894439319840716</fpage>
</bibl>

<bibl id="B306">
  <title><p>Inferring the interplay between network structure and market
  effects in Bitcoin</p></title>
  <aug>
    <au><snm>Kondor</snm><fnm>D</fnm></au>
    <au><snm>Csabai</snm><fnm>I</fnm></au>
    <au><snm>Szule</snm><fnm>J</fnm></au>
    <au><snm>Psfai</snm><fnm>M</fnm></au>
    <au><snm>Vattay</snm><fnm>G</fnm></au>
  </aug>
  <source>New Journal of Physics</source>
  <publisher>{IOP} Publishing</publisher>
  <pubdate>2014</pubdate>
  <volume>16</volume>
  <issue>12</issue>
  <fpage>125003</fpage>
  <url>https://doi.org/10.1088
</bibl>

<bibl id="B307">
  <title><p>Bitcoin network dataset</p></title>
  <aug>
    <au><cnm>MIT</cnm></au>
  </aug>
  <source>\url{https://senseable2015-6.mit.edu/bitcoin}</source>
  <pubdate>2015</pubdate>
  <note>[Online, Accessed January 11, 2020]</note>
</bibl>

<bibl id="B308">
  <title><p>Opinion dynamics in finance and business: a literature review and
  research opportunities</p></title>
  <aug>
    <au><snm>Zha</snm><fnm>Q</fnm></au>
    <au><snm>Kou</snm><fnm>G</fnm></au>
    <au><snm>Zhang</snm><fnm>H</fnm></au>
    <au><snm>Liang</snm><fnm>H</fnm></au>
    <au><snm>Chen</snm><fnm>X</fnm></au>
    <au><snm>Li</snm><fnm>CC</fnm></au>
    <au><snm>Dong</snm><fnm>Y</fnm></au>
  </aug>
  <source>Financial Innovation</source>
  <publisher>Springer</publisher>
  <pubdate>2020</pubdate>
  <volume>6</volume>
  <issue>1</issue>
  <fpage>1</fpage>
  <lpage>-22</lpage>
</bibl>

<bibl id="B309">
  <title><p>Spillover Risks on Cryptocurrency Markets: A Look from VAR-SVAR
  Granger Causality and Student Copulas</p></title>
  <aug>
    <au><snm>Luu Duc Huynh</snm><fnm>T</fnm></au>
  </aug>
  <source>Journal of Risk and Financial Management</source>
  <publisher>Multidisciplinary Digital Publishing Institute</publisher>
  <pubdate>2019</pubdate>
  <volume>12</volume>
  <issue>2</issue>
  <fpage>52</fpage>
</bibl>

<bibl id="B310">
  <title><p>Relation between volatility correlations in financial markets and
  Omori processes occurring on all scales</p></title>
  <aug>
    <au><snm>Weber</snm><fnm>P</fnm></au>
    <au><snm>Wang</snm><fnm>F</fnm></au>
    <au><snm>Vodenska Chitkushev</snm><fnm>I</fnm></au>
    <au><snm>Havlin</snm><fnm>S</fnm></au>
    <au><snm>Stanley</snm><fnm>HE</fnm></au>
  </aug>
  <source>Physical Review E</source>
  <publisher>APS</publisher>
  <pubdate>2007</pubdate>
  <volume>76</volume>
  <issue>1</issue>
  <fpage>016109</fpage>
</bibl>

<bibl id="B311">
  <title><p>Does academic research destroy stock return
  predictability?</p></title>
  <aug>
    <au><snm>McLean</snm><fnm>RD</fnm></au>
    <au><snm>Pontiff</snm><fnm>J</fnm></au>
  </aug>
  <source>The Journal of Finance</source>
  <publisher>Wiley Online Library</publisher>
  <pubdate>2016</pubdate>
  <volume>71</volume>
  <issue>1</issue>
  <fpage>5</fpage>
  <lpage>-32</lpage>
</bibl>

</refgrp>
} 





\end{backmatter}
\end{document}